# Mioara MUGUR-SCHÄCHTER

# Infra-Quantum Mechanics
## and
## critical examination of
## Bell's theorem on non locality

### The principles of a revolution of epistemology
### revealed in the descriptions of microstates
### (French text with English summary)

________________________________________________________

## L'Infra-mécanique quantique
## et
## examen critique
## du théorème de non localité de Bell

### Principes d'une révolution de l'épistémologie
### révélés dans les descriptions de microétats

# Global abstract

The first part of this book contains considerations concerning the questions of "reality" and of "knowledge", targeted on exclusively two aims: *(a)* to identify the *location* – any other more intrinsic qualification being abstracted away – of the quantum mechanical conceptualization, with respect to the two above mentioned questions; *(b)* to sensitize the reader to the role of **methodological** *decisions* when a constructive conceptualization is developed.

The second part of this book contains the construction of *Infra-Quantum Mechanics* (IQM, to be understood as: beneath the mathematical formalism of Quantum Mechanics (QM) or encrypted in it). IQM is an epistemological-physical, strictly *qualitative* discipline, elaborated **independently of the mathematical formalism of QM**. It emerges under the constraints imposed exclusively by:

* the *cognitive situation* of a human being who decides to construct communicable and consensual knowledge on 'states of microsystems' (microstates);

* general requirements of human conceptualization.

The aim of IQM is to bring into evidence how the mathematical formalism of QM manages to signify. This aim, we think, is fully achieved.

IQM brings forth a radically new type of descriptional form, *transferred* on the registering devices of macroscopic apparatuses and '**primordially** statistical' (i.e. that cannot be conceived, like in classical physics, as being in principle removable *via* a more precise way of *factually* constructing the description). QM does not yield an integrated perception of this descriptional form, nor does it permit a definition of it. Nevertheless a physicist familiar with QM clearly recognizes in the qualitative structure developed inside IQM the whole semantic essence of the quantum mechanical mathematical formalism. So it can be hoped that by utilizing IQM as *a semantic structure of reference*, the set of all the interpretation problems raised by QM will obtain a system of mutually coherent solutions.

In the third part of the book IMQ is considered globally, from its outside, and its relations with the concepts of space, time, geometry, consensus, as well as with Einstein's theories of relativity, are examined. Thereby it comes into evidence that:

**1**. There exists an *order* of progressive constructability of the inner structure of our representations of physical entities.

**2**. This order *withstands* inclusion of concepts formed inside macroscopic physics, into the primordial transferred representation of microstates.

**3**. Consequently the aim of directly 'unifying' Einstein's theories of relativity, with QM, appears to be both illusory and devoid of pertinence.

**4**. The conclusion of Bell's proof of his inequality is erroneously interpreted, possibly even by himself and certainly by many others.

**5**. The recent experiments on locality are shown to be understandable as a *factual* proof of the insensitivity of the statistical features of a primordial transferred description concerning entities defined as a *"one state of two (or several) systems"*, to manipulations tied with the current macroscopic spatial-temporal individuation.

The fourth part of the book indicates the line of thought that leads from IQM toward the general *Method of Relativized Conceptualization* (MRC), already constructed before by this author on the basis of a still not entirely explicated version of IQM. It is shown that the form of the descriptions of microstates brought forth inside IQM, has captured in it a certain sort of **universality**, thereby including the principles of a radical revolution of the theory of knowledge.

# *Remerciements*

Une version préliminaire de ce livre a été lue patiemment par Michel de Heaulme dont les commentaires ont conduit à une amélioration dans la première partie. Hervé Barreau et Francis Bailly ont accepté de lire le manuscrit initial, et François Dubois a fait une très efficace lecture de la version finale. Je les remercie tous.

Je remercie Emmanuel Malolo Dissakè pour une révision du texte dont la forme de la première partie a bénéficié notablement.

Je remercie particulièrement et très vivement Jean-Christophe Denaes pour une lecture savante et approfondie qui a suggéré une modification essentielle et a clarifié certains aspects philosophiques.

Je voudrais ajouter un remerciement d'une nature moins usuelle adressé au physicien Lee Smolin. Son dernier livre *The Trouble with Physics* commence avec un chapitre intitulé *The Five Problems in Theoretical Physics*. Le premier de ces problèmes est formulé ainsi:

> *« Problem 1. Combine general relativity and quantum theory into a single theory that can claim to be the complete theory of nature »*.

Cette formulation est suivie de quelques pages d'explicitation dont la conclusion est exprimée de la façon suivante.

> «This whole issue goes under the name *the foundational problems of quantum mechanics*. It is the second great problem of contemporary physics.
>
> *Problem 2. Resolve the problems in the foundations of quantum mehanics, either by making sense of the theory as it stands or by inventing a new theory that does make sense »*.

Il a été réconfortant de prendre connaissance d'une telle perception de la situation actuelle dans la physique théorique. Car, dans la mesure où elle n'est pas tout à fait singulière, cette perception permet d'espérer que le travail exposé ici pourrait répondre à un questionnement qui a acquis contour dans la pensée d'une certaine communauté de physiciens. En effet l'infra-mécanique quantique constitue une avancée notable vers la solution du problème n° *2* et cette avancée met en doute le problème n° *1* en tant qu'un but.

**SOMMAIRE**

**Quatrième partie**

**De l'infra-mécanique quantique vers une méthode générale de conceptualisation relativisée**

**Mot d'introduction a la quatrième partie**

**6. Les descriptions de microétats comme prémisses d'une révolution de l'épistémologie**

6.1. Des 'phénomènes'

*6.1.1. Au sens de l'épistémologie philosophique classique*

*6.1.2. Au sens des sciences cognitives*

*6.1.3. Au sens de l'infra-[mécanique quantique]*

6.2. Phénomènes et descriptions: l'idée d'un canon descriptionnel général

*6.2.1. Les sources de l'apparente singularité de la forme $D_M/G,me_G,V_M/$*

*6.2.2. L'universalité de la forme descriptionnelle $D_M/G,me_G,V_M/$*

*6.2.3. Vers une généralisation de la structure épistémologique révélée par l'infra-[mécanique quantique]*

*6.2.4. Le concept d'un canon descriptionnel général*

6.3. Vers une méthode générale de conceptualisation relativisée

**CONCLUSION GÉNÉRALE**

# *Introduction générale*

À peine quelques années sont passées depuis que Richard Feynman a frappé les esprits en déclarant: « Je crois pouvoir dire sans me tromper que personne ne comprend la mécanique quantique ». Depuis, d'autres physiciens ont exprimé avec force le même constat (Lee Smolin [2007]).

Mais *pourquoi* la mécanique quantique oppose-t-elle une difficulté si particulière aux efforts de la comprendre?

Je pense avoir identifié la réponse à cette question. Elle conduit à affirmer que le formalisme quantique a incorporé les principes d'une révolution radicale de l'épistémologie. Et le développement de ces principes engendre une refonte foncière de nos connaissances et de nos croyances concernant notre manière d'engendrer des connaissances.

Le formalisme quantique n'a introduit ce noyau révolutionnaire qu'en état diffus et cryptique, lors de la construction d'une représentation particulière, celle des états d'entités physiques microscopiques. Mais, d'une manière inattendue, j'ai pu synthétiser une forme intégrée et explicite de ce noyau où sa structure et modes de fonctionnement apparaissent au grand jour. Ensuite, par des extensions et une organisation appropriées, j'ai développé ce noyau en une *« méthode générale de conceptualisation relativisée ».*

Cette méthode est explicitement enracinée en dessous des langages, directement dans la factualité physique a-conceptuelle, comme c'est le cas, en particulier, pour les descriptions quantiques des microétats. Par la généralisation de ce trait – que l'analyse révèle être un trait *universel* de la conceptualisation humaine – la méthode de conceptualisation relativisée incorpore à sa base toute l'essence des contenus épistémologiques encryptés dans le formalisme quantique, ainsi que leurs remarquables puissances spécifiques. En outre, à la différence du formalisme quantique, la méthode de conceptualisation relativisée offre désormais une *représentation* explicite de la toute première phase des processus de construction humaine de connaissances. Cependant que la structure, et l'existence même, de cette première phase universelle de la conceptualisation, est entièrement ignorée par les épistémologies actuelles, kantienne, husserlienne, psychologique et biopsychologique, de même que par la logique, les probabilités, et toute la physique classique.

D'autre part, la méthode de conceptualisation relativisée permet d'expliciter la structure de la question métaphysique du réel, faisant ainsi corps avec la philosophie et l'unissant aux démarches de type 'scientifique'.

La méthode générale de conceptualisation évoquée plus haut a déjà été exposée dans toutes ses phases successives. D'abord dans une suite d'articles spécialisés qui a débuté en 1984 (Mugur-Schächter [1984], [1992], [1995], [2002A], [2002B]). En outre, l'état présent de cette méthode et les applications majeures élaborées à ce jour sont présentés in extenso, avec détail et aussi rigoureusement qu'il m'a été possible, dans un livre paru récemment (Mugur-Schächter [2006].

Par contre, les *sources* de la méthode n'ont jamais encore été exposées d'une manière construite et détaillée. Tout simplement parce qu'elles n'avaient pas encore acquis dans mon esprit une forme intégrée. Ces sources ont certainement agi de manière intuitive tout au cours des recherches que j'ai consacrées aux problèmes d'interprétation que le formalisme de la mécanique quantique ne cesse de soulever depuis plus de 75 ans. Très lentement, ces recherches ont développé en vrac: des brins de solution à l'une ou l'autre de ces questions d'interprétation ; les lignes propres de la méthode générale de conceptualisation relativisée ; et les différentes applications de la méthode générale. Mais les quelques compréhensions locales engendrées par ce processus indifférencié concernant spécifiquement les questions d'interprétation, brillaient éparses, sans se rattacher à une organisation conceptuelle unique, explicite, construite et cohérente.

Puis, le processus d'investigation qui se développait a atteint un point critique, par l'émergence d'une hypothèse qui suggérait une stratégie globale pour arriver à véritablement maîtriser la manière de signifier du formalisme quantique: faire table rase du formalisme mathématique et construire la description d'un microétat en termes strictement qualitatifs, telle qu'elle émerge sous les contraintes exclusives de *la situation cognitive où l'on se trouve* et des modes humains de conceptualisation. En un temps relativement très bref, la mise en œuvre de cette stratégie a précipité en effet une organisation conceptuelle qui se présente comme une *infra-mécanique quantique*, comme une réprésentation qualitative des descriptions *quelconques* de microétats (mécaniques ou autres) logée en dessous du formalisme mathématique de la *mécanique* quantique.

L'infra-mécanique quantique expose finalement à tous les regards, structurée, l'entière manière de signifier du formalisme quantique. Et son cœur même – la forme descriptionnelle spécifique qu'elle produit concernant les microétats – est cet embryon annoncé, d'une révolution de nos connaissances concernant nos manières d'engendrer des connaissances.

Dans le formalisme de la mécanique quantique, la forme descriptionnelle qui commande la structure mathématique, reste inapparente. Seuls les fragments de cette forme peuvent y être devinés, incorporés épars aux algorithmes de façon rigide et cryptique. Ils y font problème, comme les restes d'un animal aujourd'hui inconnu, éparpillés et pétrifiés dans une roche, constituent un problème pour les paléontologues. Or dans l'infra-mécanique quantique cette forme descriptionnelle se montre en état *intégré*. Elle s'y montre nue, extraite hors de toute carapace mathématique et concentrée en une structure autonome animée de modes propres de fonctionner. Il s'agit véritablement d'un *être* descriptionnel d'un type foncièrement nouveau, qui s'expose et agit sous les yeux de tous, quand *rien* ne l'annonçait, ni les grammaires, ni la logique et les probabilités classiques, ni la physique classique ou la relativité einsteinienne.

Et à la lumière des analyses subséquentes, les descriptions de ce type insoupçonné se sont révélées constituer *universellement* une toute première strate de l'entière conceptualisation humaine.

Dans la méthode de conceptualisation relativisée cette universalité est représentée dans toute sa généralité.

Si l'on considère l'infra-mécanique quantique dans son ensemble, elle constitue désormais une structure épistémologique-méthodologique de référence qui permet d'envisager un traitement de l'ensemble des problèmes d'interprétation soulevés par le formalisme quantique, en bloc et sous des contraintes de cohérence *globales* ; un traitement à accomplir par un système de comparaisons entre, d'une part le formalisme quantique mathématique, et d'autre part l'infra-mécanique quantique telle qu'elle a émergé indépendamment (Mugur-Schächter [ 2009].

Dans ce livre je présente l'infra-mécanique quantique sans nullement aborder les questions d'interprétation du formalisme mathématique de la mécanique quantique.

Ce qui est révolutionnaire est toujours d'abord très difficile à percevoir. Le déjà connu, tel qu'il est connu, attire vers ses propres formes tout ce qui pénètre dans le champ de l'attention. Pourtant dans le cas qui nous occupe cette difficulté peut être vaincue. Une traduction d'un langage mathématique en un langage courant aurait irrépressiblement appauvri et obscurci la précision et la structure de significations incorporée au formalisme quantique. Mais l'infra-mécanique quantique n'est pas une traduction de ce type. Elle est le résultat d'une construction *indépendante*, directement de la structure épistémologique-méthodologique des descriptions de microétats. Chaque pas de cette construction est marqué d'une nécessité imposée par la situation cognitive dans laquelle on se trouve et par des impératifs conceptuels-logiques. Je me suis astreinte à faire ressortir cela d'une manière très claire.
J'ai tenté de maintenir la présentation constamment accessible à un grand nombre d'intellectuels et pourtant tout à fait rigoureuse quant aux significations et, à chaque pas, de mettre en évidence ce que ce pas-là comporte de radicalement nouveau.

En dehors de son intérêt conceptuel, l'infra-mécanique quantique possède aussi une certaine importance pragmatique.
Il existe des croyances religieuses, morales, économiques, etc. Et il existe aussi des croyances épistémologiques. Celles-ci sont très profondément enracinées dans le psychisme humain. Probablement même d'une manière plus uniforme et plus agissante que les croyances religieuses. Corrélativement, elles sont enracinées dans les langages courants et y affleurent dans ses formes les plus fondamentales (par exemple, "cet arbre *est* vert" – pas "je le *vois* vert'' –, ce qui, d'emblée, *absolutise* nos perceptions humaines). Par cette voie des langages courants les croyances épistémologiques s'infusent constamment dans tous les actes de pensée. Ainsi ces actes, parce qu'ils agissent en conformité avec elles, semblent confirmer les croyances épistémologiques qu'ils ont incorporées. Cette sorte de circularité charge subrepticement les croyances épistémologiques d'une inertie très difficile à vaincre. Car ce dont on perçoit souvent l'assertion se travestit en vérité testée.
Et pourtant les croyances épistémologiques *évoluent*, irrépressiblement. L'un des moteurs de leur évolution sont les apports des flux minces mais continuels qui émanent des sciences dures et qui s'infiltrent dans la pensée publique. Mais cette évolution se produit très lentement et d'une façon plus ou moins implicite et chaotique. Ceci entraîne qu'elle reste sans contour et non contrôlée. Or il serait utile que l'évolution des croyances épistémologiques se produise d'une manière exposée aux regards, et rapide, et qu'elle puisse être guidée et optimisée. Car, à l'intérieur de la catégorie générale des croyances, les croyances épistémologiques constituent un cas tout à fait spécial, chargé de potentialités qu'il est dommage de laisser dormantes.
Bien que très fortement agissantes, les croyances épistémologiques sont le plus souvent quasi inconnues des esprits où elles agissent. Un nombre de gens relativement infime en sont avertis. Et même parmi ceux-ci, une connaissance explicite, claire et approfondie du contenu des croyances épistémologiques qui les animent, est restée jusqu'ici très rare. Ces croyances travaillent d'une manière comparable à celle des réflexes neurophysiologiques auxquels d'ailleurs elles sont liées. D'autant plus, quand il s'agit de *changements* de contenu des croyances épistémologiques, la connaissance de ceux-ci reste encore plus enfouie, plus rare et vague, presque évanescente.

D'autre part, rien ne peut empêcher les changements de croyances épistémologiques, puisque ces changements sont enracinés dans l'état des sciences de la nature, qui, lui, change irrépressiblement. De là, de cet état des sciences auquel ils sont liés, les changements de croyances épistémologiques pénètrent directement dans les sous-conscients, portés surtout par les *techniques* tirées des sciences dures, qu'un très grand nombre de gens sont amenés à s'approprier. Ainsi les changements de croyances épistémologiques se trouvent sous l'empire de contraintes d'ordre *pratique* qui contribuent à régir les relations entre l'homme et le réel physique. Par cette voie ils sont protégés à la fois des trémoussements arbitraires des modes et des inerties de l'esprit. Malgré leur état implicite et en dépit des forces inertielles qui lestent toute croyance et notamment certains paradigmes de la pensée scientifique déclarée, les croyances épistémologiques changent avec la mentalité technique.

Enfin, cependant que leur genèse et leurs contenus sont ainsi liés à la phase de développement des sciences de la nature et des techniques correspondantes, les croyances épistémologiques *réagissent* sur cette phase. Or la phase de développement des sciences et des techniques, elle, produit un impact crucial sur les évolutions économiques et sociales. En conséquence de cet enchaînement réflexif, les croyances épistémologiques – *si* elles étaient connues explicitement et à fond – pourraient libérer des forces d'orientation voulue et directe des évolutions sociales et économiques. Potentiellement, elles sont un point d'appui stratégique pour obtenir des effets de levier.

Pour ces raisons au moins, il est important d'un point de vue *pragmatique* de connaître d'une manière explicite les croyances épistémologiques induites par les sciences de la nature, dans chaque phase de leur évolution, et de contrôler l'accord évolutif de ces croyances avec les vues générales qui agissent, notamment dans la pensée scientifique.

Dès qu'on a disposé de recul il a toujours été possible d'expliciter les croyances épistémologiques induites par telle ou telle phase du développement des sciences de la nature. Mais maintenant, en ce moment précis de l'histoire de la pensée que nous sommes en train de vivre, il est d'importance très particulière de mettre cette possibilité à l'œuvre, parce que c'est la mécanique quantique qui y est impliquée et que les schémas épistémologiques encryptés dans le formalisme de la mécanique quantique fondamentale violent des croyances épistémologiques qui agissent dans la pensée *scientifique* tout autant que la pensée courante, depuis plus de trois siècles. Tous ceux qui réfléchissent à la pensée le pressentent. Ils pressentent que dans cette résistance tellement persistante à une compréhension consensuelle opposée par un formalisme qui, d'autre part, est doté d'une si remarquable efficacité, il doit se nicher l'un de ces mystères précieux qui, lorsqu'ils sont levés, font apparaître une nouvelle face des choses.

Ce livre s'adresse à tous ceux qui désirent pénétrer jusqu'aux racines de la conceptualisation humaines et former une vue sur la manière dont y émergent les descriptions de microétats incorporées à l'état cryptique dans le formalisme mathématique de la mécanique quantique fondamentale. Sa lecture aura également l'effet de les rendre immédiatement réceptifs à l'exposé de la méthode générale de conceptualisation relativisée (Mugur-Schächter [2006]), une épistémologie qualitative mais formalisée, solidement fondée dans la pensée scientifique actuelle et qui couvre les processus de génération de connaissances depuis l'encore jamais conceptualisé et jusqu'à la frontière entre le connu et le métaphysique. Mais il paraît probable que l'infra-mécanique quantique intéressera de la manière la plus intense une certaine catégorie de physiciens qui la comprendront profondément et à mi-mot. Même si cette catégorie est numériquement réduite, c'est elle qui pourrait constituer un noyau de renouvellement de la physique actuelle.

L'exposé de ce texte est organisé en quatre très brèves parties.

La première, *via* des considérations informelles concernant les questions philosophiques du réel et de la connaissance, repère la place qu'occupe la mécanique quantique dans l'évolution de la pensée.

La deuxième partie expose le processus de construction de l'infra-mécanique quantique.

La troisième partie examine très brièvement l'infra-mécanique quantique de manière globale, de l'extérieur, avec le but de spécifier ses relations avec les concepts d'espace, de temps, de géométrie, de consensus intersubjectif, et avec les théories de relativité d'Einstein. Les conclusions tirées de ces examens permettent d'éclairer le problème de localité et de mettre ainsi en évidence son caractère illusoire.

Enfin, la quatrième partie, elle aussi très brève, explicite les potentialités de novation épistémologique incorporées dans l'infra-mécanique quantique, et indique la voie qui, en réalisant ces potentialités, a conduit vers la construction d'une méthode générale de conceptualisation relativisée.

Les contenus inorganisés de l'infra-mécanique quantique ont nourri pas à pas la genèse de la méthode générale de conceptualisation relativisée, en se mêlant aux guidages sous terrains de l'intuition avant que ces deux systèmes conceptuels se fussent constitués et séparés. Mais maintenant, quand l'un comme l'autre de ces deux systèmes est déjà construit et individualisé, les relations qui les unissent et les distinguent mutuellement présentent un caractère nouveau.

L'infra-mécanique quantique peut être regardée comme *l'illustration* majeure de la méthode générale de conceptualisation relativisée, celle qui, par son produit descriptionnel propre, encore particulier, a révélé une toute première strate descriptionnelle universelle qui était ignorée.

D'autre part, l'infra-mécanique quantique peut tout autant être regardée comme un exposé systématique et approfondi des *sources* de la méthode générale de conceptualisation relativisée.

Quel que soit le choix ou les oscillations entre ces deux optiques, le texte présent et l'exposé de la méthode générale fait dans Mugur-Schächter [2006] constituent ensemble un *tout* beaucoup plus achevé que chacun de ces deux ouvrages lus séparément. Or ceux qui prendront connaissance des exposés de ces deux approches en succession immédiate, seront frappés par certaines redites. Je tiens à souligner que ces redites étaient le seul moyen de doter d'auto-suffisance chacun des deux exposés considéré séparément.

Et en tout cas, toute redondance liée à la chronologie génétique, n'a qu'une importance provisoire. Dans l'exposé final intégré de l'entière structure conceptuelle qui s'est constituée dans mon esprit – qui inclut la résolution de l'ensemble des problèmes d'interprétation du formalisme quantique ainsi qu'une certaine mathématisation de la méthode générale de conceptualisation relativisée – toute redite aura disparu sans traces. Par cette sorte de subite harmonisation géométrisée qui est la marque des touts achevés, les manifestations de l'inévitable temporalité des processus de construction s'y seront dissoutes. Les rapports essentiels, atemporels, y brilleront seuls.

Enfin, l'on constatera que la bibliographie est réduite et en grande mesure personnelle. Qu'on ne m'en tienne pas rigueur: cela est dû au fait que le contenu texte qui suit rend compte de réflexions et d'une recherche solitaires à l'extrême.

*Note technique*

**1.** La numérotation des chapitres du livre ne tient pas compte de sa division en parties. A l'intérieur de chaque chapitre la structure du sens est explicitée par des sous-titrages hiérarchisés.

**2.** Dans tout ce qui suit:
- le signe " - " indique un *concept* ;
- le signe ' -' indique le *nom d'un de concept* ou *une façon courante de dire* ;
- le signe «- » indique une *citation*.
- les italiques sont utilisées pour accentuer *l'idée* exprimée.

Les conventions posées ci-dessus ont permis de faire systématiquement des distinctions qui d'habitude sont omises et dont l'absence induit beaucoup de confusions.

*Première partie*

*Le réel et la connaissance
versus
la pensée courante.
L'épistémologie philosophique et la physique*

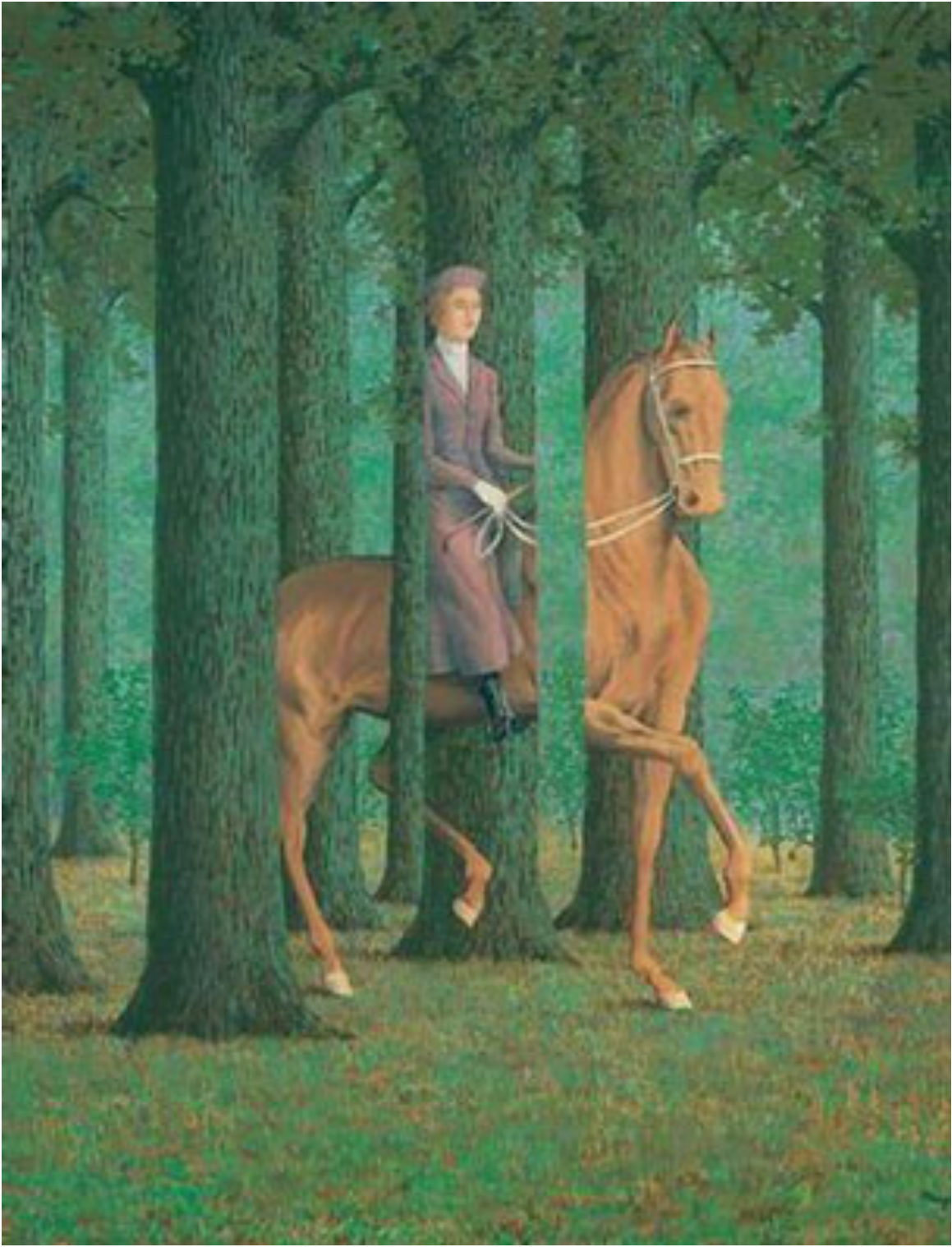

**MAGRITTE**
LE BLANC-SEIGNE

# *Mot d'introduction à la première partie*

La première partie de ce livre, constituée d'un seul chapitre, est dédiée à quelques très rapides considérations sur la question du réel et de la connaissance.

Il ne s'agit nullement d'un exposé philosophique. Il s'agit juste d'une rapide exploration introductive – tout à fait informelle – étroitement ciblée sur deux buts principaux:

\* Repérer la place qu'occupe la mécanique quantique face aux questions du réel et de la connaissance, dans le cadre d'une sommaire vue d'ensemble concernant ces questions et en perspective historique: juste la place, sans rechercher pour l'instant aucune autre spécification plus interne.

\* Sensibiliser fortement au rôle que peuvent jouer au cours d'une élaboration conceptuelle, les *décisions méthodologiques*.

# Chapitre 1

# De ce qui "existe" et de la connaissance

*"La connaissance du réel est une lumière qui projette toujours quelque part des ombres"*

*G. Bachelard*

## 1.1. Du réel dans la pensée courante et les langages naturels

### 1.1.1. Un enchevêtrement vague mais résistant

L'état actuel d'une question aussi fondamentale que celle de la spécification de ce qu'on considère comme 'existant', est très curieux. La grande majorité des gens semblent traverser leur vie entière sans jamais se demander d'une façon consciente et suivie comment s'organisent leurs propres croyances sur l'existant. Pourtant ils doivent bien être amenés de temps à autre à se dire ou à penser: « mais cela n'existe pas », ou bien « il est clair que cela existe ». Et même parmi les professionnels de la pensée et de la recherche, qui sans doute se posent tous des questions explicites sur l'existant, la réponse reste vague. En outre, elle fluctue fortement d'un penseur à un autre, et cela depuis des millénaires. À ce jour il n'y a pas de consensus sur ce sujet, ni parmi les scientifiques ni parmi les philosophes. Ceci est sans doute intimement lié au fait que les sens courants des mots 'réel' et 'existant' ne sont ni tout à fait les mêmes ni tout à fait distincts l'un de l'autre.

Si l'on posait à la ronde la question « qu'est ce qui est réel? », je crois que la réponse la plus fréquente serait, en essence, « tout ce qui est physique ». Mais si l'on demandait ensuite « le comportement de Pierre envers sa vieille mère, le fait qu'il ne lui rend jamais visite, qu'il ne lui écrit même pas, n'est-il pas réel? », la réponse risque d'être souvent du type « si, cela aussi est réel, mais d'une autre façon ». D'autre part, si l'on demandait: « Les idées existent-elles? Pas les miennes ou les vôtres, les idées elles-mêmes, l'idée de nombre, ou de 3, de beauté, de demain ou d'hier, etc. ; et est-ce que les licornes ou les naïades existent? Est-ce que votre gentillesse, ou sa méchanceté, et tous nos comportements et manières d'être (pessimisme, gaieté) existent? Est-ce que vos relations avec l'état ou avec vos concitoyens, existent? Et l'histoire et les cultures – pas des manifestations perceptibles des enregistrements qui en parlent par écrit ou par images, mais elles-mêmes et globalement – est-ce qu'elles *existent*? », alors, si l'on exigeait une réponse par 'oui' ou par 'non', il y aurait probablement encore beaucoup plus d'oscillations.

Concernant ces sujets il existe des disputes dont quelques unes sont très anciennes et persistent à ce jour, et même soulèvent des passions. L'exemple le plus frappant est le platonisme. Il est bien connu que selon les platoniciens les idées existent, par exemple l'idée de nombre en général, ou l'idée de 3. Ces idées existeraient *hors de toute conscience*, là, quelque part dehors et autour. Elles existeraient par elles-mêmes, comme on pense qu'existe ce qu'on appelle 'le réel physique'. Cependant que selon les non-platoniciens les idées humaines n'existent qu'à l'intérieur des esprits d'hommes ou dans les enregistrements de contenus de ces esprits, livres, films, etc.. Cette opinion est exprimée de façon abrégée et affligeante en disant que les idées n'existent pas, ce qui manifeste la croyance que seul ce qui est *physique* existe. Cette dispute, qui vient de l'antiquité, a traversé tout le moyen âge et toute l'époque des sciences dites classiques, et actuellement elle nourrit toujours des livres de logique (R. Haller [1985]) et des discussions nombreuses sur la philosophie des mathématiques. Il y a là un fil continu sur lequel il serait difficile de placer un repère qui puisse séparer l'ancien du moderne. Bref, pour la plupart des gens les signifiés des mots 'réalité' et 'existants' se perdent dans un non-fait comme les contours d'un bas-relief.

D'autre part les philosophes, depuis Descartes tout au moins, quasi unanimement, traitent l'existence des consciences individuelles comme une donnée première. Donc en aucun cas ils ne pensent que seul ce qui est physique existe. Ils soulignent que ce qu'on perçoit tout d'abord et le plus directement c'est qu'on est conscient, que l'on éprouve des sensations, des états affectifs, des tendances, que l'on pense, et que c'est tout cela qui constitue *l'unique* certitude absolue. Et il paraît en effet probable qu'aucun individu humain normal n'ait jamais douté que sa vie psychique existe, qu'elle est réelle, que le fait qu'il voit, entend, qu'il a mal ou est joyeux, qu'il veut ceci et abhorre cela, que tout cela existe "vraiment". Les philosophes solipsistes[1] (notamment Berkeley) sont même allés jusqu'à soutenir que rien d'autre n'existe vraiment. Ils ont réussi à soutenir cette vue sans tomber dans des contradictions logiques. Cette possibilité est remarquable en elle-même. À elle seule, elle établit une certaine sorte de priorité de la conscience face à la connaissance. Car il n'y a pas de symétrie. Il n'est pas possible de soutenir également, sans contradiction, que seuls les faits physiques existent ; ou seulement les faits non-subjectifs, soit physiques, soit ceux qui, globalement, dépassent toute conscience individuelle, comme les comportements socio-économiques, les mythes, les faits culturels, etc.. Cela nierait l'existence des faits de conscience individuelle et de pensée, donc aussi celle des faits de connaissance. Or, confronté à la question, aucun homme ne peut penser qu'il ne pense pas, ou dire qu'il sait qu'il ne sait strictement rien. Ce serait du non-sens.

Et pourtant, paradoxalement, le mot 'réel' semble avoir été assez unanimement employé surtout pour qualifier ce qui est physique, ou du moins ce qui est extérieur aux consciences individuelles, ce qui est soustrait aux subjectivités, ce qu'on estime être "objectif". Le reste n'existerait que sans être "vraiment" réel, bien que le psychique soit à la base de toute connaissance.

Nous sommes en présence d'un enchevêtrement.

Il s'agit d'un enchevêtrement vague mais très résistant qui me semble provenir de deux sources majeures. En premier lieu, il s'agit d'un effet particulier de ce que, dans mon langage privé, j'appelle *les trompe l'œil conceptuels*. Et en second lieu, il intervient aussi ce que j'appelle *la chosification de l'association d'un mot au sens de ce mot*.

### 1.1.2. Trompe-l'œil conceptuels

Chaque concept a un domaine de pertinence limité. La frontière de son domaine de pertinence est inscrite dans la genèse du concept et s'exprime dans sa définition. La plupart des concepts naissent par abstraction, à partir de *cas* individuels. Un concept qui se constitue ainsi ne peut *rien* contenir qui ne soit commun à *tous* les cas individuels qui ont participé au processus d'abstraction. Si cette chaise comporte une surface plus ou moins horizontale et 4 pieds plus un dossier et des accoudoirs, mais que d'autres chaises n'ont pas d'accoudoirs et n'ont que 3 pieds, ou 2, ou bien un seul, ou plus que 4, alors le concept tiré de ces cas divers admettra la désignation générale « chaise: une surface horizontale comportant un dossier et soutenue au-dessus du sol par un ou plusieurs supports ». Le mot 'chaise' fondé sur cette genèse, désignera l'ensemble de tous les objets sur lesquels on peut rester assis de manière commode, et rien d'autre: la conceptualisation possède certains caractères d'un calcul.

---

[1] Du latin *solus*: seul.

Ceci posé, considérons par exemple le concept général de "début", tel qu'il se constitue par abstraction. Je vois, disons, cette table, et quelqu'un me demande « sais-tu quel a été le début de cette table?». Le mot 'début' paraît d'emblée ambigu dans ce contexte ; mais précisément cela est utile dans ce qui suit: si je réponds « je l'ai achetée dans une vente publique », mon interlocuteur est en droit de remarquer « cela, c'est le début de l'époque de *ta possession* de cette table, pas le début de l'époque de l'existence de la table. La table, elle, a commencé d'exister dès que tous les éléments qui permettent d'y placer des objets lourds sur une surface maintenue à une certaine hauteur au-dessus du sol, ont été réalisés, ses pieds, son plateau, etc. Chaque entité possède son propre début, n'est-ce pas? ».

Et en effet, ni moi ni personne d'autre n'a jamais perçu un début bien spécifié qui ne soit le début de telle ou telle entité définie, et qui n'ait consisté à tirer de quelque façon cette entité-là d'autre chose de physique ou de conceptuel qui préexistait. Le concept général de début s'engendre en faisant abstraction aussi bien de l'entité particulière à laquelle est lié tel ou tel début particulier, que des entités ou substrats préexistants dont cette entité particulière a émergé. Néanmoins ce de quoi elle a émergé est à chaque fois présupposé par le processus d'abstraction. Cela fait partie intrinsèque du concept général de début.

Or, lorsque tout à coup on demande « quel a été le début de l'Univers?», le mot 'Univers' est quelquefois compris comme indiquant une entité qui est illimitée et inclut *"tout"*, donc une entité qui ne laisse rien en dehors d'elle de quoi elle ait pu émerger. Si l'on conçoit ensuite un substrat d'émergence, où peut-on le placer? Dans 'un *autre* Univers'? Ce ne serait pas une réponse, cela ne serait qu'ajourner la réponse. On sent bien que l'on amorcerait une régression non confinée.

Que fait-on alors? On se met à palier. Par exemple, on pose qu'il n'y a eu aucun substrat d'émergence. En ce cas le présupposé, pour tout début au sens courant du mot, d'un substrat d'émergence qui préexistait, est subrepticement effacé lorsqu'on soulève la question du début-de-l'Univers. On glisse hors du contenu dont le concept général de début a été doté par sa genèse, par le processus d'abstraction qui a produit ce concept. On s'en est éjecté. Au Pays des Merveilles cela pourrait s'appeler *a mad toboggan party*. On se retrouve de l'autre côté de la frontière invisible que la genèse du concept trace autour du domaine de pertinence du concept, de validité de sa définition, à savoir le domaine des entités qui peuvent avoir un début au sens courant, un début concevable, un début à partir d'un substrat préexistant, physique ou conceptuel. On se trouve désormais dans un 'no man's land' conceptuel. Mais on s'y trouve sans en avoir conscience. On se croit toujours à l'intérieur du domaine de pertinence du concept désigné par le mot 'début'. Sans s'en rendre compte, on a fabriqué un *'trompe l'œil conceptuel'*: une entité que – en l'absence de tout substrat préexistant que l'on puisse spécifier – l'on place néanmoins à l'intérieur du domaine de pertinence du concept désigné par le mot 'début'. On l'y place illusoirement, en la faisant percevoir comme ce qu'elle n'est pas, c'est-à-dire en contredisant la définition de ce concept. Sous une continuité apparente, ce trompe l'œil cache en fait une violation des séparations incorporées plus ou moins implicitement dans le langage, tel qu'il a été construit sur la base de processus d'abstraction. Et cette violation entraîne une impossibilité de concevoir. C'est une catastrophe au sens de Thom. Et au sens courant c'est l'ouverture vers des mésaventures, car cela pousse la pensée dans des stagnations, des paradoxes, des questionnements sans issue, des constructions extraordinaires.

En est-on est vraiment dupe? Je crois que oui, en général. Car sinon il n'y aurait pas autant de débats obscurs concernant le début de l'Univers. Il serait évident qu'une théorie *scientifique* ne peut statuer que concernant un concept d'Univers qui présuppose une genèse à partir de quelque substance *physique* préexistante, même s'il ne s'agit que d'un modèle, par exemple "le vide quantique". Cependant que les religions, elles, qui évidemment sont libres de ne pas supposer une substance préexistante physique, en général y suppléent en postulant une divinité dotée du pouvoir d'en émettre une. On se placerait alors clairement sur l'une ou l'autre de ces positions et on ne ressentirait aucun problème. Mais si l'on n'a pas fait un choix explicite parmi ces deux possibilités et l'on tend de manière floue et inertielle à continuer de réfléchir de la même façon que lorsqu'on recherche un début au sens courant, conforme au concept construit pour les entités usuelles, alors, dans le cas du concept d'Univers au sens de *tout*, l'esprit butte sans le savoir contre la condition implicite d'un substrat préexistant, qui dans ce cas limite n'est pas remplie ; et cela, obscurément, fait obstruction. Alors, bloqué et décontenancé par un obstacle qu'il sent agir mais dont il ne discerne pas clairement la nature, l'esprit se met à bricoler des solutions vagues et hybrides où l'on flaire de l'incohérence. Car on *sent* que les deux concepts mis en jeu, celui d'Univers au sens de tout, et celui de début au sens courant, en quelque sorte s'excluent mutuellement, ou du moins suscitent un problème nouveau, spécifique: on perçoit un trompe l'œil conceptuel, de façon implicite.

Cet exemple n'est pas unique. Il existe une véritable foule d'autres cas du même genre. Certains parmi les trompe l'œil qui sévissent dans notre conceptualisation, se manifesteront ici même, au cours des exposés de ce livre. Parmi eux se trouve le trompe l'œil colossal désigné par le mot 'objet', qui entache toute l'essence de la pensée classique, autant celle de la pensée classique courante, que celle des conceptualisations classiques logique, probabiliste, et scientifique au sens général. (Mais, remarquablement, sans contaminer aussi la pensée classique philosophique et notamment épistémologique).

Ainsi, dans le cours de nos processus implicites de construction de concepts, il arrive fréquemment que l'on se dupe en fabriquant des trompe-l'œil conceptuels. Les mathématiciens et les logiciens font barrage, à l'aide de systèmes formels. Mais les systèmes formels sont rigides. En outre, leur sémantique – qui toujours existe enfouie dans les axiomes et les règles de transformation – n'est ni donnée ni avouée. Or le domaine de pertinence d'un système formel est confiné – par construction – au domaine de pertinence de cette sémantique cachée, dont la présence est niée. Il y a là un appel muet d'une méthode générale et explicite de conceptualisation *qui soit déterminée par les contenus sémantiques*. Le règne du formel dans les approches logiques et le règne du quantitatif dans les représentations scientifiques, de concert, ont confiné les qualités en ghettos. On manque de représentations construites ouvertement et systématiquement sur la base de structurations de domaines sémantiques. Cette circonstance introduit un abîme entre la pensée libre, naturelle – avec aussi les représentations mathématisées qu'elle produit en physique théorique et même dans les disciplines de mathématiques 'pures' – et d'autre part les représentations qui consistent en systèmes formels à proprement dire.

Nonobstant la formation naturelle de trompe-l'œil conceptuels, notre pensée de tous les jours, celle qui est chargée de sémantique, ne baigne nullement dans l'illusoire. Car l'intuition, elle, perçoit les discontinuités sémantiques occultées par les trompe-l'œil conceptuels et tend à les éviter ou à les éliminer. Elle les perçoit confusément mais d'une manière lancinante qui engendre des expression indirectes, sous la forme de formulations surréalistes, provocantes, paradoxales, humoristiques, ou bien tout simplement sous la forme d'étonnements, ou de refus muets mais têtus qui enfouissent dans du silence, le temps de reconstituer une capacité d'attaque. Les métaphysiques et les religions étanchent quelquefois ces étonnements et ces refus, cependant que le rire, ou la poésie et les autres arts, les nourrissent indéfiniment. Ainsi les discontinuités que les trompe-l'œil conceptuels à la fois créent et occultent, restent indéfiniment présentes et agissantes. Les métabolismes de la pensée les empêchent de disparaître avant d'avoir produit leur propre destruction explicite. L'incompris, même s'il ne subsiste que de façon implicite, ne cesse d'agir que lorsqu'il cesse d'exister.

### 1.1.3. Données factuelles et chosification de la relation entre mot et sens

J'avais annoncé deux sources de la persistance, dans la pensée courante, de l'état enchevêtré de la question du réel. Or pour l'instant je n'ai spécifié qu'une seule de ces sources, les dérapages au delà de la frontière impartie à un concept par le processus d'abstraction qui l'a installé et qui conduisent à des trompe-l'œil conceptuels. En outre, j'ai spécifié cette source sans avoir indiqué comment un tel dérapage peut affecter la question du réel, spécifiquement. Ceci est dû au fait que la voie par laquelle les trompe l'œil conceptuels affectent la question du réel, est liée à la deuxième source de flous que j'avais annoncée, la chosification de la connexion entre le sens d'un mot, et le mot. J'indique maintenant très brièvement en quoi consiste cette deuxième source.

Les flous où flotte la question du réel sont souvent associés *à une coalescence larvée entre données factuelles et choix de langage*. Par exemple, c'est une donnée factuelle que vous ou moi, ou cette chaise et tous les autres corps tangibles qui nous entourent sur la surface de la terre ou pas trop loin de celle-ci, ont du poids. On peut changer le mot 'poids': ce n'est qu'un choix de langage, une simple étiquette verbale. Mais le *fait* que ce vase tombe par terre si je le lâche sans aucun support en dessous, n'a rien d'arbitraire ou conventionnel. Le poids des corps est remarqué et pris en compte par tout le monde. Personne ne doute de la réalité, sur cette Terre, du poids des corps. Même lorsqu'elle n'est pas conceptualisée cette prise en compte s'exprime dans le comportement de tout ce qui est vivant.

La situation est différente lorsqu'il s'agit de l'idée générale de nombre, ou du concept de licorne, ou des comportements de Marie, ou des sensations et des pensées d'un *X*. Dans tous ces derniers cas il n'y aura pas d'unanimité en ce qui concerne la réalité (ou l'existence) du désigné, parce que les bases *factuelles* d'une unanimité intersubjective concernant cette réalité, font défaut. C'est cette absence qui, dans de tels cas, laisse place à des oscillations quant à la pertinence des qualifications 'réel' ou 'existant'. Or là, cette différence statistique que l'on constate dans l'attitude face à la question de 'réalité' lorsqu'on passe de la considération du poids des corps à la considération de nombres ou de comportements, etc., se fait jour la question du rôle que joue l'association entre un mot et un signifié.

L'association entre un mot et sa signification n'est pas une donnée factuelle dans le même sens que l'est le poids. Cette association s'est installée par une assignation à caractère conventionnel. Quelquefois le caractère conventionnel de l'assignation est reconnu explicitement, comme lors d'une définition d'un concept nouveau dans une discipline des sciences dures ou des sciences humaines, ou comme lors d'un baptême. Dans ces cas il semble clair que rien n'empêche de *décider* – notamment pour des raisons de *méthode* – de modifier convenablement cette assignation, soit en introduisant un autre mot, soit en introduisant une re-définition explicite de la signification à associer au mot habituel. Par exemple, un physicien expérimentateur, ou un ingénieur, comme d'ailleurs la majorité des gens, auront tendance à dire que le mot 'réel' qualifie ce qui est *matériel*. Tandis que Teilhard de Chardin [1956] considère que ce même mot qualifie aussi l'ensemble de 'la sphère des connaissances' (la noosphère), et selon Karl Popper [1984] ce même mot qualifie 'les trois Mondes', le monde physique, les états de conscience, et les connaissances, les arts, les faits culturels. Il s'agit là de manifestations d'une liberté qu'il est très important de ne pas ignorer.

Si l'on connaît cette liberté et on l'utilise avec un certain art, on arrive souvent à réorganiser des problèmes qui paraissaient inextricables, d'une manière qui les dissout immédiatement et sans trace. Mais si, au contraire, l'on n'est pas tout à fait conscient de cette liberté, ou si tout simplement on ne l'exerce pas, on peut rester indéfiniment piégé dans une problématique faussée par une sorte de chosification illusoire de l'association d'un mot à un sens. Une sorte de confusion entre un simple *étiquetage verbal* et une existence indépendante de toute décision humaine, comme celle du désigné du mot 'soleil' ou celle du référent de l'expression plus abstraite 'la gravitation universelle'.

Il existe une tendance souvent forte à chosifier la relation entre un mot et ce qu'il signifie. Car dans l'immense majorité des cas le caractère conventionnel des actes humains par lesquels le mot s'est trouvé associé à tel signifié (désigné, référent), s'est dissous dans les inconnus des hasards et des lointains de l'histoire. Alors, depuis le début de chaque durée individuelle, depuis que je suis là ou que tel autre est là, le mot et son signifié frappent toujours l'attention *ensemble*, même si le signifié *fluctue* avec le contexte. Ainsi ils s'imposent implicitement comme une variété abstraite d'un solide rigide, et l'on est conduit à débattre si le nombre 3 '*est* réel', au lieu de se demander s'il convient – et pour quel but – de *décider* de le qualifier par le mot 'réel', en tant qu'un élément de la noosphère que l'on *déciderait* avec Teilhard de qualifier comme réelle dans son ensemble, ou en tant qu'élément de l'un des 'trois Mondes' de Popper que nous accepterions, avec lui, de *poser* comme étant tous réels.

La chosification de la connexion entre un mot et ce qu'il désigne est une séquelle de disputes nées dans le passé. Au Moyen Age un grand nombre de penseurs ont soutenu avec une force surprenante que chaque mot est *substantiellement* lié à son référent, que cette liaison est douée d'une espèce de matérialité invisible. Michel Foucault [1966] a magnifiquement décrit cette croyance et ses conséquences paralysantes.

### *1.1.4. Intermède: un dialogue imaginaire[2]*

**L.** Pourtant une chose existe ou n'existe pas, elle est réelle ou ne l'est pas. Ce n'est pas du tout arbitraire. Ce n'est pas une question de choix de mot.

**M.** Je ne conteste nullement cela. Les données factuelles intersubjectives auxquelles on *réfère* une expression verbale, c'est-à-dire *ce* qui, selon tout un groupe d'observateurs, se manifeste de telle façon et pas d'une autre, en effet n'est pas arbitraire. Mais que cette façon-là de se manifester vous la désigniez en disant c'« *est* comme ceci », « c'*est* un ceci », « c'*est* 'grand', c'*est* 'sûr', c'*est* 'insupportable', c'*est* 'une maison' » – c'« *est* [tel-MOT] » –, cela ne sera jamais plus qu'une expression verbale associée par convention tacite à la manifestation factuelle que l'on constate d'une manière intersubjective. Aujourd'hui cette convention peut elle-même nous apparaître comme une 'donnée factuelle' parce qu'elle s'est installée depuis longtemps, spontanément et implicitement, et qu'il nous semblerait 'faux', pas 'vrai', d'entendre dire « ceci est sec » tout en montrant du doigt la mer. Cette vue, dans le présent, n'est pas tout à fait contestable. Mais ici il s'agit de la genèse de l'association entre un mot et le sens de ce mot: en dernière analyse cette genèse, elle, se ramène toujours à une convention tacite. Un mot n'est qu'un étiquetage verbal qui s'est installé, et que l'on peut changer. Si l'on savait que, pour quelque raison, une nouvelle loi vient d'être votée selon laquelle tout ce qui était avant qualifié par les mots 'mouillé' ou 'humide' doit désormais, sous peine d'électrocution, être qualifié par le mot 'sec', l'on se mettrait bien à dire « ceci est sec » tout en montrant la mer.

---

[2] Dans ce qui suit 'M' se lit 'moi' et 'L' se lit 'le lecteur'.

**L**. Mais où voulez-vous en venir? Pourquoi changer un mot installé?

**M**. Cette question déplace le débat. Mais elle est bien venue et donc j'y réponds: par exemple, pour le soumettre à des exigences méthodologiques construites sous la contrainte de *buts* bien définis, notamment des buts d'intelligibilité, de cohérence, de clarté, de suppression de faux problèmes. C'est d'ailleurs exactement ce qu'on fait dans la construction de tout langage spécialisé. En géométrie euclidienne on dit qu'un 'point' est 'l'intersection dépourvue de toute dimension, de deux droites' ; tandis que dans le langage courant un 'point' est un endroit de petite étendue que l'on marque ou l'on spécifie d'une manière quelconque.

**L**. Vous entendre parler de 'construction' de langage me gêne, dans ce contexte. Et d'ailleurs la notion de "méthode" me gêne aussi. Cette façon de voir m'étonne. Surtout, je ne vois pas où vous voulez en venir quant au réel et à l'existant. Pourquoi ne dites-vous pas simplement qu'il faut rechercher avec attention le vrai sens à associer aux mots 'réel' et 'existant'?

**M**. Rechercher le 'vrai' sens de ces mots? Mais qu'est ce que cela peut vouloir dire? S'il s'agit de rechercher avec quelle signification le mot 'réel' circule dans le langage naturel, alors je viens d'indiquer précisément l'inexistence d'une signification bien défini et unique. Et si vous pensez à une 'coïncidence avec la réalité telle qu'elle est vraiment en elle-même', alors il n'y a pas de 'vrai' sens d'un mot, quel que soit ce mot. Et en plus, 'le sens vrai' du mot 'réel' – en particulier – est une expression circulaire.

La connexion d'un mot, avec une signification, avec un référent donné, s'installe par convention tacite. Point. D'ailleurs cette connexion ne s'installe pas d'une seule façon. Chaque langue, chaque dialecte, chaque langage local, entre enfants, amis, collègues, etc., installe certaines connexions qui lui sont spécifiques. Le plus souvent, le processus d'installation reste inconnu. Il est complexe, lent, lié à des analogies et des hasards, et d'habitude il s'accomplit de manière implicite. C'est un processus très étonnant concernant lequel les spécialistes des genèses des langages pourraient nous donner des renseignements fascinants.

Dans une langue naturelle, l'association d'un mot à un référent n'est presque jamais construite d'une manière délibérée et non ambiguë: ce serait contraire à l'essence même d'une langue naturelle, qui émerge et se développe spontanément, à un niveau supra individuel, et lie la signification au contexte, foncièrement, systématiquement. Il y a là une certaine finalité spontanée incorporée à tout langage naturel, un but incorporé de malléabilité des significations assurée par la possibilité de compositions de mots très diverses. Les langages courants ont besoin de certains flous, car au cours des processus de communication ceux-ci dotent l'expressivité d'élasticité, de richesse et de résilience. Mais pas n'importe quel flou. Je viens de montrer que la frontière d'un concept désigné par tel mot, bien qu'implicite, résiste au flou où la plonge un trompe l'œil conceptuel.

En outre la recherche de la possibilité de moduler le sens d'un mot par le contexte s'affaiblit dans les langages spécialisés. Dans un langage *formel* réussi elle est même éliminée entièrement. Dans ce cas limite d'un langage spécialisé de logique formelle, le but est diamétralement opposé à celui de pouvoir moduler le sens d'un mot à l'aide de son contexte: on y cherche au contraire *l'unicité stricte* de la signification de chaque signe d'étiquetage, afin d'assurer aux trajets déductifs une rigueur parfaite, à l'abri de toute ambiguïté. Dans une théorie bien construite de la physique, ou même de la philosophie, où l'on utilise des déductions soit mathématiques soit seulement logiques, sans être pour autant un langage formel au plein sens du terme, le degré de rigidité de l'association d'un mot, à un désigné, est intermédiaire ; une telle théorie participe à la fois des caractère d'un langage formel et de ceux d'un langage courant. En tout cas:

> Si dans le cadre de tel ou tel langage spécialisé on veut disposer d'un référent non ambigu du mot 'réel' alors il ne faut pas *rechercher* ce référent, il ne faut pas vouloir *le découvrir*, *il faut le **construire***.

Si l'on s'emprisonne dans l'attitude consistant à le rechercher, on ne trouve rien de clair parce que rien de clair ne préexiste. Et comme d'autre part, implicitement, on suppose une préexistence, on s'enlise. Si je dis simplement « ceci est réel », vous pouvez protester « ceci n'est pas réel », et la dispute peut ne jamais finir. Car comment *établir* si, par exemple, les idées sont réelles ou pas? Comment *trouver* cela? On s'enfonce là dans un problème sans réponse parce que c'est un problème illusoire introduit par l'idée inadéquate qu'il s'agirait d'une vérité à rechercher, à découvrir.

Aucun sens n'est lié, ou pas, au mot 'réel', ni de par la constitution intrinsèque de ce mot, ni de par la nature intrinsèque de telle ou telle signification. L'assignation d'un mot à un référent se produit hors du mot et hors du référent. Dans un langage courant, cette assignation a un caractère diffus. Mais rien n'interdit, si c'est la précision maximale qui est le but, de prolonger ce système de taches de sens diffus vers lesquelles pointent les mots d'un langage courant, par une réorganisation ciblée qui réduise à un point bien défini la tache de sens qui correspond à certains mots. Dans ce cas l'action à fournir est donc bien de nature constructive et surtout méthodologique. C'est précisément par de tels prolongements constructifs et méthodologiques, par des *décisions méthodologiques*, que tout langage scientifique s'ancre dans le langage courant, tout en remplaçant ce langage par un autre, à cohérence logique contrôlée.

**L.** Présentée ainsi je comprends votre vue. Mais du coup elle me paraît triviale.

**M.** En un certain sens vous avez raison. Pourtant il est nécessaire de l'exprimer explicitement, car dans le débat sur l'existant – à ce jour même – on se comporte comme si le sens des mots 'réel' et 'existant' était un caractère intrinsèque à ces mots, et immuable, un caractère qu'il s'agirait de rechercher et de découvrir. On réifie l'association entre les mots 'réel' ou 'existant', et leur référent, comme on le faisait au Moyen Âge. Il me paraît difficile de nier cela. Or cela bloque toute progression vers une conclusion. Les débats s'enlisent comme dans un marécage dans la tache diffuse du sens courant des mot 'réel' ou 'existant'. Pour obtenir une conclusion, il faut la construire.

**L.** Admettons. Mais comment ces deux sortes de méprises que vous incriminez, les trompe l'œil conceptuels et les chosifications de l'association entre un sens et un mot, se conjuguent-elles pour conduire au flou persistant concernant la question du réel? Cela ne ressort pas.

**M.** On le verra juste ci-dessous.

### *1.1.5. La genèse de l'enchevêtrement de la question du réel dans la pensée courante*

Dès la toute première enfance l'esprit de l'homme se met à élaborer des 'objets'. Des objets concrets, comme par exemple ceux qui correspondent aux mots 'mère', ou 'chaise', ou 'montagne' ; ou bien des objets abstraits, comme ceux qu'indiquent les mots 'beauté', 'nombre', etc.. L'élaboration d'un objet s'accomplit à l'aide du concept de 'propriétés'. Or ce qu'on appelle des propriétés émerge par l'interaction de l'esprit avec, disons, le reste. Ainsi tout objet est foncièrement dépendant des caractéristiques de la conscience humaine qui l'a construit, de ses modes de fonctionnement dans un corps muni d'un système nerveux qui inclut un cerveau et des appareils sensoriels.

Tout 'objet' est foncièrement relatif aux grilles sensorielles et aux formes de conceptualisation introduites par la conscience où il s'est constitué. Notamment, il est relatif aux seuils de perception comportés par les grilles sensorielles mises en jeu. Si l'homme pouvait distinguer des dimensions spatiales de l'ordre de 10 à la puissance −10 d'un centimètre et des dimensions temporelles de l'ordre de 10 à la puissance −30 d'une seconde, les objets construits par les consciences humaines seraient tout à fait autres, ou au moins différents de ce qu'ils sont. Par exemple une cuillère apparaîtrait comme un nuage à bords mouvants. Et les objets formés par les appareils sensoriels d'autres espèces d'êtres, un oiseau, un insecte, sont en général différents des objets que forment les hommes.

Les philosophes, les physiologistes, les psychologues, et de plus en plus l'homme cultivé, sont clairement conscients de cette foncière dépendance de tout 'objet' face aux fonctionnements biopsychiques par lesquels il s'est formé, et des relativités qui en découlent. Les chercheurs dans le domaine des sciences cognitives attestent de cette dépendance et la décrivent avec un détail et une précision de plus en plus remarquables.

Mais l'intuition les occulte, avec une force tyrannique. Son travail d'occultation s'accomplit en dessous du conscient. Les processus d'interaction de la conscience, avec son extérieur, par lesquels l'esprit constitue ce qu'on appelle des objets, sont en général réflexes et ataviques. Or les intuitions pragmatiques (au sens darwinien) que ces processus réflexes installent dans l'inconscient, poussent d'une manière irrépressible à croire que les objets désignés par les mots de nos langages existent indépendamment de toute conscience ; que cette chaise, par exemple, est là, au dehors, *avec* toutes *ses* propriétés *telles* que je les perçois et qu'elle posséderait tout à fait indépendamment du travail de toute conscience ; qu'elle *est* marron, rigide, etc. et qu'elle existe ainsi indépendamment de vous ou moi ou de nous tous. Il s'agit là d'un immense glissement inconscient hors du domaine de validité du concept général d'objet tel que ce domaine s'est constitué par la genèse biopsychique factuelle du concept. Ce glissement réflexe s'accompagne de la fabrication d'un trompe l'œil conceptuel résistent comme un diamant, et omniprésent dans la pensée courante.

D'une part chaque élément de la classe du concept général d'objet n'émerge que par des interactions entre les consciences-et-corps humains et l'univers 'extérieur' à ces consciences – qu'elles postulent – et il porte la marque indélébile de ces interactions fondatrices, auxquelles il reste foncièrement relatif. La classe du concept général d'objet ne contient *que* des entités produites par cette sorte de genèse.

Mais d'autre part, ce même élément, dès qu'il a été construit, est conçu implicitement comme étant sans aucune relation avec les consciences-et-corps humains, comme étant dépourvu de toute relativité à des actions réflexes de lieux vivants d'émergence, dépourvu de toute relativité à la structure de grilles de qualification comportées par ces lieux d'émergence. Ainsi, subrepticement, tout 'objet' au sens courant est éjecté de la classe à laquelle il appartient par sa genèse. Il acquiert ainsi le caractère illusoire de quelque chose qu'il n'est pas: il se produit une altération en trompe l'œil conceptuel du référent du nom du concept, du mot 'objet'.

Cette altération introduit une immense tache aveugle dans le champ de notre connaissance intuitive de nos propres processus cognitifs. Et c'est là, dans cette tache aveugle, que prennent source des flous et des inconsistances dans nos croyances concernant le sens des mots 'réel' ou 'existant', en s'y mêlant à des chosifications de l'association du mot 'réel', avec tout mot-nom d'objet *physique*.

Quand elle s'accomplit à l'aide de délibérations *conscientes*, la genèse d'une action constructive est explicite dans l'esprit où elle s'est produite. En ce cas les conséquences de cette genèse concernant son produit, sont comprises. La genèse délibérée du concept de mètre étalon, ou celle d'une équation mathématique, définit explicitement, dans l'esprit où elle s'est produite, la nature du construit qui en a résulté, ses frontières et ses potentialités fonctionnelles. Alors ce construit, avec sa nature, ses limites et ses potentialités, est *ressenti* comme pas tout à fait extérieur à l'esprit, pas tout à fait imposé du dehors ; et *'donc'* comme moins réel qu'une montagne, par exemple, ou que la gravitation. C'est en conséquence du fait que dans la genèse du concept "montagne" dans tel ou tel esprit, le rôle de cet esprit a été entièrement non conscient, réflexe, que ce concept y est ressenti comme un simple enregistrement qui s'impose du dehors aux sens biologiques, tout fait, entièrement, avec sa nature et ses caractéristiques qui ne nous doivent rien. Et c'est cette manière de ressentir qui se reflète dans le fait que la réponse à la question « cette montagne est-elle réelle? » sera en général un *oui* immédiat et décidé. Cet *oui* est soutenu par la coalescence implicite entre ce que nous percevons lorsque nous percevons de quelque manière le désigné du mot 'montagne', et d'autre part le mot-qualification 'réel': on *ressent* que la montagne ″est″ réelle, car sinon comment s'imposerait-elle avec tant de force à nous qui n'avons aucunement contribué à ce qu'elle soit là et telle qu'on la perçoit? En radicalisant, on a même tiré une soi-disante définition du réel: « le réel » est ce qui résiste aux représentations » (ce qui peut s'y avérer faux).

L'association entre le mot-qualification 'réel' et ce que nos sens nous font percevoir lorsqu'on dit un nom d'objet *physique*, comme 'montagne', s'est chosifiée implicitement, intuitivement. Cependant que lorsqu'il s'agit d'une équation ou de l'idée de bonté, ou bien d'un comportement individuel ou social, la connaissance explicite du rôle des hommes dans la genèse de l'entité considérée se traduit en réticences et fluctuations quant à la *vérité*, à leur égard, du mot-qualification *réel*. D'où il vient que, plus une entité est extérieure à notre volonté et nos actions, plus *grand* est le degré de réalité qu'on a tendance à lui assigner. Cela nonobstant le fait que les états de conscience d'un être humain, qui lui sont foncièrement intérieures, soient les uniques données dont l'existence, pour lui, soit certaine.

L'enchevêtrement de nos vues concernant le "réel" émane de forces intuitives tellement impérieuses que seule une méthodologie solidement structurée pourrait le dissoudre.

## 1.2. De la connaissance dans la pensée courante et les langages naturels

Dans la pensée courante les cohérences conceptuelles qui sous-tendent les mots et la syntaxe, ne sont pas explicites. Elles ressortent de l'ensemble des contextes. Mais elles existent et elles dirigent les assentiments et les désaccords. En effet il est profondément étonnant de réaliser, sur la base des études des familles de mots, des étymologies, des grammaires, à quel point ces cohérences entre concepts, souvent étranges, existent, bien constituées, et sont mises en évidence à tour de rôle par les variantes des syntaxes. Qui a surveillé l'accomplissement de ces cohérences et leur maniement, en dehors des consciences individuelles, à l'intérieur de cette noosphère supra individuelle? Comment du rationnel si étroitement lié aux significations, arrive-t-il à se former en dehors de tout esprit individuel? Telle est la question qui, en relation avec un autre étonnement du même type, reviendra à la base même de la construction de l'infra-mécanique quantique accomplie dans la deuxième partie de cet ouvrage.

Mais pour l'instant, en ce qui concerne notamment la cohérence entre la question « que connaît-on » qui met l'accent sur "le réel", et la question « comment arrive-t-on à connaître » qui souligne le processus de génération de "connaissances", ce qu'une analyse triviale de la pensée courante et des langages naturels fait ressortir au premier abord est, en gros, ceci. On cherche à connaître 'ce qui *est*' ou 'ce qui est *réel*', c'est-à-dire ce qui est 'vrai'[3]. Et on y arrive par les perceptions directes de ce qui 'est' *via* nos sens biologiques et leurs prolongements (humains, techniques, méthodologiques, etc.), ces perceptions étant quelquefois soutenues par certains hasards et, plus ou moins systématiquement, par des raisonnements

Voilà ce qu'on trouve dans la pensée et les langages naturels. On pourrait détailler, mais ce serait dépourvu d'intérêt pour le but poursuivi ici.

## 1.3. Du "réel" et de la "connaissance" en physique

La physique fait usage de langages spécialisés qui réduisent beaucoup le flou des langages naturels. En conséquence de cela la cohérence du tout constitué par les concepts utilisés devient beaucoup plus apparente qu'elle ne l'est dans un langage naturel. Notamment, les concepts "réel" et "connaissance" y apparaissent très explicitement comme reliés. Mais ce n'est pas pour autant qu'en physique cette relation est *construite*. Ni d'ailleurs les définitions des contenus séparés de ces deux concepts. Car la plupart des physiciens pensent qu'il est nécessaire de *séparer* la physique de toute autre démarche, notamment psychologique ou philosophique, par souci de sauvegarder un domaine clairement délimité de haute compétence spécialisée: la physique aurait vocation de statuer concernant les phénomènes physiques, *point*. Concernant le "réel" c'est aux philosophes de statuer, selon leurs propres critères. Quant aux processus d'émergence des "connaissances", ce sont l'épistémologie et les sciences cognitives qui doivent statuer à leur égard.

Dans ces conditions, ce qui suit n'exprime que des explicitations opérées *concernant* la physique, mais en dehors d'elle, à partir d'une part de la structure des sciences physiques, et d'autre part à partir d'assertions et analyses faites par tel ou tel physicien intéressé aussi par la philosophie en général et notamment par l'épistémologie philosophique, ou, plus récemment, par les sciences cognitives. On trouve couramment de telles explicitations dans la philosophie des sciences. Le résultat de ces explicitations est bien connu. O peut le résumer de la façon qui suit.

---

[3] Voir la note technique qui suit l'introduction générale: "." met l'accent sur le concept cependant que '.' met l'accent sur un mot (ou une façon de dire).

### 1.3.1. Le "réel" selon la physique classique

La pensée scientifique classique a nourri une sorte de réalisme à degrés, selon lequel seul ce qui est de nature physique existe vraiment, pleinement. Bien entendu, l'existence, aussi, des connaissances, n'est pas niée, ni l'existence de faits socio-économiques, culturels, etc. Mais la pensée scientifique classique et notamment la physique classique, traite plus ou moins explicitement toute entité dont la nature n'est pas exclusivement physique, y compris les consciences elles-mêmes, comme des *produits* des structures de substance physique. C'est en ce sens adouci que la pensée scientifique classique postule que seuls les faits ou entités physiques existent: elles seules existent au degré le plus fort, en tant que matière *première*. Toutes les autres sortes d'existences peuvent en principe en être déduites – et sans aucune perte – de l'existence et des modes d'être et de fonctionner des entités physiques. C'est ce qu'on appelle le réductionnisme matérialiste.

### 1.3.2. La "connaissance" selon la physique classique

En ce qui concerne l'émergence et le contenu des connaissances, la physique classique reproduit, en essence, la conception impliquée dans la pensée courante: on cherche à connaître 'ce qui est réel', le 'vrai' ; et on y arrive par, d'abord, les perceptions directes de ce qui 'est' *via* nos sens biologiques humains et leurs prolongements matériels techniques (des appareils), et ensuite, par des développements déductifs fondés sur ces perceptions. La différence majeure avec la vue de la pensée courante réside dans les exigences bien connues de *méthode scientifique,* que l'on impose autant à l'enregistrement de données observables, qu'aux élaborations déductives fondées sur celles-ci. Toutes ces actions de recherche sont soumises à des conditions explicites d'observation, rigueur, et vérifiabilité, conçues de manière à conduire à des consensus intersubjectifs aussi larges que possible. Mais ces conditions de la méthode scientifique ne spécifient rien quant à la structure psychophysique des processus impliqués dans les actions de recherche. L'étude de cette structure, ceci a déjà été noté, est laissée aux sciences cognitives et, éventuellement, à l'épistémologie philosophique.

### 1.3.3. "Réel" et "connaissance" et physique moderne

La physique moderne fondamentale – la relativité einsteinienne et la mécanique quantique – prolongent d'une façon floue et muette les implications de la physique classique en ce qui concerne le réel: c'est le réel physique qui existe fondamentalement, au premier ordre, le but de la connaissance étant d'arriver à le *représenter* tel qu'il est vraiment.

Mais d'autre part, les disciplines de la physique moderne introduisent foncièrement et explicitement, dans l'élaboration des représentations qu'elles affirment, les actions cognitives de l'observateur humain, et des *relativités* des représentations élaborées, à ces actions cognitives, ainsi qu'à *l'état d'observation* dans lequel ces actions sont accomplies. *Cela trouble la cohérence de la situation conceptuelle.* Car il semble clair que cela introduit des contradictions dans la vue implicite concernant *ce* qu'on connaît, donc concernant ce qu'on appelle 'connaissance'. Cette vue ne peut plus rester la même que celle admise dans la physique classique. Toutefois l'établissement de la nouvelle relation entre "réel" et "connaissance" qui est impliquée, n'est toujours pas considéré comme étant l'affaire de la physique, en dépit du fait marquant qu'Einstein, Bohr, Schrödinger, de Broglie, et un nombre croissant d'autres physiciens, lui aient consacré des analyses dont quelques une sont devenues célèbres.

Néanmoins, d'ores et déjà l'impression s'est fortement imposée que la physique moderne comporte une mutation qui y a fait apparaître comme une sorte de dendrites tendues vers l'épistémologie philosophique, et peut-être même vers les sciences cognitives : la possibilité de maintenir la séparation de la physique face au reste des démarches de la pensée, semble désormais compromise.

Cependant, dans la physique moderne aussi la relation entre la question du réel et les représentations de la physique moderne, n'est toujours pas construite. De l'intérieur des théories modernes du réel physique, cette relation s'impose comme une grande question et quelques uns de ses traits s'ébauchent ; mais pour l'instant elle n'est pas élaborée.

## 1.4. Kantisme et néo-kantisme, "réel" et "connaissance"

Dès le XVIII$^{ème}$ siècle Emmanuel Kant [1781], pour la première fois dans l'histoire de la pensée et à sa manière bien entendu, a associé au mot 'réel' un système de significations clair et cohérent. Il a réalisé cela en *re*-définissant un grand nombre parmi les connexions entre mots et les signifiés correspondants, qui dans la pensée courante et les langages naturels flottent dans le flou. Kant a construit son système philosophique à coup de décisions méthodologiques. Il a *méthodologisé*, comme on le fait aussi afin de construire un domaine des sciences dures.

Le système kantien de mots-et-significations est fondé sur la mécanique newtonienne, qui avait été publiée environ un siècle plus tôt. Il s'agit donc d'un système conceptuel conçu par un très grand génie mais qui est fondé sur les tout premiers débuts de la pensée scientifique au sens moderne de cette expression. Dans ces conditions on peut s'attendre à ce que la science de nos jours contienne quelques éléments qui ne trouvent pas une place pour se loger dans le schéma kantien, même si les grandes lignes de ce schéma perdurent remarquablement.

### 1.4.1. Du "réel"chez Kant

Selon le système philosophique de Kant il existe à la fois – en tant que données premières – du réel physique et du réel psychique, des consciences individuelles. La conception kantienne domine encore la philosophie.

### 1.4.2. De la "connaissance" chez Kant

Effleurer par quelques mots seulement ce chef d'œuvre de la pensée humaine qu'est l'épistémologie kantienne, peut sembler sacrilège. Pourtant je le ferai, car exclusivement une certaine quintessence joue un rôle dans ce que je voudrais mettre en évidence ici[4].

Selon l'épistémologie kantienne, l'émergence de *connaissances* concernant le réel physique s'accomplit par l'interaction des deux sortes d'existants admis, le réel physique lui-même et les consciences individuelles. Cette émergence est représentée à l'aide d'une structure complexe de concepts re-définis ou définis et dénommés par Kant. L'organisation de cette structure ne laisse place à aucune contradiction logique. Et elle assigne une position centrale à l'assertion que toute conscience individuelle introduit deux 'formes a priori de l'intuition', l'espace et le temps, dont cette conscience est foncièrement inséparable et dans le moule desquelles elle dépose tous les produits de ses interactions avec le réel physique.

---

[4] On peut trouver tout détail souhaité dans l'exposé de Kant lui-même, ainsi que dans un très grand nombre d'exégèses excellentes écrites par des philosophes.

Selon Kant l'espace et le temps ne sont donc *pas* du réel physique, ni des 'propriétés' du réel physique. Ce sont des réceptacles innés où se logent les perceptions et les connaissances humaines concernant le réel physique.

L'épistémologie kantienne opère *une transmutation des absolus physiques de l'espace et du temps newtoniens, en traits du psychisme humain*.

Les effets immédiatement perceptibles des interactions entre une conscience humaine et le réel physique, s'élaborent dans cette conscience comme des 'phénomènes' (ou 'apparences phénoménales') – au sens de l'épistémologie philosophique, *pas* au sens des physiciens, ni au sens courant. Les phénomènes au sens de l'épistémologie philosophique naissent inscrits dans les deux formes *a priori* de l'intuition, l'espace et le temps. Tels qu'ils émergent d'abord dans les consciences, ce sont encore des faits strictement subjectifs. Afin qu'ils deviennent communicables et donc susceptibles de consensus intersubjectifs et notamment *scientifiques*, les phénomènes doivent être soumis à certaines opérations de 'légalisation' à la suite desquelles ils se trouvent 'objectivés'. Selon Kant l'objectivité scientifique consiste donc en consensus intersubjectifs concernant des apparences phénoménales qui au départ sont foncièrement subjectives, mais qui sont par la suite 'légalisées', notamment selon des méthodes scientifiques (J. Petitot [1997]). Mais cette légalisation ne donne nullement accès à la connaissance du réel *tel qu'il est en lui-même*. Car – par définition de langage – nous ne percevons directement que des phénomènes, subjectifs, formés dans des moules biopsychiques humains, jamais du *réel-en-soi*. Toutes nos conceptualisations partent de tels phénomènes subjectifs. Les conceptualisations qui tissent l'entière épaisseur des connaissances humaines, sont toutes fondées, selon Kant, sur des *apparences* phénoménales qui, elles, sont l'effet d'interactions entre du réel physique inconnu et des consciences humaines génétiquement dotées de moules *a priori*, notamment de ceux de l'intuition que sont l'espace et le temps. Le réel physique lui-même, le réel physique *en soi*, échappe à la connaissance, foncièrement, définitivement.

Selon la vue kantienne, les connaissances humaines n'informent donc pas sur le réel physique seul, séparé des formes et des modes de fonctionnement introduits par les consciences humaines individuelles. Entre le réel physique en soi et nos connaissances, les consciences individuelles interposent l'écran de leurs propres formes et fonctionnements, que rien, jamais, ne pourra abolir dans la structure de qu'on appelle une *connaissance*. Tout ce qui est *connu* est marqué d'une façon indélébile par ces formes et modes de fonctionnement de nos consciences. « Comment est le réel en soi? » est une question impossible. On ne peut pas concevoir de 'réponse' à une question qui est impossible en tant que question. Si néanmoins on en fabrique une, on glisse subrepticement hors du domaine de pertinence du concept de connaissance, et l'on engendre du non-sens.

Ces assertions sont à la fois surprenantes et assez évidentes. En effet, il est clair qu'une fourmi perçoit le réel – quoi que cela peut vouloir dire – autrement qu'un homme ou qu'un aigle ; que si je regarde la tête d'un chaton sous une grande loupe elle m'apparaît comme celle d'un tigre, cependant qu'au seul toucher, sa patte continue de me sembler minuscule ; que – selon les modèles de la microphysique actuelle – pour un neutron qui serait doté d'une capacité de perception consciente de type humain, le corps humain dont il fait partie apparaîtrait comme un nuage de petits maxima de densité de substance grave très éloignés les uns des autres ; etc. Toutes ces relativités aux modes de percevoir crèvent l'entendement dès qu'on analyse la genèse des perceptions et l'on met en jeu les vues de la science actuelle. 'Le réel en soi' ne peut *être* de toutes ces *façons* différentes à la fois. Il ne peut que, soit être tout court, caché dans une stricte absence de toute apparence, donc de toute cognoscibilité, soit paraître tel ou tel, à travers telle ou telle grille de perception ou de représentation et donc être connu à travers cette grille.

Bien après Kant, Husserl s'est attaché dans sa phénoménologie à approfondir et à détailler les processus par lesquels la rencontre d'une conscience avec du réel en soi, engendre les données premières de la connaissance. D'autres penseurs et chercheurs, notamment les biologistes modernes et les 'cognitivistes' ont également contribué au développement d'une vue néo-kantienne et néo-husserlienne sur ce sujet. Mais l'essence de l'épistémologie kantienne a perduré parmi les philosophes. En ce sens il n'y a pas de solution de continuité classique-moderne dans le point de vue philosophique concernant la genèse des connaissances.

## 1.5. Physique, philosophie, "réel" et "connaissance"

### *1.5.1. Philosophie kantienne et physique classique*

Considérons maintenant la pensée classique scientifique. Là non plus, rien n'a compromis l'essence de la pensée kantienne. Dans le domaine de la physique classique notamment, ni la thermodynamique, ni l'électromagnétisme, ni la mécanique statistique, ni la progression des mathématiques et de la logique incorporées dans ces théories, n'ont altéré cette essence.

Et pourtant la pensée scientifique des physiciens classiques ne s'est jamais imprégnée du schéma kantien. Même s'il a dû y avoir à toute époque, ici ou là, quelques chercheurs en sciences exactes familiers de la conception de Kant, la pensée vivante de la physique classique considérée dans son ensemble, n'a tout simplement fait aucun contact déterminant avec cette conception, *pendant plus de 200 ans !* On peut étendre cette affirmation aux autres démarches scientifiques. Pendant tout ce temps la pensée scientifique classique a nourri un réalisme sans aucune épistémologie, un réalisme affirmé de manière abrupte mais non construit, réductionniste au seul réel physique sans que la question de la réductibilité ait été examinée.

Or la structure de la philosophie kantienne-husserlienne, largement dominante, est foncièrement non-réductionniste. Elle pose l'existence des consciences individuelles comme une donnée première et elle spécifie des processus biologiques-psychiques d'émergence des représentations scientifiques consensuelles, à partir des phénomènes subjectifs perçus dans les consciences individuelles, avec les formes *a priori* que celles-ci comportent. C'est dans la spécification de ces processus que consiste l'entière épistémologie kantienne qui s'est prolongée dans la phénoménologie de Husserl.

Ainsi la pensée scientifique classique s'est développée sans contact déterminant avec les démarches philosophiques kantienne et néo-kantienne et l'épistémologie que celles-ci affirment. Ces démarches sont ignorées. Même le concept de 'phénomène', qui chez Kant et Husserl est un concept foncièrement subjectif, a été 'objectivé' d'office et transformé en "phénomène *physique*" auquel on assigne une existence *en soi* qui serait telle qu'on la perçoit : un trompe l'œil conceptuel du même type que celui d'objet' au sens de la pensée classique, et aussi énorme, qui a fait glisser le concept désigné par le mot 'phénomène' jusqu'à *l'opposé* du concept que ce même mot désigne chez les philosophes kantiens et husserliens.

La pensée physique classique a totalement occulté la question épistémologique.

Elle a laissé un trou épistémologique qui prolonge le vide philosophique concernant l'existence même des consciences individuelles avec leurs subjectivités et leurs formes *a priori*. Corrélativement, la physique classique pose que l'on *découvre* les lois 'objectivement vraies' du réel physique ; que, au fur et à mesure que la physique progresse, elle nous rapproche de manière asymptotique de la connaissance du réel physique tel-qu'il-est-en-lui-même *'vraiment', 'objectivement'*, en entendant par ces mots : *indépendamment de toute perception ou action cognitive humaine*. La physique classique ne se soucie pas de spécifier comment les physiciens – des hommes qui pour connaître agissent nécessairement à travers leurs consciences individuelles liées à leurs corps – peuvent arriver à 'découvrir les lois objectives' de la réalité physique telle qu'elle est indépendamment de toute conscience humaine, de toute perception ou action cognitive humaine.

### *1.5.2. Philosophie kantienne et néo-kantienne versus physique moderne*

Dans la physique moderne l'observateur, avec son état d'observation et ses actions cognitives conscientes, joue un rôle explicite et essentiel dans la construction des représentations mathématiques des phénomènes physiques perçu[5]. En conséquence de cela ces représentations, je l'ai déjà souligné, contiennent une sorte de dendrites abstraites tendues dans la direction de la théorie kantienne de la connaissance. Néanmoins il semble exclu qu'une élaboration de mise en cohérence de la physique moderne avec l'épistémologie kantienne et néo-kantienne, puisse conserver inaltéré le schéma fondamental de cette épistémologie qui a été conçue sous contrainte de cohérence avec la physique macroscopique newtonienne.

Considérons d'abord les deux théories de la relativité d'Einstein. Commençons avec les vues concernant l'espace-temps. Selon l'entière physique classique l'espace et le temps sont des donnés *absolus*. Tandis que les deux théories einsteiniennes de la relativité – élaborées elles aussi en tant que disciplines macroscopiques – assignent une validité *relativisée* aux résultats des *mesures* de distances et de durées.

Dans la relativité restreinte, la géométrie – euclidienne – de Minkowski-Einstein est juste une classification synthétique de l'ensemble des coordonnées d'espace-temps possibles pour un événement, selon le type de distance d'espace-temps entre un observateur 'inertiel' *donné* et tel ou tel événement. Cette classification divise l'espace-temps en deux 'cônes de lumière' *de l'observateur considéré* et deux 'ailleurs' *de cet observateur* : l'observateur 'inertiel' considéré peut interagir avec les événements de l'intérieur de ses deux cônes de lumière – où, à la limite, avec ceux qui se trouvent sur la surface de ces cônes – mais il n'a aucune possibilité d'interaction avec les événements intérieurs à ses ailleurs. Donc les contenus d'événements de *ses* deux cônes de lumière sont, *pour* l'observateur considéré, des entités-objet-de-connaissance possibles ; cependant que les contenus d'événements de *ses* ailleurs échappent à l'obtention, par *cet* observateur, de toute connaissance les concernant. Cette multiplicité de géométries subjectives, non intégrées en un seul tout, *ne permet pas une organisation causale de l'ensemble des événements*. Plus radicalement, elle n'affirme *rien* sur la *nature* de l'espace-temps lui-même et considéré globalement. Notamment, aucune connexion n'est élaborée avec le concept de formes *a priori* de l'intuition de Kant. D'ailleurs Einstein a déploré le 'mal' fait par les philosophes en introduisant ce concept (A. Einstein [1955]) (ma traduction):

> « L'unique justification des concepts et des systèmes de concepts est le fait qu'ils permettent de comprendre en un seul regard des complexes de vécus ; il n'en existe pas une autre justification. Je suis convaincu que les philosophes ont accompli un acte des plus nocifs en transposant certains concepts fondamentaux des sciences de la nature, du domaine de l'empirique et de l'utile, accessibles au contrôle, dans les hauteurs inattaquables de la nécessité logique, de *l'a priori*. Car, bien qu'il soit certain que les concepts ne peuvent être déduits de manière logique (ou de quelque autre manière) de l'expérience et dans un certain sens ils sont des créations libres de l'esprit humain (sans lesquelles aucune science n'est possible), néanmoins les concepts sont aussi peu indépendants de nos vécus que, par exemple, les habits de la forme du corps humain. Ceci est particulièrement valide concernant nos concepts de temps et d'espace, que les physiciens, contraints par des réalités, ont dû descendre de l'Olympe de *l'a priori* afin de les remanier et les mettre en état d'être utilisés [6].

---

[5] Ici j'emploie le langage des physiciens. Je prie le lecteur d'excuser le pléonasme que ce langage entretient. Mais le danger de confusion par des oscillations entre 'phénomène' au sens de l'épistémologie philosophique ou bien au sens des physiciens, l'impose. John Archibald Wheeler a souvent cité Bohr qui disait: « Un phénomène n'est phénomène que si c'est un phénomène *perçu* ».

En relativité générale le type de structure géométrique assigné à l'espace-temps, est riemannien. Sa métrique est définie – en termes différentiels – en fonction de la distribution de *masses* qui est supposée, et donc de la structure du *champ gravitationnel* créé par cette distribution de masses: cette métrique est *variable*. Dans la relativité générale on semble poser (implicitement) l'existence d'un désigné de l'expression 'espace-temps' qui serait *indépendant* des psychismes humains, comme dans la mécanique classique, et à cette existence indépendante l'on y associe une structure géométrique qui semble n'être déterminée que par des entités (masses) et faits (champs) qui sont physiques, extérieurs eux aussi à tout psychisme. D'autre part, implicitement, dans la détermination de cette structure géométrique se trouvent incorporés un principe de 'relativité' et un principe d''équivalence' qui sont *observationnels*. Ces principes s'appliquent à la subjectivité des observateurs humains et ils servent un *but*, à savoir la représentabilité consensuelle de la forme de la loi de mouvement en champ gravitationnel.

Ainsi dans la relativité générale tout autant que dans la relativité restreinte, la question métaphysique de la nature assignée à l'espace-temps reste non traitée (car refusée[7]). *A fortiori* ce concept d'espace-temps non plus ne se rattache d'aucune façon élaborée aux formes *a priori* de l'intuition posées par Kant.

Considérons maintenant *l'attitude descriptionnelle* qui marque les théories einsteiniennes de la relativité.

La démarche d'Einstein est fortement méthodologique, normative, constructive: on y poursuit un *but* d'unité et de consensus, et l'on construit de manière à atteindre ce but. Dans le relativité restreinte, le but est d'unifier l'électromagnétisme macroscopique et la mécanique macroscopique. À cette fin l'on postule la constance universelle de la vitesse observée de la lumière, les transformations Lorentz-Einstein des coordonnées d'espace et de temps lorsqu'on passe d'un observateur 'inertiel' à un autre, et le principe de relativité lié à ces transformations ; sur cette base l'on reconstruit les concepts de masse et d'énergie. Dans la relativité générale, le concept – non définissable d'une façon effective – d'observateur 'inertiel', est abandonné, et la démarche est élaborée sous l'empire du but que *tous* les observateurs qui observent un *même* 'mobile' macroscopique, quels que soient leurs états individuels d'observation, soient toujours d'accord sur le fait que *la loi de mouvement constatée est une géodésique de la*

---

[6] Il est surprenant de constater qu'Einstein regardait les *a priori* comme des « nécessités *logiques* » et les concepts d'espace et de temps comme des concepts fondamentaux « des sciences de la nature, du domaine de l'empirique et de l'utile ».
[7] Les physiciens qui, actuellement, utilisent dans leurs recherches les théories einsteiniennes de la relativité, en pâtissent gravement, car ils pensent qu'Einstein a décrété (ou même démontré) que la *nature* de l'espace-temps est *physique* – et ils se soumettent à ce dictat supposé – ce qui mine leur manière de concevoir.

*structure géométrique posée pour l'espace-temps*. De nouveau il s'agit d'une démarche à caractère délibéré et ciblée sur un but de *consensus*. Notons que ce but ne concerne qu'un fait descriptionnel circonscrit, bien que doté d'un grand degré de généralité.

Il est difficile d'imaginer une attitude de recherche qui s'éloigne plus de l'assertion de la physique classique selon laquelle les lois physiques se 'découvrent', et 'telles qu'elles sont vraiment en elles-mêmes'. Mais ce n'est pas pour autant que cette attitude se rapproche de celle pratiquée dans l'épistémologie kantienne et néo-kantienne. Cependant que la démarche finalisée d'Einstein est volontariste, méthodologique, normative, l'épistémologie kantienne et néo-kantienne cherche à identifier de manière 'neutres' les processus spontanés, naturels, de constitution de connaissances.

Quant aux représentations de la mécanique quantique, elles concernent foncièrement des entités microscopiques, étrangères à la physique de Newton et à la philosophie kantienne et néo-kantienne.

Dans cette théorie les concepts d'espace et de temps sont atteints d'un degré notable de désagrégation et évanescence, à cause de la nature primordialement[8] probabiliste des descriptions.

En outre, le mot 'phénomène', on le verra, lorsqu'il est utilisé pour désigner les manifestations observables liées à un microétat, introduit un contenu sémantique qui s'écarte aussi bien de celui que l'on assigne à ce même mot en physique classique, que de celui qu'on lui assigne dans l'épistémologie kantienne-husserlienne.

Dans ces conditions, nonobstant le fait que l'essence de l'épistémologie kantienne ait subsisté à ce jour dans la pensée philosophique, il paraît exclu que les implications épistémologiques de la relativité einsteinienne d'une part, et d'autre part celles de la mécanique quantique, se révèlent être entièrement compatibles avec la vue kantienne qui a été construite en cohérence avec la physique newtonienne. D'autant plus que ces deux disciplines restent elles-mêmes *non* unifiées.

*L'épistémologie scientifique moderne reste à construire* en cohérence explicite non seulement avec la biologie moderne, comme d'ors et déjà on le tente  dans les sciences cognitives, mais aussi avec la physique actuelle.

### *1.5.3. Métaphysique et pensée physique*

La frontière entre ce qui reste définitivement hors d'atteinte *via* une démarche expérimentale-déductive et ce qui, en principe, est accessible à des progressions par une telle démarche, reste floue. Quant à cette accessibilité on ne distingue pas une ligne claire de contour, on perçoit toute une zone d'incertitude. Par exemple, le grand débat qui oppose le 'monisme matériel' (Jean-Paul Baquiast [2007]) au 'dualisme matière esprit' – se place dans la zone d'incertitude. Mais je crois certaines questions, comme par exemple celles du 'réel' ou de la *nature* du temps et de l'espace, apparaissent à tous ceux qui y pensent comme ayant un caractère indiscutablement et définitivement métaphysique.

Jusqu'à présent, la pensée scientifique et notamment la pensée physique autant classique que moderne, n'incorpore aucune recommandation ouverte et élaborée concernant le traitement des questions métaphysiques. La morale scientifique est largement discutée et même institutionnalisée par l'installation de comités d'éthique.

---

[8] Ce terme est justifié dans la deuxième partie de ce livre.

Mais la question des relations entre recherches expérimentales-déductives et problèmes possiblement ou certainement métaphysiques, traîne sur le plan des comportements scientifiques enfermée dans des boules subjectives étanches qui s'entrechoquent de manière sourde dans une sorte de billard insensé, sans révéler la structure de leurs contenus e jeu se poursuit dans une sorte de brume d'interdiction des interrogations métaphysiques.

Pourtant rien n'empêchera jamais l'esprit de l'homme de s'envoler dans ce qui transcende la conceptualisation expérimentale-déductive. La pensée humaine engendre irrépressiblement et continuellement des questions métaphysiques incontournables concernant les énigmes de la vie et de la mort. Tout être humain, même très jeune ou très primitif, se sent quelquefois frappé de stupeur et submergé d'une impression d'étonnement face à l'existence de tout ce qui l'entoure, qui lui apparaît comme miraculeuse. La condition de l'homme, la condition du vivant, l'être universel, les « pourquoi est-ce comme cela et pas autrement » concernant ce qu'on constate, le début et la fin de Tout, *la question du sens* – bien au delà de celle du vrai – hantent l'entendement avec une généralité, une constance et une force qui dominent de très haut celles des interrogations scientifiques. Ceci est un *fait*.

Alors pourquoi devrait-on soumettre les vues scientifiques à un décret d'arrêt qui empêche *a priori* l'émergence de constructions où l'on puisse voir *jusqu'au bout* comment l'esprit humain fabrique ses grandes cohérences, et *étudier* comment chez tel biologiste ou mathématicien ou physicien, la structure de sa pensée de recherche scientifique et celle de ses interrogations métaphysiques, se tissent l'une à l'autre? Pourquoi favoriser la subsistance de blancs qui cachent la texture de ces mises en relation qui certainement existent dans l'esprit de tout chercheur qui est un penseur, au lieu d'étaler des cheminements de pensées qui faciliteraient pour tous des choix métaphysiques moins vagues, ou moins dogmatiques, plus construits et plus consensuels, et qui même, de temps à autre, révèlerait sans doute quelque avancée notable?

Par exemple, je peux déclarer: « je choisis d'admettre qu'il existe du réel physique, et de là il découle… », etc.. Ou bien, au contraire, je peux déclarer « je choisis d'admettre qu'il n'existe pas du réel physique, et donc… » (comme le font les solipsistes), etc. Rien ne peut *bannir* l'un ou l'autre de ces choix d'une manière dotée d'une *nécessité* empirique-déductive. Alors selon quels critères faire un tel choix? Si dans l'expression publique des démarches scientifiques les critères de cette sorte restent secrets, des progressions concernant ces critères ne seront disponibles que dans les œuvres des philosophes et des théologiens. Or celles-ci, quelle que soit l'érudition scientifique et la profondeur de la réflexion que souvent elles manifestent, restent plus ou moins étrangères aux réalités psychologiques intimes des processus de création scientifique, où agissent des éléments techniques chargés d'une signifiance forte.

Notons aussi un autre aspect dont on sous-estime l'importance. La séparation de toute question métaphysique, dépourvoit les théories expérimentales-déductives d'une véritable 'clôture'. En langage topologique l'on pourrait dire que ces théories restent dépourvues de points d'accumulation qui en tracent une frontière définie. Cette lacune est particulièrement frappante dans le cas de la vue exprimée par une grande théorie de physique mathématique. Une telle théorie incorpore une certaine axiomatique logique exprimée mathématiquement. Cette axiomatique n'est jamais étanche face au conceptuel métaphysique. Or le conceptuel métaphysique *est là*, tout autour, inévitablement, et sa force de pénétration est grande. Si la théorie physique considérée est dépourvue d'une mise en relation *explicite* et logiquement *réglementée* avec son au-delà conceptuel métaphysique[9], elle reste exposée à la pénétration *non* réglementée de celui-ci. Et les mélanges d'expérimental-déductif, avec du métaphysique, lorsqu'ils sont incontrôlés, altèrent la pureté de la structure de *sens* incorporée dans la théorie. Cela se manifeste par du flou dans les critères d'applicabilité, ce qui engendre des impasses.

Je risque une fiction. Imaginons, qu'Einstein, au lieu de vilipender les philosophes, aurait déclaré clairement que la *nature* – psychique ou physique – de ce qu'on appelle 'espace' et de ce qu'on appelle 'temps', lui semblait *inaccessible* à toute vérification de type scientifique, et que par conséquent il choisissait de *partir* d'un *fait* d'observation: tout être humain loge dans l' 'espace' et le 'temps' toute entité 'physique' (évènement, corps, association de corps et événements) qu'il perçoit ou conçoit ; et qu'il entreprenait de construire une représentation mathématique de, exclusivement, ces *actes humains de location* – tels qu'ils sont accomplis dans le domaine *macroscopique*, à l'aide de signaux lumineux – la représentation recherchée étant *a priori* soumise aux contraintes du *but* d'assurer un *consensus observationnel* aussi large que possible parmi tous les observateurs qui décrivent une même entité physique. *Cela n'aurait rien changé aux deux théories de la relativité, ni au passage de l'une à l'autre*. Mais cela aurait *évité* ce concept ambigu – *l*'espace-temps 'physique' (au singulier) – qui nargue l'entendement parce qu'on se sent enclin à penser que, s'il varie avec le champ, il devrait au moins avoir la *même* définition de structure quelle que soit la nature du champ considéré (gravitationnelle, électromagnétique, etc.); cependant que, d'autre part, on n'arrive pas à construire une représentation formelle où cette unicité soit exprimée en accord avec les exigences de cohérence, de vérifiabilité et d'*intelligibilité* que l'on aimerait voir satisfaites.

Chaque physicien, s'il avait été clairement débarrassé de ce problème d'unicité, se serait alors senti libre de considérer chaque variante d'espace-temps 'physique' einsteinien, comme un construit conceptuel-physique particulier relatif à un ensemble de contingences physiques. Il aurait alors pu envisager d'éventuellement associer aux théories de la relativité d'Einstein, l'*a priori* kantien des formes de l'intuition, par un postulat métaphysique concernant spécifiquement la – *seule* – *nature* de l'espace et du temps ; et il aurait pu tenter de relier explicitement ce postulat métaphysique, aux théories d'Einstein, en termes conceptuels-mathématiques. *Cela non plus n'aurait rien changé à ces théories*. Et cela les aurait dotées d'une *clôture métaphysique*.

On aurait alors, peut-être, conçu que dans la relativité restreinte l'*ensemble* des perceptions de *locations* distinctes d'*entités physiques*, dans l'espace-temps 'a priorique' général et qualitatif, conduit, pour des observateurs inertiels, à la géométrie euclidienne de Minkowski-Einstein, *via* ce genre de processus d'abstraction signalé en détail par Henri Poincaré [1898] (et réaffirmé par Einstein lui-même) qui, à la fin, *élimine* les éléments physiques complètement. Et en partant de cette première base acquise, l'on aurait peut-être délimité de manière plus stricte le domaine spécifique de signifiance de l'espace-temps 'physique', variable, de la relativité générale ; donc on aurait aussi délimité mieux ses domaines de *non* signifiance. En ces conditions l'on aurait peut-être conçu depuis longtemps déjà un mode de chercher à 'unifier' la physique, différent de celui qui depuis quelque 50 ans résiste à toutes les tentatives.

Mais quittons la fiction et revenons au plan général. Le vaste domaine de la pensée scientifique ne peut être ressenti comme véritablement *accompli*, il ne peut permettre à l'entendement de s'y mouvoir d'une façon libre, avertie, sûre, que s'il offre

---

[9] Par les définitions de l'espace et du temps qu'elle introduit, la mécanique newtonienne déclare d'emblée sa relation au métaphysique

d'une part une démarcation claire entre zone scientifique et zone métaphysique, et d'autre part aussi une *mise en relation* explicite de ces deux zones. Un mur d'interdits à la place d'une mise explicite en relation, dissout l'entendement en prudences et témérités sans guidage qui se réfléchissent sur les constructions expérimentales-déductives de la zone scientifique et y répandent des confusions.

Pour les raisons exposées je tiens que chaque domaine de pensée scientifique fondamentale, et notamment de physique fondamentale, aurait avantage à se munir, par postulats métaphysiques déclarés, d'une clôture métaphysique bien définie et associée d'une façon réglementée aux contenus de connaissances du domaine considéré.

Une telle clôture métaphysique est nécessairement à double face. Une face interne où s'inscrivent les postulats métaphysiques, que la rationalité scientifique consensuelle, expérimentale-déductive, peut juste *atteindre* en développant ses conceptualisations jusqu'au bout, et s'y *accoler*, mais qu'elle ne pourra jamais transpercer ou dépasser ; et une autre face, externe, qui reste à jamais inaccessible. Mais les yeux de l'esprit, eux, perçoivent à leur manière cette face externe inaccessible à la rationalité scientifique expérimentale-déductive. Ils la perçoivent intuitivement, comme on perçoit tout ce qu'on appelle 'des données premières'. Et ils exigent l'incorporation de cette face également, dans une unité globale physique-métaphysique, en l'absence de laquelle ils restent frustrés comme notre regard scientifique est frustré s'il ne perçoit pas une unité factuelle-déductive suffisamment étendue, consensuelle et convenablement construite.

### 1.5.4. Conclusion sur les relations entre philosophie et physique

Entre la philosophie et la physique classique s'est installé un clair hiatus concernant le réel. Quant aux processus de génération de connaissances, à leur égard dans la pensée physique classique il subsiste un pur néant. Il est surprenant qu'en dépit d'une totale occultation de l'épistémologie, la physique classique ait atteint le degré d'efficacité qu'on connaît, dans toutes les disciplines qu'elle a produites. Le système de la physique classique s'est bien plus effrité à cause de déficiences formelles, de cohérence logique interne, qu'à cause de déficiences d'efficacité pratique. L'efficacité pratique semble se trouver sous protection séparée. Les osmoses non explicites entre l'homme et le réel physique où il est immergé, sont si puissantes qu'elles assurent un remarquable degré d'indépendance de l'efficacité de nos représentations de connaissances, face aux principes et aux croyances épistémologiques déclarées qui accompagnent ces représentations.

Quant à la physique moderne, elle ébauche une ouverture vers une re-construction actualisée de l'unité philosophie-physique accomplie par Kant pour la physique newtonienne. Mais cette ouverture reste à être élaborée.

Cette situation conceptuelle est synthétisée dans les schémas 1, 2 et 3.

# FACE - À - FACE  [ PHILOSOPHIE ↔ PHYSIQUE ]  CONCERNANT *LE RÉEL*
# (*QUOI* EXISTE)

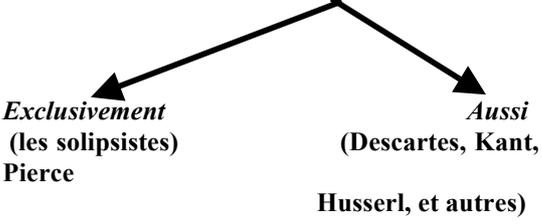

| PHILOSOPHIE | PHYSIQUE *CLASSIQUE* |
|---|---|
| **1. Les Univers Intérieurs Individuels** *existent*<br><br>*Exclusivement* (les solipsistes) Pierce    *Aussi* (Descartes, Kant, Husserl, et autres) | **(a) Consciences individuelles**<br><br>*VIDE PHILOSOPHIQUE*<br><br>**(b) Le réel 'de base'** est exclusivement *physique*. Toutes les autres formes du réel *dérivent* du réel physique. |
| **2. L'univers extérieur,** avec le réel *physique*, *existe* aussi.<br><br>**On recherche les relations 1↔2** | **PHYSIQUE MODERNE**<br><br>*Les subjectivtés existent et contribuent à déterminer les représentations scientifiques.*<br><br>**Ouverture** *non élaborée* **vers la philosophie.** |
| **L'accent tombe sur une méthode philosophique** *fondée sur la mécanique newtonienne* | **Globalement,** la physique admet de façon plus ou moins explicite qu'il existe aussi bien du réel *physique* que des consciences individuelles, mais on ne cherche *pas* à spécifier *jusqu'au bout* leurs relations générales. |

Schéma 1

**FACE - À - FACE  [ PHILOSOPHIE ↔ PHYSIQUE *CLASSIQUE* ]
CONCERNANT LA *CONNAISSANCE*
(*QUOI ET COMMENT* CONNAISSONS-NOUS?)**

| PHILOSOPHIE | PHYSIQUE CLASSIQUE |
|---|---|
| Les données premières de la connaissance sont les apparences phénoménales disposées d'emblée dans les formes *a priori* de l'intuition: *l'espace et le temps* <br><br> **LE RÉEL EN SOI EST *À JAMAIS NON-CONNAISSABLE*** <br><br> Toute connaissance scientifique *se construit* à partir de phénomènes 'légalisés', rendus susceptibles de *consensus intersubjectif*: c'est l'unique forme d'objectivité qui est accessible à l'homme. <br><br> **La question « comment est ce qui existe , '*en soi*',  ? »** **est une question mal formée.** <br><br> La question bien formée est: **Comment *construire* des connaissances consensuelles?** <br><br> La réponse à cette question est fondée sur *la mécanique newtonienne.* | On *découvre* de plus en plus, asymptotiquement, comment le réel physique est vraiment, *en soi*, de façon '*OBJECTIVE*', indépendamment de toute perception ou action humaine. <br><br> LE SUJET CONNAISSANT ; COM.M.ENT IL CONNAÎT 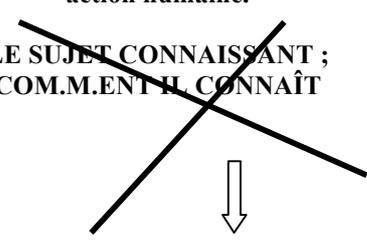 <br><br> *TROU ÉPISTÉM.IQUE* <br><br> La physique prescrit des méthodes scientifiques pour connaître le réel tel-qu'il-est-*en-soi*. *Ces méthodes ne disent rien concernant l'effet sur les  représentations, des actions cognitives de l'observateur-concepteur.* |
| **L'accent tombe sur une méthode philosophique fondée sur la mécanique newtonienne** | **L'épistémologie déclarée par la physique classique perd tout contact avec l'épistémologie philosophique.** <br><br> *L'évolution de la physique classique ignore et brise l'unité construite par Kant entre l'épistémologie philosophique et celle déclarée par la physique* |

Schéma 2

**FACE - À - FACE [ PHILOSOPHIE ↔ PHYSIQUE *MODERNE* ]**
**CONCERNANT *LA CONNAISSANCE***
**(*QUOI ET COMMENT* CONNAISSONS-NOUS?)**

| PHILOSOPHIE | PHYSIQUE *MODERNE* |
|---|---|
| Les données premières de connaissance sont des apparences 'phénoménales' (au sens des philosophes) disposées d'emblée dans les formes a priori de l'intuition, l'espace et le temps.<br><br>*Le réel en soi n'est pas connaissable*.<br><br>La connaissance scientifique *se construit* à partir de phénomènes 'légalisés' susceptibles de *consensus intersubjectif*: c'est *l'unique forme d'objectivité accessible à l'homme* et elle est fondée sur des subjectivités légalisées.<br><br>Mais les procédures de légalisation fondées sur la mécanique newtonienne sont désuètes*:*<br><br>L'évolution de la physique a rendu *insuffisante* la théorie philosophique actuelle de la connaissance.<br><br>***IL FAUT RECONSTRUIRE L'EPISTEMOLOGIE PHILOSOPHIQUE*** | La relativité einsteinienne et la mécanique quantique fondent explicitement leurs représentations mathématiques sur les actions cognitives de l'observateur-concepteur humain.<br><br>*Les lois formulées dans les théories physiques modernes dépendent explicitement des contenus et des buts des actions cognitives de l'observateur-concepteur humain*.<br><br>**LA PHYSIQUE MODERNE CONTIENT L'EMBRYON D'UNE NOUVELLE THÉORIE DE LA CONNAISSANCE**<br><br>**MAIS CET EMBRYON Y EST INCORPORÉ EN ÉTAT *CRYPTIQUE, IMPLICITE, NON-ELABORÉ*.** |
| Au niveau *explicite*, la réponse amorcée dans la physique moderne à la question « comment construire des connaissances scientifiques consensuelles en compatibilité avec une métaphysique explicite? », reste *inachevée*.<br>**UNE THÉORIE DE LA CONNAISSANCE EXPLICITE, ACHEVÉE, CAPABLE D'UNIFIER DANS UN TOUT COHÉRENT LA PHYSIQUE MODERNE ET LA PHILOSOPHIE, *RESTE À FAIRE*.** ||

Schéma 3

# 1.6. En conclusion à la première partie

Kant avait élaboré une unification cohérente entre une épistémologie philosophique explicite et l'épistémologie impliquée dans la physique newtonienne. La physique classique post-newtonienne a ignoré cette unification et l'a brisée. La physique moderne contient des éléments d'amorce d'une restauration de l'unité philosophie-physique fondée sur l'épistémologie impliquée dans les théories actuelles.

Mais je tiens que c'est l'épistémologie impliquée dans *la mécanique quantique* qui apparaît comme la base de départ appropriée.

En effet la relativité d'Einstein a été conçue originellement comme une discipline *macroscopique* dont les caractéristiques ont été déterminées par des exigences de consensus aussi large que possible entre des groupes d'observateurs ayant des états d'observation distincts, mais qui, tous, peuvent percevoir directement un même 'mobile' ou événement, en interagissant avec lui *via* des signaux lumineux. Or *ces exigences sont dépourvues de pertinence concernant des entités microscopiques.* Il n'est pas surprenant que les tentatives d'utiliser les théories de la relativité d'Einstein afin d'associer des *modèles* aux microsystèmes et microphénomènes, rencontrent des résistances à ce jour même.

Tandis que la mécanique quantique fondamentale – qui n'est pas relativiste – introduit une conceptualisation qui, d'emblée et spécifiquement, concerne des états de microsystèmes. L'approche épistémologique impliquée dans cette théorie est radicalement novatrice face à *l'ensemble* des théories macroscopiques, les deux relativités d'Einstein inclusivement.

En ces conditions, afin d'accomplir une unification générale entre le noyau dur de l'épistémologie philosophique kantienne et néo-kantienne, et d'autre part l'épistémologie qui régit la physique moderne, il convient de comprendre d'abord à fond le mode de conceptualisation impliqué dans la mécanique quantique fondamentale.

On peut espérer que les traits non classiques de la relativité pourront être incorporés eux aussi, cas par cas, sur la base d'un examen de pertinence spécifique. Mais il faudra tenter cela *après* avoir bien établi les distinctions et les relations entre la conceptualisation classique en général et la conceptualisation de la mécanique quantique fondamentale, qui sont de types radicalement différents. Cependant qu'il semble illusoire d'espérer, inversement, qu'en commençant avec la relativité, les contraintes provenant de la mécanique quantique puissent être introduites *à la fin*: dans la refonte d'une unité entre philosophie et physique, les contenus épistémologiques de la mécanique quantique fondamentale jouent certainement un rôle tellement novateur qu'il est nécessaire leur réserver un terrain d'expression pur de toute contrainte émanant d'une autre source que, exclusivement, ces contenus.

Ainsi nos regards se trouvent dirigés vers les contenus épistémologiques encryptés dans la mécanique quantique.

*Deuxième partie*

*L'infra-mécanique quantique*

# *En guise d'introduction à la 2ème partie*

Cette partie est dédiée à la recherche du mode implicite de conceptualisation qui a conduit au formalisme mathématique de la mécanique quantique.

L'approche dont on rend compte plus loin s'est fait jour lentement. Au premier abord elle pourrait paraître inattendue. Le résultat en est ce que j'ai dénommé *l'infra-mécanique quantique*. A l'intérieur de la conceptualisation globale où elle s'intègre, l'infra-mécanique quantique constitue une étape dont le rôle, bien que nodal, n'est qu'intermédiaire.

- D'une part cette étape fonde la possibilité future d'une approche *unifiée* de l'ensemble des problèmes d'interprétation soulevés par la mécanique quantique (Mugur-Schächter [ 2009].

- D'autre part cette étape fonde rétroactivement de la manière la plus détaillée et construite que j'ai pu élaborer à ce jour, la méthode générale de conceptualisation relativisée (Mugur-Schächter [2006]).

Dans cet ouvrage, c'est la construction de l''infra-mécanique quantique *elle-même*, isolée de ces conséquences, qui constitue le but central.

# 1

# Naissance d'un projet

« Les mots techniques que l'on est obligé d'utiliser apportent, dans ces discours inhabituels, beaucoup d'obscurité sur le sens général »

Vitruve

## 1.1. Sur le processus d'émergence de la mécanique quantique

La façon dont la mécanique quantique s'est constituée comporte un certain caractère qui est unique dans l'histoire des théories physiques : elle a émergé – longuement, entre 1900 et 1935 environ – d'une vraie petite *foule* de contributions d'auteurs différents. Bohr, Plank, Einstein, de Broglie, Schrödinger, Heisenberg, Born, Pauli, von Neumann, Dirac. Et j'en oublie un bon nombre. Or ces contributions (dont quelques unes ont même émergé d'une façon parallèle, pratiquement sans interactions) qui toutes ont été essentielles, chacune vraiment originale et de grande envergure créative, mais aussi fortement dissemblables, se sont finalement assemblées dans un tout parfaitement cohérent. Pourtant il serait difficile d'attribuer cette mise en cohérence à une personne déterminée dont on puisse imaginer qu'elle l'a surveillée à l'intérieur de son esprit individuel, par un type de processus que chacun peut imaginer par introspection. Comme, par exemple, on est porté à attribuer à Newton la mise en cohérence des données connues à son époque concernant le mouvement des corps macroscopiques, ou à Maxwell la mise en cohérence des données connues à son époque concernant les phénomènes électriques et magnétiques, etc.

Qu'est-ce qui assure donc la cohérence logique de cette foule de contributions qui se sont fondues dans le formalisme de la mécanique quantique ?

## 1.2. Une hypothèse

Le phénomène et la question mentionnés suggèrent une hypothèse: Ceux qui se sont attelés à la tâche de représenter les microsystèmes et leurs états d'une façon qui puisse être tolérée à la fois par l'essence de la mécanique newtonienne et par la théorie macroscopique des champs électromagnétiques, se sont trouvés confrontés à une situation cognitive qui, à l'époque, était sans doute inusuelle à un point tel, que l'effort d'innovation nécessaire dépassait de loin les facultés d'une seule intelligence. Et même les capacités d'un seul génie. Mais d'autre part cette situation cognitive singulière imposait plus ou moins implicitement des contraintes tellement contraignantes, que celles-ci ont agi comme un moule commun qui a assuré un grand degré d'unité entre les résultats des différentes approches. C'est la situation cognitive qui a orchestré la construction de la mécanique quantique.

Placée sur un niveau supra individuel, intersubjectif, cette situation cognitive a remplacé d'une manière implicite le contrôle unificateur conceptuel-logique qui d'habitude fonctionne explicitement à l'intérieur d'un seul esprit novateur. Omniprésente d'une manière extérieure et neutre, elle a agi comme un organisateur et un co-ordinateur.

Le formalisme qui s'est constitué ainsi n'exprime *pas* explicitement la situation cognitive qui l'a déterminé. Toutefois il a dû sans doute incorporer en état cryptique les contraintes qui l'ont modelé, puisqu'il est performant. Mais cela est courant. Pour toute théorie mathématique d'un domaine du réel physique, les choses se passent plus ou moins ainsi.

L'inhabituel, dans le cas de la mécanique quantique doit donc consister dans la nature particulièrement nouvelle et contraignante de la situation cognitive impliquée. Celle-ci, après avoir fait obstacle à la conception du formalisme par un seul physicien, et après avoir ensuite orchestré la construction du formalisme par tout un ensemble de physiciens, doit être aussi, par l'*extériorité* dans laquelle elle s'est maintenue face à tout esprit individuel, la raison pour laquelle, jusqu'à ce jour, le formalisme quantique est ressenti comme si peu compréhensible, même par les physiciens et théoriciens qui l'ont longuement pratiqué et y ont réfléchi à fond. Parmi les fondateurs eux-mêmes il serait difficile de trouver deux qui aient été entièrement d'accord sur les 'significations' incorporées dans le formalisme qu'ils ont contribué à créer.

Aucune autre théorie physique, pas même la relativité d'Einstein, n'a soulevé des débats aussi résistants concernant les significations. Ces débats subsistent et *évoluent* depuis plus de 70 ans, sans trouver des solutions qui soient acceptées de manière unanime. Le formalisme est là, cohérent et puissant. Mais aucun consensus n'a pu s'installer sur *sa façon de signifier*, ni, *a fortiori*, sur la question de savoir pourquoi cette façon de signifier est comme elle est et pas autre.

### 1.3. Un projet

#### 1.3.1. Formulation du projet

L'hypothèse formulée suggère un projet : Faire abstraction du formalisme quantique et s'attacher à construire soi-même une représentation uniquement qualitative, mais qui mérite clairement d'être appelée une "description" – et en termes *mécaniques* (position, vitesse, etc.) – de ce qu'on conçoit comme correspondant à l'expression 'états de microsystèmes'. Si l'on se place dans ces conditions de départ de pénurie sévère, on sera forcé d'utiliser à fond le peu qui reste disponible et agit inévitablement. À savoir la situation cognitive justement. Et bien sûr aussi les traits qui caractérisent les modes humains de conceptualiser, nos capacités opératoires, les exigences de *communicabilité* et de consensus interjubjectif (sans quoi ne peut exister aucune activité consensus 'scientifique'), et le *but* de construire en termes qualitatifs l'essence sémantique d'une catégorie donnée de 'descriptions'.

En ces conditions, si l'hypothèse mentionnée, d'une détermination des spécificités du formalisme quantique, par les contraintes qu'impose la situation cognitive, est correcte, ce qui émergera devra, avec une évidence indiscutable, être l'expression qualitative de l'essence sémantique des algorithmes quantiques tels qu'on les connaît.

Et une fois connue, cette essence sémantique offrirait un solide élément de référence pour comprendre les choix plus ou moins implicites qui ont conduit à la représentation mathématique des microétats, et leur manière de signifier. Ce qui devrait permettre de dissoudre les problèmes d'interprétation.

### *1.3.2 Nouveauté du projet*

On a énormément discuté la situation cognitive impliquée dans telle ou telle expérience particulière, souvent magistralement et très en détail. Einstein, Bohr, Schrödinger, de Broglie, Bell, Wigner, ont fait des analyses de cas très profondes dans le lit desquelles se sont ensuite précipité des torrents de gloses, de spécifications, ou simplement de fantaisies. Mais toutes ces analyses font intervenir le formalisme quantique tout autant que la situation cognitive spécifique du cas considéré. Les problèmes et les analyses considérés ont en général renvoyé *plus* à l'expression mathématique des algorithmes quantiques, qu'à la situation cognitive qui a engendré l'essence sémantique de ces algorithmes. Ce mélange a piégé l'entendement. Il l'a empêché de s'extraire radicalement du formalisme et d'en pouvoir percevoir, isolément, les sources, la structure des racines qu'il a implantées dans la factualité physique, et les modes humains d'opérer et de conceptualiser qui ont agi. Bref, il semble utile de construire *a posteriori* la genèse du formalisme quantique.

Le moment actuel y est peut-être beaucoup plus propice. Dans le passé, ceux pour qui le mot *comprendre* pointait malgré tout vers autre chose que l'application automatique d'un algorithme dont l'efficacité est établie, et qui ont recherché une approche qui puisse doter d'un caractère de nécessité les algorithmes quantiques et leur fonctionnement, sont tous restés isolés, même les plus grands comme Einstein, Schrödinger, et de Broglie.

Ce n'est que dans les années 1980 – exactement à partir du théorème de Bell concernant la question de 'localité'[10] – que les débats ont pris de la densité. Ce théorème impliquait des *expérience*s *réalisables* qui pouvaient établir si oui ou non les microphénomènes sont 'non locaux', comme on acceptait que l'affirme le formalisme quantique. Tout à coup, tout le monde s'est mis à vouloir comprendre, ouvertement et avec une sorte de désespoir. La tension de bizarrerie introduite par ce qu'on appelait le caractère non local du formalisme quantique était brusquement ressentie comme insupportable. Cette nouvelle phase, imprévue, était comme la tombée d'un masque officiel. Ceux qui avaient soutenu qu'il n'était pas nécessaire de comprendre, dans le fond d'eux-mêmes avaient été convaincus qu'il n'y avait rien de vraiment *important* à comprendre : le formalisme n'est pas local parce qu'il est newtonien, non relativiste. D'accord. On y pense et puis on oublie. Car la réalité physique, elle, est certainement locale comme il se doit selon la relativité d'Einstein. Un point c'est tout. Tout est en ordre. Un formalisme utile à 99% est bon. Il faut être pragmatique. Mais les expériences ont été réalisées et elles ont *confirmé* les prévisions du formalisme quantique dans ce cas, spécifiquement. Elles l'ont confirmé là où l'on ne voulait pas et on ne s'attendait pas qu'il soit confirmé! Du coup, il devenait urgent de comprendre cette situation. Ainsi le problème de l'interprétation du formalisme s'est finalement trouvé officialisé, institutionnalisé par un succès inattendu du formalisme.

A partir de là a commencé une phase nouvelle de communauté des questionnements et celle-ci a conduit à des bilans globaux qui se poursuivent. Désormais s'est installée une attention plus ou moins générale aux questions d'interprétation. Seule la persistance des questionnements de départ et l'accumulation de questionnements nouveaux ont pu conduire – dans une génération renouvelée de physiciens et penseurs – vers la claire perception d'un problème *global* de la signification du formalisme quantique ; un problème porté et mûri par le cours du temps (Sclosshauer [2003]).

---

[10] Si un événement qui se produit en un point donné d'espace-temps, en un ici-maintenant donné *A*, influence un événement qui se produit en un point d'espace-temps *B* qu'aucun signal lumineux issu de *A* ne peut atteindre, alors on dit qu'il y a un 'non-localité' au sens de la relativité d'Einstein.

### 1.3.3. Intermède: une conférence sur la question de localité soulevée par Bell

Je pense que le document reproduit ci-dessous – extrait du volume *Einstein 1879-1955 (6-9 juin 1979), colloque du centenaire, Collège de France*, Editions du Centre National de la Recherche Scientifique – est intéressant et utile pour montrer l'état des choses en 1979 et qui en grande mesure *perdure* à ce jour.

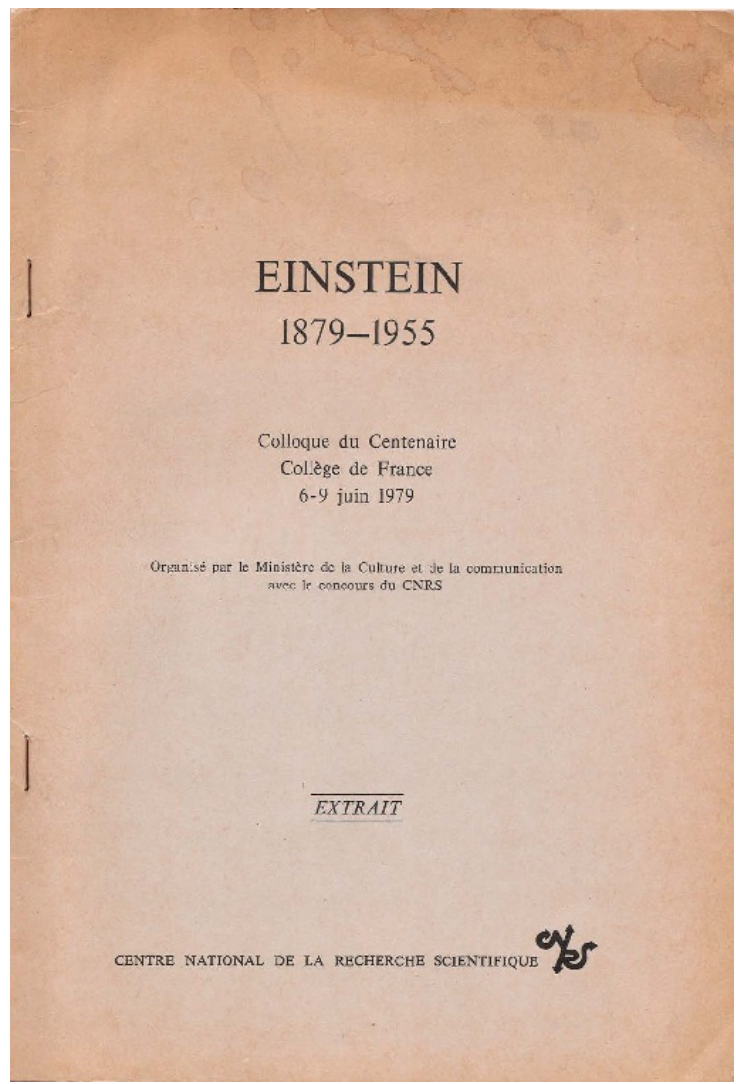

## REFLEXION SUR LE PROBLEME DE LOCALITE

**M. Mugur – Schächter**
**UNIVERSITE DE REIMS**
**B.P 347  51062 REMS CEDEX**

**(EXTRAIT)**

<u>But</u>

Depuis huit ans ce que l'on appelle le problème de localité retient de plus en plus l'attention. Des théoriciens, des expérimentateurs, des penseurs pluridisciplinaires investissent des efforts importants pour élucider ce problème. Les aspects techniques – mathématiques et expérimentaux – ont été déjà examinés dans un grand nombre de travaux et ils sont bien connus de ceux qui font des recherches spécialisées. Mais la configuration conceptuelle qui est en jeu me paraît avoir des contours beaucoup moins définis. Le but de l'exposé qui suit est d'examiner cette configuration conceptuelle. J'essaierai de procéder à cet examen d'une manière aussi simple et frappante que possible, presque affichistique, à l'aide de schémas et de tableaux. Ces moyens me paraissent être les plus adéquats pour donner le maximum de relief aux insuffisances que je perçois dans la définition même du problème de localité.

<u>Bref rappel.</u>
<u>Le paradoxe EPR (1935).</u> Le problème de localité est soulevé par un théorème bien connu de J. Bell (1) qui se rattache à un raisonnement formulé en 1935 par Einstein, Podolsky et Rosen (2). Ce raisonnement, connu sous la dénomination de "paradoxe EPR", et été construit pour démontrer que le formalisme de la Mécanique Quantique ne fournit pas une description complète des microsystèmes individuels. Les hypothèses qui constituent la base de départ du paradoxe EPR sont indiquées dans le tableau suivant (où des notations abrégées leur sont associées):

| | |
|---|---|
| Toutes les prévisions de la Mécanique Quantique sont vraies. | $\forall$ MQ |
| La Mécanique Quantique fournit une description complète des microsystèmes. | C (MQ) |
| La réalité physique existe indépendamment de l'observation. Elle est "déterministe" et locale (ou "séparable") * | $\exists$ (r.d.l.) |

Le "paradoxe EPR" consiste dans la démonstration du fait que les hypothèses énumérées ne sont pas compatibles.

L'interprétation proposée par Einstein, Podolsky et Rosen, de cette démonstration, a été la suivante:

Les prévisions du formalisme quantique se montrent correctes. Il n'existe donc aucune base pour abandonner l'hypothèse $\forall$MQ. L'hypothèse $\exists$(r.d.l.) exprime un credo métaphysique que l'on est libre d'accepter ou de rejeter. Mais *si* on l'accepte, alors il faut l'adjoindre aux prévisions de la Mécanique Quantique. En ce cas la démonstration de l'incompatibilité du système d'hypothèses [$\forall$MQ + C(MQ) + $\exists$(r.d.l.)] oblige à abandonner hypothèse de complétude C(MQ). En d'autres termes cette démonstration oblige alors à accepter la possibilité d'une théorie déterministe et locale (TDL) des microphénomènes, où le formalisme quantique sera complété par des éléments descriptifs additionnels, des paramètres cachés (par rapport au formalisme quantique) déterministes et locaux (p.c.d.l.) qui permettent d'accomplir une description complète des microsystèmes individuels. Cette description complète fournie par TDL doit être compatible avec la Mécanique Quantique – en vertu de l'hypothèse $\forall$MQ – et avec la Relativité, en vertu de l'hypothèse $\exists$(r.d.l.) qui se trouve intégrée dans la théorie de la relativité. Cette structure d'idées peut être représentée par le schéma suivant:

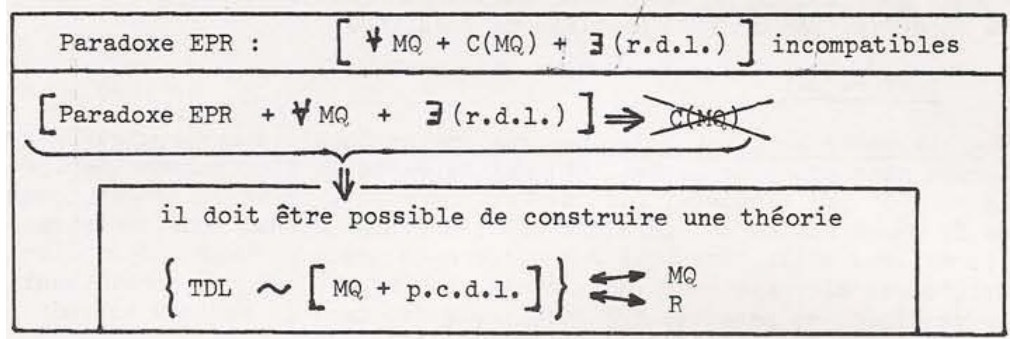

Les réactions pendant 30 ans. Les réactions ont été diverses. Pourtant la note dominante a été nettement celle du positivisme: l'hypothèse "réaliste" ∃(r.d.l.) est dépourvue de toute signification opérationnelle. Elle est donc essentiellement métaphysique, extérieure à la démarche scientifique. L'incompatibilité dénommée "paradoxe EPR" n'existe que par rapport à cette hypothèse non scientifique, et donc elle ne constitue pas un problème scientifique. Pour la science il s'agit là d'un faux problème.

Le théorème de J. Bell (1964). Trente années plus tard J.Bell a démontré un théorème qui semble contredire l'interprétation associée par Einstein Einstein, Podolsky et Rosen à leur propre démonstration. La conclusion du théorème de Bell peut s'énoncer ainsi (ou de manières équivalentes): il n'est pas possible, à l'aide de paramètres cachés déterministes et locaux, d'obtenir dans tous les cas les mêmes prévisions que la Mécanique Quantique ; en certains cas, de tels paramètres conduisent à d'autres prévisions. Si alors on veut rétablir l'accord avec les prévisions de la Mécanique Quantique, il faut supprimer le caractère local des paramètres cachés introduits, ce qui contredira l'hypothèse ∃(r.d.l.), que la théorie de la Relativité incorpore. Par conséquent la théorie déterministe TDL compatible à la fois avec la Mécanique Quantique et la Relativité, dont Einstein Podolsky et Rosen ont cru avoir établi la possibilité, est en fait impossible.

La démonstration repose sur la production d'un exemple. On considère deux système $S_1$ et $S_2$ à spins non nuls et corrélés, créés par la désintégration d'un système initial S de spin nul. On envisage des mesures de spin sur $S_1$ selon trois directions a, b, c, à l'aide d'un appareil $A_1$, et des mesures de spin sur $S_2$ selon ces mêmes directions, à l'aide d'un appareil $A_2$ qui peut se trouver à une distance arbitrairement grande de $A_1$. L'hypothèse ∃(r.d.l.) est ensuite formalisée: des paramètres cachés sont introduits et ils sont soumis à des conditions telles qu'elles fournissent une traduction mathématique des qualifications de "déterministes" et "locaux". Ainsi la conceptualisation introduite auparavant au niveau d'une sémantique claire, mais qualitative, est élevée jusqu'à un niveau sémantique syntaxisé. Un tel pas est souvent important, car il peut permettre des déductions mathématiques à conclusions quantitatives. Et en effet Bell a démontré que l'hypothèse ∃(r.d.l.) ainsi formalisée entraîne nécessairement une certaine inégalité concernant les corrélations statistiques entre les résultats de mesures de spin enregistrés sur les appareils $A_1$ et $A_2$. Or, cette inégalité n'est pas satisfaite par les corrélations statistiques prévues par la Mécanique Quantique. On pourrait retrouver les corrélations quantiques en supprimant la condition qui traduit mathématiquement le caractère "local" des paramètres cachés introduits, c'est-à-dire en renonçant à une partie de l'hypothèse ∃(r.d.l.). On exprime ceci en disant que, dans la circonstance considérée, "la Mécanique Quantique est non-locale" ou "implique des effets non-locaux" qui la rendent incompatible avec ∃(r.d.l.). Schématiquement, on peut résumer l'apport de Bell ainsi (en notant (p.c.d.l.)$_B$ les paramètres cachés soumis aux conditions de Bell).

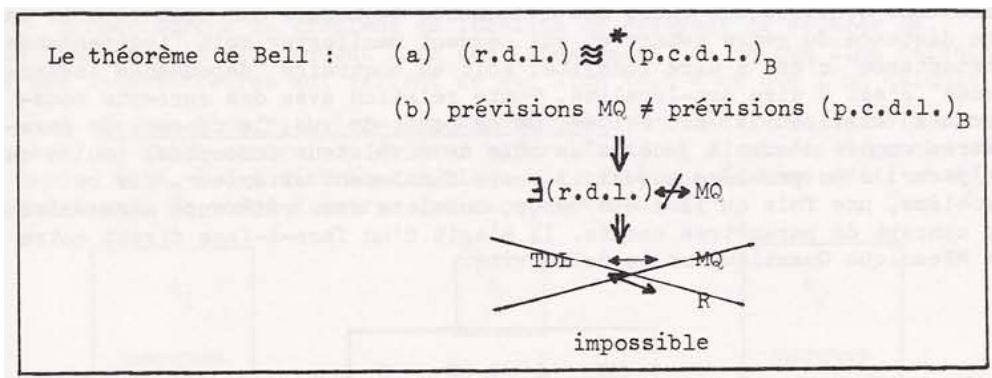

Comme les statistique dont il s'agit sont observables, il est en principe possible d'établir expérimentalement si les faits physiques correspondent aux prévisions de la Mécanique Quantique ou à celles entraînées par les paramètres cachés déterministes et locaux au sens de Bell. C'est l'un des traits les plus forts du théorème de Bell.

Si l'expérience infirmait la Mécanique Quantique, la situation conceptuelle créée paraîtrait claire. On devrait admettre la possibilité d'une théorie déterministe et locale des microphénomènes, mais différente de celle envisagée par Einstein, Podolsky et Rosen, car elle n'obéirait pas à l'exigence d'identité prévisionnelle avec la Mécanique Quantique, pour tous les cas.

Mais un certain nombre d'expériences de vérification a déjà été fait et il se trouve que les résultats obtenus à ce jour – bien qu'ils ne tranchent pas encore définitivement – étayent fortement la supposition que la prévision de la Mécanique Quantique s'impose comme correcte.

Il s'agit donc de comprendre la situation conceptuelle qui semble s'établir et que l'on dénomme "problème de localité".

<u>Interprétations</u>

Le problème de localité est ressenti diversement. Je distinguerai en gros trois interprétations, en omettant ou en bousculent beaucoup de nuances.

I- <u>Interprétations de refus.</u> Un certain de nombre de physiciens semble considérer cette fois encore qu'il s'agit d'un problème métaphysique qui n'existe que par rapport au concept non opérationnel de paramètre cachés, mais qui se dissout dès qu'on refuse ce concept. D'autres physiciens considèrent que le problème n'existe parce qu'il est faussement posé (3).

2- <u>Interprétation minimale.</u> Selon d'autres physiciens[*] (4), (5), (6), (7), etc.…, le problème satisfait cette fois aux normes positivistes les plus draconiennes, parce qu'il conduit à des testes expérimentaux. Toutefois, ils refusent de conceptualiser au-delà de ce que ces tests mettent en jeu. Ils ne prennent en considération strictement que des corrélations statistiques entre des évènements de mesure qui sont séparés par une distance du genre espace et qui peuvent manifester soit "indépendance instantanée" c'est-à-dire localité, soit au contraire "dépendance instantanée" c'est-à-dire non-localité. Toute relation avec des concepts sous-jacents "explicatifs" est évitée. De ce point de vue, le concept de paramètres cachés n'aurait qu'un rôle de révélateur conceptuel (ou de catalyseur) d'un problème auquel il reste finalement extérieur. Car ce problème, une fois qu'il a été perçu, subsiste sans référence nécessaire au concept de paramètres cachés. Il s'agit d'un face à face direct entre la Mécanique Quantique et de Relativité.

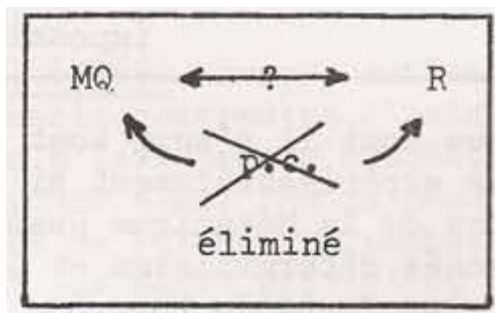

3- <u>L'interprétation épistémologique.</u> Il existe enfin une tendance (8) à connecter le problème de localité à notre conceptualisation la plus courante de la réalité, qui postule l'existence d'objets isolés possédant des propriétés intrinsèques et permanentes. La violation des inégalités de Bell serait incompatible avec ces suppositions. Il s'agirait donc en dernière essence d'un face-à-face entre la Mécanique Quantique et – à travers le concept de paramètres cachés et à travers la Relativité – des postulats épistémologiques fondamentaux.

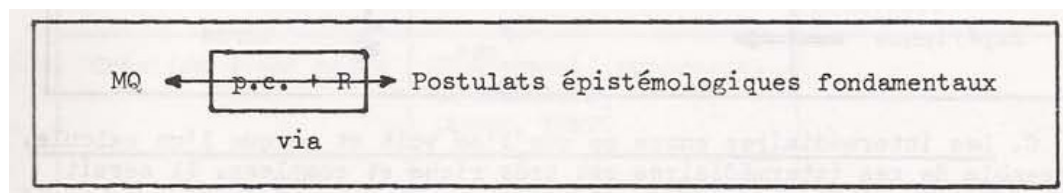

Je n'examinerai pas l'interprétation de refus, car elle ne peut conduire à aucun élément nouveau.

---

[*] Je prie à l'avance ceux qui n'estimeraient pas appartenir à cette catégorie, de m'excuser.

Quant aux deux face-à-face impliqués par les deux autres interprétations, aucun d'eux ne me semble s'imposer dans la phase actuelle du débat. Seule une question ressort clairement:

Qu'est ce qui est en jeu – au juste – dans le problème de localité?

L'examen qui suit montrera que, pour fixer une réponse, les conceptualisations existentes et les tests sur l'inégalité de Bell ne peuvent pas suffire. Inévitablement d'autres conceptualisations encore, et les tests correspondants, devront être abordés. Sinon, aucune conclusion définitive ne pourra être tirée, même si l'inégalité de Bell est clairement violée.

### Le problème de localité et le terrain conceptuel sous-jacent

Reconsidérons le problème de localité en essayant de séparer ce que l'on perçoit directement lors des expériences, de ce que l'on calcule, et des intermédiaires qui relient ce que l'on voit à ce que l'on calcule.

A. Ce qu'on voit lors des expériences. On voit (tous les détails mis à part) un objet central $A_o$ et deux appareils $A_1$ et $A_2$ placées à gauche et à droite de $A_o$ à des distances égales. Sur certaines parties de $A_1$ et $A_2$ apparaissent de temps à autres des marques visibles.

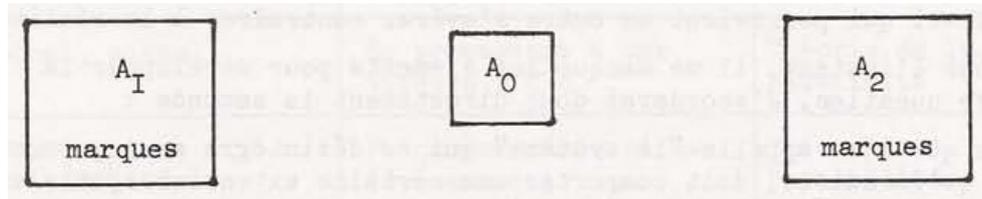

B- Ce qu'on calcule. On calcule des corrélations statistiques en employant trois sortes de distributions de probabilités conduisant à trois fonctions de corrélation, une fonction $F_{(TDL)B}$ caractéristique d'une théorie déterministe locale au sens de Bell, une fonction $F_{MQ}$ obéissant aux algorithmes de la Mécanique Quantique, et une fonction $F_{obs}$ correspondant aux statistiques observées. L'inégalité de Bell distingue $F_{(TDL)B}$ de $F_{MQ}$. L'expérience doit montrer si la réalité observée reproduit $F_{MQ}$ ou $F_{(TDL)}$.

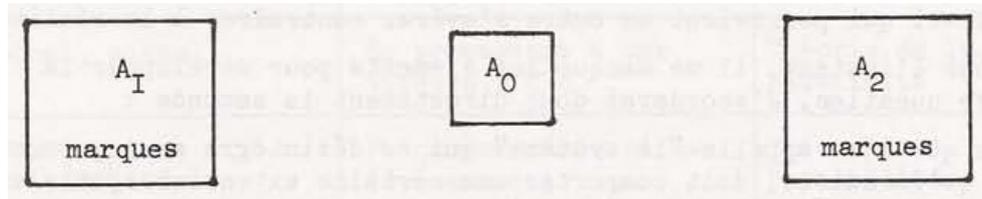

C- Les intermédiaires entre ce qu'on voit et ce qu'on calcule. L'ensemble de ces intermédiaires est très riche et complexe. Il serait insensé de vouloir donner une énumération et une caractérisation exhaustive. Je vais donc opérer une sorte d'échantillonnage, mais en distinguant entre les mots que l'on emploie, les concepts reliés à ces mots, et les organisations syntaxiques dans lesquelles ces concepts se trouvent intégrés.

(Voir tableau page suivante)

La colonne centrale de ce tableau est peut-être quelque choquante d'un point de vue positiviste. Mais de toute façon les paramètres cachés déterministes et locaux de Bell violent la pudeur sémantique dictée par le positivisme. Alors autant aller jusqu'au bout et avouer l'ensemble des questions sémantiques liées aux interprétations 2 et 3 du problème de localité telles que je les ai distinguées plus haut.

Je commence par l'interprétation minimale. Je perçois deux questions.

En premier lieu, les contenus sémantiques assignés aux qualificatifs "déterministes" et "locaux", tels qu'ils sont impliqués par la modélisation mathématisée de Bell, permettent-ils la représentation la plus générale concevable d'un processus d'observation d'un "microétat" à l'aide d'un "appareil" macroscopique?

En second lieu, en supposant que la modélisation de Bell d'un processus d'observation n'introduit vraiment aucune restriction non nécessaire, quelle sorte de non-localité, exactement, la violation des inégalités de Bell démontrerait-elle? La non-localité que la théorie de la Relativité interdit clairement, ou bien des prolongements spontanés et encore flous de celle-ci qui pourraient en outre s'avérer contraire à la réalité?

Pour l'instant, il me manque les éléments pour développer la première question. J'aborderai donc directement la seconde:

Ce qu'on appelle "le système" qui se désintègre en $A_o$, pour autant qu'il existe, doit comporter une certaine extension spatiale non nulle de départ $\Delta x_s(t_o) \neq 0$ (ce qui peuple ce domaine d'espace, est-ce un "objet" ou un "processus", ou les deux à la fois? les définitions même manquant pour répondre). Ce qu'on désigne par les termes "désintégration" ou "création d'une paire $S_1$ et $S_2$", comment le concevoir? Les mots indiquent dans le substrat conceptuel l'hypothèse d'un processus, d'une entité réelle en cours de changement. Pour exister, ce processus doit se produire quelque part et il doit durer, il doit occuper un certain domaine non nul d'espace-temps $\Delta s_c(t).\Delta t_c \neq 0$ (l'indice c: création) à l'intérieur duquel "le système de départ S" existe encore mais change, cependant que $S_1$ et $S_2$ n'existent pas encore mais se forment.

| MOTS | CONCEPTS | ORGANISATIONS SYNTAXIQUES |
|---|---|---|
| ↑ système | Macroobjet, objet | Logique des classes d'objets et des prédicats |
| Création d'une paire | Changement, processus, SUCCESSIONS, DUREE, TEMPS | |
| 2 systèmes corrélés | Objets ⇄ corrélés isolés | |
| Appareil | Macroobjet capable de réagir d'une manière significative avec un microsystème | |
| Spin | Propriété d'observation | |
| Mesure | Processus d'observation Evénément de connexion entre microobjets et macroobjets | |
| Paramètres cachés | Propriétés intrinsèques et permanentes (d'objets ou de processus) | |
| Déterministe | Prévisible? A partir de quoi? | |
| Local, signal | Se propageant à une vitesse $v < c$ | Théorie de la Relativité |
| Statistiques Probabilités | Phénomènes aléatoires (événements, processus, objets) | Théorie des Probabilités |
| Prévisions quantiques | Vecteurs d'états, algorithmes quantiques | Mécanique Quantique |
| Influence à distance | Changement transporté à vitesse $v < c$ | ? |

Dans l'écriture qui désigne ce domaine d'espace-temps, le facteur de durée $\Delta t_c = t_{12,o} - t_o$ s'étend – par définition – d'une certaine "valeur initiale de temps" $t_o$ où le changement de création commence, jusqu'à une "valeur finale de temps" $t_f \equiv t_{12,o}$ à partir de laquelle "la paire $S_1$, $S_2$ de systèmes corrélés" commence à exister (des objets? des processus eux aussi? les deux à la fois?). Quand au facteur d'extension spatiale $\Delta s_c(t)$, il semble obligatoire de concevoir, puisqu'il s'agit d'un processus, qu'il change en fonction de la "valeur de temps" t, avec ($t_o < t < t_f$), en restant toutefois métastablement connexe tant que $t < t_f$ (c'est-à-dire tant que S subsiste encore et que $S_1$ et $S_2$ ne sont pas encore créés). Pour tout $t > t_f$, toutefois, ce domaine spatial devrait être devenu non connexe *via* une scission plus ou moins "catastrophique" conduisant à cette nouvelle forme de stabilité à laquelle on rattache l'expression "la paire $S_1$, $S_2$ de deux systèmes corrélés".

Je m'arrête un instant et je regarde ce que je viens d'écrire. Quel mélange de "nécessités" et d'arbitraire, de signes et de mots qui ont l'air de pointer vers un désigné précisé et sous lesquels pourtant on ne trouve que des images floues et mouvantes accrochées à ces mots et ces signes de manière non séparée. J'écris entre guillemets "valeur de temps", par exemple, parce qu'à chaque fois que je réfléchis au degré d'inexploration où se trouvent encore les concepts de durée et de temps et leur relation, je ressens une réticence à écrire quoi que ce-soit en dehors d'un algorithme qui fixe une règle du jeu. La paramétrisation de la propriété fondamentale de durée à l'aide de la variable de temps t, telle que cette paramétrisation est pratiquée dans les théories existantes et même dans la Relativité, est encore certainement très simplificatrice et souvent falsificatrice, rigidifiante, mécanisante en quelque sorte. Les changements ne sont pas toujours des déplacements d'entités stables intérieurement. Pour pouvoir rendre compte pleinement de l'entière diversité des types et des intensités de changements, il faudrait une sorte de grandeur vectorielle, un *champ* de temps processuel défini en chaque point de l'espace abstrait encadré par l'axe de durée et par les axes des changements envisagés.

Mais le temps se transformerait-il selon Lorentz? Quel rôle joue la vitesse d'un "signal" lumineux face aux vitesses de propagation "d'influences" (?) dans un tel espace processuel? Qu'est ce que la Relativité impose véritablement aux processus *quelconques* et qu'est-ce qu'elle laisse en blanc? Lorsqu'il s'agit de processus très "intenses" localement, "catastrophiques", comme c'est probablement la "création d'une paire", que devient "le temps"?

En théorie relativiste générale de la gravitation, par exemple, un gradient non nul du champ de gravitation est lié à une impossibilité de définir un temps unique, pour les observateurs d'un même référentiel, si ces observateurs sont spatialement distants l'un de l'autre. Quand à l'invariance de la vitesse *de la lumière* elle-même (et non la vitesse d'autres sortes "d'influences") lorsqu'on passe d'un référentiel à un autre, elle n'est postulée que localement, car il n'existe aucune définition uniforme des distances et des temps dans des champs gravitationnels variables (9) (espace-temps courbes). Comment savoir quelle sorte de "courbure" locale de l'espace-temps est produite (ou non) par un processus – essentiellement variable – de création d'une paire?

Enfin, la Relativité n'introduit aucune quantification au sens de la Mécanique Quantique, sa description est continue. Lorsqu'on écrit [vitesse = distance/temps], le temps est un paramètre continu.

Si ensuite on se demande comment on trouve la valeur de t, on s'aperçoit qu'elle est de la forme $NT_H$ où N est un entier et $T_H$ une période "d'horloge" (supposée constante !) ce qui ramène au discret. En macroscopie cela peut être négligeable aussi bien sur le plan du principe que sur le plan numérique. Mais lorsqu'on considère des processus quantiques et relativement très brefs, quel est le degré de signification d'une condition comme

$$v = \frac{\text{distance}}{\text{temps}} = \frac{\text{distance}}{NT_H} = \text{const?}$$

Quelle horloge faut-il choisir, avec quel $T_H$, et comment par ailleurs s'assurer que lorsqu'on écrit $\Delta t = 10^{-x}$, on fait plus qu'un calcul vide de sens?

On comprend, devant de telles questions, les prudences positivistes et les normes qui conseillent de se maintenir dans la zone salubre de l'opérationnellement défini et du syntaxisé, où la pensée circule sur des voies tracées et consolidées. Au dehors, on s'enfonce dans une véritable boue sémantique. Pourtant ce n'est que là, dans cette boue, et lorsqu'on force le regard à discerner les formes mouvantes, que l'on peut percevoir du nouveau. De toute façon le problème de localité nous y force: c'est un problème très fondamental où tout comportement inertiel, inanalysé ou approximatif, conduit inéluctablement à l'arrêt de la capacité de raisonnement, ou à des problèmes et perspectives illusoires. On ne peut pas cette fois suivre un chemin parce qu'il est construit. On est obligé de choisir et de suivre la direction qui convient.

Je reviens donc sur la création d'une paire corrélée $S_1$, $S_2$. J'imagine ce processus comme ayant des analogies avec la formation de gouttes. (Ceci peut être faux, mais ce n'est pas *a priori* impossible, et <u>je n'ai</u>

besoin que d'un exemple de possibilité). Je dessine donc ainsi la projection spatiale (en deux dimensions) du domaine d'espace- temps $\Delta s_c(t).\Delta t_c$, $t_0 < t < t_f$, pour 4 époques:

* $t = t_0$ ;
* $t_0 < t < t_f$
* $t_0 < t < t_f^-$ (c'est-à-dire immédiatement *avant* $t_f$) ;
 et $t = t_f^+$ (c'est-à-dire immédiatement *après* $t_f$)

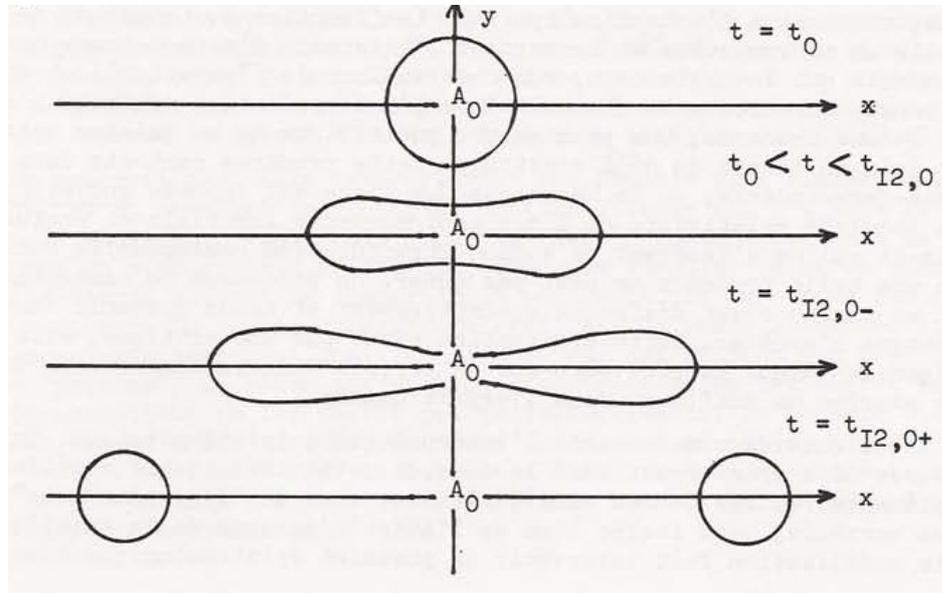

Supposons maintenant que la distance $d_{l2}$ entre les appareils $A_1$ et $A_2$ est plus petite que la projection maximale de $\Delta s_c(t)$ sur l'axe x correspondant à $t = t_f^-$.

Les appareils $A_1$ et $A_2$ seront donc atteints non pas par "$S_1$" et "$S_2$" respectivement, mais par "S en cours de désintégration", qui <u>peut</u> néanmoins déclencher des enregistrements sur $A_1$ et $A_2$. Supposons encore que la durée des évènements de mesure se trouve être telle par rapport à $d_{l2}$, que la distance d'espace-temps entre les événements de mesure soit du genre espace. Enfin, supposons que, en dépit de tout cela, les évènements de mesure ne soient pas indépendants au sens de Bell, c'est-à-dire supposons qu'un changement de $A_2$ puisse agir à une vitesse v > C sur le résultat de l'un des enregistrements de $A_2$. Les statistiques de résultats d'enregistrements sur $A_1$ et $A_2$ seront alors "non localement corrélées" et l'inégalité de Bell sera violée. Mais serait-il en ce cas justifié de conclure qu'on a démontré une contradiction avec la théorie de la Relativité? La théorie de la Relativité ne statue <u>que</u> sur des "signaux" (quelle est exactement la définition?) se propageant "dans le vide". Elle ne statue rien du tout concernant la transmission "d'influences" (définition?) à travers un "système" (objet? processus?). En particulier, elle  n'impose rien du tout concernant "l'ordre temporel" (?) ("causal" ou "non causal") (?) d'événements placés à des endroits spatiaux différents d'"un même système". L'exemple imaginé – un modèle de "création d'une paire" – n'appartient tout simplement pas au domaine de faits que la Relativité décrit. Aucune théorie constituée ne le décrit encore. Pourtant cet exemple, quelles que soient ses inadéquations face à la réalité inconnue, caractérise certainement d'une manière en essence acceptable ce qui mérite la dénomination de processus de création d'une paire: un tel processus doit occuper un domaine non nul d'espace-temps, dont la projection spatiale, connexe au départ, évolue, devenant non connexe.

Cet exemple de possibilité me semble suffire comme base pour la conclusion suivante: les tests destinés à vérifier l'inégalité de Bell, même s'ils violaient définitivement l'inégalité, ne pourront jamais établir à eux seuls que le principe <u>relativiste</u> de localité a été enfreint. Pour préciser ce qui est en jeu, la modélisation de Bell et le test correspondant devront être associés à d'autres modélisations et à d'autres tests, concernant l'extension d'espace-temps des évènements qui interviennent, non observables ("création") et observables (mesures). La minimalité de l'interprétation minimale n'est en fait qu'une prudence, une peur encore positiviste de se laisser entraîner trop loin en dehors du déjà construit. Cette prudence cantonne dans un face-à-face indécis, où la Mécanique Quantique est opposée <u>indistinctement</u> à la localité relativiste et à des prolongements inertiels et confus de celle-ci qui ne s'insèrent en aucune structuration conceptuelle constituée. Mais une telle prudence ne peut pas durer. Un processus de conceptualisation en chaîne s'est déclenché subrepticement et aucun obstacle factice ne pourra l'arrêter. Cette affirmation n'est pas une critique, elle désigne la valeur la plus sûre que je perçois dans la démarche de Bell, et elle exprime ma confiance dans l'esprit humain.

Je considère maintenant l'interprétation épistémologique. Celle-ci s'avance déjà précisément dans le sens de cette inéluctable modélisation supplémentaire. Les termes considérés sont ceux de "1 système" et "2 systèmes corrélés mais isolés l'un de l'autre" (au sens de la Relativité). La modélisation supplémentaire mentionnée fait intervenir le postulat épistémologique courant d'existence de propriétés intrinsèques pour des entités réelles isolées. On déduit de ce postulat au même type que celle de Bell, concernant des statistiques de résultats de mesures sur des entités supposées isolées. On établit donc une connexion entre des tests sur des inégalités observables d'une part, et d'autre part le postulat épistémologique d'existence de propriété intrinsèque pour des objets isolés au sens de la Relativité. Sur cette base on admet (il me semble?) que la violation de l'inégalité de Bell infirmerait à elle seule la signifiance de la conceptualisation en termes d'entités isolées possédant des propriétés intrinsèques. Or j'ai montré ailleurs (10) (en termes trop techniques pour être reproduits ici) que cela n'est pas possible. Ici je ne ferai à ce sujet que quelques remarques qualitatives.

Tout d'abord, les considérations faites plus haut concernant la création d'une paire peuvent aussi se transposer d'une manière évidente au cas de l'interprétation épistémologique. Mais prolongeons encore autrement ces considérations: plaçons-nous cette fois d'emblée à l'instant t = $t_o$ où $S_1$ et $S_2$ sont créés. Pour t > $t_o$, $S_1$ et $S_2$ occupent maintenant deux domaines d'espace disjoints $\Delta s_1(t)$ et $\Delta s_2(t)$, qui s'éloignent l'un de l'autre et qui rencontrent ensuite respectivement les appareils $A_1$ et $A_2$, produisant des interactions de mesure. L'interaction de mesure de $S_1$ avec $A_1$ est elle-même un évènement qui occupe un domaine non nul d'espace-temps $\Delta s_{m1}(t_{m1}).\Delta t_{m1} \neq 0$ (l'indice m se lit: mesure) où $t_{m1} \in \Delta t_{m1}$ et le facteur de durée $\Delta t_{m1}$ dépend de l'extension spatiale $\Delta s_{m1}(t_{m1})$ liée à l'époque $t_{m1} \in \Delta t_{m1}$ (en supposant que cette extension spatiale reste constante au cours de l'époque $t_{m1} \in \Delta t_m$). Il en va de même pour l'évènement de mesure sur $A_2$ dont l'extension d'espace-temps est $\Delta s_{m2}(t_{m2}).\Delta t_{m2} \neq 0$. Comment définir maintenant la distance d'espace-temps entre ces deux évènements de mesure? Quelle que soit la distance spatiale fixée entre $A_1$ et $A_2$, comment savoir si la distance correspondante d'espace-temps entre les évènements de mesure est ou non du genre espace? Car c'est cela qui décide si oui ou non la condition cruciale "d'isolement" réciproque de ces évènements de mesure, se réalise, et c'est sur la base de cette condition que l'on s'attend à l'inégalité de Bell pour les statistiques des résultats enregistrés. Que la distance d'espace-temps entre les évènements de mesure soit ou non du genre espace, cela dépend évidemment (entre autres) des facteurs d'extension spatiale $\Delta s_{m1}(t_{m1})$ et $\Delta s_{m2}(t_{m2})$. Or, que savons-nous de la valeur de ces facteurs? $S_1$ et $S_2$ se déplacent-ils "en bloc", "mécaniquement", comme le suggèrent le modèle de de Broglie et le concept récent de soliton, ou bien s'étalent-ils comme le suggère le concept quantique courant de paquet d'ondes à évolution linéaire Schrödinger?

On pourrait peut-être espérer avoir une réponse plus claire dans le cas où $S_1$ et $S_2$ seraient des photons "dont la vitesse est C". Mais la vitesse de quoi? Du front de l'onde photonique, oui, mais que penser du "reste" du photon? Comment est fait un photon, comme un microsystème de Broglie, avec une singularité et un phénomène plus étendu autour? Le comportement manifesté par des ondes radio le laisse supposer. De quelle extension alors? Dans la phase actuelle, que savons nous, exactement et individuellement sur ces entités que l'on dénomme "photons"? La Mécanique Quantique newtonienne ne les décrit pas ; l'électromagnétisme ne les décrit pas individuellement. La théorie quantique des champs a été marquée, au cours des années récentes, par des essais "semi-classiques" dont le but est d'éliminer tout simplement la notion de photon afin d'éviter les difficultés conceptuelles liées aux algorithmes de renormalisation (11).

On peut

d'espace-temps du genre espace, il faudrait connaître (entre autres) l'extension spatiale des états de ces microsystèmes, en fonction du temps.

Sans dé

À eux seuls, les tests de l'inégalité de Bell ne permettront jamais de conclure concernant la signifiance de l'assignation de propriétés intrinsèques à des entités réelles isolées au sens de la Relativité d'Einstein. Donc pour l'instant aucun face-à-face n'est encore défini entre la Mécanique Quantique et les postulats épistémologiques de notre conceptualisation courante de la réalité. Seule une direction de pensée est tracée, qui suggère l'intérêt de recherches nouvelles sur la structure d'espace-temps de ce que l'on appelle les microsystèmes individuels. Cette direction de pensée me paraît courageuse et très importante, mais dans la mesure où elle se reconnaît et s'assume. Elle s'associe alors naturellement à des recherches récentes sur l'extension des microsystèmes à masse non nulle au repos (12), (13) et sur le concept de photon (11). Il est très remarquable de voir que toutes ces recherches se concentrent sur les phénomènes et concepts d'interférence. En effet c'est là qu'à travers la statistique peut apparaître l'individuel. C'est là que peut se trahir – si on l'y cherche – la confusion entre des interférences mathématiques de statistiques standard et d'autre part des statistiques d'interférences physiques d'une entité individuelle qui se superpose avec elle-même (14), (15).

salubrement d'algorithme en algorithme, accrochés à des cordes de mots, me paraît mériter d'être connue de plus près. Il faudra bien y plonger pour forger les concepts nouveaux qui manquent et en fixer les contours d'une manière qui permette de s'élever jusqu'à des syntaxisations.

A trave

Le conc

maximale. Mais cette théorie est <u>foncièrement</u> inapte à une description non restreinte des changements. En effet, la logique des classes d'objets et des prédicats est fondée sur la relation d'appartenance ∈ : si pour l'objet x le prédicat f est vrai, alors x appartient à la classe $C_f$ définie par f: $f(x) \rightarrow x \in C_f$. Mais cette relation fondamentale d'appartenance ∈ est conçue au départ d'une manière statique, hypostasiée. Aucun aménagement ultérieur ne peut compenser les rigidités introduites ainsi au départ. La théorie des probabilités d'une part et d'autre part les différentes théories physiques (la mécanique, la thermodynamique, les théories des champs, la Mécanique Quantique, la Relativité) sont arrivées à combler cette lacune à des degrés différents. Mais chacune pour une catégories particulière de faits et par des méthodes implicites et diversifiées. Une théorie générale et spécifique des événements et des processus, une logique des changements absolument quelconques, à méthodologie explicite et unifiées, n'a pas encore été construite[*].

Considé

d'une mesure de probabilité définie sur les classes. Pourtant, à ce jour, une telle quantification numérique de la logique n'a pas pu être accomplie. Les "quantificateurs" logiques ∃, ∀, Ø, sont restés <u>qualitatifs</u> !

Complé

tribu d'événements τ, définie sur l'univers U={$e_i$, i=1,2,....} d'événements élémentaires $e_i$. Cette tribu peut refléter, en particulier, une classification des événements élémentaires $e_i$ commandée par un prédicat f et en ce cas des propriétés spécifiques "logiques" s'ensuivent pour l'espace [U,τ, p(τ)]. *Via* ces propriétés classificatrices, la connexion entre logique et probabilités pourrait être amorcée. Mais ceci n'a pas été tenté, et la connexion reste pour l'instant non élaborée.

Considé

D'autre

des probabilités et la Mécanique Quantique reste pour l'instant elle aussi très obscure.

concepts tels que l'on pourrait vouloir les imaginer en dehors de l'observation. Même la probabilité de présence n'est qu'une probabilité de résultats d'interactions d'observation: il est permis par la Mécanique Quantique d'imaginer qu'un "système" qui fait une marque sur un écran à un moment t, se trouvait, en lui-même, aussi loin que l'on veut de cette marque, aussi peu que l'on veut avant le moment t. La Mécanique Quantique laisse parfaitement non conceptualisée en elle-même, "la réalité" dont elle codifie de manière si riche et détaillée les manifestations observables à travers les interactions de mesure.

Considé

<u>observationnels</u> et quantifiés de la Mécanique Quantique, soulève des problèmes bien connus et très résistants.

Ainsi n

les bases disparaissent. Quand à l'ensemble des concepts liés à la propriété fondamentale de durée, les concepts de processus, d'événement, de changement, de permanence, de succession, de TEMPS, ils n'agissent librement qu'à l'état épars, primitif et subjectif, tels que l'expérience et le langage les a diversement induits dans les esprits. Car les organisations auxquelles ces concepts ont été soumis à l'intérieur de la théorie de la Relativité, de la théorie des probabilités, ou à l'intérieur de telle ou telle autre théorie physique, sont toutes particularisantes et amputantes. La situation est encore telle que la décrivait Bergson: «La déduction est une opération réglée sur les démarches de la matière, calquée sur les articulations mobiles de la matière, implicitement donnée, enfin, avec l'espace qui sous-tend la matière. Tant qu'elle roule dans l'espace ou dans le temps spatialisé, elle n'a qu'à se laisser aller. C'est la durée qui met des bâtons dans les roues' » (17).
    Je résume une fois encore par un schéma:

---

[*] J'ai pu prendre connaissance d'une tentative originale et courageuse de formaliser la durée (16). Jusqu'ici seules les valeurs associables à la durée ("le temps") ont fait objet de certaines formalisations.

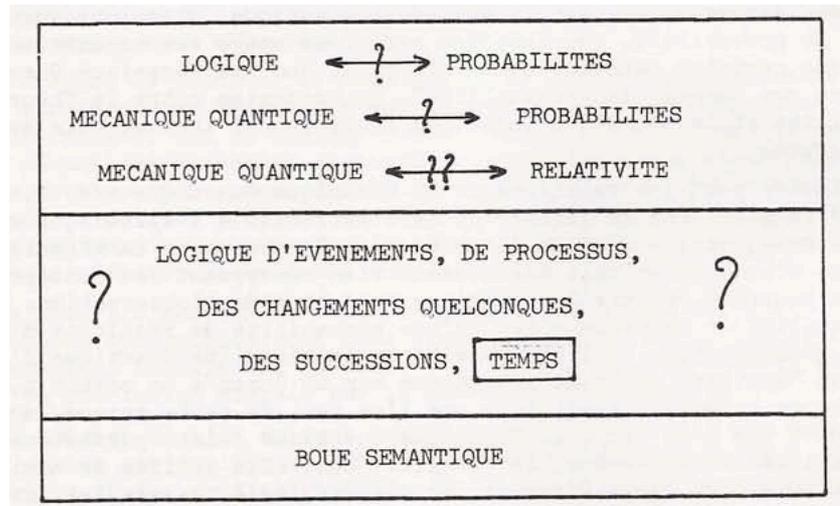

Quand il n'existe encore aucune unification entre la démarche statistique, discrète, observationnelle, orientée vers le microscopique, de la Mécanique Quantique, et d'autre part la démarche individuelle, continue, réaliste, orientée vers le cosmologie, de la Relativité, quant tout ce qui touche à la durée et au temps est encore si peu élucidé, quand tout ce qui touche à la manière d'être de ces entités que l'on appelle des microsystèmes – ou plus encore, de microétats – est encore tellement inexploré, quel sens cela peut-il bien avoir d'affirmer qu'on se trouve – sur la base de tests de "non-localité" – devant un face-à-face <u>contraignant</u>, direct ou pas, entre la Mécanique Quantique et la Relativité? Ou bien entre la Mécanique Quantique et notre conceptualisation du réel?

<div align="center">Conclusion</div>

Je ne puis qu'écarter, pour ma part, les face-à-face que les autres physiciens pensent percevoir. Pour moi la valeur du théorème de Bell réside ailleurs: ce théorème, et l'écho qu'il soulève, illustrent d'une manière frappante la puissance d'action des modélisations <u>mathématisées</u>, lorsqu'elles sont connectables aux tests expérimentaux. Pendant des dizaines d'années, les tabous positivistes ont fait obstacle aux modèles. Le résultat est ce vide vertigineux de modèles syntaxiques, et même seulement qualitatifs, que l'on découvre maintenant sous les algorithmes quantiques. Or, la modélisation de Bell a déclenché une dynamique de conceptualisation et de syntaxisation. Cette dynamique atteindra peut-être l'attitude positiviste. Elle ébranlera peut être la Mécanique Quantique et la Relativité. Car elle attire et maintient longuement l'attention sur l'état du milieu conceptuel dans lequel les théories actuelles sont immergées. De ce contact prolongé sortiront peut-être des théorisations nouvelles, plus unifiées, plus étendues et plus profondes. Je perçois (ici comme en théorie de l'information) les premiers mouvements de formalisation de l'épistémologie, les premières ébauches, peut-être, d'une méthodologie mathématisée de la connaissance. Et cela pourrait s'avérer plus fertile que toute théorie particulière d'un domaine donné de réalité.

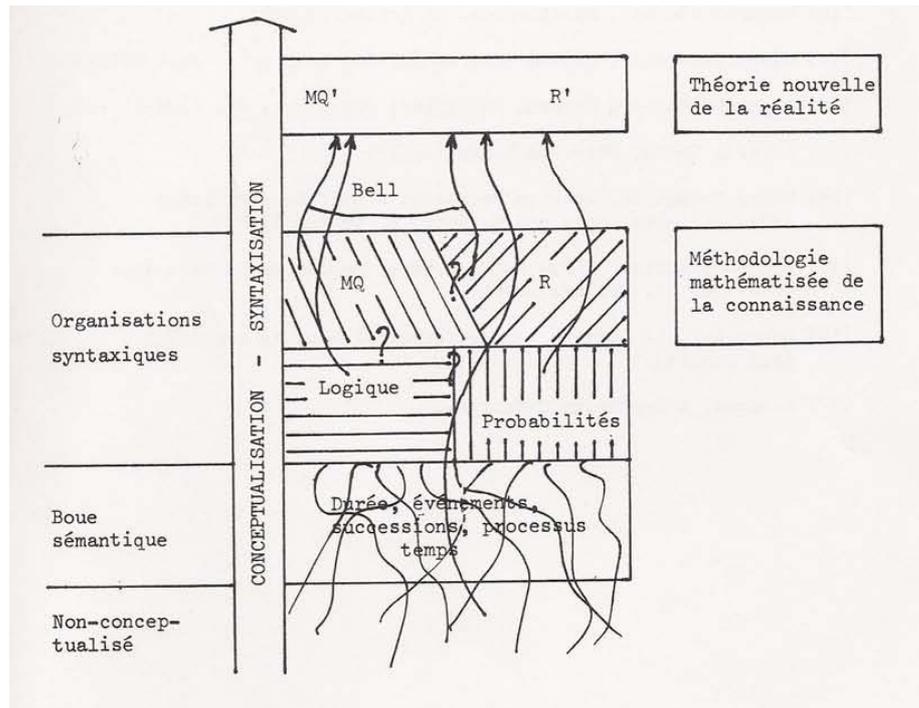

**************

Depuis juin 1979 et à ce jour je n'ai jamais arrêté de travailler à la réalisation du programme esquissé dans la conclusion de cette réflexion sur la question de localité, et surtout dans le schéma qui la clôt. Ce qui suit en est une étape importante vers la réalisation de ce qui y est dénoté MQ': une refonte de la mécanique quantique en relation explicite avec ses substrats sémantiques et apte à être insérée dans une méthode épistémologique générale formalisée et même mathématisée. D'ores et déjà, ici même, les relations entre mécanique quantique et probabilités et entre mécanique quantique et relativités d'Einstein se trouveront considérablement éclairées.

Quant aux relations *générales* entre probabilités et logique, elles ont été élucidées dans la méthode de conceptualisation relativisée, cependant que s'y trouve également construite une représentation du concept de temps.

# 2

# L'infra-mécanique quantique

## Une représentation qualitative des microétats construite *indépendamment* du formalisme quantique

**"Conçois bien le sujet, les mots suivront"**
**Cato le Vieux (234 BC-149 BC)**

### 2.1. Préalables

Je fais donc abstraction du formalisme quantique : Je me trouve devant une table rase où les seuls éléments dont il est permis de faire usage sont : Les conditions physiques-cognitives qui contraignent l'approche ; les impératifs introduits par les modes humains de conceptualisation ; les impératifs introduits par le *but* choisi.

L'entière démarche restera ***indépendante de la mécanique quantique***[11]. Elle conduira à une discipline d'un type nouveau que l'on pourra qualifier de "représentation épistémo-physique des microétats"

Dans chaque phase j'essaierai de faire apparaître les caractéristiques de cette phase là:

- La façon dont, dans la phase considérée, interviennent les modes humains généraux de conceptualiser (notamment le fait que le nouveau qu'on veut construire est irrépressiblement rattaché (par continuité ou par opposition) à ce qui a été construit avant.

- Comment, lors de chaque nouvelle avancée, s'entrelacent inextricablement concepts, opérations physiques, données factuelles, et mots et signes d'étiquetage qui assurent la communicabilité.

- Comment émergent progressivement des structures de penser-et-dire liées à des modes de faire.

- *Comment on peut se heurter à un obstacle qui arrête toute progression vers le but choisi et qui ne peut être contourné* **que** *par une* **décision méthodologique** *adéquate pour pouvoir continuer, mais qui écarte de manière abrupte de la structure de pensée 'classique' acquise historiquement au cours de l'interaction avec la factualité physique avec laquelle l'homme interagit naturellement via ses sens biologiques et qui, en partie du moins, a été intégrée dans les structures et les fonctionnements biopsychologiques des corps humains.*

---

[11] De temps à autre il sera fait allusion au formalisme de la mécanique quantique – presque exclusivement dans des notes – afin de préparer progressivement dans l'esprit du lecteur la comparaison constructive développée dans la deuxième partie de ce travail, qui engendre la "2ème mécanique quantique" annoncée dans le titre et l'introduction. Mais la construction de la première partie de ce travail restera rigoureusement indépendante du formalisme quantique.

En conséquence d'une rupture introduite par une décision méthodologique qui se sera imposée dès le départ d'une manière incontournable, le résultat final qui émergera sera foncièrement non-"classique", nonobstant le fait que, sous l'empire de contraintes biopsychologiques et de buts de continuités historiques, l'observateur-concepteur humain aura, lui, pensé et agi selon ses modes courants de penser et agir. Et je montrerai que", en un certain sens bien défini, ce résultat foncièrement non-classique *peut* être regardé comme une "description de microétat. Cet aboutissement, lié à cette scission [(actions cognitives classiques)-(résultat non-classique)], illustrera le rôle essentiel des décisions méthodologiques au cours de toute action délibérée dominée par un *but*.

Bref, j'essaierai de mettre au jour pas à pas le développement du processus constructif délibéré par lequel peut émerger un système cohérent de manières de penser et d'exprimer, faisant unité organique avec un système correspondant de procédures physiques et conceptuelles, et qui fonde le concept de descriptions de microétats.

Afin de réussir une communication sans flou, je serai obligée d'introduire quelques notations : des dénominations par symboles au lieu de mots, afin d'éviter des formulations verbales répétées à outrance et trop longues à chaque fois. Mais ces notations n'auront rien de mathématique. L'exposé restera rigoureusement qualitatif, afin de mettre en évidence ce qui, véritablement, s'impose au niveau primordial des structures de faits-et-concepts.

## 2.2. Comment introduire un microétat en tant qu'objet de description ?

### 2.2.1. Une opération de génération d'un micro-état

Toute *description* implique une entité-objet[12] définie et des qualifications de cette entité-objet, et elle doit être *communicable*. Les descriptions scientifiques sont en outre soumises à l'exigence de permettre un certain consensus intersubjectif. Le but propre de la mécanique quantique est d'offrir des descriptions *mécaniques* d'*états* de micro-*systèmes*, de microétats[13] (désormais nous écrirons la succession de ces deux mots sans tiret interne, sauf quand il faudra accentuer la différence entre leurs désignés). Ceux-ci sont des entités hypothétiques qu'aucun humain ne perçoit. En outre, ce sont des entités conçues comme étant en général instables, comme tout état mécanique. Nous voulons expliciter la stratégie descriptionnelle qui a conduit à la description quantique des microétats. La première question qui s'impose est donc :

**Q1 : « Comment fixer un microétat en tant qu'objet d'étude ? »**

---

[12] Dans tout ce qui suit l'expression 'entité-objet' est une abréviation de l'expression 'entité-(objet d'étude)' où 'objet d'étude' spécifie un *rôle descriptionnel*. (Par conséquent le pluriel s'écrit : entités-objets). Cette spécification est très importante d'un point de vue conceptuel : en général l'objet d'étude n'est *pas* aussi un 'objet' au sens de la pensée et le langage classique, ni au sens de la logique classique fondée sur la pensée classique.

[13] Les caractéristiques stables d'un micro-*système* – qu'on peut baptiser 'objectités' – sont étudiées par la physique atomique et la physique des particules. La mécanique quantique fondamentale présuppose connues les objectités des micro-systèmes qu'elle considère, et elle s'assigne la tâche spécifique d'étudier leurs *états mécaniques*, des microétats, c'est à dire leurs caractères mécaniques (position, vitesse, énergie) instables qui dépendent du temps et de l'environnement. (langage introduit par Dirac)

Pourquoi ce mot, 'fixer' ? Parce qu'en l'absence de toute espèce de stabilité – récurrence ou reproductibilité – on ne peut pas commencer à conceptualiser. Ce qui est dépourvu de *toute* stabilité peut être étudié aussi, bien sûr, mais en tant que changements subsumés sous quelque chose de durable ou de récurrent ou de répétable que l'on place en position fondamentale. Il s'agit là d'une manière de hiérarchiser qui est inhérente au fonctionnement de l'esprit humain.

Les microétats qui ont la plus grande importance pragmatique sont conçus comme étant liés par des forces d'attraction au sein de microstructures stables, atomes, noyaux, molécules, etc., et là ils sont stables eux aussi. C'est dire que, bien que non perceptibles, ils sont conçus comme étant déjà disponibles pour être étudiés, en état naturel et doté de certaines stablilités. C'est pourquoi les toutes premières ébauches de pensée 'quantique' ont concerné les microétats liés.

Mais quelle que soit leur importance particulière, les microétats liés n'épuisent pas la totalité des microétats concevables. Il convient d'admettre que l'espace fourmille de microsystèmes en états non liés, libres ou 'progressifs'.

La question *Q1* concerne un microétat libre dont la manière d'être nous est totalement inconnue.

Puisqu'un microétat n'est pas perceptible, il est clair qu'on ne peut pas le fixer en tant qu'objet d'étude en le sélectionnant tout simplement dans un ensemble d'entités préexistantes, comme un caillou, par exemple, que l'on ramasserait par terre et que l'on poserait sur la table du laboratoire. Il faut trouver un autre procédé.

D'autre part, il est clair également que si nous – des humains liés par nos corps et nos capacités perceptives à un certain niveau d'organisation de la matière, celui qu'on dit 'macroscopique' – voulons étudier des 'microétats' non perceptibles par nous, alors nous aurons besoin d'appareils enregistreurs, c'est-à-dire, d'objets macroscopiques eux aussi, qui soient aptes à développer *sur eux*, à partir d'interactions supposées avec les microétats présupposés, des marques qui, elles, soient perceptibles par nous. Mais tant qu'on n'a pas introduit un procédé qui puisse fixer en tant qu'objet de l'étude à faire, un microétat défini, il n'y a aucun moyen de savoir *quelle* marque correspond à *quel* microétat : une médiation introduite exclusivement par un appareil enregistreur ne peut suffire. Cet appareil manifesterait juste une foule de marques (visibles ou audibles, etc.), rien d'autre. Ces marques ne seraient pas des descriptions de microétats, elles seraient dépourvues de signification. Elles ne seraient que des événements connaissables mais tout à fait déconnectés du concept hypothétique de microétat. Or, de par sa définition courante, le concept de description concerne un objet spécifié. Ce problème consistant à fixer un microétat en tant qu'objet d'étude, est incontournable et primordial.

Pour amorcer une dynamique de solution, considérons une opération impliquant des objets et des manipulations *macroscopiques* et qui soit telle que, sur la base de certaines données dont nous disposons par la voie historique du développement de la physique, nous puissions imaginer que cette opération engendre un microétat. (Par exemple, on a été conduit à admettre qu'une plaque métallique chauffée suffisamment perd, par agitation thermique intensifiée, des électrons qui circulent librement sur sa surface. Ce procédé semble donc pouvoir être conçu comme engendrant des états libres d'électrons).

Mais on se demande aussitôt: Comment peut-on savoir que l'opération considérée engendre vraiment un tel microétat ? Et tout d'abord, quel sens cela a-t-il de parler de 'microétats' avant de savoir ce que c'est ?

Eh bien, on ne peut pas d'emblée savoir quels contenus sémantiques associer au mot 'microétat', justement. On produit au départ juste un lieu conceptuel blanc désigné par ce

'microétat' tel qu'il s'est formé dans l'histoire de la physique. Mais on peut essayer de *construire* un sens à associer à ce mot, sur la base de données préexistantes, de raisonnements et de conventions: il s'agit de construire une stratégie cognitive. Pour commencer cette construction on peut supposer, sur la base invoquée plus haut, que telle opération macroscopique met en jeu ce qu'on appelle un électron, et en détermine un certain état, inconnu.

D'ailleurs, on n'a pas de choix.

Si l'on veut initier une étude des microétats, on doit investir de quelque façon, car de rien on ne peut rien tirer. Et pour cela on doit puiser dans le réservoir dont nous disposons. Or ce réservoir ne contient que la pensée et le langage courant, les modes humains généraux de conceptualisation, et les systèmes de connaissances pré-existants, avec les représentations qui y sont affirmées et les opérations macrocopiques qu'ils suggèrent, réalisables à l'aide d'appareils macroscopiques. On investit donc du savoir conceptuel préconstitué qui induit la supposition qu'un microétat déterminé mais inconnu de telle ou telle sorte de microsystème, a été produit par telle opération macroscopique réalisée à l'aide de tels appareils macroscopiques, afin de se donner ainsi une base pour tenter de construire à partir d'elle un savoir nouveau, défini et *vérifiable* concernant ce microétat inconnu.

### 2.2.2. Étiquetage et communicabilité

La question *Q1*, toutefois, subsiste toujours : en quel sens peut-on fixer ce microétat hypothétique et inconnu, en tant qu'objet stable d'une étude subséquente ? Le fait d'avoir reconnu qu'il faut créer le microétat par une opération macroscopique, n'efface pas cette question. A la réflexion, l'*unique* réponse – qui s'impose – est frappante :

*En étiquetant le microétat à étudier par l'opération qui l'a engendré.*

Mais un tel étiquetage ne peut être utile que si l'opération de génération d'état mise à l'œuvre est *reproductible*. Si elle l'est, et seulement dans ce cas, alors nous pouvons en effet **convenir** de dire qu'à chaque fois qu'on réalise cette opération, elle fait émerger 'le microétat correspondant'. Sur la base de cette convention on peut maintenant étiqueter l'opération de quelque manière, et attacher la même étiquette au microétat correspondant.

Il est en effet strictement nécessaire d'étiqueter, car sinon on ne peut pas communiquer ce qu'on a obtenu, même pas à soi-même. Cependant que ce que nous voulons réaliser, est une procédure librement communicable (sans limitations de distance, etc.) et en plus consensuelle, conduisant à une description consensuelle d'un microétat. En outre, sans étiquetage, sans notation symbolique, cet exposé écrit non plus ne pourrait continuer sans s'enfoncer dans des vertiges de mots. Convenons donc d'une notation. Par exemple, notons *G* une opération donnée de génération de microétat. Le micro-état correspondant pourra alors être symbolisé $me_G$, ce qui se lit : *le micro-état engendré par l'opération de génération dénotée G.*

L'opération *G* de génération de microétat est supposée être bien définie par la spécification de paramètres opérationnels macroscopiques, mais par ailleurs elle est quelconque (de même que lorsqu'on dit « pensez à un nombre », l'interlocuteur pense à un nombre défini, *3* ou *100*, etc., mais n'importe lequel). Si la nécessité se présente de considérer deux opérations de ce type, l'on pourra, par exemple, introduire les notations *G1* et *G2*.

Mais aussitôt on se demande : « En quoi ces précisions nous avancent-t-elle concernant le problème de fixer un microétat spécifié en tant qu'objet d'étude ? Des notations ne sont pas des faits physiques. Écrire '$me_G$' ne peut pas fixer physiquement un microétat en tant qu'objet d'étude ». Or, si, justement. Car l'opération de génération dénotée *G*, elle, est un acte physique, et qui se trouve sous notre contrôle puisque c'est nous qui le concevons et l'accomplissons. Et les associations [(opérations)-(symbolisations)] ont une grande puissance d'organisation dans un processus de conceptualisation. On peut s'en rendre compte le plus clairement précisément lorsqu'on se trouve en situation de pénurie conceptuelle.

Pour y voir clair, procédons par contraste. Imaginons d'abord une entité physique macroscopique qu'il nous est loisible d'examiner directement, par exemple un morceau de tissu. Supposons la tâche de 'fixer l'état' de ce morceau de tissu. Comment s'y prendrait-on ? On l'examinerait et l'on enregistrerait quelque part (dans sa mémoire, sur papier, dans un ordinateur, etc.) des mots ou des notations indiquant les propriétés constatées et qui paraissent suffire pour l'individualiser : genre de substance, dimensions, forme, couleur, degré d'usure, défauts, etc. L'on assignerait aussi un nom à ce morceau de tissu, ou une étiquette, ou les deux à la fois, et ensuite on le rangerait en se disant que désormais il sera possible de le retrouver tel qu'il était, ou bien au pire de le reproduire.

Mais un procédé de ce type n'est évidemment pas applicable à un microétat qui n'est pas directement perceptible. Tout ce dont on dispose pour fixer ce microétat en tant qu'objet d'étude, c'est qu'il a été engendré par l'opération de génération d'état qui a été dénotée $G$ : *c'est le microétat généré par l'opération G et étiqueté me$_G$*. Point. Rien de plus, au départ. Or – et cela est remarquable – déjà cela suffit comme bout de fil. En effet – selon *notre* rationalité humaine – l'entité étiquetée *me$_G$*, puisqu'elle a émergé par l'opération physique $G$, a dû émerger imprégnée de certaines marques physiques relatives à cette opération. Des marques non connues, mais dont on conçoit qu'elles ont singularisé cette entité factuellement à l'intérieur du continuum du réel physique. Désormais cette entité 'existe' donc, en ce sens qu'elle a cessé de se fondre dans le reste du réel physique. Elle existe d'une façon spécifique qui porte le sceau de l'opération $G$ de génération.

En outre puisque l'opération dénotée $G$ est reproductible, le microétat dénoté *me$_G$* est désormais 'capturé', en *ce* sens qu'on peut désormais le re-produire lui aussi autant de fois qu'on voudra.

Les traits minimaux de stabilité et de communicabilité sans lesquels on ne peut pas démarrer un processus de conceptualisation publique, viennent d'être acquis. Une association [(opération physique)-(symbolisations)] a permis de franchir ce premier pas. Bref : Selon notre manière usuelle de penser et de dire, on peut fixer un microétat en tant qu'objet d'étude en le *créant* physiquement à l'aide d'une opération de génération reproductible, et en l'étiquetant par l'étiquette associée à cette opération.

### *2.2.3. Une décision méthodologique inévitable*

Mais à nouveau on doute. On se dit : « Si l'on veut aboutir à décrire des microétats *il faut* en effet partir d'une opération macroscopique accomplie à l'aide d'un appareil macroscopique, car nous sommes rivés au niveau macroscopique pour initier une action dans le réel microscopique ; *il faut* en effet poser l'hypothèse que ce qui résulte de l'opération accomplie est ce qu'on appelle un microétat, sinon on ne peut pas tenter de construire une démarche qui mérite le nom d'étude des microétats : un échantillon de réel purement factuel ne peut être hissé dans le réseau de conceptualisation qu'en l'incluant *a priori* dans un réceptacle conceptuel. En outre *il faut* aussi se référer à ce microétat hypothétique en l'appelant *le microétat qui correspond à l'opération de génération G* et en le dénotant de quelque façon qui inclue l'étiquette $G$, car on ne dispose d'aucune autre caractérisation, cependant que, pour communiquer, il faut bien nommer. Tout ceci paraît en effet incontournable. Mais qu'est ce qui prouve que lorsqu'on effectue ce qu'on appelle reproduire l'opération de génération étiquetée $G$ c'est vraiment la même opération qui se réalise ? Et, en supposant qu'il en serait ainsi, qu'est-ce qui prouve en outre qu'à chaque fois qu'on réalise *G* c'est le même microétat qui émerge ? ».

Eh bien, *rien ne prouve ces deux 'mêmetés', ces deux invariances, celle du désigné de G et celle, corrélative, du désigné de me_G.* D'autre part elles sont cruciales. Car comment pourrait-on étudier un 'microétat' qui pour l'instant n'est que juste un nom d'un concept posé par nous à l'avance, dont le désigné n'est pas perceptible, et dont on ne saurait même pas si l'on arrive ou non à le re-créer ? Cela est au-delà des limites du concevable. Et il est évident que vouloir d'abord savoir si chaque réalisation de l'opération de génération *G* aboutit ou pas à re-produire le même microétat, afin de décider sur la base de ce savoir si oui ou non on peut 'logiquement' se lancer à commencer de construire les connaissances concernant les microétats, c'est se laisser piéger dans un cercle vicieux et *renoncer au but initial.*

La pensée scientifique classique nous a profondément habitués à poser l'existence simultanée d'entités matérielles ou conceptuelles préconstituées, à partir desquelles, par des appositions de symboles (notamment verbaux) et des mises en successivités purement discursives appropriées, on peut assurer de la *déductibilité.* Mais ici il est évident qu'il ne s'agit pas de déduire et qu'on ne dispose pas d'un réservoir de simultanéités préconstituées, disponibles. *Nous sommes dans une phase radicalement initiale de pure construction,* une construction primordiale, à la fois physique et conceptuelle, au cours de laquelle seul l'ordre approprié des actions cognitives peut engendrer les éléments qui, successivement, s'imposent comme strictement nécessaires pour assurer la possibilité de progression. Or dans le cas qui nous occupe, l'arrêt de la constructibilité ne peut être évité que par cette sorte de mortier abstrait qu'est la ***DÉCISION MÉTHODOLOGIQUE*** suivante :

***DM. Ce*** qui émerge lors d'une réalisation quelconque de l'opération de génération *G* telle qu'elle est spécifiée à l'aide de paramètres macroscopiques – les seuls dont nous disposons – sera dénommé par définition 'le microétat *me_G* correspondant à *G*', *quel qu'en soit le contenu factuel non connaissable.* Et l'on posera que la relation entre *G* et *me_G* est une relation est une relation de un-à-un que l'on symbolisera par $G \leftrightarrow me_G$.

Sans cette décision méthodologique il n'est tout simplement pas possible d'assigner une définition claire à ce qu'on veut désigner comme 'le microétat *me_G* correspondant à *G*'. L'on ne pourrait donc pas s'y appuyer pour construire. Si donc l'on *veut* tenter de développer une démarche qui permette d'aboutir à des connaissances consensuelles concernant les 'microétats', alors il n'y a pas de choix: on est obligé de commencer par *admettre* pour les désignés des symboles *G* et *me_G*, les 'mêmetés' corrélées posées plus haut, sans s'immobiliser dans la question de savoir si elles sont 'vraies' ou non. Car on *manque de tout accès direct de contrôle de la vérité de ces 'mêmetés'.* Cependant dans la phase actuelle de la progression amorcée, rien n'interdit cette décision, et elle seule peut permettre d'avancer. Nous verrons bien ce qui en découle. Plus tard, enrichis de la construction accomplie, il deviendra possible de discerner des conséquences de la décision méthodologique posée à la base. Nous serons alors armés pour conclure si finalement nous devons revenir en arrière et abandonner cette décision, avec la construction fondée sur elle, ou si au contraire nous pouvons adhérer définitivement à sa pertinence. *Pas à sa vérité, mais à sa pertinence méthodologique.*

Pour l'instant nous construirons donc sur la base de la décision méthodologique qui vient d'être posée.

Enregistrons et soulignons cette première entrée en scène spectaculaire de la nécessité inévitable de décisions méthodologiques au cours d'un processus explicite de conceptualisation sous la contrainte de buts. Les buts – si l'on exclut d'y renoncer –, imposent des décisions méthodologiques, en tant qu'actions inévitables et prioritaires.

### 2.2.4. Une catégorie particulière d'opérations de génération d'un microétat : Opérations de 'génération composée'

Il existe une catégorie particulière d'opérations de génération de microétats qui est liée à un fait d'expérience bien connu, celui que l'on désigne par l'expression 'interférence corpusculaire' (interférence avec des microsystèmes lourds, i.e. à masse non nulle).

Soit un microsystème lourd d'un type donné, disons un électron. Dès qu'on a spécifié une opération de génération $G1$ qui engendre pour un tel microsystème un microétat correspondant $me_{G1}$, et également un autre opération de génération $G2$ qui pour le même type de microsystème engendre un autre microétat correspondant $me_{G2}$, on peut spécifier pour ce type de microsystème, une opération de génération *composée* qui combine $G1$ et $G2$ et engendre un microétat tel que, en un certain sens détaillé plus bas, on peut le considérer comme étant un état d' "interférence" des microétats $me_{G1}$ et $me_{G2}$.

Historiquement, l'étude des états de microsystèmes à masse non-nulle a été abordée en connaissant déjà, d'une part le comportement des mobiles lourds tel que celui-ci est décrit par la mécanique newtonienne, et d'autre part le comportement macroscopique des ondes (sans masse), tel que celui-ci est décrit par l'électromagnétisme classique. Or – lorsqu'on les découvre avec ces acquis historique dans l'esprit – les microétats d'interférence corpusculaire frappent l'attention. Car dans ce cas, face aux repères constitués par ces acquis, on se trouve devant un système surprenant de similitudes et de différences quant aux manifestations observables. En effet, avec des ondes, une configuration – étendue – d' "interférence" se forme *d'un seul coup*, pas comme cela se passe avec des microétats, au fur et à mesure, *via* des impacts localisés qui sont observables individuellement. Cependant que selon la mécanique classique, les impacts isolés que l'on obtient par les réitérations d'un même état d'un mobile donné (un 'corpuscule', si ce mobile est très petit), devraient ne pouvoir jamais engendrer une configuration finale étendue ayant la même forme qu'une figure d'interférence obtenue avec une onde[14].

La possibilité des phénomènes d'interférence corpusculaire apparaît donc comme une spécificité surprenante de certains états de microsystèmes lourds. Cette spécificité a joué un rôle central dans la formulation mathématique de la mécanique quantique[15]. Toutefois l'importance des phénomènes d'interférence corpusculaire reste majeure, également, face aux buts *propres* de la construction tentée cette première partie. Car ces phénomènes comportent des implications concernant les opérations de génération de microétats qu'il est vital de formuler explicitement. Arrêtons-nous donc un instant sur les phénomènes d'interférence corpusculaire. On peut le faire le plus simplement à l'aide d'un compte rendu de la célèbre 'expérience des trous d'Young', fait dans les termes que nous avons construits jusqu'ici.

* Afin de fixer les idées supposons qu'on travaille avec des électrons. On utilise un écran *opaque*. Du côté droit de cet écran on place une plaque couverte d'une substance sensible (où se forme une marque observable lorsque la surface de la plaque est atteinte[16]

---

[14] Les "mobiles" au sens de la mécanique classique, lorsqu'ils rencontrent des obstacles, des impacts distribués de manière à constituer une figure de "diffusion", pas une figure d' "interférence".

[15] Dans la première partie de ce travail ce rôle n'est pas perceptible, mais il sera examinée en détail dans la deuxième partie.

[16] Conformément à la note technique de la fin de l'introduction générale, les expressions verbales qui suggèrent des images importées de la pensée et le langage courants sont écrites sous la forme '…..' où les points indiquent expression verbale. Cette précaution permet de maintenir ces expressions sous contrôle, et finalement les éliminer.

par un microsystème lourd). Du côté gauche de l'écran opaque, on produit, par quelque opération de génération préliminaire $G_p$, un microétat préliminaire d'électron, et ensuite on laisse passer du temps. On constate alors que jamais, à la suite de la génération du microétat préliminaire, il n'apparaît une marque observable sur la plaque sensible de la *droite* de l'écran opaque. On exprime ce fait en disant que le microétat préliminaire ne traverse pas l'écran opaque.

    \* On perce maintenant l'écran opaque d'un trou. Dénotons-le par le chiffre *1*. Dans ces nouvelles conditions on constate que, à la suite de chaque acte préliminaire de génération $G_p$ accompli à la gauche de l'écran opaque, si les paramètres sont convenablement chisis il se produit systématiquement un impact observable sur la plaque sensible de la droite de l'écran opaque. On exprime ce fait en disant que le microétat préliminaire correspondant à $G_p$ passe par le trou *1*, devenant de ce fait un *nouveau* microétat. Si l'on élimine toute image suggérée par cette façon de dire, on reste avec l'affirmation que, dans le contexte expérimental considéré, l'existence du trou *1* agit comme une nouvelle opération de génération *G1* qui, à partir de l'état préliminaire produit par $G_p$, crée du côté droit de l'écran un microétat correspondant $me_{G1}$.

    En répétant la même procédure un grand nombre de fois, on obtient sur la plaque sensible une certaine configuration étendue *c(1)* de marques observables (pas une seule même marque réitérée) qui est *compatible* avec la figure de *diffusion* que la mécanique newtonienne affirme, dans des conditions similaires, pour un corpuscule au sens classique.

    \* Appliquons la même procédure décrite ci-dessus, mais en utilisant un écran opaque percé d'un trou *2* qui est placé à un *autre* endroit que celui où était placé le trou *1*. On observe alors de nouveau, à la suite de chaque opération préliminaire de génération $G_p$, une marque sur la plaque sensible de la droite de l'écran. On exprime ce fait en disant que cette fois le microétat préliminaire passe par le trou *2* ; ce qui, en faisant abstraction de toute image suggérée par le langage employé, revient à considérer que, dans ce nouveau contexte expérimental, l'existence du trou *2* agit comme une opération de génération *G2* qui, à partir de l'état préliminaire $G_p$, produit du côté droit de l'écran un microétat correspondant $me_{G2}$, différent de $me_{G1}$.

    En répétant $G_p$ un grand nombre de fois, on obtient cette fois sur la plaque sensible une nouvelle configuration étendue de marques observables, *c(2)*, consistant dans une figure de *diffusion* compatible avec la mécanique newtonienne et qui est déplacée face à celle obtenue avec le trou *1*.

    \* On utilise maintenant un seul écran opaque percé de *deux* trous, *1* et *2*. A gauche de cet écran opaque on répète un grand nombre de fois l'opération préliminaire de génération $G_p$. Dans ces conditions nouvelles on constate qu'à la suite de chaque réalisation de $G_p$ à la gauche de l'écran opaque, l'on obtient sur la plaque sensible de droite *une* seule marque observable. Et – à condition que les trous *1* et *2* sont assez rapprochés – la configuration finale de l'ensemble des marques observables, *diffère* cette fois de la figure que l'on devrait obtenir selon la mécanique classique (à savoir une figure ayant le même aspect qu'une juxtaposition (addition) des deux configurations de diffusion *c(1)* et *c(2)* où l'on ne conterait qu'une seule fois deux marques produites à un même endroit : *la configuration finale inattendue que l'on obtient – dénotons-la c(1+2) – reproduit par des impacts individuels successifs, l'aspect de la figure d'interférence que, avec une onde, l'on obtiendrait d'un seul coup.*

    On peut exprimer cette situation par la représentation qui suit.

*(a)* On considère que lors de chaque réalisation de la procédure impliquée, l'existence simultanée des deux trous *1* et *2* agit comme *une* seule nouvelle opération de génération où les deux opérations précédentes *G1* et *G2* se 'composent'. Dénotons cette nouvelle opération de génération par *G(G1,G2)* et baptisons-la *opération de génération composée*. Notons *me$_{G(G1,G2)}$* le microétat correspondant généré à la droite de l'écran opaque et nommons-le *le microétat correspondant à une opération de génération composée G(G1,G2)*.

*(b)* Convenons de dire que, dans le microétat à génération composée *me$_{G(G1,G2)}$*, les deux microétats *me$_{G1}$* et *me$_{G2}$* que produi*raient*, respectivement, l'opération *G1 seule* et l'opération *G2 seule*, 'interfèrent': l'introduction de cette façon de dire permet de faire une référence verbale aux figures d'interférence ondulatoire étudiés dans la théorie macroscopique des ondes électromagnétiques, mais qui apparaissent d'emblée (i.e. pas par la composition progressive d'impacts successifs observables individuellement).

On vient de voir comment s'est forgé dans ce cas un système de représentation comportant:
- des opérations physiques seulement *conçues* mais pas effectivement accomplies (*G1*, *G2*), ou bien *effectuées* (*G$_p$*, *G(G1,G2)*); des objets et procédures macroscopiques (ceux qui interviennent dans l'expérience d'Young décrite plus haut) ;
- des faits observables (les marques individuelles sur la plaque sensible et les configurations globales constituées de ces marques) ;
- des façons de dénoter et de dire où interviennent des références à des conceptualisations accomplies auparavant ;
- et enfin, certains prolongements commodes de la pensée et du langage courants qui est essentiellement modélisante, mais dont par la suite *on peut faire abstraction*.

On peut concevoir un nombre illimité de microétats distincts correspondant à des compositions distinctes d'une seule et même paire d'opérations de génération de départ, *G1* et *G2*. En effet on peut complexifier la manière de composer *G1* et *G2* : En posant un 'filtre' sur le trou *1* on peut affaiblir l'intensité de 'ce' qui passe par le trou *1* relativement à ce qui passe par le trou *2*, quoi que cela puisse vouloir dire (car en fait on n'en sait strictement rien) ; ou *vice versa*. Il existe également des procédés pour 'retarder la progression de ce qui passe par l'un des trous, face à la progression de ce qui passe par l'autre trou'. L'ensemble des opérations que l'on exprime en ces termes modélisants d'affaiblissements ou retardements relatifs, engendre une nombre illimité de micro-états dont les manifestations observables sont en effect mutuellement distinctes (sans qu'on sache en fait 'pourquoi'), mais qui tous – chacun *via* une opération composée correspondante *G(G1,G2)* – sont liés à un même couple de deux opérations élémentaires de génération de départ, *G1* et *G2*.

En outre, on pense que les constatations formulées plus haut pour le cas de deux opérations de génération *G1* et *G2*, peuvent être généralisées à tout nombre fini d'opérations de génération (de trous). En tout cas à ce jour on n'a pas signalé des restrictions concernant ce nombre. Acceptons donc *l'hypothèse* de la possibilité d'opérations de génération composées *G(G1,G2,…Gn)* où *n* est fini mais quelconque et appelons cette hypothèse *le principe de composabilité des opérations de génération de microétats*. Etant donnée une opération de génération *G(G1,G2,…Gn)* bien définie, le microétat à génération composée correspondant sera dénoté *me$_{G(G1,G2,…Gn)}$*. Ce symbole dénote donc un seul microétat engendré par l'unique *opération de génération composée* mais qui, dans la façon d'en parler et de le représenter dans sa genèse et ses manifestations observables, est référé aux microétats *possibles* mais *pas réalisées* $me_{G1}$, $me_{G2}$,.... $me_{Gn}$ correspondants, respectivement, aux opérations $G1,G2,...Gn$ considérées chacune séparément.

### 2.2.5. Mutation du concept de "définition" d'une entité-objet-d'étude

D'ores et déjà la dynamique de construction, qui tout simplement s'est *imposée*, entraîne un premier pas dans le *no man's land* situé en dehors du domaine de la pensée classique. Mettons sous loupe la dynamique de cette transgression.

Le produit de l'opération de génération $G$ s'installe dans la démarche amorcée en tant que morceau – imaginé – de pure factualité physique, un morceau de factualité conçu *par le concepteur-observateur humain* comme portant des spécificités physiques relatives à $G$, mais qui sont tout à fait inconnues. Car dire que « ce qu'on étiquette $me_G$ est le microétat engendré par $G$ » ne renseigne nullement sur la façon d'être *spécifique* de l'entité (hypothétique) *particulière* elle-même qui – parmi toutes les entités subsumées sous le concept *général* de microétat – est la seule qui est étiquetée $me_G$. C'est dire exclusivement comment cette entité dénotée a été produite, pas comment elle 'est' ou comment elle se comporte.

De même, les expressions à sonorité modélisante qui interviennent dans l'introduction du concept d'opération de génération composée (comme dire d'un microétat $me_G$ qu'il 'passe' par tel trou, etc.), ne disent strictement *rien* concernant le mode d'être et le comportement du microétat correspondant dénoté $me_G$. Et en outre, on vient de le voir, ces expressions à sonorité modélisante peuvent être éliminées à la fin de la construction.

Bref, le fragment de conceptualisation accompli jusqu'ici a donc véritablement contourné l'impossibilité, au départ, d'utiliser des prédicats conceptuels pour définir un microétat (comme, pour définir une chaise particulière ou un certain couteau, on dirait respectivement «la chaise marron qui se trouve.....etc. » ou « le couteau très coupant, à poignée rouge...etc. »), ou de montrer du doigt, ou de *pointer vers*... à l'aide de contextes verbaux. Il a véritablement contourné le fait que rien, aucun moyen pratiqué dans la conceptualisation classique, n'est disponible pour que, à l'aide de cela, le microétat à étudier étiqueté '$me_G$' soit individualisé au sein de la catégorie générale dénommée 'microétat'; pour qu'il y soit individualisé d'une manière qui s'applique directement à lui *spécifiquement*. Malgré cette totale pénurie de façons usuelles de définir, nous avons pu produire pour le microétat à étudier, au niveau de connaissance *zéro* sur lequel nous nous sommes placés par hypothèse, *une définition a-conceptuelle, strictement **non-qualifiante***.

Ce qui importe dans cette réussite, c'est que dès que cette sorte de définition est acquise, elle permet de continuer le processus de construction d'une description, cependant que sans elle ce processus restait bloqué.

Bien sûr, il existe d'innombrables autres entités-objet que l'on ne définit qu'opérationnellement. Par exemple, une robe est elle aussi un objet que (souvent) on ne définit que par une suite d'opérations. Mais pendant qu'on fabrique une robe, on perçoit tous les substrats sur lesquels s'appliquent nos actions de génération, ainsi que les processus que ces actions produisent. Et quand la confection est terminée, on voit la robe, on la porte, etc.

Tandis que la définition d'un microétat, par l'opération physique qui produit ce microétat factuellement, descend jusqu'à un substrat physique placé par hypothèse à l'intérieur d'un domaine de factualité physique où règne un inconnu total. A partir d'un tel substrat :

On ne peut disposer d'aucune autre sorte de connecteur au domaine du connu, qu'un réceptacle conceptuel préfabriqué – toute une classe dénommée 'microétat' – que l'on immerge mentalement jusque dans ce domaine d'inconnu total, pour y accueillir comme dans un ascenseur conceptuel, le microétat particulier étiqueté $me_G$, et le hisser jusque dans *du conçu et dicible* où il soit rendu disponible pour être qualifié donc 'conceptualisé'.

Or sans *aucun* tel connecteur il n'est tout simplement pas possible de hisser un fragment de factualité entièrement non connu, jusqu'au contact avec le volume du conceptualisable. Si l'on veut accomplir cette sorte de contact, il *faut* parler, il *faut* écrire – avec des mots qui existent et que chacun peut comprendre – il *faut* investir du connu commun déjà constitué, si l'on veut le prolonger par du connu nouveau. Mais l'utilisation d'un réseau de concepts-et-mots ('microsystème', 'microétat', 'génération de microétat', etc.), qui conditionne une telle prolongation, ne dit pas plus sur le microétat particulier dont on parle, que ne dit la structure du filet utilisé par un pêcheur, sur le 'poisson' particulier qui y a été attrapé.

Sur la base de ces considérations, nous admettons désormais que *toute* opération *G* de génération d'un microétat comporte aussi – en dehors d'une opération physique – un connecteur conceptuel qui, à partir du connu déjà constitué, émet un filet d'accueil conceptuel-verbal qui guide pour spécifier l'opération physique à utiliser et offre un lieu d'accueil conceptuel du résultat de cette opération.

Cette spécification supplémentaire du concept d'opération de génération *G*, permet d'affirmer maintenant en toute rigueur qu'une telle opération introduit comme objet-étudier un morceau de factualité physique encore strictement non-qualifié, lui, spécifiquement. Qu'elle l'introduit d'une manière qui est entièrement indépendante de toute action de qualification passée ou future de cette entité-objet particulière. Toute qualification à proprement dire – *spécifique* de l'entité-objet *particulière* dénotée *$me_G$*, qui puisse, non pas la singulariser de manière *a-cognitive* au sein du factuel physique inconnu, mais la caractériser cognitivement *elle, en particulier*, au sein de la classe (posée) de tous les microétats – devra donc être réalisée ***ensuite***.

Cependant que toute "définition" au sens classique requiert *à la fois* l'introduction d'une entité-objet-de-description *et* certaines qualifications caractéristiques de celle-ci.

Ainsi, dans le cas d'un microétat encore strictement non connu – lui, *spécifiquement* – le concept classique de "définition" se trouve radicalement scindé en deux étapes *indépendantes* l'une de l'autre : Une première étape de *génération non qualifiante* qui absorbe en elle la fonction de mettre à disposition un 'microétat' particulier en tant qu'entité-objet-d'étude, et une étape de *qualification* de cette entité-objet-d'étude (de son étude à proprement parler), qui reste à construire.

*Le concept de définition de l'entité-objet **se sépare** du concept de qualification de cette entité.*

Cela est foncièrement nouveau par rapport à la conceptualisation classique. La pensée courante et les langages qui l'expriment – le langage courant mais aussi les grammaires, la logique, les probabilités, toute la pensée scientifique classique – n'impliquent pas une structure de cette sorte pour initier un processus de création de connaissances. Les entités-objet-d'étude sont partout introduites par des actions qui, d'emblée, sont plus ou moins *qualifiantes* : des gestes ostentatoires qualifiants (là, ici, celui-ci, etc.), ou par des référence à des contextes qualifiants, ou carrément par des qualifications seulement verbales. Ouvrons un dictionnaire. On y trouve « *chat* : un petit félin domestique, etc. ». Et dans les manuels scientifiques il en va souvent de même. Nous sommes profondément habitués à ce qu'une entité-objet et ses qualifications nous soient données *dans la même foulée*. Les grammaires définissent en général l'entité-objet d'une assertion descriptionnelle (d'une *proposition* au sens grammatical) en introduisant un nom d'*objet* au sens courant classique, 'maison', 'ciel', 'montagne' etc., suivi des nom(s) de *propriétés*, de prédicats (au sens grammatical) qui – eux – définissent en le qualifiant l'objet indiqué par le nom considéré. Tout cela est souvent exclusivement verbal et suppose la *pré-existence* des entités-objet, que ce soit au sens descriptionnel ou au sens grammatical général. Ceux-ci ne doivent qu'être sélectionnées dans le champ de l'attention, pour usage, étude, etc. Pas question de les créer physiquement. Et la sélection d'entités-objet, ce qui introduit dans le champ de l'attention telle ou telle parmi ces entités pré-existantes, ce sont des qualifications qui l'opèrent, des prédicats, non pas une opération physique non-qualifiante. La logique classique entérine cette façon de faire. Toute la pensée classique flotte dans le nuage de trompe-l'œil conceptuels que nous désignons par les mots 'objet' et 'propriété'.

Mais un 'microétat' n'est pas un "objet" au sens classique. Ce n'est qu'une entité physique hypothétiques, conçue à l'avance par prolongation de la physique classique, et que nous devons d'abord produire dans le rôle d'entité-(objet-*de-description*), si l'on veut pouvoir tenter ensuite de l'intégrer au domaine du conceptualisé. Et ce tour difficile exige impérativement que le référent courant du mot 'définition' soit soumis à une mutation de son sens classique ; une mutation telle qu'elle vide ce référent de tout contenu sémantique propre en le séparant radicalement du référent du mot 'qualification'.

### *2.2.6. Scission*

Pour clore cette étape **2.2** je rappelle ce qui suit, qui a été déjà mentionné, mais qui mérite d'être extrait et très fortement souligné.

La pensée classique du concepteur-observateur *humain* – qui est irrépressiblement modélisante, et, néc*é*ssairement, modélisante *dans le cadre des 'formes a priori kantiennes de l'intuition' d'espace et de temps* – a agi au cours du processus de construction comme un simple catalyseur qui ne se fixe pas dans le produit descriptionnel final. Le produit final ne consiste que *exclusivement* dans le fait essentiel que ce qu'on dénomme a priori 'microétat', quoi que ce soit, se trouve désormais crée et connecté à un réseau de conceptualisation préexistante, comme un poisson dans un filet que l'on n'a pas encore remonté et ouvert.

Entre les caractères de ces deux catégories descriptionnelles – d'une part *les actions cognitives du concepteur-observateur*, fondées sur des représentations classiques, et d'autre part le résultat non-classique de ces actions – *il s'est installé spontanément une scission*.

Cette sorte de miracle épistémologique est la conséquence directe de la décision méthodologique qui pose la relation de un-à-un symbolisée par $G \leftrightarrow me_G$ : C'est cette relation qui permet d'*isoler* de l'intérieur du concept classique de définition, la fonction vitale de mettre stablement à la disposition des concepteurs-observateurs, une entité-objet-d'étude à qualifier *plus tard*, et d'absorber cette fonction dans l'opération de génération $G$. Il s'est amorcé une chimie épistémologique dont il sera utile de noter les développements.

### 2.3. Qualifier un microétat

Sur la base acquise à partir de la question *Q1* il est désormais possible d'envisager d'étudier le microétat étiqueté $me_G$, c'est-à-dire d'essayer d'acquérir quelques données communicables sur sa manière propre, particulière, de se manifester. Il s'est créé une ouverture vers une éventuelle acquisition d'un savoir nouveau et *non*-hypothétique lié à des caractéristiques spécifiques au microétat $me_G$ créé par l'opération de génération $G$ considérée, et par aucune autre.

Cette ouverture, toutefois, n'aura été utilisée que lorsqu'on aura réussi à en tirer des manifestations observables par le concepteur-observateur humain, des manifestations qui impliquent le microétat $me_G$ et dont on puisse dire en quelque sens précisé qu'elles *qualifient $me_G$*. Mais quelles sortes de manifestations? Comment les concevoir? Comment les réaliser ? *Comment leur associer **des significations**?* Bref, on se trouve maintenant en présence de :

La question *Q2 :* Comment qualifier un microétat $me_G$ ?

Un chemin pour aborder la question *Q2* s'ébauche lorsqu'on commence par examiner comment nous qualifions habituellement.

### 2.3.1. Comment qualifions-nous habituellement ?
### Grille normée de qualifications communicables et consensuelles

Lorsqu'on qualifie un objet on le fait toujours relativement à quelque point de vue, quelque biais de qualification, couleur, forme, poids, etc. Une qualification dans l'absolu n'existe pas. Supposons alors que l'on recherche une qualification de couleur pour une entité-objet macroscopique. Notons tout de suite que le mot couleur lui-même n'indique pas une qualification bien définie. Il indique une nature commune à tout un ensemble ou spectre de qualifications, rouge, vert, jaune, etc. Il indique une sorte de dimension, ou terrain abstrait, bref un réceptacle ou *support sémantique* où l'on peut loger *toute* qualification qui spécifie une couleur bien définie. On pourrait alors dire, par exemple, que rouge est une 'valeur' (non numérique dans ce cas) que la dimension sémantique de couleur peut loger ou manifester, et que la couleur au sens général ne peut, elle, se manifester que par des valeurs de couleur. Ce langage peut paraître inutilement compliqué. Mais il apparaîtra vite que les distinctions introduites sont toutes nécessaires. Il en va de même pour ce qu'on appelle forme, poids, position, énergie, bref, pour tout ce qui indique un biais de qualification: Plus ou moins explicitement, mais toujours, une qualification fait intervenir deux paramètres de qualification, hiérarchisés : *(a)* une dimension 'sémantique' de qualification ; *(b)* un spectre de 'valeurs' qualifiantes porté par cette dimension ; *(c)* une (au moins) opération 'de mesure' (d'estimation) qui permet d'associer à certaines entités (celles qui peuvent être qualifiées via la dimension sémantique considérée) une valeur du spectre *(b)*. Bref, un genre le plus proche, des différences spécifiques, et au moins une modalité d'estimation de différence spécifique à l'intérieur de ce genre le plus proche, dans le cas d'entités qui admettent ce genre. Cela vient de la nuit des temps, porté sans doute par notre agencement biopsychologique.

Comment apprend-on quelle est la valeur de couleur d'un objet que l'on veut étudier du point de vue de la couleur? On le regarde. Cela peut s'exprimer aussi en disant qu'on assure une *interaction de mesure* de couleur entre l'objet et notre *appareil* sensoriel visuel. L'avantage de cette façon de dire est qu'elle introduit d'emblée un langage qui pourra convenir aussi bien aux examens scientifiques délibérés, qu'aux examens plus ou moins spontanés de la vie courante. Cette interaction (biologique) de mesure de couleur produit une sensation visuelle, une *quale* (qualité (latin), pluriel : qualia) que seul le sujet connaissant éprouve, donc peut connaître et reconnaître. Mais on admet que cette sensation visuelle se trouve en corrélation stable avec l'objet étudié, dans la mesure où cet objet et son état sont eux-mêmes stables, cependant que l'état du sujet percepteur est stable également, et 'normal'. En outre, par des apprentissages préalables qui impliquent des processus de comparaison et d'abstraction, le sujet finit par distinguer, dans sa sensation visuelle globale d'un lobjet-d'étude, cette dimension sémantique particulière dont le nom public est 'couleur', ainsi que les valeurs particulières de couleur par lesquelles cette dimension se manifeste à lui. Une valeur de couleur, telle qu'elle est ressentie par le sujet, est essentiellement indicible quant à sa qualité, sa *quale*, sa nature subjective intime. Mais par l'apprentissage de la correspondance entre cette *quale* et le *mot* que les autres prononcent en regardant la même entité physique (correspondance qui est constante dans le cas d'un groupe d'observateurs normaux et en situation observationnelle normale) le sujet arrive à *étiqueter* lui aussi cette *quale* particulière non communicable, par ce même *mot*, son nom public qui, lui, est communicable et consensuel : 'rouge', 'vert', etc. Cela lui permet de s'entendre avec les autres sujets humains en ce qui concerne les valeurs de couleur[17]. Bref, dans le cas de notre exemple, le sujet annoncera par *un mot consensuel* que le résultat de la mesure (de l'estimation de valeur) de couleur qu'il vient de réaliser à l'aide de ses yeux, est telle valeur de couleur, disons la valeur 'rouge'.

Mais un aveugle peut-il répondre à une question concernant la couleur d'un objet ? Ce n'est pas impossible. Il peut mettre l'objet d'étude dans le champ d'un spectromètre de couleurs connecté à un ordinateur vocal qui annonce en noms publics de couleurs les résultats de l'analyse spectrale qu'il opère. Ainsi le résultat produit par une interaction de mesure de couleur entre l'objet d'étude et un appareil de mesure de couleur qui est différent des appareils sensoriels biologiques de l'aveugle, est perçu par l'aveugle *via* un appareil sensoriel biologique dont il dispose, son ouïe. Cela le met en possession – directement – de l'expression publique du résultat de l'interaction de mesure considérée. La perceptibilité sensorielle visuelle, dont il manque, a été court-circuitée.

En outre, nonobstant l'absence de *toute* perception sensorielle visuelle, l'aveugle peut néanmoins se construire progressivement une certaine 'perception intellectuelle' subjective de la signification des noms de couleur, 'rouge', 'vert', etc. Cela est possible à l'aide de certains contextes (verbaux ou d'autres natures) qu'il est capable de percevoir. À partir de ces contextes, le total vide d'intuition qui, pour lui, se cache sous le mot publique 'rouge' qu'il a appris à utiliser, est osmotiquement pénétré d'une sorte de brume de qualités, de *qualia*, peut-être associée à un mélange d'images fugaces, comme un 'modèle' vague et changeant.

Cette dernière remarque n'est pas dépourvue d'importance parce que l'expérimentateur scientifique est comparable à l'aveugle en ce qui concerne le type de signification qu'il peut associer à certaines qualifications qu'il réalise sans aucun autre support perceptif que des marques 'sans forme' ni signification propre, ou même à l'aide seulement d'annonces par symboles recueillies sur des enregistreurs d'appareils qui sont extérieurs à son corps (pensons aux indications que l'on lit sur un écran de surveillance des phénomènes qui se passent dans un accélérateur du CERN). A la différence des qualifications déclenchées exclusivement par ses perceptions sensorielles biologiques, les significations associées à de telles marques ou annonces ne parviennent plus à la conscience de l'expérimentateur sous la forme de *qualia* liées à l'entité-objet-d'étude, il ne les perçoit *QUE* par l'intermédiaire d'étiquetages publics de résultats obtenus opérationnellement: *la dimension sémantique qui porte la 'valeur' impliquée par ces*

---

[17] L'essence de ce procédé est la même que dans l'entière physique macroscopique, les relativités d'Einstein inclusivement : Un ensemble d'observateurs perçoivent et examinent tous une même entité physique qui leur est extérieure. Ils étiquettent les résultats subjectifs de leurs observations selon des règles publiques conçues de façon à assurer certaines *formulations* qualifiantes finales qui sont invariantes d'un observateur à un autre.

*résultats, n'est plus **sensible***. Mais après coup, comme l'aveugle, l'expérimentateur se construit une certaine perception intellectuelle subjective de cette dimension sémantique. Quand il dit, par exemple : «j'ai mesuré une 'différence de potentiel électrique'», il associe à cette expression un certain mélange flou d'images portées par les formulations verbales qui ont formé ce concept dans son esprit et qui, au plan intuitif, le relient à son expérience sensorielle de départ.

On vient de détailler comment se constitue une *grille normée de qualifications communicables* dans le cas d'une couleur ou d'autres dimensions sémantiques. Une conclusion analogue vaut pour toute autre sorte de qualification physique communicable et normée, de forme, poids, etc., et même pour des qualifications communicables et normées abstraites.

Le schéma général affirmé au départ pour "une grille de qualification", peut donc maintenant être précisé : Une telle grille de qualification normée comporte toujours *(a)* une dimension sémantique ; *(b)* portant un ensemble de valeurs posées sur cette dimension sémantique ; *(c)* une procédure d'interaction de 'mesure' mettant en jeu, soit d'emblée et exclusivement un ou plusieurs appareils sensoriels biologiques du concepteur humain, soit d'abord un appareil de mesure artificiel sur les enregistreurs duquel s'enregistrent des marques physiques qui, elles, sont directement observables par le concepteur *via* ses appareils sensoriels biologiques (il doit *toujours* y avoir un effet *final* d'un acte de mesure, que le concepteur humain perçoive directement par ses sens biologiques) ; *et* en outre *(d)* une procédure de traduction de l'effet *final* perçu par le concepteur humain, en termes communicables et publiquement organisés désignant une, et une seule, parmi les *valeurs* portées par la dimension sémantique introduite.

Telle est l'essence du schéma qui fonctionne lors des actions de qualification de notre vie courante et, en général, lors des qualifications effectuées dans le cadre de la pensée classique. Au premier abord, les éléments de ce schéma et les phases de son édification peuvent n'apparaître que d'une façon confuse. Mais la présence de chaque élément et de chaque phase est toujours identifiable par analyse.

## 2.3.2. De la  grille usuelle de qualifications communicables, à une 'condition-cadre générale' pour la qualifiabilité d'un microétat

Imaginons un microétat *me*$_G$ spécifié par une opération de génération *G*. On veut le qualifier. Il est clair d'emblée que pour ce cas le schéma classique ne fonctionne plus tel quel, ne serait-ce que parce qu'un micro-*état* n'est pas disponible tout fait et stable, de manière à ce qu'on puisse lui 'appliquer' une grille qualifiante déjà disponible et stable elle aussi. Mais il peut y avoir d'autres difficultés, insoupçonnées. Il faut donc identifier explicitement, une à une, les spécificités qui apparaissent, et faire face systématiquement aux contraintes qu'elles comportent.

### 2.3.2.1. Préalables : Spécificités d'une opération de qualification d'un microétat

Les considérations qui suivent constituent une digression pour mise en contexte général, et qui reste extérieure au processus de construction entrepris ici.

Partons de l'exemple des couleurs considéré plus haut. Dans cet exemple on suppose que l'entité-objet préexiste et que, d'emblée, elle est apte à produire, lors d'une interaction avec nos yeux, l'effet de couleur dénommé 'rouge'. Ce résultat est supposé se produire en vertu d'une *propriété* dont on conçoit que d'ores et déjà elle est actuelle, réalisée en permanence *dans* l'entité-objet considérée, d'une façon intrinsèque, indépendante de toute interaction d'estimation de la couleur : la propriété d'émettre constamment des radiations électromagnétiques d'une longueur d'onde comprise dans l'intervalle étiqueté par le mot 'rouge'.

L'aveugle admet lui aussi que l'entité-objet dont il ne peut voir la couleur, préexiste avec constance, puisqu'il peut à tout instant la toucher, l'entendre tomber, etc. Et il admet également que cette entité-objet 'possède' de par elle-même une propriété qui, par interaction avec un appareil récepteur adéquat, produit le genre d'impression qu'on appelle 'couleur'. C'est sur cette double base qu'il soumet l'entité-objet, telle qu'elle est, à une interaction avec un spectromètre, c'est à dire avec un simple détecteur de couleur qui ne fait qu'enregistrer des effets de la propriété, préexistante dans l'entité-objet, d'émettre des radiations dans la bande du visible par l'homme normal.

Or dans le cas d'un microétat, la supposition de propriétés intrinsèques préexistantes dans l'entité-objet d'étude, ne vaut plus. Une telle supposition serait *inconsistante* avec la situation cognitive dans laquelle on s'est placé par hypothèse via le processus de conceptualisation que nous avons entrepris ici. Concernant – spécifiquement – un microétat particulier qui, via une opération de génération *G*, a été rerndu 'disponible' pour être étudié, mais qui n'a encore jamais été qualifié, on ne sait *rien* à l'avance en dehors de la manière dont il a été généré. (J'emploie de manière répétée ces curieuses précisions – spécifiquement, particulier – afin de constamment distinguer le savoir nouveau que l'on veut gagner, du type de 'savoir générique d'accueil' porté par l'affirmation *posée a priori* et absorbé dans *'G '*, que l'effet de l'opération de génération mise à l'œuvre est de la catégorie qu'on convient d'appeler 'un microétat'). Dans la phase initiale du processus de construction de connaissances concernant un microétat, celle de création de l'entité-objet-d'étude, nous n'assignons aucune propriété qui soit propre (si l'on peut dire) au microétat que nous avons étiqueté *'me<sub>G</sub>'* : il n'est doté que des propriétés générales assignées *à l'avance* au concept d'accueil d''un microétat'. Nous n'avons même pas encore d'indices directs que ce microétat existe. On *pose* qu'il existe en une relation de un-à-un avec l'opération *G* qui l'a produit) mais on ne le *sait* pas.

Dans ces conditions, afin d'arriver à associer au microétat *me<sub>G</sub>* telle valeur de telle dimension de qualification, il faudra *construire* tout un chemin. Il faudra, en général tout au moins, commencer par *changer* ce microétat hypothétique de telle manière qu'il produise quelque effet *observable*. C'est-à-dire, il faudra soumettre le microétat à étudier, à une *interaction* qui, selon des critères explicitement définis, justifie l'assertion qu'il s'agit bien d'une opération de 'mesure' de, précisément, *ce qu'on veut mesurer sur ce microétat*, donc d'une 'valeur' logée sur une dimension sémantique *définie*. Et il faudra que l'effet observable que l'on a produit puisse – de quelque façon bien définie elle aussi – *signifier une valeur précisée*, parmi toutes celles considérées comme possibles pour la dimension de qualification voulue (ce qui, si accompli, étayera aussi l'hypothèse que le microétat *me<sub>G</sub>* existe). Et enfin, il faudra rester vigilant concernant *l'entité* que – exactement – la valeur ainsi obtenue qualifie. Bref, en l'occurrence, il faudra *créer, conceptuellement et physiquement*, une qualification par une valeur du type sémantique recherché, et en outre il faudra construire un sens pour l'affirmation que cette valeur là peut être associée à un microétat. Explicitons ces mises en garde.

L'affirmation de nécessité de vigilance signale d'emblée un piège vers lequel le langage pousse subrepticement la pensée, et auquel il faut échapper. Lorsqu'on dit « je veux qualifier ceci » (cette pierre, ce lac), ce qu'escompte automatiquement un concepteur-observateur humain doté de la manière classique de penser, est un renseignement sur la manière d'être *de 'ceci' même*. Or dans notre cas, puisqu'on doit changer le microétat à étudier afin de le qualifier de manière observable, et peut-être même le changer radicalement, ce qui implique un processus de durée non-nulle, le résultat *final* observé impliquera déjà un *autre* microétat, différent du microétat à étudier $me_G$, celui qui a été initialement engendré par l'opération de génération $G$ en tant qu'entité-objet-d'étude. En outre ce résultat final ne qualifiera même pas *exclusivement* cet autre microétat changé non plus, il ne qualifiera que, globalement, *l'interaction de mesure* entre un appareillage et le microétat $me_G$, qui aura produit le changement de $me_G$ *et la manifestation finale observée*. En effet la valeur indiquée par la manifestation observée aura été produite par cette interaction de mesure *considérée globalement*, pas exclusivement et directement par ce que nous appelons le microétat $me_G$ à étudier. Il faudra surveiller de près cet ordre d'idées et notamment expliciter s'il existe un sens, dans ces conditions, dans lequel le résultat des qualifications construites peut être regardé comme une 'description du microétat étudié $me_G$' lui-même. Et si un tel sens semble définissable, il faudra l'expliciter avec rigueur.

Une autre remarque s'impose. Le formalisme mathématique de la mécanique quantique concerne spécifiquement les déplacements des microétats dans l'espace, puisqu'il s'agit d'une *mécanique*. Les dimensions sémantiques de qualification qui y interviennent sont indiquées par les mots 'position', 'quantité de mouvement', 'énergie', moment de la quantité de mouvement[18], espace, temps. Comment a-t-on pu arriver à associer de telles qualifications, à des microétats non perceptibles et encore strictement non connus, dont on ne savait même pas si, et en quel sens, on peut dire qu'il se *déplacent* ? Cela, sans se fonder sur aucun modèle? Car n'oublions pas que le modèle du genre bille en mouvement, que l'on associait à un microétat avant la construction de la mécanique quantique, par transfert direct de la conceptualisation newtonienne de la mécanique classique, a échoué, et que c'est à la suite de cet échec qui a laissé un *vide de modèle* que l'on a conçu la nécessité d'une autre conception sur les microétats. *Les acquis historiques qui au début 20$^{ème}$ siècle appelaient une nouvelle 'mécanique' des microétats, lançaient cet appel avec un mot, 'microétats', juste un mot qui s'était vidé de tout contenu établi.*

Un nouveau départ avait été marqué par le célèbre modèle onde-particule proposé dans la thèse de Louis de Broglie (de Broglie, L., [1924] et [1963]). Mais ce modèle n'a pas clairement survécu dans la formalisation finale acceptée généralement. En tout cas il n'y a pas survécu avec la même signification que dans la thèse et les travaux ultérieurs de Louis de Broglie. Il n'en est pas vraiment absent non plus. Il y subsiste implicitement dans les écritures mathématiques et dans le langage qui accompagne ces écritures, où il a instillé *une foule* de traces. Mais celles-ci ont diffusé et ont perdu les marques de leur origine. À tel point que, depuis Bohr et Heisenberg et jusqu'à ce jour, on affirme couramment que la mécanique quantique actuelle n'introduirait aucun modèle, ni de microsystème ni de microétat.

*Or cela est **certainement** inexact.*

Le but d'élaborer une mécanique des microétats a *dû* présupposer la signifiance de 'grandeurs mécaniques' pour ces entités inobservables et hypothétiques étiquetées par le mot 'microétats'. Dans la mécanique classique les grandeurs mécaniques n'ont été définies que *que* pour des mobiles macroscopiques. Leur signifiance pour des 'microétats' également n'a pu être qu'un postulat posé *a priori*. Un postulat admis sur la base de tout un ensemble d'indices expérimentaux et conceptuels-historiques, mais un postulat *a priori* tout de même. Ce postulat ne pouvait être justifié que d'une manière constructive, par la réalisation effective d'une représentation des grandeurs mécaniques à l'intérieur d'un tout nouveau doté d'efficacité prévisionnelle, constitué par un système d'algorithmes mathématiques et d'opérations physiques et appareils associés à ces algorithmes. On a *décidé* d'induire des 'grandeurs mécaniques' conçues initialement dans une discipline macroscopique, dans une représentation nouvelle liée à des dimensions d'espace et de temps dont les ordres de grandeur dépassent à un degré gigantesque les seuils de perception des organes sensoriels biologiques de l'homme. On a construit selon cette *décision* et la construction s'est justifiée a posteriori. Mais afin de pouvoir construire l'on a forgé des éléments formels – des '*opérateurs dynamiques'* – tirés d'une reformulation mathématiques des modèles classiques de mobiles qualifiables par des grandeurs mécaniques newtoniennes (à savoir la reformulation hamiltonienne-lagrangienne). Les algorithmes mathématiques liés à ces opérateurs ont été *adaptés* à des actions descriptionnelles conceptuelles, et à des opérations physiques, d'un type foncièrement différent de celui des actions descriptionnelles et des opérations physiques associées aux mesures classiques de grandeurs mécaniques[19]. Or la forme mathématique de ces éléments formels nouveaux, et *surtout* la structure de l'opération physique de mesure qui est juste **affirmée** de façon *in*-formelle comme 'correspondant' à tel ou tel opérateur formel, portent quasi systématiquement et ouvertement les marques de leur origine historique et d'une franche *modélisation* sur la base du modèle onde-particule de de Broglie.

---

[18] À ces qualifications fondamentales, ont été ensuite ajouté d'autres (spin, parité).

Afin de concrétiser, nous donnons immédiatement un exemple d'une telle opération de mesure, celle de la grandeur dynamique fondamentale de quantité de mouvement. Cet exemple est paradigmatique. Il relie explicitement ce que nous avons déjà exprimé ici, aux questions générales qui seront traitées plus loin.

Soit la grandeur – vectorielle – $X \equiv$'*quantité de mouvement'*, dénotée $p$[20]. Un acte de mesure sur un microétat, d'une valeur $Xn$ du spectre de cette grandeur fondamentale (dénotons-la $Xn \equiv p_n$), doit s'accomplir par la méthode 'time of flight'[21], de la manière suivante.

Soient, respectivement, $\delta E(G)$ et $\delta t(G)$ les domaines d'espace et de temps qui seront peuplés par une réalisation de l'opération $G$ de génération d'un exemplaire du microétat $me_G$ à qualifier. On place un écran sensible $\mathcal{E}$ – très étendu – suffisamment loin du domaine d'espace $\delta E(G)$ pour que ce domaine puisse être considéré comme quasi ponctuel par rapport à la distance $O\mathcal{E}$ entre $\delta E(G)$ et $\mathcal{E}$ mesurée le long d'un axe $Ox$ partant – en gros – de $\delta E(G)$ et tombant perpendiculairement sur $\mathcal{E}$ ; cependant que $\delta t(G)$ puisse être regardé comme négligeable par rapport à la durée moyenne qui s'écoule entre le moment $t_o$ de la fin de la réalisation de l'opération $G$ et le moment $t$ où l'on enregistre un impact sur $\mathcal{E}$.

*(a)* On accomplit effectivement une opération $G$ en notant le moment $t_o$ assigné à *sa fin*, i.e. celui assigné au début de l'existence du microétat $me_G$. (Notons que $t_o$ est une *donnée* de départ qui est caractéristique *de G*, ce n'est pas un enregistrement obtenu par l'acte de mesure sur $me_G$ qui doit suivre).

---

[19] Il s'agit de la définition d'un 'opérateur différentiel' associé à chaque grandeur mécanique classique, avec ses 'états propres' et ses 'valeurs propres', et de la spécification de l'opération physique par laquelle on doit réaliser l'interaction de mesure correspondante.

[20] Ici l'écriture en gras '$X$' n'indique que la nécessité, dans un formalisme mathématique non spécifié conçu pour décrire des microétats, de définir une *composition*, pour la dimension sémantique considérée, de trois autres dimensions sémantiques; une composition reliée à celle qui, pour un vecteur au sens classique, unit ses trois 'composantes', eu un 'vecteur').

[21] Feynman a souligné qu'une mesure de quantité de mouvement n'est compatible avec la théorie quantique des mesure **que** si elle est accomplie selon la méthode 'time of flight' (il s'ensuit que les mesures 'par trace' pratiquées souvent, ne sont pas réglementaires).

*(b)* Si *G* a comporté des champs, au moment $t_o$ *on les éteint*. Si entre le support d'espace-temps *δE.δt(G)* et l'écran *E* il préexiste des champs extérieurs ou des obstacles matériels, *on les supprime*. Sur la base de ces précautions l'*évolution de mesure* assignée à l'exemplaire du microétat $me_G$ créé par l'opération *G*, est posée être 'libre' (dépourvue d'accélérations) (notons que ces précautions se rapportent avec évidence à la présupposition, ***dans $me_G$***, d'une 'quantité de mouvement' *dont toute accélération **modifierait** la valeur vectorielle*).

*(c)* Après quelque temps il se produit un impact $P_n$ sur l'écran *E*. L'aiguille d'un chronomètre lié à *E* aqcuiert alors une position, disons $ch_n$, qui marque le moment $t_n$ de cet événement (l'ensemble des données qui concernent l'acte de mesure considéré ici, et qui pourraient changer de valeur dans un autre acte de mesure accompli sur un autre exemplaire du microétat $me_G$, sont indexés par un numéro d'ordre *n*). On dit que 'la durée de vol' (time of flight) – **mais *de qui? de quoi?*** – a été $\Delta t_n = t_n - t_o$.

*(d)* La valeur de la distance – vectorielle – ***$d_n$*** parcourue entre *δE(G)* et le point d'impact $P_n$, est ***$d_n = OP_n$***. Le carré de la valeur absolue de cette distance est $|d_n|^2 = d_{xn}^2 + d_{yn}^2 + d_{zn}^2$ où $d_{nx} \equiv O\!E$ est mesurée sur l'axe *Ox* et $d_{yn}$, $d_{zn}$ sont mesurés sur deux axes placés dans le plan de *E* et qui, avec *Ox*, déterminent un système de référence cartésien droit.

*(e)* On définit, respectivement, la valeur ***$p_n$*** mesurée pour la grandeur ***X≡p***, et sa valeur absolue $|p_n|$, selon les formules

$$p_n = m(d_n/\Delta t_n) \qquad\qquad |p_n| = m(\sqrt{(d_{xn}^2 + d_{yn}^2 + d_{zn}^2)}/\,\Delta t_n)$$

où *m* est la masse associée au microsystème dont on étudie le microétat $me_G$ (définie dans la physique atomique ou la théorie des particule élémentaires).

Ceci clôt l'acte de mesure considéré. Notons maintenant ce qui suit.

Dans le cas exposé plus haut les *manifestations physiques observables* produites par l'acte de mesure sont : le point $P_n$ et la position $ch_n$ de l'aiguille du chronomètre lié à l'écran *E*. Ces manifestations ne sont *pas* directement des valeurs numériques, ni n'en 'possèdent'. Ce sont seulement des *marques physiques* perceptibles, disons $\mu_{1n}$ et $\mu_{2n}$, respectivement, produites par l'acte de mesure, sur les deux 'enregistreurs' de 'l'appareil' de mesure qui a été conçu pour accomplir cet acte de mesure (l'appareil étant constitué du chronomètre associé à l'opération *G*, de l'extincteur de champs extérieurs, de l'écran *E* et du chronomètre lié à l'écran).

Les *significations* associées aux manifestations observables enregistrées, ainsi que les valeurs numériques associées à ces significations, à la fois, sont définies par: la manière de *concevoir* un acte de mesure de la grandeur *X* ≡ 'quantité de mouvement' assignée à un microétat et dénotée ***p*** et par les relations posées ***p=m(d/Δt)=mv***, *Δt=t-t_o* et $|d| = \sqrt{(d_x^2 + d_y^2 + d_z^2)}$, $\Delta t_n = t_n - t_o$ et $|d_n| = \sqrt{(d_{xn}^2 + d_{yn}^2 + d_{zn}^2)}$ dont les trois premières définissent une fonction générale '*f*' de structure interne de '***p***', à savoir ***p** = f(m, d, Δt)*, et les autres permettent de calculer la valeur ***$p_n$*** mesurée pour ***p*** en cohérence avec la fonction de structure '*f*' et sur la base des deux manifestations physiques observables $\mu_{1n} \equiv P_n$ et $\mu_{2n} \equiv ch_n$.

On voit clairement sur l'exemple donné que :

- La quantité de mouvement ***p*** associée au microétat à étudier $me_G$, est définie de la manière *classique*, par ***p=mv***.

- Les prescriptions pour calculer la valeur numérique ***$p_n$*** sont exactement celles que l'on devrait suivre pour un mobile classique libre.

- Cela *seul* justifie que l'évolution de mesure posée pour $me_G$ exige d'éteindre tout champ et d'éliminer tout 'obstacle', i.e. de supprimer tout ce qui – selon la mécanique *classique* – modifierait par des 'accélérations' la valeur **p** 'de départ' et que c'est cette exigence qui constitue la base sur laquelle on conçoit (plus ou moins explicitement) qu'un acte de mesure 'time of flight' est *'convenable'* par ceci précisément *qu'il change le microétat étudié* $me_G$ *d'une manière qui n'altère pas aussi la valeur à mesurer de* **p**[22].

Bref, l'entière procédure qui vient d'être exposée serait entièrement arbitraire – et même *inconcevable* – en l'absence d'un modèle macroscopique classique pour lequel on a décidé d'assigner, par prolongement, une signification, aussi, pour les microétats. Et le prolongement est accompli via l'acceptation non-déclarée du modèle one-particule de Louis de Broglie.

Tout ce qui vient d'être dit plus haut concerne le formalisme mathématique de la mécanique quantique. Mais, comme j'ai déjà souligné, cela reste extérieur à la démarche constructive tentée dans la première partie de ce travail : si l'on supprimait tout ce qui vient d'être dit sur la mesure 'time of flight', le travail constructif resterait le même et il n'en pâtirait pas dans sa structure, il n'en pâtirait que dans son degré d'intelligibilité immédiate. Car la démarche qui est en cours d'être développée ici *est soumise délibérément à des contraintes plus fondamentales et conceptuellement plus ascétiques que celles qu'a subies la construction du formalisme de la mécanique quantique*. La construction de la mécanique quantique avait un autre **but**, à savoir de représenter *mathématiquement* une **mécanique** des microétats. Ce but-là, d'emblée, plaçait les bases de la recherche sur un niveau de conceptualisation subséquent à celui posé ici, à savoir un niveau où ce concernant quoi il fallait construire des connaissances avait déjà été doté d'un *modèle* dont l'efficacité s'était manifestée et qui, même si on ne le plaçait pas à la base de l'action descriptionnelle, néanmoins guidait cette action fortement ; cependant que les qualifications recherchées avaient déjà été conceptualisées dans la mécanique classique.

Cependant qu'ici l'on recherche une structure descriptionnelle qualitative qui découle de contraintes imposées – *exclusivement* – par les modes humains généraux de conceptualiser et la situation cognitive où se trouve un observateur-concepteur qui veut construire des connaissances *absolument quelconques* concernant des 'microétats', en partant du niveau de connaissance zéro où – hormis le concept général de 'microétat' – rien n'est donné, ni la manière de se *doter* d'un microétat à qualifier, ni le concept général de qualification définie pour un microétat, ni des modalités d'obtenir telle ou telle sorte de qualification concernant un microétat dont par avance on se serait doté. Ce but là, et l'ascèse qu'il exige, sont commandés par l'objectif d'ériger un milieu conceptuel d'immersion et de référence, maximalement général mais rigoureusement structuré, qui permette ensuite de percevoir, par comparaison et entièrement tiré hors des brouillards, le statut conceptuel de chaque élément descriptionnel qui intervient dans le formalisme quantique. En ces conditions, l'utilisation – **ici** – de modèles de prolongement de la mécanique classique, déborderait notre règle du jeu. Et surtout elle continuerait de *cacher les sources et les implications de la représentation mathématique, dans la mécanique quantique, des mesures de grandeurs mécaniques accomplies sur des microétats*. Or, parmi les problèmes soulevés par le formalisme quantique, celui des mesures quantiques est probablement celui qui a nourri le plus grand nombre de questionnements et de controverses.

---

[22] Cette condition, généralisée, est imposée dans la mécanique quantique à toute mesure d'une grandeur mécanique.

Ces spécifications soulignent que rien de ce qui vient d'être mis en évidence n'entraîne que l'insertion d'emblée, dans le formalisme quantique, de modèles de prolongement de la mécanique classique, soit répréhensible de quelque point de vue : il est bien clair que *sans de telles insertions il n'y aurait simplement pas de **mécanique quantique**.*

### *2.3.2.2. La condition-**cadre** et la grille primordiale de qualification pour un microétat*

Le contenu de ce paragraphe est crucial[23]. Il exige une attention particulière. Seulement s'il est assimilé à fond il pourra permettre une réelle compréhension de la solution du problème des mesures quantiques proposée dans la deuxième partie de ce travail.

J'ai mis en évidence que selon une définition classique générale plus ou moins explicite une "grandeur qualifiante" doit introduire une "dimension sémantique" et un "spectre" formé de l'ensemble de toutes les "valeurs" de cette dimension sémantique qui sont prises en considération[24]. Mais dans le cas d'un microétat $me_G$ correspondant à une opération de génération $G$ tout simplement on ne dispose d'aucune dimension sémantique préconstruite dont on sache qu'elle peut s'appliquer à $me_G$, et en quel sens. Comment concevoir une "grandeur qualifiante" face au but de qualifier *un 'microétat'*, et *avant toute modélisation, au niveau **zéro** de **connaissances** pré-acquises concernant le **concept d'accueil** désigné par ce mot introduit par prolongement du langage de la microphysique classique? COMMENT ?*

Dans les conditions cognitives où l'on se trouve en ce qui concerne ce but, tout ce qu'un concepteur-observateur humain peut faire est de tenter de soumettre le microétat $me_G$ qui correspond à une opération de génération $G$, à une "interaction-test" physique – dénotons-la **𝗫** – réalisable à partir du niveau macroscopique à l'aide de manipulations d'un groupement d'appareil macroscopiques. Dénotons globalement $A(\textbf{𝗫})$ un tel groupement. Il apparaîtra dans ce qui suit que, malgré la non perceptibilité d'un microétat lui-même, cela permet néanmoins de fonder sur des interactions-test **𝗫** une toute première strate de grilles de qualification et une première strate descriptionnelle correspondante, une strate proprement **primordiale** ; mais si et seulement si l'on peut assurer les deux conditions suivantes:

* Chaque réalisation par $A(\textbf{𝗫})$ d'une interaction-test **𝗫** avec un exemplaire du microétat à étudier doit produire un groupe de $m$ marques physiques directement observables *{$\mu_k$}, k=1,2,...m,)* (*m*: un entier qui en général est petit, rarement plus grand que *4* et souvent égal à *1*)) sur/par les enregistreurs de l'appareil $A(\textbf{𝗫})$ : marque sur un

---

milieu sensible à des micro-impacts, un son émis par un compteur lors d'interaction avec une entité microscopique invisible, position d'aiguille d'un chronomètre, etc.[25].

Cette première condition est triviale.

** L'on doit arriver à définir un *codage* de tout groupe de marques $\{\mu_k\}$, $k=1,2,...m$, au sens suivant. Soit une interaction-test $\boldsymbol{X}$ entre le microétat à étudier $me_G$ et l'appareil $A(\boldsymbol{X})$. Soit $n=1,2,....$ un indice qui, dans une suite *indéfinie* de répétitions de la succession $[G.\boldsymbol{X}]$ d'une réalisation de l'opération $G$ de génération de $me_G$ suivie immédiatement par une réalisation de l'opération-test $\boldsymbol{X}$, singularise *une* de ces répétitions, fixée mais quelconque. Et soit $\{\mu_{kn}\}$, $k=1,2,...m$ le groupe des $m$ marques physiques directement observables produit par la réalisation d'indice $n$ de la succession $[G.\boldsymbol{X}]$, dénotée $[G.\boldsymbol{X}]_n$. Si a *tout* groupe $\{\mu_{kn}\}$, $k=1,2,...m$ de marques produit par la réalisation d'une succession $[G.\boldsymbol{X}]_n$, distinguée par un indice $n$ quelconque, l'on peut de quelque manière faire correspondre une "valeur" $vj$ – numérique ou *non* – *et une seule*, appartenant à un "spectre" $(v1, v2,....vj...vJ)$ avec $j=1,2,...J$ et $J$ *fini*, qui puisse caractériser l'interaction-test $\boldsymbol{X}$, alors nous dirons que l'on a défini pour les groupes de marques $\{\mu_k\}$, $k=1,2,...m$ produits par l'interaction-test $\boldsymbol{X}$ un codage

$$C(\{\mu_k\}, k=1,2,...m) \leftrightarrow vj \quad \text{avec } j=1,2,....J$$

qui permette de communiquer concernant les effets de cette interaction, faire des confrontations intersubjectives, rechercher des consensus, bref, les intégrer à la base d'une démarche scientifique [26, 27].

Cette deuxième condition, à la différence de la première, est loin d'être triviale. Au contraire, dès qu'on la considère elle soulève des questions qui peuvent sembler insurmontables. En effet, mettons sous loupe la situation conceptuelle.

Un groupe de marques physiques directement observées $\{\mu_k\}$, $k=1,2,...m$, – *en tant que marques* – déclenchent bien dans l'esprit d'un observateur humain un "phénomène" au sens psychophysiologique de Husserl. Mais les *qualia* liées à *ce* phénomène psychique là, *n'établissent aucune connexion **avec le microétat $me_G$**, auquel pourtant elles sont conçues comme étant reliées certainement via l'interaction-test $\boldsymbol{X}$ qui est posée avoir produit les marques ; ce sont des qualia propres* – **exclusivement** – *aux marques observées* : Entre le microétat $me_G$, juste présupposé exister, et le phénomène psychophysiologique déclenché par les marques enregistrées, il y a une radicale coupure sémantique observable.

Et plus en amont, un ensemble de marques physiques $\{\mu_k\}$, $k=1,2,...m$, **n'indique même pas quelque catégorie sémantique dénommée**.

Il est évidemment essentiel de dépasser ce hiatus qui sépare la sémantique sensorielle *propre aux groupes de marques $\{\mu_k\}$, $k=1,2,...m$*, de la notion de microétat '$me_G$' qui, elle, telle qu'elle a été introduite initialement, est par construction encore entièrement vide de

---

[25] Le microétat $me_G$ *lui-même*, isolément, ne 'manifestera' rien. *Tout* ce qui sera observable sera effet de *l'interaction* entre $A(\boldsymbol{X})$ et le microétat $me_G$.

[26] Ce cardinal $J$ est posé être *fini*. Ceci correspond à ce qui se passe toujours factuellement – par construction – car toujours on définit **une zone d'observation finie** et des grilles d'observation munies d'unités, notamment d'espace et de temps, qui **discrétisent.** Ceci devient très clair et fondamental dans le cadre de la méthode générale de conceptualisation relativisée (MMS [2006]) où l'entière construction est développée sous condition d'effectivité.

[27] L'exemple donné dans le paragraphe précédent concernant la grandeur 'quantité de mouvement' définie dans la mécanique quantique, peut aider à vite comprendre cette condition de codabilité.

*toute* sémantique propre, tant qu'on n'a pas encore assuré les conditions de possibilité d'engendrer une telle sémantique, précisément en associant à '*me_G*' des significations, i.e. des qualifications.

A fortiori, dans cette phase de construction de connaissances absolument premières où nous place par hypothèse la démarche amorcée ici, on ne dispose évidemment pas non plus de représentations *formelles* de "grandeurs" qui sont *conçues* et *représentées* à l'avance [28].

D'autre part, en absence de tout codage, comment pourrait-on communiquer concernant les résultats factuels $\{\mu_k\}$, $k=1,2,...m$ de l'interaction-test **X**, chercher des consensus, en développer des conséquences, l'insérer dans la conceptualisation scientifique intersubjective ? Que faut-il conclure de cette situation conceptuelle ? Sommes nous piégés dans un situation circulaire inextricable?

Non, la réponse qui se révèle et *s'impose* n'est pas celle la. Elle est du même genre que celle exprimée par la décision méthodologique *DM1*: Il faut se secouer avec violence et se débarrasser définitivement de cette tendance insidieuse qui a subrepticement pris possession de l'esprit des concepteurs-observateurs humains dont la pensée scientifique s'est forgée selon le postulat implicite que l'on se trouverait à la recherche de vérités préexistantes et que l'action à dérouler en recherche scientifique serait une action de pure découverte. Une fois de plus, il ne s'agit nullement de cela. Il s'agit d'organiser une structure de manières d'agir et de dire qui, dans le contexte considéré, permette de *construire* des connaissances qui permettront de prédire et de modéliser. Il s'agit – sans s'emprisonner dans des buts impossibles de 'découvertes' de 'faits vrais' qui en ce cas tout simplement ne préexistent pas – de trouver *les posés méthodologiques* qui permette d'organiser *un pont qui unisse ce dont on dispose, à ce qu'on veut atteindre.* Un codage $C(\{\mu_k\}, k=1,2,...m) \leftrightarrow (vj, j=1,2,....J)$ est précisément un tel pont et il faut expliciter et poser des conditions suffisantes pour le construire. Point. Voilà le problème tout nu.

Mais *en quoi* une telle condition suffisante pourrait-elle consister ? Au premier abord la question est déroutante. Si l'on considère d'abord le cas d'une marque $\mu_{kn}$ physique et *directement* perceptible qui est isolée, alors – intrinsèquement – une telle marque est en général beaucoup *trop largement* catégorielle : juste un point d'impact sur un milieu sensible, comme tout autre tel point ; ou un son produit par un appareil de comptage, comme tout autre tel son. C'est dire qu'une telle marque ne comporte *pas* d'aspects *intrinsèques, propres,* qui soient à la fois perceptibles et *spécifiques de cette **seule** marque-là.* De par ses caractères *propres*, une seule marque physique directement perceptible $\mu_k$ n'offre aucune prise à partir de laquelle il soit possible de lui faire correspondre une "valeur" $Xj$ – numérique ou *non* – *et une seule*, appartenant à un "spectre" *(v1,v2,....vj...vJ)* avec $j=1,2,....J$ et *J fini*, qui puisse caractériser l'interaction-test **X**,

Si l'on considère alors un cas où se forme un groupe de *plusieurs* marques, et on considère ce groupe comme un tout, on peut chercher une *caractéristique globale, observable et **spécifique***, que *tout* tel groupe posséderait *nécessairement* quel que soient **X** et l'indice d'ordre *n* de la succession *[G.**X**]_n* considérée. Or sur cette voie il apparaît qu'une telle caractéristique existe : deux groupes de marques $\{\mu_{kn}\}$, $k=1,2,...m$ et $\{\mu_{kn'}\}$, $k=1,2,...m$, $n'\neq n$, possèdent – nécessairement et pour tout $\mathbf{X}$ et tout $(n,n')$ – des configurations d'espace-temps, ou d'espace seulement ou de temps seulement – qui peuvent être mutuellement différentes ou identiques. Car, étant *une* entité *physique* perceptible par le concepteur-observateur humain, *une marque* $\mu_k$ *est nécessairement perçue avec une location d'espace-temps*. Dans le cas d'une marque isolée cette location reste *extrinsèque* à la marque, et en général elle n'en est pas spécifique. Mais lorsqu'il s'agit d'un groupe de plusieurs marques, et face à un référentiel d'espace-temps doté d'une origine et d'unités définies, les locations individuelles constituent une certaine *configuration* d'espace-temps propre à ce groupe, intériorisée par ce groupe, qui *peut* en être spécifique[29].

---

[28] Comme la grandeur dynamique quantité de mouvement $p=f(m,d,\Delta t)=m(d/\Delta t)=mv$ (qui dans le formalisme quantique, via les paranthèses de Poisson, acquiert une autre représentation mathématique, $(ih/2\pi)\partial/\partial x$, tout en conservant la même représentation conceptuelle qu'en physique classique).

Suivons ce fil.

Associons à l'appareil $A(\mathbf{X})$ un référentiel d'espace-temps *Réf(ET)* doté d'une origine d'espace-temps fixée et d'unités spécifiées. *Lors du début de chaque nouvel enregistrement d'une durée, l'origine de temps dans Réf(ET) est remise à zéro, comme dans le cas des tests sportifs*. Une réalisation de la succession *[G.$\mathbf{X}$]* correspondante au microétat étudié $me_G$, produira un groupe de marques $\{\mu_k\}$, $k=1,2,...m$ ayant une configuration d'espace-temps donnée relativement à *Réf(ET)* ; dénotons-la *Config.ET($\mathbf{X}$,Réf(ET))*. (En conséquence du zéro du temps renouvellé pour chaque succession *[G.$\mathbf{X}$]* l'estimation de temps n'aura qu'une signification *locale*, relative à la succession considérée).

Soit maintenant une suite *[G.$\mathbf{X}$]$_n$*, $n=1,2,....N$ de réitérations de la succession d'opérations *[G.$\mathbf{X}$]*, avec $N$ un entier très grand. Elle produira $N$ groupes $\{\mu_{kn}\}$, $k=1,2,...m$. Dénotons par *Config.ET($\mathbf{X}$,n,Réf(ET))* la configuration d'espace-temps face à *Réf(ET)*, du groupe produit par la succession *[G.$\mathbf{X}$]$_n$*[30].

Si l'on veut éviter des particularisations arbitraires on doit concevoir qu'une 'bonne' interaction-test $\mathbf{X}$ produit en général, lors de répétitions, des groupes de marques dont les configurations *Config.ET($\mathbf{X}$,n,Réf(ET))* ne seront pas toutes identiques. Mais nous admettons d'autre part qu'en général chaque telle configuration ne sera pas non plus différente de *toutes* les autres. Car si c'était le cas, l'interaction-test $\mathbf{X}$ serait une source de parfait hasard (randomness), ce qui dans d'autres contextes serait précieux, mais *ici* conduirait à éliminer $\mathbf{X}$ de la catégorie des candidats inintéressants pour être répertoriés en tant que 'bonnes' interactions-test. En effet ici nous recherchons à discerner les conditions à imposer à un processus de "qualification" dont les résultats soient dotés de *régularités statistiques (en général) capables de caractériser le microétat étudié via l'interaction-test considérée, et cela exige un nombre **fini** de qualifications possibles, stables et mutuellement distinguables sans ambiguïté*. Nous exigeons donc explicitement que, avec une 'bonne' interaction-test $\mathbf{X}$, l'on voie se réaliser tantôt une configuration d'espace-temps *Config.ET($\mathbf{X}$,n,Réf(ET))* du groupe de marques $\{\mu_{kn}\}$, $k=1,2,...m$ enregistré, tantôt une autre, parmi un nombre *fini H* de telles configurations possibles, stables et mutuellement distinguables sans ambiguïté. (En particulier le nombre *H* peut se réduire à *1*, mais en général il doit être conçu comme étant différent de *1* si l'on ne veut pas introduire une exigence arbitrairement restrictive). Alors, étant donné que pour *toute* valeur de l'indice *n* l'on obtient l'une ou l'autre des *H* configurations possibles pour le groupe *{μ<sub>kn</sub>}*, *k=1,2,...m* de marques observables enregistrées, nous pouvons réindexer les configurations d'espace-temps par un indice *h=1,2,...H* au lieu de l'indice *n*. Donc désormais nous écrivons *Config.ET($\mathbf{X}$,h,Réf(ET))*.

---

[29] Penser au cas paradigmatique d'une mesure de quantité de mouvement par la méthode du "temps de vol" exposée dans *2.3.2.1*. Dans ce cas la valeur recherchée de la grandeur mesurée (quantité de mouvement) a été identifiée à l'aide de **la durée** $\Delta t_n=t_n-t_o$ et la **distance** $|d_n|=\sqrt{(d_{xn}^2+d_{yn}^2+d_{zn}^2)}$ fournies par l'appareillage **sur la base des marques physiques directement observées "$t$" et un point sur l'écran sensible**. L'identification a été faite par un calcul dicté par la structure fonctionnelle du concept qualifiant *préconstruit* de 'quantité de mouvement' $p=f(m,d,\Delta t)=m(d/\Delta t)=mv$. On perçoit l'analogie, même si *la situation considérée ici est plus fondamentale que celle qui concerne toute mesure de la mécanique quantique*, et notamment celle pour 'time of flight', puisqu'ici aucun concept préconstruit de grandeur qualifiante n'est donné.

[30] L'exemplaire du microétat $me_G$ qui avait été mis en jeu aura, en général, été 'consommé' après chaque succession *[G.$\mathbf{X}$]*, il sera changé, devenu non réutilisable.

L'on peut donc assigner à chaque groupe de marques *{μ<sub>k</sub>}, k=1,2,...m* rencontré dans une suite *{μ<sub>k</sub>}, k=1,2,....m, n=1,2,....N* d'enregistrements de tels groupes, une valeur donnée de l'indice *h* et une seule de la configuration d'espace-temps *Config.ET($\mathbf{X}$,Réf(ET)* correspondante. Ainsi l'on aura finalement établi un codage du type *C({μ<sub>k</sub>}, k=1,2,...m)* ↔ *vj* avec *j=1,2,....J*, à savoir un codage fondé sur les localisations face aux dimensions-"cadre" d'espace et de temps (cf. MMS [2002A], [2002B], [2006]). Dénotons par *Cod.cadre(ET)* un tel codage :

$$\textbf{Cod.cadre(ET) :} \quad C(\{\mu_k\},\ k=1,2,...m) \leftrightarrow Config.ET(\mathbf{X},h,Réf(ET)),\ avec$$
$$h=1,2,...H$$

Le mot 'cadre' symbolisé dans la dénomination de ce codage souligne qu'il n'y intervient strictement aucune référence à tel ou tel **contenu** qualifiant classique connu à l'avance ("grandeur mécanique" (énergie, vitesse, quantité de mouvement, position) ou quelque "grandeur" classique d'une autre nature) qui soit assignable au microétat qualifié lui-même.

Notons maintenant que :

*(a)* Le concept dénoté *Config.ET($\mathbf{X}$,Réf(ET)* peut être regardé comme une dimension sémantique *'liée'* au microétat étudié *me<sub>G</sub>* (bien que pas perçue à la manière classique comme étant "possédée" par lui) ;

*(b)* Cette dimension sémantique porte un spectre de valeurs, *{Config.ET($\mathbf{X}$,h,Réf(ET))}, h=1,2,...H* ;

*(c)* Toute réalisation *[G.$\mathbf{X}$]<sub>n</sub>* d'une succession *[G.$\mathbf{X}$]* produit un groupe corrspondant *{μ<sub>kn</sub>}, k=1,2,...m* de marques physiques observables sur les enregistreurs de l'appareil *A($\mathbf{X}$)* ;

*(d)* Ce groupe de marque est codable via le codage *Cod.cadre(ET)* en termes d'une valeur *Config.ET($\mathbf{X}$,h,Réf(ET))* portée par la dimension sémantique *Config.ET($\mathbf{X}$,Réf(ET)*.

Donc toutes les données sont réunies afin d'associer à un microétat quelconque la suivante *grille de qualification primordiale liée à l'interaction-test $\mathbf{X}$ et définie pour un microétat, gq.primord.($\mathbf{X}$).me<sub>G</sub>* :

**Def.(gq.primord.($\mathbf{X}$).me<sub>G</sub>).** L'association

*[Config.ET($\mathbf{X}$,Réf(ET)), {Config.ET($\mathbf{X}$,h,Réf(ET))} où h=1,2,...H, [G.$\mathbf{X}$],*
*Cod.cadre(ET)]*

constitue *une grille de qualification primordiale définie pour le microétat me<sub>G</sub>.*

Cette transposition au cas d'un microétat, du concept classique de grille de qualification *vide donc ce concept de tout contenu sémantique construit et dénommé préalablement et que l'on puisse considérer comme étant 'possédée' exlusivement par le microétat considéré lui-même*. Elle crée de toutes pièces un type nouveau de contenu sémantique explicitement et spécifiquement lié aussi à l'interaction-test $\pmb{X}$. Et nous avons montré sous loupe pourquoi – inévitablement – cette transposition sépare à son intérieur ce vide de toute sémantique préexistante de type classique. Nous introduisons donc le nouveau langage approprié suivant.

Parce que *l'unique* effet directement perceptible d'un acte d'interaction-test $\pmb{X}$ consiste en un groupe *{μ$_{kh}$}, k=1,2,...m* de marques physiques observables produites sur/par les enregistreurs de l'appareil $A(\pmb{X})$, dépourvues de toute qualia assignable au microétat examiné lui-même, nous dirons qu'il s'agit d'un groupe de marques *"de transfert"* sur les enregistreurs de $A(\pmb{X})$ (ou des marques "transférées").

Parce que la genèse de ce groupe de marques implique de manière non séparable le microétat étudié *et* l'appareil $A(\pmb{X})$, nous parlerons d'une manifestation observable *reliée* au microétat étudié – *pas* d'une manifestation "de" ce microétat.

Et parce que l'entité qualifiée n'avait jamais encore été conceptualisée auparavant, nous parlerons de qualification transférée *primordiale* ou *de base*.

Nous venons de définir *une strate de qualifications primordiales transférées liées à un microétat me$_G$*. C'est une strate de connaissances extrêmales qui, pour tout microétat donné, établit une référence fixe valide pour toute autre qualification de ce microétat.

L'on pourra désormais confronter de façon intersubjective ces qualifications de référence fixe, établir des consensus les concernant, les développer vers le haut de la verticale des conceptualisations via d'autres qualifications : Nous venons de poser une petite dalle de fond sur le plancher du volume du *conceptualisé*. Car en effet il s'agit d'un premier accès au conceptualisé, au connu. La décision méthodologique *DM1* par laquelle nous avons posé la relation de un-à-un $G \leftrightarrow me_G$ afin de spécifier un microétat en tant qu'objet d'étude, ne nous avait pas encore extrait de ce vide de toute connaissance de, spécifiquement, *me$_G$*, où nous avons placé délibérément le début de la démarche tentée ici. *On ne "connaît" que par des qualifications*. Ce qui n'a jamais été qualifié n'est pas "connu". L'opération de génération de $G \leftrightarrow me_G$ était encore strictement non qualifiante. Mais le codage-cadre *C({μ$_k$}, k=1,2,...m) $\leftrightarrow$ h, avec h=1,2,...H* établit enfin un brin de connaissance d'un type strictement primordial lié *spécifiquement* au microétat *'me$_G$'*. Un enchaînement

$$(\pmb{POC})_h \equiv [G\text{-}me_G\text{-}DM1\text{-}\pmb{X}\text{-}\{μ_{kh}\},\ k=1,2,...m], \quad h=1,2,...H$$

constitue un petit *pont opérationnel-conceptuel* qui, pour la valeur *h* de l'indice de configuration d'espace-temps des marques observables, relie l'accomplissement d'une opération de génération *G*, à l'effet observable *{μ$_{kh}$}, k=1,2,...m* de la performance immédiatement successive à *G* d'une opération d'interaction-test $\pmb{X}$, *conceptualisé en termes de cette goutte de conceptualisation primordiale dénotée Config.ET($\pmb{X}$,h,Réf(ET))*, *h=1,2,...H*. Dans une suite de successions *[G.$\pmb{X}$]$_n$, n=1,2,....N* nous disposons désormais, pour chaque valeur de l'indice *n*, d'un tel petit pont jeté au-dessus de ce vide de perceptibilité directe et de *qualia* qui caractérise la situation cognitive dans laquelle on se trouvait encore à la suite de la performance de l'opération *G* seulement, liée à *me$_G$* via *DM1*.

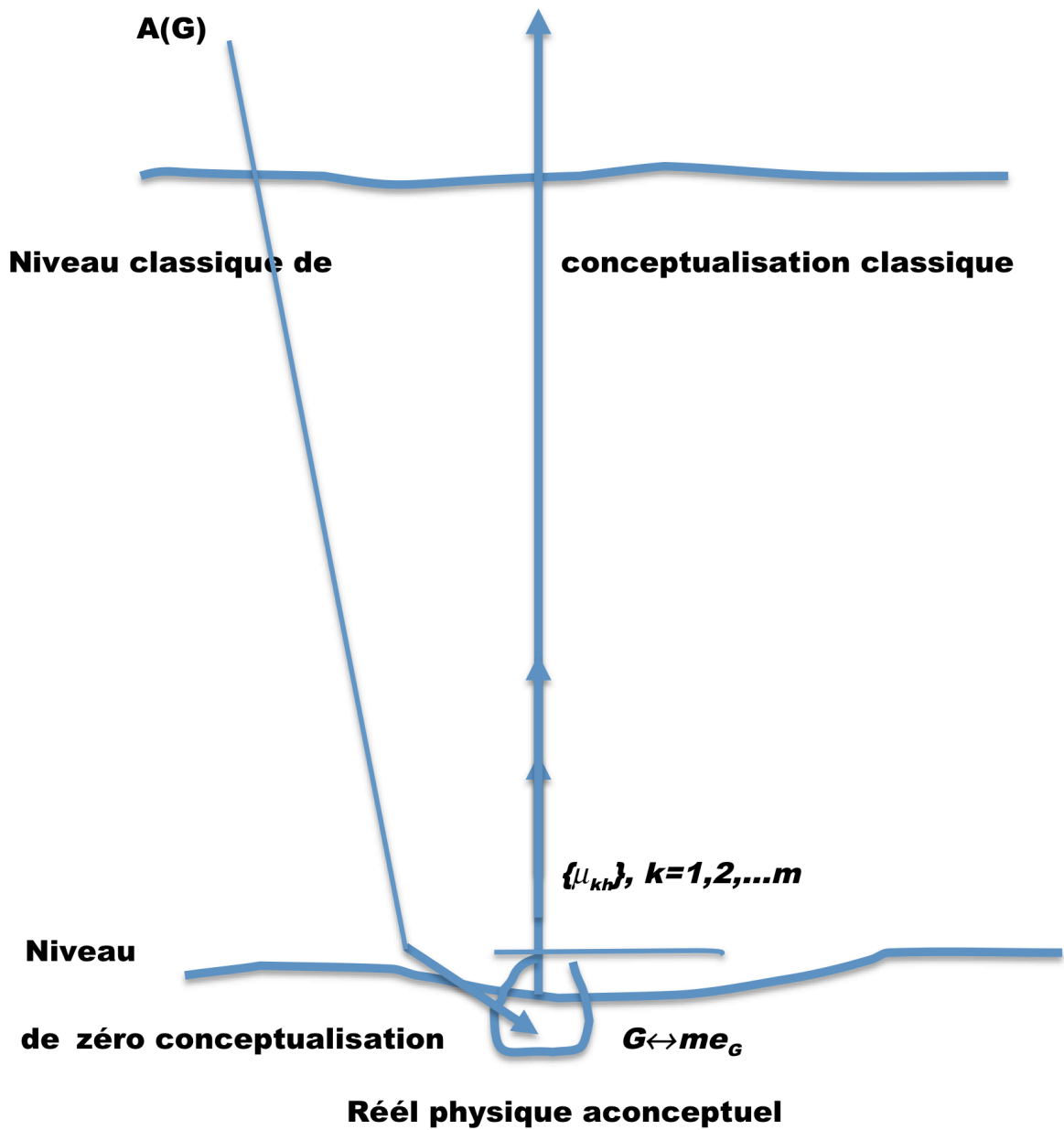

**Fig. 1.** Opération de qualification transférée primordiale

En dessous de ce pont s'opère à chaque fois un remarquable et curieux *changement de sens de progression* de notre action cognitive, le long de la verticale de nos conceptualisations: Au cours de la phase *'G'* cette action s'avance – à partir d'un appareil classique "de génération" *A(G)* – du 'haut' dominé par la conceptualisation classique, vers le non-classique zéro local de connaissance qui marque le sol (évolutif) de la verticale de nos conceptualisations ; elle *atteint* ce sol et s'y enfonce dans du réel physique aconceptuel ; là elle engendre-et-capture en tant qu'entité-à-qualifier, un fragment de réel physique baptisé '$me_G$' ; puis cette action **inverse** son sens et, du 'bas' extrémal non-classique, elle s'avance de nouveau vers le 'haut' classique devenu lointain, en accomplissant une qualification **toute première** du fragment de réel physique engendré, qui désormais est capturé dans la reproductibilité posée par *DM1*.

Là, dans cette inversion représentée dans la *figure 1* il s'agit d'un geste qui, à partir du réel encore a-conceptuel, tisse un fond au volume du "connu". Sur ce fond du connu, bien que l'on sache déjà quelque chose, on ne peut pas encore **comprendre**. Car *com*prendre implique relier (prendre ensemble). Le *"sens"* et la "compréhension" sont des concepts internalisés par un groupement de données, comme une configuration d'espace-temps, comme une "distance", comme une "comparaison". Face à une donnée isolée ces concepts simplement n'existent pas.

Cependant que le codage-cadre $C(\{\mu_k\}, k=1,2,...m) \leftrightarrow h$, *avec h=1,2,...H* du résultat directement observable $\{\mu_{kh}\}, k=1,2,...m$ d'une succession *[G.**X**]* ne relie ce résultat à rien d'autre. Entre le connu signifiant et le résultat $\{\mu_{kh}\}, k=1,2,...m$ d'une succession d'opérations *[G.**X**]* codé selon *Cod.cadreET* il subsiste un hiatus radical. L'on est en présence d'un connu *pur* nouveau, d'une donnée première au sens strict de cette expression. Il en est ainsi parce qu'une interaction-test **X** définie pour des microétats est absolument non spécifiée de quelque point de vue *sémantique* déjà installé : Ce n'est qu'un 'test' *quelconque* qui peut s'appliquer à des boîtes noires dénotées '$me_G$' et qui produit des données strictement premières liées à ces boites noires.

### 2.3.2.3. Descriptions transférées de base fondées sur des représentations de "grandeurs" importées de la conceptualisation classique

Or il est possible de réaliser également des qualifications transférées primordiales des microétats, à l'aide non pas d'interactions-test encore non conceptualisées mais de "grandeurs" $X$ importées de la conceptualisation classique, à structure conceptuelle préconstruite et déjà dénommée ; notamment des "grandeurs *mécaniques*" redéfinies pour des microétats : On peut l'affirmer parce que ceci a été réalisé dans le cas du formalisme de la mécanique quantique.

Une telle grandeur $X$ – *de par sa construction initiale, classique* – comporte une dimension sémantique dénommée (qui soit est introduite en tant qu'une grandeur fondamentale (masse, durée, longueur, charge, etc.) soit est structurée conceptuellement en termes de telles grandeurs fondamentales (comme $p=md/dt(\Delta x)$) ainsi qu'un spectre de valeurs *(X1,X2,....Xj,...XJ)*. Mais cette dimension sémantique et ce spectre de valeurs sont conçus comme des "propriétés" des entités physiques elles-mêmes qu'ils concernent. Afin de les redéfinir pour des microétats il faut changer cette manière de les concevoir. Corrélativement, il est – probablement – nécessaire toujours de modifier leurs représentations formelles, et il est certainement nécessaire de redéfinir l'opération physique de mesure *Mes(X)* qui convient. Donc à la place de successions *[G.**X**]* interviendront des successions *[G.Mes(X)]*. Mais exactement comme dans le cas d'une interaction-test **X**, le résultat directement observable d'une succession *[G.Mes(X)]* appliquée à un microétat *me_G* consiste en un groupe *{μ_k}, k=1,2,...m* de marques physiques transférées sur les récepteurs de l'appareil *A(X)* utilisé. Il faut donc pouvoir définir un codage, dénotons-le *Cod(X)* qui associe à tout tel groupe de marques une valeur du spectre des valeurs considérées, et une seule telle valeur (cf. l'exemple de *2.3.1*).

En supposant que toutes ces conditions ont été remplies pour une grandeur classique *X* donnée, on peut définir une grille de qualification *gq.primord.(X).me_G* d'une manière tout à fait analogue à celle mise en œuvre pour une interaction-test **𝑿** :

**Def.(gq.primord.(X).me_G).** L'association *[X, {X1,X2,....Xj,...XJ}, Mes(X), Cod.X]* constitue *une grille de qualification primordiale liée à une grandeur classique X redéfinie pour le microétat me_G.*

Les différences face au cas d'une interaction-test **𝑿** considéré dans *2.3.2.2* concernent exclusivement *le codage* et le processus de *Mes(X)* :

Le codage *Cod.X* des groupes de marques observables produit par un acte de mesure *Mes(X)* doit être accompli dans les termes de valeurs *Xj* du spectre *(X1,X2,....Xj,...XJ)* de la grandeur classique *X* redéfinie pour des microétats :

$$\textbf{Cod.X}: \quad C(\{\mu_k\}, k=1,2,...m) \leftrightarrow Xj, \quad avec \ j=1,2,...J$$

*- Ce codage introduit donc a priori des significations* **préconstruites** *Xj qui connectent directement le microétat examiné me_G à la conceptualisation* **classique**.
**- LE PROCESSUS DE MESURE IMPLIQUÉ Mes(X) EST SOUMIS À LA CONDITION D'ASSURER DE QUELQUE MANIÈRE LA POSSIBILITÉ DU CODAGE Cod.X.**

Comment assurer la possibilité d'un codage, ne peut pas être spécifiée en termes généraux, car cela dépend de la représentation conceptuelle-formelle de la grandeur *X* et des implications physiques-opérationnelles de cette représentation. Toutefois, notons bien ce fait :

*Les groupes de résultats physiques observables* $\{\mu_k\}$, *k=1,2,...m produits par les successions [G.Mes(X)]) où X désigne une grandeur classique redéfinie pour un microétat me_G,* **RESTENT, EUX, DES QUALIFICATIONS TRANSFÉRÉES PRIMORDIALES**.

Ces qualifications laissent en dessous d'elles ces mêmes **vides** de qualités *perçues* **sur** *l'entité à qualifier elle-même,* surmontés par de petits ponts qui unissent du réel physique aconceptuel, à du conceptualisé intersubjectif, qui sont la signature de toute qualification transférée primordiale. Les significations classiques *Xj, j=1,2,...J* introduites par le *Cod.X* sont superposées *en l'absence de toute qualité perçue,* via la représentation conceptuelle-formelle abstraite en laquelle consiste la redéfinition pour des microétats de la grandeur *X*. Toutefois la spécificité introduite par des codages en termes de valeurs d'une grandeur classique redéfinie pour des microétats, est loin d'être mineure, car :
*- Les codages – en ce cas – dotent les résultats de mesure, de sens, d'intelligibilité, ils en font du connu compréhensible.*
*- Les qualifications de microétats qui interviennent dans les algorithmes de la mécanique quantique se trouvent précisément dans ce cas où les qualifications recherchées concernent une grandeur définie dans la pensée classique, à savoir les grandeurs dynamiques de la mécanique macroscopique.*

*2.3.2.4. Conclusion sur 2.3.2*

La démarche développée dans ce travail concerne la catégorie *générale* des descriptions transférées de base des microétats.

Le domaine propre des qualifications quantiques des microétats est *contenu* dans le domaine général des qualifications transférées de base des microétats.

Les qualifications transférées primordiales qui correspondent à de simples interactions-test 𝗫 encore dépourvues de toute relation avec des "grandeurs" classiques, notamment mécaniques, se trouvent donc à *l'extérieur* du domaine de la mécanique quantique. Elles restent emprisonnées dans la strate primordiale où elles s'engendrent. Mais leur codage en termes de configurations d'espace-temps du groupe de marques enregistré définit une bande frontalière extrêmale de la strate des qualifications transférées primordiales, et par cela elles établissent un élément de *référence* éclairant :

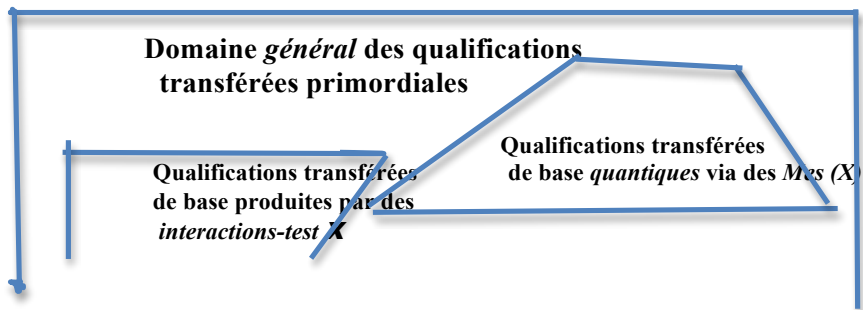

**Domaine *général* des qualifications transférées primordiales**

**Qualifications transférées de base *quantiques* via des *Mes (X)***

**Qualifications transférées de base produites par des *interactions-test 𝗫***

*Fig.2 : Qualifications transférées de base produites par des interactions-test, ou qualifications quantiques produites via des mesures de grandeurs classiques redéfinies pour des microétats*

### 2.3.3. Deux conséquences du concept de grilles de qualification applicables à des microétats

L'élaboration de **2.3.2** implique deux traits notables, l'un – déjà signalé – concerne le concept de 'propriétés' ; l'autre concerne la relation entre qualification et modèle. Dans ce qui suit nous focalisons brièvement l'attention sur ces deux traits.

#### 2.3.3.1. Qualifications de microétats et 'propriétés'

Afin d'établir le codage-cadre *Cod.cadre(ET)* et les deux definitions des grilles de qualification *Def.gq.primord.(𝗫).me*$_G$ et *Def.gq.primord.(X).me*$_G$ il n'a pas été nécessaire de supposer l'existence permanente, pour un microétat, de 'propriétés'. En fait, à la faveur des exigences comportées par l'introduction des concepts mentionnés, une véritable mutation s'est introduite subrepticement en ce qui concerne les propriétés *'de'* l'entité-objet étudiée. Par exemple il est clair que si l'on veut finalement réaliser pour les microétats des qualifications en termes de grandeurs *mécaniques*, il est obligatoire d'utiliser un certain concept de *position*, car c'est l'un des deux concepts de base, avec la vitesse, de ce qu'on appelle une mécanique. Le concept de 'position' veut dire 'ici' ou 'là', à tel endroit localisé de l'espace. Il répond à la question courante 'où?' Or il est apparu que concernant un microétat un tel renseignement ne peut être obtenu que *via* une manifestation physique observable *transférée* sur des enregistreurs d'appareils. Cela n'**exclut** pas que cette manifestation physique, marque ou déclic, puisse se traduire en termes d'une valeur bien définie d'une *propriété* permanente de *position* 'possédée' par le microétat étudié. Au premier abord il peut même paraître que le seul fait que la manifestation observable, elle, se soit produite à un endroit d'espace bien défini, *suffit* pour permettre de parler en termes de position possédée par le microétat étudié, d'une manière *séparée* de l'interaction de mesure[31]. Mais en fait ceci n'est nullement le cas. Car rien ne prouve que *ce* qui, lorsque l'enregistrement de la marque s'est produit, a agi de telle manière qu'il se soit engendré une marque quasi ponctuelle sur l'écran sensible d'un appareil, existait dès *avant* l'émergence de l'enregistrement ; ou que cela ne se trouvait pas ailleurs et s'est ramassé là, etc. Si l'on ne pose *a priori* vraiment aucune ombre de modèle, il n'existe aucune base pour exclure que l'enregistrement final d'une marque observable localisée étiquetée par le mot 'position' ait été *créé de toutes pièces* par l'interaction-test *X* ou par l'interaction de mesure *Mes(X)* considérée. Cette éventualité a été évoquée et discutée (de Broglie, L., [1956]). Par exemple, Einstein a objecté qu'elle impliquerait la possibilité de processus *physiques* de localisation quasi instantanée à l'intérieur de l'étendue assignée au 'microétat' – qui peut être illimitée – ce qui, pensait-il, pouvait être contraire à la théorie de la relativité. (Notons que – en rigoureuse absence de tout modèle – cette objection ne peut être reçue elle non plus). Or, on l'a vu, à l'intérieur d'une description de base transférée, l'interprétation d'une marque localisée en termes d'une valeur d'une 'propriété' de position possédée de manière permanente par le microétat étudié, serait un saut au-dessus d'un précipice de conceptualisation.

Cet exemple permet d'estimer l'ampleur de la mutation que subit le concept classique de propriété '*de*' une entité-objet, en la stricte absence de toute connaissance préconstruite concernant ces fragments de pure factualité que nous avons dénommés 'microétats'.

Bref, en toute rigueur, on doit raisonner en admettant que les manifestations observables produites par une interaction-test *X* ou un acte de mesure *Mes(X)* pourraient être *entièrement* créées par l'interaction de l'appareil avec l'entité-objet strictement inconnue dénotée $me_G$, et en tant que propriétés émergentes *des enregistreurs de l'appareil*, pas du microétat étudié ; le microétat-objet-d'étude n'est pas qualifié *isolément* par *X* ou *Mes(X)*.

Les considérations qui précèdent, associées au contenu de *2.3.2*, permettent de réaliser quelle structuration conceptuelle complexe – à savoir celle correspondant aux ponts *(POC)$_h$* – a été reléguée dans le non-examiné lorsqu'on a décidé, juste décidé, quelles opérations physiques sont à accomplir *sur un microétat* lors d'un acte de 'mesure' de telle ou telle 'grandeur *mécanique'* dont on a posé à l'avance la *signifiance* pour des microétats (comme dans le cas de la méthode du temps de vol).

### 2.3.3.2.  Spécification sur qualification de microétats versus *modèle*

Nous avons d'abord souligné dans *2.3.2.1* que les qualifications mécaniques spécifiées par le formalisme quantique n'ont pu être définies qu'à partir de modèles classiques prolongés pour le cas des microétats, par la voie d'éléments descriptionnels mathématiques et de définitions correspondantes d'opérations physiques de mesure.

Il est apparu ensuite dans *2.3.2.2.* que le codage-cadre *Cod.cadre(ET)*, parce qu'il ne fait aucune référence à quelque modèle, a dû laisser subsister un vide de spécification sémantique qui entraîne une radicale coupure avec les contenus de la conceptualisation classique.

---

[31] En mécanique quantique on parle souvent, même pratiquement toujours, de la position *de* la 'particule', ou de la position *du* 'système' (même pas du micro-*état*), et l'une comme l'autre de ces façons de parler conduit à une cohue de confusions entre 'micro*état*' et *modèle* de micro-*système*. En physique des particules élémentaires, par contre, le fait que la 'position' est représentée elle aussi par un 'opérateur', introduit explicitement une certaine distance face au concept classique de "propriété".

Et il est apparu dans *2.3.2.3* que le recours à des grandeurs classiques liées à des modèles intrinsèques à la conceptualisation classique, n'excluent nullement que les qualifications produites soient des qualifications primordiales transférées, non attribuables au microétat étudié lui-même, dépourvues de qualia et même d'une structure définie d'espace-temps.

On sent que l'on est là en présence de faits conceptuels importants et reliés. Les remarques suivantes détaillent brièvement la structure de ces faits.

Les fonctionnements biologiques et psychiques de l'être humain, comme aussi ses intuitions et sa pensée-et-langage, se sont forgées par des interactions directes avec du réel macroscopique, *via* les appareils biologiques neurosensoriels humains qui engendrent dans l'esprit des *qualia*. Les structures mentionnées sont si profondément et inextricablement inscrites dans l'être matériel de chaque humain normal, qu'il est *vain* de vouloir les 'dépasser entièrement' par des entraînements de familiarisation avec des concepts et structures auxquels ont abouti des élaborations mathématiques ou méthodologiques réalisées dans des théories scientifiques modernes. Les injonctions qui exigent un tel dépassement 'total' sont simplement irréalistes. Il subsiste *irrépressiblement* un cordon ombilical plus ou moins évident qui rattache tout construit conceptuel, au réseau d'intuitions et de conceptualisation issu de nos interactions sensorielles directes avec du réel macroscopique. Il n'est pas *pensable* de *qualifier* du non percevable dans le sens courant du terme, d'une façon qui, à la fois, ne fasse intervenir strictement aucun élément induit par la conceptualisation naturelle macroscopique, et qui néanmoins soit ***intelligible***.

On peut ressentir quelquefois une surprise totale, un manque total de compréhension, face à certains faits d'observation qui apparaissent spontanément, comme lors de la découverte des rayons-*X*. Mais à ce jour, de manière plus ou moins explicite, toute qualification d'entités physiques inobservables qui a été *organisée délibérément* à été fondée sur des éléments tirés de la conceptualisation classique, *qui est modélisante dès qu'elle est explicite* et qui, toujours, dotent d'un certain sens.

Dans l'approche présente toutefois, tout tel élément classique a été déclaré et le sens qu'il transporte a été réduit au minimum possible. A savoir, à juste des *mots* dénommant des concepts *catégoriels* utilisés comme des ascenseurs envoyés jusqu'au niveau zéro sur la verticale de nos conceptualisations, afin de les y charger de fragments de réel physique encore jamais conceptualisés – eux, spécifiquement – et afin de tirer de ces fragments des données strictement premières placées sur le fond du volume du conceptualisé.

Cet investissement classique minimal – vidé de tout contenu sémantique attribuable isolément à l'entité-objet particulière concernant laquelle on veut construire des connaissances – nous permettra dans ce qui suit de poursuivre jusqu'au bout et explicitement cette phase de conceptualisation transférée primordiale qui jusqu'ici avait été occultée dans des coalescences avec des phases ultérieures, déjà modélisantes. Au terme de ce processus constructif il sera devenu possible d'analyser les processus de conceptualisation en *trois* strates clairement distinctes:

- Une strate de conceptualisation transférée *primordiale (de base)* accomplie au-dessus d'un vide radical de *qualia*, consistant exclusivement en des marques perçues sur des enregistreurs d'appareils et codées en termes de leur configuration d'espace-temps et dépourvues de toute 'signification' préconstituée, coupées radicalement de la conceptualisation classique.

- Une strate de conceptualisation transférée *primordiale 'quantique'*, accomplie elle aussi au-dessus d'un vide radical de *qualia* et consistant elle aussi exclusivement en des marques perçues sur des enregistreurs d'appareils, mais qui – via des grandeurs classiques redéfinies pour des microétats – est dotée par construction de significations en termes de valeurs de grandeurs classiques redéfinies pour des microétats, qui la relient à la conceptualisation classique *et qui sont subrepticement modélisantes*.

- Et enfin, toute une hiérarchie illimitée de chaînes de *modélisations* explicites successives fondées au départ sur des qualifications transférées primordiales, quantiques *ou non* (MMS [2002B], [2006]).

Cela permettra d'économiser une foule d'échecs aveugles, de débats circulaires, de stagnations.

### 2.3.4. Conclusion globale sur les qualifications transférées primordiales de microétats, et modélisation de celles-ci

En examinant quels principes permettent de concevoir des qualifications associables à des microétats il est apparu que la démarche développée ici, parce qu'elle part d'une base fondamentale au sens extrême de ce mot, englobe les qualifications de microétats produites dans la *mécanique* quantique, dans la catégorie plus vaste des qualifications transférées primordiales quelconques. Cela permet de percevoir avec un relief maximal deux faits conceptuels notables :

D'une part, l'on peut percevoir l'abîme qui sépare le concept général de qualification applicable à des microétats, du concept classique de 'propriété possédée' par l'entité-à-étudier, qui émerge par un 'oui' ou un 'non' résultant de la simple comparaison entre un 'prédicat' qui préexisterait dans l'air du temps en état pérenne et indépendant, avec d'autre part un 'objet' au sens du langage courant qui lui aussi subsisterait dans l'air du temps, pérenne et indépendant et prêt à être qualifié.

Et d'autre part l'on peut percevoir aussi l'intérêt de construire une continuité avec le système classique de penser et de dire, qui est foncièrement modélisant, mais qui – pour nous, peuple de la région de conceptualisation 'classique' – introduit du *sens*.

Si la construction explicite des descriptions transférées primordiales qui est entreprise ici réussit à assainir le formalisme de la mécanique quantique, une modélisation unanimement acceptable de la représentation des microétats pourra être tentée avec des chances notablement accrues.

## 2.4. Description qualitative de microétats

### 2.4.1. Annonce générale

Nous venons de dépasser les deux difficultés majeures qui surgissent lorsqu'on veut transposer à des microétats le concept classique de "description" : la mise à disposition pour des opérations de qualification, d'un microétat dans le rôle d'entité-à-qualifier, et l'accomplissement de toutes les étapes exigées afin de qualifier ce microétat. Désormais nous pouvons aborder la construction de *descriptions de microétats*. Une telle description fait intervenir toute une *classe* de qualifications (comme d'ailleurs c'est déjà le cas lorsqu'on veut définir le codage qui intervient dans la réalisation d'une qualification isolée) : Nous tâcherons de mettre en évidence les conditions et les caractères spécifiques de ces classes de qualifications.

Il vient d'être souligné que l'approche constructive amorcée ici délimite une catégorie générale de qualifications transférées primordiales qui se sépare en deux sous-catégories, une sous-catégorie fondée sur des interactions-test qui reste coupée de la conceptualisation classique, et une sous-catégorie fondée sur des 'grandeurs mécaniques' redéfinies pour des microétats qui relient à la conceptualisation classique.

Afin de faciliter la comparaison future avec les algorithmes quantiques, qui constitue le but majeur de ce travail, nous parlerons désormais exclusivement en termes de "*grandeurs **mécaniques**" redéfinies pour des microétats* (en bref, grandeurs mécaniques). Mais, notons-le bien :

Les considérations qui suivent ne sont nullement restreintes à des grandeurs *mécaniques*, ni à des 'grandeurs' tout court. Elles sont ***indépendantes** de tout modèle* de microétat et valides pour ***tout** acte physique de qualification défini opérationnellement pour un microétat, que ce soit une intéraction physique de mesure *Mes(X)* d'une grandeur *X* quelconque[32], ou une interaction-test physique **X** quelconque.

Nous caractériserons donc la catégorie ***générale*** des descriptions transférées primordiales de microétats. Seul le *langage* employé se calquera sur le cas spécial du langage utilisé dans la mécanique quantique[33].

Et nous ne ferons aucun usage des représentations mathématiques utilisées dans la mécanique quantique. La démarche continuera de rester rigoureusement non mathématique, qualitative. Même lorsque nous écrirons des chiffres liés à des dénombrements, aucun calcul ne sera impliqué.

En outre la démarche restera *indépendante* de celle poursuivie dans la construction de la mécanique quantique : Cette dernière a obéi à des buts et des contraintes différentes qui seront examinées dans la deuxième partie de ce travail.

### 2.4.2. Quelques définitions fondamentales

Nous allons parler d'interactions de *Mes(X)* opérées sur *un* exemplaire d'un microétat *me_G* engendré par une opération de génération d'état *G*, et dont l'effet est codé en termes d'une valeur *Xj* de la grandeur mesurée *X*. *Cela implique-t-il qu'un seul acte de mesure sur un seul exemplaire du microétat étudié me_G, ne peut produire que l'enregistrement d'une seule valeur Xj d'une grandeur dynamique définie pour un microétat ?* Voulons-nous admettre cette implication?

La réponse est *non*, car 'un micro-*état*' est une expression qui ne désigne pas la même chose qu''un micro-*système*'[34]. Et un micro-état – selon le langage installé en microphysique – peut concerner un *ou plusieurs* micro-systèmes, selon le substrat physique sur lequel a été appliquée l'opération de génération *G* : *C'est dans ce point précis que se concentre la nécessité de **distinguer** entre les deux concepts, de microsyst*ème et de micro-état.

En outre, nous avons signalé que selon le principe de composabilité des opérations de génération de microétats (cf. *2.2.4*), plusieurs opérations de génération *Gk, k=1,2,...n,* qui peuvent opérer séparément sur des exemplaires distincts du microétat 'initial' (possiblement non connu) utilisé comme matière première, peuvent toujours s'appliquer également toutes à la fois sur un seul exemplaire d'un tel microétat-initial-matière première. Mais comment compte-t-on ces divers 'un' et 'deux'? Quelles présuppositions sont ici à impliquer afin de rester en accord avec l'emploi que l'on fait de ces termes afin d'accompagner verbalement les écritures du formalisme quantique?

---

[32] Les qualifications purement formelles-conceptuelles, comme celles de 'projection' ou de 'parité', échappent peut-être au domaine de pertinence des considérations qui suivent.

[33] Nous agirons d'ailleurs d'une manière analogue en ce qui concerne le concept de probabilité: Nous utiliserons le mot, mais nous soulignerons que la signification que nous lui assignons est différente de celle posée actuellement dans la mécanique quantique (MMS [2009], [2013].

[34] Dans ce paragraphe, pour clarté de lecture, nous écrivons 'micro-système' et 'micro-état', avec tiret à l'intérieur. Ensuite nous reviendrons à l'écriture en bloc, sauf dans les cas ou l'on accentue de nouveau la distinction entre "système" et "état".

Lorsqu'on considère ces questions on perçoit à quel point la démarche entreprise ici n'est pas de la nature d'une recherche passive de faits 'naturels', ni ne pourrait se déployer en tant que telle. On perçoit à quel point cette démarche est une construction délibérée soumise à des buts et à d'autres contraintes aussi bien factuelles que conceptuelles.

Soit donc ce qu'on appelle 'un micro-*état*'. Ce micro-*état* comporte nécessairement quelque micro-*système*, ou plusieurs, dont il est le (micro)-*état* (l'entière conceptualisation humaine associe tout 'état', à quelque support plus stable, que l'on peut désigner par le terme 'système').

La décision méthodologique *DM1* pose qu'à *une* opération de génération *G* correspond *un* microétat $me_G : G \leftrightarrow me_G$. Ceci est inébranlable.

Par ailleurs, sous l'empire des questions énoncées et des contraintes qui agissent, nous avons identifié l'organisation suivante de penser-et-dire (quelque peu obscure et mouvante) que nous adoptons désormais explicitement parce qu'elle nous paraît être l'unique qui permet, à la fois, continuité face aux façons de dire courantes et cohérence globale relativement aux implications de l'entière microphysique.

### Définitions [(un micro-système) et (un micro-état de un micro-système)]

Soit un micro-état tel qu'une seule opération de mesure accomplie sur un seul exemplaire de ce micro-état *ne* **peut** *produire qu'une seul groupe* $\{\mu_k\}$, *k=1,2,...m de marques observables*. On dira que ce micro-état met en jeu ce qu'on appellera **un** *micro-système* et que par conséquent c'est *un micro-état de* **un** *micro-système*.

### Définition [un *micro état de* n *microsystèmes*]

Soient maintenant $n>1$ micro-systèmes, c'est à dire deux ou plusieurs micro-systèmes d'un type dont on sait que, pour chacun séparément, on peut engendrer un micro-état au sens de la définition précédente. Mais soit *Gn* une seule opération de génération qui, agissant sur quelque support physique initial utilisé comme matière première, a engendré **à la fois** n micro-*états* de tels *n* micro-systèmes[35]; ou bien, qui a même engendré *à la fois* tous ces *n* micro-*systèmes* eux-mêmes, *avec* leurs micro-états respectifs[36]. *Dans les deux cas on dira que le micro-état correspondant est* **un** *micro-état de* **n** *micro-systèmes, et nous le dénoterons par* $me_{Gn}$.

L'unicité de l'opération de génération *Gn* peut être conçue a priori comme source possible de spécificités du comportement global d'un microétat de *n* microsystèmes *face* au cas de "*n* microétats de *n* microsystèmes" *générés chacun séparément*, i.e. au sens de la définition précédente.

Cette circonstance sera approfondie plus tard.

### Définition [mesure complète sur un *micro-état de* n *micro-systèmes*]

Une opération de mesure *complète* accomplie sur un seul exemplaire d'un micro-état $me_{Gn}$ de *n* micro-systèmes, produit *n* groupes distincts de marques observables, un groupe pour chaque micro-système impliqué, les grandeurs et les valeurs auxquelles ces marques sont liées pouvant être identiques ou différentes.

---

[35] C'est le cas, par exemple, lorsque *Gn* consiste en une "interaction" entre deux électrons préexistants, ou un électron et un proton préexistants, etc.
[36] Il s'agit alors d'une 'création' au sens de la physique des particules.

***Définition [mesure incomplète sur** un *micro-état de* n *micro-systèmes]***

Une opération de mesure accomplie sur un seul exemplaire du micro-état $me_{Gn}$ et qui produit moins de *n* marques observables – c'est-à-dire qui ne tire pas de $me_{Gn}$ une qualification pour chaque micro-système impliqué et par conséquent ne qualifie pas cet exemplaire de ce micro-état, par cette mesure, aussi exhaustivement qu'on peut le concevoir conformément à la définition précédente – est *une mesure incomplète* sur $me_{Gn}$.

***Remarque***

D'autre part, en conséquence des définitions explicitées plus haut, un acte de mesure *Mes(X)* opéré sur un seul exemplaire d'un micro-état $me_G$ de *n* micro-systèmes ne peut – par construction des concepts – produire que tout au plus *n* valeurs observables *Xj*.

***Définition [un** *micro-état à génération composée]***

Soit un micro-état d'un micro-système ou de *n>1* micro-systèmes, indifféremment. Si ce micro-état a été engendré par une opération de génération *G(G1,G2,..Gk)* qui – sur un *autre* micro-état de départ utilisé comme matière première afin de générer le micro-état considéré – a composé les effets de deux ou plusieurs opérations de génération *G1,G2,...Gk* (où *k* est un entier) dont on sait que chacune aurait pu agir séparément, alors on dira que le micro-état considéré est *un micro-état à génération composée*[37].

## 2.5. Construction d'une *'description'* de microétat

### 2.5.1. Le caractère primordialement statistique des qualifications 'mécaniques' d'un microétat

Dans ce qui suit immédiatement nous supposerons qu'il s'agit d'un micro-état d'*un* seul micro-système. Le cas d'un micro-état de *n>1* micro-systèmes sera évoqué de nouveau plus tard.

Revenons sur un caractère des *qualifications* de microétats qui s'est déjà fait jour dans *2.3.2.2* et qui, ici, est repris à la base de la caractérisation des *descriptions* de microétats :

Considérons l'exemplaire du microétat $me_G$ qui a été produit par *une* seule réalisation d'une opération de génération *G*. Supposons qu'il est soumis à une opération de mesure d'une grandeur mécanique particulière, bien précisée. Notons *B* cette grandeur (afin de la distinguer clairement, aussi bien d'une grandeur définie mais quelconque qui est dénotée *X*, que d'un appareil de mesure qui est dénoté *A(X)*). Nous considérons donc un acte de *Mes(B)*. Pour éviter des restrictions arbitraires *a priori* il faut admettre que, en général tout au moins, l'acte de *Mes(B)* doit *changer* le microétat $me_G$ de départ, et de telle manière que l'on obtienne un groupe {$\mu_k$, *k=1,2,...m)}* de *m* marques physiques observables relevées sur les enregistreurs de l'appareil *A(B)*. *Via* le codage *Cod.B* correspondant (*2.3.2.3*), ces marques interviendront dans l'identification de l'une parmi les différentes valeurs possibles du spectre de *B*, disons *B4*. Au bout de cette séquence *[G.Mes(B)]* de réalisation d'une opération de génération *G* suivie d'un acte de *Mes(B)*:

* Les manifestations observées – *sur l'appareil* – qui signifient la valeur *B4* de la grandeur *B*, incorporent une inamovible *relativité au processus Mes(B)* qui a permis d'obtenir ces manifestations.

---

[37] Cette dernière définition – déjà introduite avant pour le cas d'un micro-état d'un seul micro-système – place maintenant tout micro-état à génération composée (i.e. lié au principe de composabilité des opérations de génération), dans le contexte de l'ensemble des définitions de ce paragraphe.

    * L'exemplaire individuel d'un microétat $me_G$ qui a été soumis à l'acte de *Mes(B)* en général n'*existe plus* tel qu'il avait été engendré par l'opération de génération *G*. En général ce microétat de départ a été d'abord changé par l'évolution de *Mes(B)*, et en outre, souvent, sa transformée finale reste capturée dans l'un ou l'autre des objets macroscopiques qui constituent les enregistreurs de l'appareil *A(B)*.

    Cette dernière circonstance oblige, si l'on veut *vérifier* le résultat *B4*, d'engendrer d'autres exemplaires du microétat $me_G$ et d'autres successions *[G.Mes(B)]* d'une opérations *G* et un acte de *Mes(B)*. Or les sciences physiques accordent une importance majeure à la condition de vérifiabilité des résultats annoncés: c'est cette condition qui garantit la possibilité d'un consensus intersubjectif, sans quoi il n'y aurait pas ce qu'on appelle objectivité. Bref: afin de faire face à la condition centrale de vérifiabilité, il faudra faire usage de tout un ensemble de réalisations d'une succession *[G.Mes(B)]* et de l'entier *ensemble* correspondant d'exemplaires du microétat $me_G$.

    Imaginons alors que l'on ait effectivement répété un grand nombre de fois la succession *[G.Mes(B)]*. Si à chaque fois l'on retrouvait le résultat *B4* que l'on avait trouvé la première fois, on se dirait: « j'ai trouvé une petite loi: si un microétat $me_G$ engendré par l'opération de génération *G* est soumis à un acte de *Mes(B)*, l'on obtient le résultat *B4* ».

    On pourrait se demander ensuite si *toute* mesure de toute autre grandeur *C, D*, etc., effectuée de manière répétée sur des exemplaires du microétat $me_G$, produit stablement une et même valeur de la grandeur mesurée, disons *C17* pour *C*, *D154* pour *D*, etc. Et s'il s'avérait qu'effectivement c'est le cas, on se dirait: « j'ai trouvé une nouvelle loi plus importante que la précédente: un microétat $me_G$ introduit un groupe bien déterminé de valeurs observables des grandeurs *B,C,D,...* considérées, une valeur et une seule pour chacune de ces grandeurs mécaniques re-définie pour ces microétats ».

    Mais il se trouve qu'en fait les choses se passent autrement. Lorsqu'on répète une succession *[G.Mes(B)]* qui comporte un acte de *Mes(B)* accompli sur un microétat $me_G$ engendré par une opération de génération *G*, en général on n'obtient *pas* à chaque fois une même valeur de la grandeur *B*. En général – nonobstant le fait qu'à chaque fois il s'agit de la 'même' opération *G* et du 'même' acte de *Mes(B)*, on obtient une fois telle valeur de *B* et une autre fois telle autre valeur. Et lorsque le nombre d'essais s'accroît, l'ensemble des valeurs obtenues ainsi tend à couvrir progressivement tout le spectre *{B1, B2,....Bj,...}* de *B*. Et même s'il arrive que, pour le microétat $me_G$, ce soit à chaque fois la même valeur de *B* qui apparaît, disons *B4* – et l'expérience montre la possibilité d'un tel cas –, alors on trouve *toujours* d'autres grandeurs différentes de *B* pour lesquelles, face à $me_G$, les résultats sont dispersés: *aucun* microétat ne s'associe avec un ensemble de résultats de mesure qui soit dépourvu de dispersion pour *toutes* les grilles de qualification mécanique définies pour des microétats: si un microétat $me_G$ donné est tel qu'il conduit à un ensemble de résultats de mesure dépourvu de dispersion pour une parmi ces grilles – ce qui n'est pas le cas général – alors il existe *toujours* d'autres grilles de qualification pour lesquelles $me_G$ conduit à un ensemble dispersé de résultats de mesure.

    Il n'y a jamais stabilité de la valeur produite par la répétition d'un acte de mesure donné opéré sur un microétat, pour *toutes* les grilles de qualification définissables pour un microétat.

    Ceci est un fait d'expérience, une donnée factuelle. Ainsi *les faits nous éjectent sur un niveau statistique*.

    Dans ces conditions il est clair d'emblée qu'une valeur donnée du spectre de la grandeur *B*, disons *B4*, peut apparaître, par un acte de *Mes(B)*, pour une infinité de

microétats différents produits par des opérations de génération différentes. Une valeur d'une grandeur $X$ n'est donc jamais spécifique à un microétat donné.

Or le caractère statistique auquel on se trouve confronté ici est *primordial*, en *ce sens* qu'on ne peut pas l'assigner à quelque ignorance. Car – par construction – la structure de connaissances concernant des microétats dont on surveille la genèse ici, émerge *foncièrement première*. Elle émerge d'un inconnu qui est posé être total. *Notamment, elle émerge en dehors du postulat déterministe de la possibilité de principe, toujours, d'une formulation fondamentale des lois mécaniques, et en général des lois tout court, en termes individuels (non statistiques et a fortiori non probabilistes).*

### 2.5.2. Exigence de stabilité des manifestations observées

Ce n'est donc *que* sur un niveau statistique qu'on peut encore rechercher un invariant observationnel lié à un microétat donné. Or sans invariants il n'y a pas de lois, pas de science prévisionnelle. Comment procéder?

De ce qui vient d'être dit il découle que pour chaque paire *(G,X)* il faudra, pour tout $X$, accomplir un grand nombre $N$ de fois la succession d'opérations *[G.Mes(X)]* correspondante, enregistrer à chaque fois la valeur codante $Xj$ du spectre *{X1,X2,…Xj……}* de $X$ qui a été obtenue et rechercher l'ensemble des $N$ valeurs obtenues. Ce sera alors forcément un invariant non individuel ; et si c'*est* en effet un invariant de la statistique constatée ce sera un invariant 'probabiliste'[38], car on ne connaît pas une autre sorte d'invariant sur le niveau de conceptualisation statistique.

Soit donc un grand nombre $N$ de répétitions de la succession *[G.Mes(X)]*. Soit *{n(G,X1)/N, n(G,X2)/N,…n(G,Xj)N…… n(G,Xk)/N}* l'ensemble des fréquences relatives enregistrées (où $n(G,Xj)$ désigne le nombre des réalisations de la valeur $X(j)$ pour le microétat $me_G$, $N$ est le nombre total d'essais faits, et $n(G,Xj)N$ est la fréquence relative du résultat $Xj$ parmi les résultats des $N$ essais *[G.Mes(X)]* accomplis). Cet ensemble de fréquences relatives est 'la distribution statistique' des $Xj$.

On dira que la situation s'avère être 'probabiliste'[39] *si et seulement si*, lorsqu'on mesure la distribution statistique *{n(G,X1)/N, n(G,X2)/N,…n(G,Xj)N…. n(G,Xk)/N}}* pour des nombres d'essais $N$ qui s'accroissent autant qu'on veut, l'on constate que ces distributions semblent manifester une tendance de *converger*. Si donc une telle tendance se manifeste, alors la distribution – statistique – constatée sera dénommée dans ce qui suit une *distribution de probabilité* et on dira qu'elle constitue *la loi factuelle de probabilité*

$$\{p(G,Xj)\}=\{ p(G,X1, p(G,X2),…. p(G,Xj)….. p(G,Xk)\}$$

---

[38] En fait un tel invariant ne peut pas exister en conditions *naturelles*, c'est un *artefact opérationnel-conceptuel* construit délibérément sous la contrainte du but de pouvoir prévoir avec le degré de certitude maximal concevable au niveau statistique de conceptualisation (MMS [2009], [2013]). La proportion étant gardée avec les ordres de grandeur des durées et des distances impliquées, ceci vaut même concernant des phénomènes cosmiques.

[39] Ici – provisoirement – j'introduis des définitions nécessaires pour pouvoir employer les manières courantes de parler. Mais en fait les concepts de 'convergence probabiliste' et de 'loi factuelle de probabilité' soulèvent des problèmes conceptuels très sérieux. Ceux-ci ont déjà été exposés et traités dans MMS [2006], mais de façon insuffisante. Dans MMS [2009] on trouve une version plus avancée du traitement, et dans MMS [2013] le problème est enfin résolu d'une manière qui semble être stabilisée. L'on y conclut que'une *'loi de probabilité' est un artefact conceptuel déterminé par des buts prévisionnels, qui n'est constructible – en certaines conditions – que sur des domaines délimités d'espace-temps. Or les conditions mentionnées NE SONT PAS RÉALISABLES pour des descriptions primordiales transférées : ELLES EXIGENT UNE MODÉLISATION.* Dans la deuxième partie de ce travail nous reviendrons sur cette question, très brièvement..

où $p(G,Xj)=lim.N\rightarrow\infty [(n(G,Xj)/N]$ dénote par définition la limite de convergence de la fréquence relative $n(G,Xj)/N$, supposée exister.

### 2.5.3. Exigences de spécificité face à $me_G$ d'une loi de 'probabilité' $p(G,Xj)$, versus *grandeurs mutuellement incompatibles*

#### 2.5.3.1. Questions concernant la spécificité des observations

Lorsqu'on dit qu'on recherche la description du microétat $me_G$ il est sous-entendu que l'on cherche un ensemble de qualifications de $me_G$ qui soit spécifique à $me_G$, c'est-à-dire tel qu'aucun autre microétat généré par une opération de génération différente de celle, $G$, qui produit $me_G$, ne puisse faire apparaître exactement le même ensemble de qualifications. Ce qu'on cherche en fait ainsi est une redéfinition de $me_G$ qui soit à la fois qualifiante et vérifiable par l'expérience et qui puisse *in fine* remplacer par quelque sorte de *description* sa 'définition a-conceptuelle' de départ via l'étiquetage par l'opération $G$ correspondante, qui ne nous donne *rien* à connaître concernant spécifiquement le microétat $me_G$.

Or comment savoir si la loi de probabilité *{p(G,Xj)}* obtenue est spécifique du microétat étudié $me_G$?

*Si* elle l'était, alors d'ores et déjà, à elle seule, elle pourrait peut-être être regardée comme une 'description' du microétat $me_G$, nonobstant le fait qu'elle ne concerne pas $me_G$ isolément, mais seulement les interactions de $me_G$ avec l'appareil $A(X)$ utilisé pour accomplir les *Mes(X)*. Mais si, même avec cette réserve de taille, cette distribution de probabilité n'est pas spécifique au microétat $me_G$, alors il est clair d'emblée qu'en aucun cas il ne s'agit là d'une 'description' de $me_G$: l'appeler ainsi trahirait radicalement la définition du concept. Or aucun fait ni aucun argument n'entraîne avec nécessité qu'une distribution de probabilité $p(Xj)$ établie pour une paire $(G,X)$ donnée (i.e. pour un microétat $me_G$ et avec une grandeur $X$ donnés), ne peut se réaliser également pour un paire *(G',X)* où $G'\neq G$, c'est à dire avec le même $X$ mais pour un autre microétat $me_{G'}\neq me_G$. Il faut donc chercher un autre critère de spécificité observationnelle de $me_G$.

Cela conduit à la nouvelle question suivante. Que se passe-t-il si l'on considère *deux* grandeurs dynamiques $X$ et $Y$ différentes, au lieu d'une seule? Ne pourrait-on pas, à l'aide d'une paire de deux lois de probabilités *{p(G,Xj)}* et *{p(G,Yq)}* définir avec certitude une spécificité observationnelle face à $me_G$? Ou bien, puisque peu de choses peuvent être certaines lorsque d'emblée il s'agit de probabilités, du moins agrandir la chance *a priori* d'avoir identifié une spécificité observable de $me_G$?

#### 2.5.3.2. Compatibilité mutuelle de deux aspects de qualification d'un microétat[40] versus spécificité des observations

Quand il s'agit de microétats, deux grandeurs $X$ et $Y$ ne sont vraiment 'différentes' que lorsqu'elles sont mutuellement incompatibles, c'est-à-dire, lorsqu'elles s'excluent mutuellement en **ce** sens précis *qu'un acte de mesure de Mes(X) est conçu comme changeant le microétat étudié **autrement** qu'un acte de mesure de Mes(Y)*. Ce qui évidemment est impossible à réaliser sur **un** seul *exemplaire* du microétat à étudier. Deux grandeurs qui sont incompatibles au sens spécifié, ne peuvent donc pas qualifier par un seul acte de mesure, un et *même* exemplaire du microétat à étudier, *si* celui-ci – comme il est supposé ici – est un microétat d'un seul microsystème[41,42].

---

[40] il apparaîtra que les considérations qui suivent ne s'appliquent qu'au cas d'un micro-état de **un** micro-système.

Si deux grandeurs mécaniques *X* et *Y* ne sont *pas* incompatibles au sens spécifié, alors, par opposition aux conditions impliquées par la définition d'incompatibilité mutuelle, le mode de changement du seul exemplaire du microétat étudié *me*$_G$, qui se réalise lors d'une interaction de *Mes(X)*, *peut* (avec un choix convenable d'appareil) être le même que celui qui se réaliserait pour ce même exemplaire lors d'une interaction de *Mes(Y)*: c'est la définition de la compatibilité mutuelle de deux grandeurs qualifiantes définies pour des microétats. Donc en cas de compatibilité, rien n'empêche de procéder de la façon suivante: Un seul exemplaire du microétat à étudier est soumis au type unique de changement qui convient – à la fois – comme processus de changement par un acte de *Mes(X)* et comme processus de changement par un acte de *Mes(Y)*. On peut donc introduire une nouvelle notation: *un acte de Mes(XY)*. Cette unique interaction de *Mes(XY)* ne peut évidemment produire qu'un seul ensemble de marques physiques observables sur les enregistreurs de l'appareil *A(XY)*. Mais l'unicité de cet ensemble de marques physiques observées permet néanmoins de qualifier le microétat étudié, à la fois, par une valeur correspondante *Xj* de *X* et par une autre valeur correspondante *Yq* de *Y*. En effet, ces deux *valeurs* peuvent, *elles*, être distinguées l'une de l'autre, si la définition *conceptuelle* de la grandeur *X* – superposée à la définition opérationnelle de *'Mes(XY)'*, introduit, disons, un super-codage en termes d'une valeur *Xj* de *X* qui est différent du super-codage introduit par la définition *conceptuelle* de la grandeur *Y*, mais ce dernier étant un super-codage *relié* au précédent, identifiable ou même calculable *à partir* de celui-ci. Ceci revient à dire que, dans ces conditions, on peut considérer que les noms '*X*' et '*Y*' ne désignent en fait *pas* deux dimensions de qualification qui sont distinctes *physiquement*. Qu'ils ne désignent que deux utilisations conceptuelles-formelles différentes mais corrélées, d'une seule dimension physique de qualification. Disons, en employant une image, que d'un point de vue physique opérationnel les deux grandeurs compatibles qui interviennent dans *Mes(XY)* peuvent être regardées comme deux '*directions de qualification colinéaires*' que l'on peut superposer dans une seule dimension de qualification[43].

---

[41] C'est l'essence de ce qui, en mécanique quantique, est lié au *principe de complémentarité*. Mais souvent (sinon toujours) la condition d'**unicité de l'exemplaire** du microétat de *un* microsystème que l'on considère, est oubliée lorsqu'on parle de complémentarité en mécanique quantique. Cela conduit à beaucoup de confusions. Car c'est précisément cette condition d'unicité de l'exemplaire mis en jeu lors d'un seul acte de mesure, qui est essentielle (et qui en outre est douée d'universalité (MMS [2002A], [2002A], [2006])).

[42] Il apparaîtra plus loin que cette définition de l'incompatibilité mutuelle de deux grandeurs définies pour un microétat, posée ici pour le cas d'un micro-système, *cesse d'être valide dans le cas d'un micro-état de deux ou plusieurs micro-systèmes: Ce qui décide finalement de la 'compatibilité', est la possibilité, ou non, de mesurer les grandeurs considérées simultanément **sur un seul exemplaire** du micro-état considéré.*

[43] Par exemple, imaginons que *X* est la re-définition pour le cas d'un microétat, de la grandeur classique 'quantité de mouvement' selon une seule dimension d'espace (dont en mécanique classique la valeur s'écrit alors *p=mv* où *m* est la masse du mobile et *v* dénote sa vitesse selon la dimension d'espace considérée), cependant que *Y* est la re-définition conceptuelle de la grandeur classique 'énergie cinétique' selon une seule dimension d'espace (dont en mécanique classique la valeur s'écrit $T=(p^2/2m)=(mv^2/2)$). En transposant ces définitions conceptuelles classiques, dans les termes du formalisme quantique (sur la base d'une construction abstraite qui prolonge les concepts dénotés '*p*' et '*T*' dans la mécanique classiques), leur relation de dépendance conceptuelle et formelle se maintient et elle conserve sa forme. En ces conditions il est évident que selon le formalisme quantique il suffit qu'un seul acte de *Mes(PT)* produise, à partir d'un seul exemplaire du microétat étudié, un unique ensemble de manifestations physiques observables commun à *P* et à *T*, puisque le super-codage comporté par la re-définition conceptuelle de *P* permet d'associer à cet ensemble de marques un *sens* en termes d'une valeur *pj* de la quantité de mouvement *P*, cependant que le super-codage comporté par *T* consiste à simplement calculer ensuite aussi un autre sens, énergétique, $Tj=((pj)^2/2m$. Bref, on peut procéder selon la méthode 'time of flight' pour obtenir un seul ensemble de marques physiques observables, et ensuite, faire les *deux* calculs mentionnés: En général maintenant, dire que deux grandeurs *X* et *Y* sont 'compatibles' ne veut dire *que* ceci, précisément, que *X* et *Y* ne diffèrent l'une de l'autre *que* conceptuellement, par la manière d'associer à un ensemble donné *unique* de marques physiques observées, les deux valeurs *Xj* et *Yq* différentes qui correspondent à cet ensemble unique selon deux codages différents mais reliés que l'on peut transformer l'un dans l'autre.

### 2.5.4. La 'description' versus *sa genèse produite par le concepteur-observateur*

Il ne semble donc pas impensable que l'on obtienne la même loi de probabilité *{p(G,Xj)}* pour deux microétats distincts (produits par deux opérations de génération différentes). Bien que cela ne semble pas pouvoir être fréquent, on aurait des réticences à en affirmer l'impossibilité. Si alors, afin d'agrandir le degré de vraisemblance d'avoir établi une spécificité de $me_G$, on considéré deux grandeurs qualifiantes au lieu d'une seule, il résulte des considérations du paragraphe précédent que l'effet de spécification recherché ne se produira pas si les deux grandeurs choisies sont mutuellement compatibles, car dans ce cas elles n'introduisent qu'une seule dimension *physique* de qualification.

Par contre, sur la base de ces mêmes considérations il paraît suffisamment sûr d'admettre que deux groupes non compatibles de grandeurs mutuellement compatibles, agissent comme deux 'directions de qualification' distinctes qui, en s''intersectant', fournissent une spécificité du microétat étudié $me_G$ ; c'est à dire, qu'aucun autre microétat engendré à l'aide d'une autre opération de génération différente de l'opération $G$, ne conduit exactement au même couple de deux lois de probabilité liées à ces deux grandeurs non compatibles, que le couple de lois trouvées avec $G$ et $me_G$.

D'autant plus, alors, l'ensemble de *toutes* les lois de probabilité obtenues avec **une** *opération G fixée* et **tous** les groupes non compatibles de grandeurs compatibles qui sont redéfinies pour les microétats, peut être tranquillement considéré comme spécifique à $me_G$. En *ce* sens, et en ce sens seulement, il semble possible d'affirmer que :

L'ensemble des lois de probabilité évoqué – bien qu'il ne concerne *pas* le microétat $me_G$ isolé des interactions de mesures qui l'ont engendré – *constitue néanmoins une "description" **de** ce microétat lui-même, en ce sens et ce sens seulement que cet ensemble de lois de probabilité constitue une redéfinition de ce microétat en termes – non pas de l'opération de génération G qui l'a produit – mais en termes de **connaissances** qui **concernent** ce microétat **lui-même et spécifiquement** et qui sont communicables, vérifiables et consensuelles.*

La 'description' – dans le sens précisé – que nous venons d'obtenir est ***foncièrement relative*** à :
 * l'opération de génération $G$ qui agit uniformément dans toutes les successions réitérées *[G.Mes(X)]*, pour *toute* grandeur $X$ définie pour le microétat étudié;
 * l'effet $me_G$ de l'opération $G$;
 * la grille de qualification constituée par l'*ensemble* des grandeurs (mécaniques ou d'autre nature) définies pour un microétat.

Mais une fois que le processus descriptionnel est accompli, l'effet final de ce processus – globalisé sur le niveau de conceptualisation statistique, pour toutes les grandeurs mécaniques redéfinies pour des microétats – consiste en, ***EXCLUSIVEMENT***, des marques physiques transférées via de interactions de mesure sur des enregistreurs – épars – d'appareils divers, des marques codées en termes de 'valeurs' $Xj$ de 'grandeurs' $X$ et dont on a dénombré les fréquences relatives de réalisation: il s'agit de *'descriptions transférées' primordiales 'probabilistes'* au sens de statistiques stables *face* à l'entier $N$ choisi pour le nombre de répétitions des successions *[G.Mes(X)]* qui interviennent.

*Or là, dans le **PRODUIT** final – transféré et primordial – que les actions cognitives du concepteur-observateur ont déposé, toute organisation **globale** d'espace-temps est absente*[44].

Afin de mettre en évidence ce fait épistémologique fondamental nous introduisons les dénominations et les notations suivantes :

La grandeur *X* correspond à une *vue $V_X$* où le symbole $V_X$ dénote globalement la grille de qualification *gq.primord.(X).me$_G$* définie dans *2.3.2.3*. Dénommons cette symbolisation *vue-aspect mécanique correspondant à la grandeur mécanique X redéfinie pour des microétats*. Soit *{$V_X$}* l'ensemble des 'vues-aspect' $V_X$. Cet ensemble sera dénoté $V_M$ et sera dénommé la vue mécanique globale définie pour des microétats.

**L'action** descriptionnelle **globale** face à $V_M$ ou à $V_X$ – **avec sa genèse et son résultat final** – sera symbolisée, respectivement,

$$D_M/G, me_G, V_M/ \quad \text{ou} \quad D_X/G, me_G, V_X/$$

Les désignés de ces symboles sont dotés d'une organisation définie d'espace-temps **des actions descriptionnelles du concepteur-observateur**.

L'ensemble *{p(G,Xj)}*, $\forall V_X \in V_M$, de – **exclusivement** – toutes les distributions de probabilité obtenues pour le microétat *me$_G$* qualifié par $V_M$ sera dénommé *la description mécanique **proprement dite** de me$_G$* et sera dénotée par

$$D_M \equiv \{p(G,Xj)\}, \, j=1,2,...J, \quad \forall V_X \in V_M$$

Si en particulier l'on a $V_M \equiv V_X$ l'on dénotera la description mécanique correspondante proprement dite par

$$D_X \equiv \{p(G, Xj)\}, \, j=1,2,...J$$

et l'on dira qu'elle n'est reliée à l'entité-objet *me$_G$* que face à l'unique vue-aspect mécanique *X*.

**Les descriptions proprement dites $D_M$ et $D_X$ sont dépourvues de toute structure globale d'espace-temps**.

Voilà les contenus de la forme *qualitative* générale des descriptions de microétats impliquées dans le formalisme quantique. Cette forme est maintenant là, devant nos yeux, dénudée de toute adhérence à des éléments mathématiques, *intégrée*, à fonctionnalités bien définies tout autant en ce qui concerne le contenu de chacune de ces fonctions, qu'en ce qui concerne leur succession : Une forme descriptionnelle comme vivante.

L'hypothèse faite au départ, selon laquelle le contenu épistémologique qualitatif du formalisme quantique est déterminé par la situation cognitive dans laquelle on se trouve lorsqu'on veut décrire des microétats, semble se confirmer. Mais il reste à parfaire cette confirmation. Car nous venons de parler de "description probabiliste" $D_M \equiv \{p(G,Xj)\}$,

---

[44] En conséquence des *non*-compatibilités mutuelles (*2.5.3.2*) entre certaines grandeurs mécaniques distinctes, les configurations d'espace-temps des groupes de marques *{μ$_k$}, k=1,2,...m* produits par les différents actes de mesure impliqués, ne sont en général *pas* intégrables dans une définition d'une structure d'espace-temps définie du résultat **global** d'une description d'un microétat, face à **toutes** les grandeurs mécaniques redéfinies pour lui.

*j=1,2,...J,* $\forall V_X \in V_M$, et – en un certain sens – il n'y a pas de doute qu'il semble adéquat en effet de qualifier cette description comme 'probabiliste'. Mais en ce cas le terme 'probabiliste' possède un sens *nouveau* et il faut expliciter à fond en quoi cette nouveauté consiste. Je le ferai ci-dessous, sur la base d'une explicitation de la structure d'espace-temps de l'action descriptionnelle $D_M/G,me_G,V_M/$ ou $D_X /G,me_G,V_X /$ déployée, respectivement, afin d'accomplir une description proprement dite $D_M \equiv \{p(G,Xj)\}$, *j=1,2,...J,* $\forall V_X \in V_M$ ou $D_X \equiv \{p(G,Xj)\}$, *j=1,2,...J.*

## 2.6. L'arbre de probabilité de l'opération de génération *G* d'un microétat *:*

### *2.6.1. Le cas fondamental d'un microétat de* un *microsystème et à opération de génération* simple

Dans ce qui précède, la *genèse* de la description $D_M/G,me_G,V_M/$ d'un microétat – **opérée par le concepteur-observateur humain** – a été indiquée d'une manière très morcelée et étalée, argumentée pas à pas. L'on ressent le besoin d'une reformulation plus rapide et synthétique qui permette de percevoir en un seul coup d'œil la structure globale de cette genèse. Sur le terrain conceptuel affermi dans les paragraphes précédents, nous élaborerons maintenant – d'abord pour le cas fondamental de *un* microsystème, et à opération de génération simple (non-composée) – une représentation d'espace-temps synthétisé, 'géométrisé', de *l'action descriptionnelle* comportée par $D_M/G,me_G,V_M/$. Cette représentation permettra de jauger mieux les spécificités et les remarquables perspectives nouvelles ouvertes par le type descriptionnel dénoté $D_M/G,me_G,V_M/$). Car il apparaîtra que l'organisation qualitative du concept de 'probabilités' *primordiales* (de statistiques primordiales relatives stables) qui s'est constitué, *se distingue foncièrement du concept qu'avait dans son esprit Kolmogorov ([1933]) lorsqu'il a formulé sa théorie classique des probabilités* [45, 46].

### *2.6.1.1. Construction et résultat global*

Ce qui suit immédiatement concerne donc le cas fondamental d'un micro-état d'*un* seul micro-système (*2.4.2*) et à opération de génération simple. Les autres cas seront considérés ensuite *via* des généralisations de ce premier cas.

Soit une opération de génération *G* qui engendre un microétat $me_G$. Dénotons ici *B*, *C*, *D*, etc. les grandeurs mécaniques redéfinies pour des microétats qui engendrent la description $D_M/G,me_G,V_M/$ de $me_G$ via des mesures *Mes(B)*, *Mes(C)*, *Mes(D)*, etc. Tout ce qui est essentiel concernant $D_M/G,me_G,V_M/$ – avec son entière genèse – peut être représenté d'une manière intuitive visuelle, à l'aide d'un schéma d'espace-temps qui *globalise* la réalisation et les résultats d'un grand nombre de répétitions de la succession d'opérations *[G.Mes(B)]*, ainsi que de la succession d'opérations *[G.Mes(C)]*, ainsi que de la succession d'opérations *[G.Mes(D)]*, etc. Il ne s'agit pas d'un modèle. Il s'agit juste d'une *représentation* accomplie *dans l'espace-temps de l'observateur-concepteur humain*.

---

[45] Faute des guidages complexes qu'offre la méthode générale de conceptualisation relativisée, l'exposé qui suit est très simplifié. Dans MMS [2002] pp. 256-291 et [2006] pp.193-257 on peut trouver des exposés beaucoup plus accomplis.

[46] Comme annoncé dans la note 2, nous accompagnerons les résultats qui émergeront, de brèves indications anticipées concernant leurs correspondants dans le formalisme mathématique de la mécanique quantique : Ces commentaires, toutefois, resteront strictement *extérieurs* à la construction élaborée dans la première partie de ce travail, de simples pointage du doigt qui préparent à une compréhension rapide de la comparaison constructive qui constitue la deuxième partie de ce travail.

Détaillons, mais en simplifiant au cas essentiel de seulement deux grandeurs dynamiques mutuellement incompatibles, $B$ et $C$.

Considérons d'abord le processus de génération $G$ du microétat $me_G$. Ce processus commence à un moment initial, disons $t_o$, et il finit à un moment ultérieur, disons $t_G$. Il possède donc une durée $(t_G$-$t_o)$. Il balaye aussi un certain domaine d'espace, disons $d_G$. Donc il couvre un domaine global d'espace-temps $[d_G.(t_G$-$t_o)]$. Au moment $t_G$ quand le processus de génération $G$ est accompli – donc le microétat $me_G$ peut déjà être supposé exister – on commence aussitôt un acte de $Mes(B)$. Ainsi l'on accomplit une succession $[G.Mes(B)]$. Cette succession finit avec l'enregistrement par l'appareil $A(B)$, d'un certain groupe de marques physiques observables. Au moment où l'enregistrement de ce groupe de marques est accompli, l'opération physique de $Mes(B)$ est terminée et l'entière succession $[G.Mes(B)]$ est elle aussi accomplie. Notons $t_B$ ce moment final. Le processus physique $Mes(B)$ aura donc couvert dans le temps du concepteur-observateur une durée $(t_B$-$t_G)$. Il aura également balayé un certain domaine d'espace, disons $d_B$. Il se sera donc produit sur un domaine d'espace-temps $[d_B.(t_B$-$t_G)]$. Quant à l'entière succession $[G.Mes(B)]$, elle aura couvert le domaine d'espace-temps $[d_G.(t_G.t_o)+d_B.(t_B$-$t_G)]$.

Après avoir accompli le processus physique $Mes(B)$, on accomplit en outre une opération supplémentaire, *conceptuelle* cette fois: le groupe de marques physiques observables enregistré par l'appareil $A(B)$ doit être *codé* conformément à la règle de codage associée à la re-définition conceptuelle de la grandeur $B$ pour le cas des microétats, et ce codage fournit une traduction du groupe de marques physiques enregistrées, en termes d'une valeur $Bj$ du spectre $\{B1, B2,… Bj,…Bk\}$[47] de la grandeur $B$, comme l'exige la définition d'une grille de qualification de $me_G$ via la grandeur $B$.

Répétons un très grand nombre $N$ de fois la réalisations de la même succession $[G.Mes(B)]$, en remettant à chaque fois le chronomètre à $0$ comme pour des épreuves sportives. Idéalement, le domaine total d'espace-temps couvert sera à chaque fois le même: $[d_G.(t_G$-$t_o)]$. Quant au domaine d'espace-temps $[d_B.(t_B$-$t_G)]$, en général il variera car l'enregistrement final d'une manifestation observable aura des coordonnées d'espace-temps aléatoires (puisque la situation est en général probabiliste, donc à la base, statistique). L'on considérera donc le domaine d'espace-temps *global* couvert par l'ensemble des actes de mesure et on le désignera par la même notation. Si le nombre $N$ est assez grand, l'ensemble de toutes les $N$ répétitions de la succession $[G.Mes(B)]$aura progressivement matérialisé *toutes* les valeurs du spectre $\{B1, B2, B3,....Bj,... Bk\}$, $j=1,2,....k$, de la grandeur $B$, puisque aucune parmi ces valeurs n'a une probabilité *a priori* nulle. Chacune de ces valeurs aura été obtenue avec une certaine fréquence relative. Et si $N$ est très grand, alors l'ensemble $\{n(G,B1)/N, n(G,B2)/N,.... n(G,Bj)/N.....n(G,Bk)/N.\}$ des fréquences relatives obtenues sera assimilable à la loi correspondante de probabilité

$$\{p(G,Bj)\} = \{p(G,B1), p(G,B2),....p(G,Bj),...p(G,Bk)\}, \qquad j=1,2,....k,$$

présupposée 'existante'. Bref, au bout de ces $N$ réalisations d'une succession $[G.Mes(B)]$ le 'plafond' du domaine d'espace-temps $[d_G.(t_G$-$t_o)+d_A(t_A t_G)]$ se trouvera finalement garni par toutes les valeurs du spectre $\{B1, B2, B3,....Bj,.... Bk \}$ de la grandeur $B$, et – sur un niveau descriptionnel supérieur – l'on pourra inscrire l'entière loi factuelle de probabilité $\{p(G,Bj)\}$ constatée sur ce spectre. Or le couple

---

[47] Nous supposons un spectre fini, pour simplifier.

$$[(B1, B2, B3, ..Bj \ ..Bk), \quad \{p(G,Bj)\}]$$

est l'essence[48] de ce qui, dans la théorie moderne des probabilités formulée par Kolmogorov ([1933]), est dénommé *un espace de probabilités*: le résultat qui s'est constitué est donc, en termes simplifiés, l'espace de probabilité qui inclut la loi de probabilité *{p(G,Bj)}*.

Considérons maintenant la grandeur *C* – qui par hypothèse est *in*compatible avec la grandeur *B* – et refaisons concernant *C* un chemin strictement analogue à celui qu'on vient d'indiquer concernant la grandeur *B*. Au bout d'un très grand nombre *N* de répétitions de la succession de deux opérations *[G.Mes(C)]* couvrant un domaine global d'espace-temps *[d_G.(t_G-t_o)+d_C.(t_C-t_G)]* qui cette fois correspond à *Mes(C)*. Au dessus du plafond de ce nouveau domaine d'espace-temps l'on aura finalement inscrit un autre espace de probabilité

$$[(C1, C2, C3, ....Cq, ..Cm), \ \{ \ p(G,Cq) \ \}]$$

Puisque les grandeurs *B* et *C* sont incompatibles, les deux domaines d'espace-temps *[d_B.(t_B-t_G)]* et *[d_C.(t_C-t_G)]* couverts respectivement par des *Mes(B)* et des *Mes(C)*, seront *différents*. Mais le domaine d'espace-temps *[d_G.(t_G-t_o) ]* couvert par l'opération de génération *G* est le *même* dans les successions *[G.Mes(B)]* et les successions *[G.[Mes(C)]* parce que les mesures de la grandeur *B* et celles de la grandeur *C* ont été effectuées toutes sur des exemplaires du même microétat *me_G* engendré par l'opération de génération *G*. Ainsi la structure globale d'espace-temps de tout l'ensemble de successions de mesure accompli, est arborescente, avec un tronc commun couvrant le domaine d'espace-temps *[d_G.(t_G-t_o)]* et deux branches distinctes qui couvrent globalement deux domaines d'espace-temps *[d_B.(t_B-t_G)]* et *[d_C.(t_B-t_G)]* différents. Chaque branche est surmontée d'un espace de probabilité spécifique à la branche. C'est pour cette raison qu'il est adéquat d'appeler cette structure *un arbre de probabilité du microétat me_G correspondant à l'opération de génération G* (en bref, d'un microétat). On peut désigner cet arbre par le symbole *T[G,(V_M(B)∪V_M(C)] (T:* tree ; *(V_M(B)∪V_M(C)*: la vue (grille) qualifiante constituée par *l'union* des grandeurs mécaniques redéfinies pour des microétats qui ont été dénotées par *B* et *C*).

Si l'on avait représenté *toutes* les branches possibles impliquées dans la description *D_M/G,me_G ,V_M/*, correspondant à toutes les grandeurs mécaniques mutuellement incompatibles définies pour un microétat, l'on aurait dû employer l'article défini et dire '*l'arbre de probabilité* .......'. Dans ce cas général la dénotation aurait été *T(G,V_M))*[49].

---

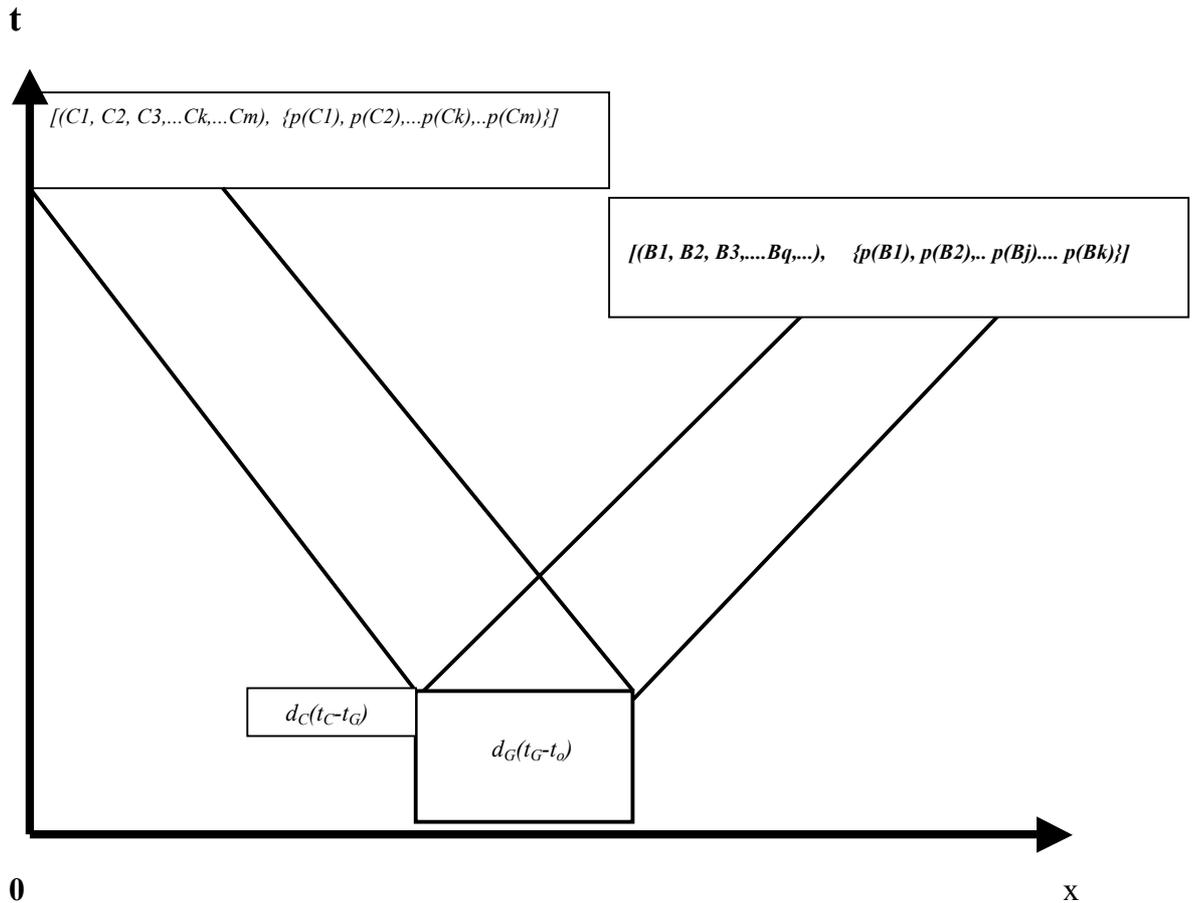

**t**

$[(C1, C2, C3,...Ck,...Cm),\ \{p(C1),\ p(C2),...p(Ck),..p(Cm)\}]$

$[(B1, B2, B3,....Bq,...),\ \ \ \{p(B1),\ p(B2),..\ p(Bj)....\ p(Bk)\}]$

$d_C(t_C{-}t_G)$

$d_G(t_G{-}t_o)$

**0**          x

***Fig. 3:*** *Un arbre de probabilité T[G, (V$_M$(B)∪V$_M$(C)] de l'opération de génération G d'un microétat de **un** microétat.*

Concernant le cas général notons maintenant ce qui suit. L'arbre de probabilité $T(G,V_M)$ étale, géométrisée, l'entière structure d'espace-temps de la genèse opérationnelle accomplie par le concepteur humain afin d'établir une description transférée primordiale de microétat, $D_M/G,me_G,V_M$.

*L'effet global observable* d'une telle genèse opérationnelle, auquel, à strictement parler la 'description' proprement dite $D_M ≡ \{p(G,Xj)\}$, $j=1,2,...J$, $\forall V_X \in V_M$, *se réduit*, est lui aussi explicitement représenté dans ce schéma: c'est l'ensemble des *symbolisations* des espaces de probabilité qui *coiffent* les branches de $T(G,V_M)$, *écrites* mais *a-morphes*[50] : En effet il saute maintenant aux yeux que cet effet global observable ne peut pas être représenté géométriquement lui aussi, car étant dépourvu de structuration d'espace-temps on ne peut pas en évacuer le temps et en garder une géométrisation purement spatiale, comme nous venons de le faire pour la représentation de la genèse opérationnelle accomplie par le concepteur humain.

---

[50] Un physicien de la mécanique quantique perçoit tout de suite que les espaces de probabilité symbolisés *au dessus* des branches – à eux *seuls* et *tous ensemble* – sont l'équivalent qualitatif de la description du microétat $me_G$ offerte par le formalisme quantique *via* le vecteur d'état ǀΨ> associé à $me_G$ (qui permet, à l'aide aussi d'autres éléments de l'algorithme quantique, de calculer l'ensemble de lois de probabilité concernant $me_G$ ). Mais l'arbre lui-même, tout ce qui, dans le schéma de la *Fig.3* est en *dessous* de la couronne de lois de probabilité, *n'est pas représenté* dans le formalisme mathématique de la mécanique quantique.

Qu'on ne se laisse pas porter par des impressions inertielles de trivialité. L'expression 'arbre de probabilité' est employée dans des contextes divers où elle n'implique aucune nouveauté conceptuelle. Mais le concept d'arbre de probabilité *de l'opération de génération G d'un microétat* est un concept probabiliste foncièrement nouveau qui *déborde* de plusieurs manières le concept d'un espace de probabilité de Kolmogorov. Pour s'en rendre compte il suffit de considérer les précisions qu'il comporte concernant les notions de phénomène aléatoire, de dépendance probabiliste et de 'méta-loi de probabilité, qui sont exposées plus bas.

### 2.6.1.2. Le phénomène aléatoire isolé

La théorie classique des probabilités est dépourvue d'une expression formalisée de la notion de phénomène aléatoire. Cette notion y est posée comme engendrant un espace de probabilité, mais elle n'intervient que dans le substrat *verbal* de la théorie, de façon vague. On ne conceptualise et l'on ne représente mathématiquement que l'espace de probabilité engendré par le phénomène aléatoire, dont les éléments, par contre, sont étudiés en détail. On peut caricaturer par le schéma:

[? 'phénomène aléatoire'? ] → [espace de probabilité]

Les signes d'interrogation soulignent le vide de conceptualisation laissé en ce qui concerne la genèse d'un espace de probabilité.

Cependant que l'arbre de probabilité d'un microétat inclut au contraire une représentation minutieusement explicitée de, précisément, les *phénomènes aléatoires-branches* qui engendrent les événements élémentaires à partir desquels on construit ensuite les espaces de probabilité qui coiffent les deux branches. *Tout* ce qui dans la *Fig. 3* se trouve en dessous de l'écriture qui dénote l'espace de probabilités d'*une* branche, est phénomène aléatoire correspondant. On y *voit*, on comprend en détail, on *sent* toute la montée créatrice d'une "situation probabiliste": L'opération de génération *G crée* l'entité-objet $me_G$ à étudier, de l'intérieur de la factualité physique a-conceptuelle ; l'acte de *Mes(X)*, où, à tour de rôle, X=B,C, etc., s'empare de cette entité-objet – en la *changeant* par interaction avec l'appareil *A(X)* – et porte cette entité incorporée dans l'interaction de mesure jusqu'au *bord* de la conceptualisation, en produisant les groupes *{μk}, k=1,2,...m* de marques physiques observables sur les enregistreurs de *A(X)* ; la loi de codage qui traduit ces groupes de marques en valeurs *Xj* de *X*, constitue *un prolongement – conceptuel – du processus physique de mesure*, qui dépose cette fois *dans* la toute première couche du conceptualisé, une 'valeur' de l'univers *{X1, X2,...Xj...Xk}* de telles 'valeurs' possibles assignées à *X*, en lequel consiste le spectre des événements élémentaires ; et l'espace de probabilité correspondant *[{X1, X2....,Xj,...Xk}, {p(G,Xj)}]* (où l'algèbre posée sur l'univers d'événements élémentaires est ici supprimée pour simplification) est finalement construit sur cet univers d'événements élémentaires : Le phénomène aléatoire-'branche' qui engendre cet espace est défini dans toutes ses phases.

Ainsi, à la faveur du cas particulier des microétats, le concept probabiliste *général* de phénomène aléatoire est en cours d'acquérir une représentation, qualitative mais à signifiance non restreinte (MMS [2006]). Or il apparaît que cette représentation est elle aussi (comme la description proprement dite de $me_G$) explicitement et foncièrement relative à l'opération de génération *G* de l'entité-objet-d'étude $me_G$, à cette entité $me_G$ elle-même, et à la vue qualifiante $V_X$ utilisée:

*Cette sorte d'éléments descriptionnels G, $me_G$, $V_X$, et cette sorte de relativités à ceux-ci n'interviennent pas dans la conceptualisation probabiliste classique, et pourtant il* *semble fortement vraisemblable qu'un examen attentif révélerait qu'ils **sont implicitement présents et agissent en toute description probabiliste**.*

### 2.6.1.3. Dépendances probabilistes non-classiques

Jusqu'ici, pour simplifier, nous avons fait abstraction de l'algèbre d'un espace de probabilité au sens de Kolmogorov. Maintenant nous allons remédier à cette lacune afin de pouvoir faire des remarques importantes sur le concept de *dépendance probabiliste* concernant lequel Kolmogorov a écrit ([1950], p.9):

«…..one of the most important problems in the philosophy of the natural sciences is – in addition to the well known one regarding the essence of the concept of probability itself – to make precise the premises which would make it possible to regard any given real events as independent.»

Nous rappelons donc ce qui suit.

Un espace de probabilité *[U, τ, p(τ)]* de Kolmogorov contient: un *univers* (ensemble) *U d'événements élémentaires $e_i$*, une *algèbre τ d'événements e définie sur U*, et une *mesure de probabilité p(τ) qui est définie **sur l'algèbre d'événements τ**.* Un événement au sens des probabilités est constitué par tout *ensemble* d'événements élémentaires. Une algèbre τ d'événements définie sur l'univers *U* d'événements élémentaires est un ensemble de sous-ensembles *e* de *U (un ensemble de $e_i$)* – contenant *U* lui-même ainsi que l'ensemble vide *Ø* – et qui est tel que si les sous-ensembles d'événements élémentaires *A* et *B* sont contenus dans *U* alors *U* contient également la réunion *A∪B* et la différence *A-B*. Enfin, *une mesure de probabilité définie sur τ* consiste en un ensemble de nombres réels *p(A)* dont chacun est associé à un événement *A* de *τ* et qui satisfont aux conditions suivantes: *0≤p(A)≤1, p(U)=1* (normation), *p(Ø)=0*, et *p(A∪B)≤p(A)+p(B)* où l'égalité se réalise ssi *A* et *B* sont disjoints (n'ont aucun événement élémentaire $e_i$ en commun *(A∩B=Ø)*.

Deux événements *A* et *B* de l'algèbre *τ* sont posés par définition être mutuellement *'indépendants'* si le produit numérique *p(A)p(B)* de leurs probabilités est égal à la probabilité *p(A∩B)* de l'événement-produit-ensembliste *A∩B*. (Cette définition est étendue au cas de deux (ou plusieurs) algèbres distinctes *τ* et *τ'≠τ*, mais en présupposant la possibilité d'une réalisation *conjointe* de *A∈τ* et *B∈τ')*[51].

Ces définitions restent pertinentes à l'intérieur – isolément – de *chaque* espace de probabilité qui coiffe une branche de l'arbre de probabilité *T(G,$V_M$)* du microétat *$me_G$*.

Considérons maintenant un événement *A(X)* de l'algèbre d'événements *τ(U(Xj))* posée sur l'univers *U(Xj)* d'événements élémentaires de l'espace de probabilité

$$[U(Xj), τ(U(Xj)), p[τ(U(Xj))]]$$

qui coiffe la branche de *T(G,$V_M$)* correspondant à des *Mes(X)*. Et soit un événement *B(Y)* de l'algèbre d'événements *τ(U(Yk))* posée sur l'univers *U(Yk)* d'événements élémentaires de l'espace de probabilité

---

[51] La théorie classique des probabilités ne définit pas directement une éventuelle dépendance entre deux (ou plusieurs) événements *élémentaires*. Une telle dépendance ne peut être appréhendée que par *comparaison* avec la loi de probabilité posée sur un univers produit cartésien qui implique *U* en tant que l'un des facteurs, et qui est lié à un phénomène aléatoire différent de celui qui engendre l'espace *[U, τ, p(τ)]*.

*[U(Yk)}, $\tau$(U(Yk)), p[ $\tau$(U(Yk))]]*

qui coiffe une *autre* branche de *T(G,V$_Q$)* correspondant à des *Mes(Y) in*compatibles avec les *Mes(X)*.

Comment associer une définition mathématique à une dépendance – au sens *courant* du terme – entre deux événements de ce type? Car selon la pensée naturelle deux tels événements ne sont certainement *pas* 'in-dépendants', puisqu'ils concernent le *même* microétat *me$_G$* produit par une et même opération de génération *G*. Pourtant la notion d'occurrence 'conjointe' qui intervient dans la définition probabiliste classique de la dépendance entre deux événements, n'a pas de sens dans ce cas, car – par construction de *T(G,V$_M$)* – les deux événements *A(Xj)* et *B(Yk)* ne se produisent *jamais* 'simultanément', i.e. *pour un même exemplaire du microétat me$_G$*: deux branches distinctes de *T(G,V$_M$)* sont mutuellement incompatibles dans tous leurs éléments.

On pourrait se dire que l'incompatibilité entre *A(Xj)* et *B(Yk)* est une forme limite de dépendance maximale, car ces deux événements s'excluent systématiquement, et donc, puisque les probabilités séparées *p(A(Xj))* et *p(B(Yk))* ne sont pas nulles, on *peut* exprimer la dépendance en question en écrivant *p(A(Xj).B(Yk))=0*, ce qui – comme il est exigé par la définition classique de la dépendance probabiliste – est bien différent de *p(A(Xj))* x *p(B(Yk))*.

Mais on flaire un glissement, un abus d'extension de la syntaxe probabiliste classique.

La réticence se confirme lorsqu'on considère deux événements de deux *arbres* différents, fondés sur deux opérations de génération différentes *G* et *G'*, donc correspondant à deux microétats différents : Dans ce cas *aussi* la probabilité conjointe considérée plus haut est toujours nulle (si par 'conjoint' on entend 'pour un même exemplaire d'un microétat *donné*') cependant que les deux probabilités considérées isolément ne sont pas nulles. Mais d'autre part, se dit-on, pourquoi, dans ce nouveau cas, y aurait-il systématiquement *dépendance?* Avec *G* et *G'* différents on se trouve dans deux schémas probabilistes distincts, pas dans un seul schéma, comme avec une seule opération *G* de génération d'un microétat.

Et en effet le point important est là : dans le cas des microétats le tout probabiliste qui s'impose est celui de l'ensemble des espaces de probabilité liés à une opération de génération *G* donnée, et une seule. Tandis que dans la conceptualisation classique du concept de probabilité l'opération de génération de l'entité impliquée dans les événements élémentaires et les événements, n'intervient pas.

Il semble clair que par cette absence – au moins – *le domaine de validité de la conceptualisation probabiliste classique est confiné à un seul espace de probabilité, arbitrairement*. Car lorsqu'on considère l'ensemble des espaces de probabilité appartenant à l'arbre de probabilité d'un microétat, une certaine notion de dépendance *probabiliste* s'impose intuitivement – en conséquence de l'existence du tronc commun de l'arbre. Or cette notion tout simplement dépasse la notion classique de dépendance probabiliste.

*Il faut élargir la conceptualisation probabiliste classique[52].*

Cette conclusion se trouve renforcée juste ci-dessous.

### 2.6.1.4. Méta-phénomène aléatoire et méta-dépendances probabilistes

Chaque loi de probabilité *{p(G,Xj), j=1,2,...n}* de l'arbre $T(G,V_M)$ d'un microétat est contenue dans l'espace de probabilité qui coiffe une seule branche, et cette loi de probabilité est relative à la triade *(G,me$_G$,V$_M$(X))* spécifique de cette branche-là. Mais le fait qu'un et même couple *(G,me$_G$)* est impliqué dans le tronc commun de tous les phénomènes aléatoires de toutes les branches de l'arbre, ainsi que dans toutes les lois de probabilité que ces phénomènes aléatoires produisent, conduit irrépressiblement à poser, par exemple, qu'entre la probabilité *p(Yk)* d'un événement élémentaire de la branche des *Mes(Y)* de $T(G,V_M)$ et la loi de probabilité d'une autre branche de $T(G,V_M)$ – *considérée globalement, car comment détailler?* – il existe une certaine relation 'méta-probabilistes'. On pose donc que l'on a

$$p(Yk) = F\{p(G,Xj)\}^{53}$$

Ici *'F'* est une relation fonctionnelle dont la structure reste à être spécifiée. (Dans le cas particulier de la *Fig.3* on écrirait donc *p(Ck) = F{p(Bj)}* où *{p(Bj)}* désigne l'entière loi de probabilité sur les valeurs observables *Bj* de la grandeur *B*, qui est incompatible avec la grandeur *C*).

Il existe un *fait d'expérience* (dont ici, par notre règle de jeu, on ne peut énoncer que l'essence qualitative) qui fonde plus concrètement le postulat avancé plus haut: A l'intérieur d'un arbre $T(G,V_M)$ donné, si la dispersion des valeurs observables *Yk* de la grandeur *Y* distribuées selon la loi *{p(Yk)}*, est grande, alors la dispersion des valeurs observables *Xj* de la grandeur *X* – incompatible avec *Y* – distribuées selon la loi *{p(Xn)}*, est petite. Et *vice versa* [54] :

Quelle que soit la forme mathématique de la relation fonctionnelle *'F'* de méta-dépendance probabiliste, ce fait d'expérience implique que cette relation *existe*, et d'une façon qui comporte des manifestations physiques observables.

Selon le postulat posé, l'ensemble de tous les phénomènes aléatoires liés à toutes les grandeurs mécaniques *X* redéfinies pour des microétats et correspondant à *un même couple (G,me$_G$)*, peut être conçu comme *un méta-phénomène aléatoire global spécifique du microétat me$_G$* [55]: l'opération de génération *G* commune qui peuple le tronc de l'arbre fonde factuellement ce méta-phénomène aléatoire global.

---

[52] Ceci a été accompli de manière explicite et en termes tout à fait généraux, dans MMS [2002A], [2002B], [2006], [2013]. Mais sous une forme implicite et – en plus – cryptique, c'est déjà aussi accompli à l'intérieur du formalisme de la mécanique quantique, depuis quelque 80 ans.

[53] Un physicien de la mécanique quantique reconnaîtra dans l'assertion posée le trait fondamental de la *théorie des 'transformations'* de Dirac (de la base choisie dans l'espace Hilbert du système). Il notera également que *dans le formalisme quantique on n'associe **pas** à ce trait une signification probabiliste définie (qui, si elle était posée, devrait être reconnue comme non-classique)*.

[54] Un physicien reconnaît là tout de suite l'assise qualitative du *théorème* dit 'd'incertitude' établi dans le formalisme mathématique de la mécanique quantique (à ne pas identifier au *principe* d''incertitude' de Heisenberg, qui est extérieur au formalisme quantique).

[55] Les logiciens emploient quelquefois le mot 'méta' pour désigner le langage *d'immersion* du langage considéré. J'attire l'attention qu'ici, au contraire, ce mot est employé comme signifiant *'au dessus'*, 'qui, d'un point de vue constructif, se place à un niveau conceptuel plus élevé que celui où émerge le concept classique d'un seul phénomène aléatoire'.

Considérons maintenant *exclusivement* l'ensemble de toutes les lois de probabilité qui coiffent les branches de l'arbre d'un microétat $me_G$ – séparément du méta-phénomène aléatoire global que l'on vient de définir et qui engendre cet ensemble de lois – muni de la relation méta-probabiliste inconnue $p(Yk)=F\{p(G,Xj)\}$ posées entre ses éléments.

Cet *ensemble* de lois de probabilité $\{p(G,Xj)\}{\equiv}D_M, j=1,2,...J, \forall V_X{\in} V_M$, reliées l'une à l'autre peut être regardé comme constituant *une, 'la', méta-loi de probabilité spécifique de $me_G$*[56,57].

Pour cette raison nous allons désormais dire que $D_M$ est *la méta-loi de probabilité de l'arbre de probabilité $T(G,V_M)$ engendré par G* est nous écrirons symboliquement

$$\{p(G,Xj)\}{\equiv}D_M{\equiv}\mathcal{M}\ell(G)$$

La suite des remarques qui viennent d'être faites peut être synthétisée en disant que l'arbre de probabilité $T(G,V_M)$ d'un microétat $me_G$ constitue un *tout* probabiliste nouveau où les différents phénomènes aléatoires et les différents espaces de probabilité qui interviennent, s'unissent de manière organique.

On pourrait avoir tendance à subsumer les relations méta-probabilistes posées plus haut, au concept probabiliste classique de 'corrélation probabiliste'. Mais en fait les caractéristiques spécifiques du type de relations signalées ici, n'ont aucun correspondant explicitement élaboré dans le calcul classique des probabilités. Ici on est en présence de méta-relations de 'dépendance' probabilistes d'un type bien défini, foncièrement liée au fait particulier mais très fréquent que les lois de probabilité $\{p(Yk)\}$ et $\{p(Xn)\}$ concernent le même microétat. La conceptualisation probabiliste classique n'individualise *pas* cette sorte particulière de corrélations et la condition spécifique de leur émergence.

### 2.6.1.5. La logique de l'ensemble des événements d'un arbre de probabilité d'un microétat

Si au lieu de chercher plus ou moins artificiellement une 'logique quantique de *propositions'* fondée directement sur la structure mathématique des algorithmes quantiques Hilbert-Dirac (von Neumann, Birkhoff, Jauch, Piron), l'on descend au niveau plus profond et général de la conceptualisation des microétats où se placent les arbres de probabilité des opérations de génération *G*, et si l'on y définit *la logique des événements d'un arbre de probabilité d'un microétat donné*, alors se révèle un terrain conceptuel beaucoup plus naturel et plus clairement pré-organisé. Ceci permet *d'unifier la logique de l'ensemble des événements liés à un microétat donné, avec les probabilités de ces événements* (MMS [1992C].

Cela conduit ensuite à une refonte *générale* du concept de probabilité et à une *unification générale entre les conceptualisations probabiliste et logique* [2002B], [2006]).

---

[56] Le singulier souligne *l'unité instillée par l'unicité de la paire (G,me$_G$)* qui intervient dans toutes ces lois de probabilité.
[57] Mackey [1963], Gudder [1976], puis Suppes, Beltrametti, et d'autres (probablement à ce jour même), ont recherché – par des voies purement mathématiques – une formulation satisfaisante pour une méta-loi de probabilité associable à l'entier vecteur d'état $|\Psi>$ correspondant à un microétat $me_G$. Ici *le fondement qualitatif d'une telle formulation se fait jour naturellement et il permettra peut-être d'identifier la forme mathématique pertinente de la fonctionnelle dénotée F*.

### 2.6.1.6. Conclusion sur l'arbre de probabilité de l'opération de génération
### de *un* microétat de *un* microsystème

Le concept qualitatif d'arbre de probabilité d'un micro-état $me_G$ de un micro-système qui déploie la structure globale d'espace-temps des actions cognitives du concepteur-observateur qui opère la genèse de la forme descriptionnelle $D_M/G,me_G,V_M/$, met au grand jour des implications non-classiques de cette forme descriptionnelle. Celles-ci – avec l'inclusion des compatibilités et incompatibilités mutuelles entre des processus de mesure et les conséquences en termes de méta-dépendance probabiliste, et avec la signification méta-probabiliste de l'ensemble des lois de probabilité qui coiffent les branches d'un arbre $T(G,V_M)$ – constituent un apport nouveau au concept général de probabilité ([MMS 2006]).

### 2.6.2. Deux généralisations du concept
### d'arbre de probabilité de l'opération de génération G d'un microétat

Dans *2.6.1* nous avons considéré seulement le cas fondamental de l'arbre de probabilité d'un microétat de *un* microsystème et à opération de génération 'simple' (pas composée). Que devient ce concept dans le cas d'un micro-état de plusieurs micro-systèmes, ou dans le cas d'un microétat à génération composée?

#### 2.6.2.1. L'arbre de probabilité d'un micro-état de deux ou plusieurs micro-systèmes

Le concept d'arbre de probabilité établi pour un micro-état d'un seul micro-système se transpose d'une manière évidente à un micro-état de deux ou plusieurs micro-systèmes (cf. les définitions de *2.4.2*).

Pour fixer les idées, considérons un micro-état de *2* micro-systèmes, *S1* et *S2*, engendré par une opération de génération d'état $G_{12}$ qui soit a seulement impliqué les deux micro-systèmes *S1* et *S2*, soit les a même engendrés. Symbolisons ce micro-état par $me_{G12}$. Soit $T(G_{12},V_M)$ l'arbre de probabilité correspondant. Dans ce cas *une* opération de mesure opérée sur *un exemplaire* du micro-état $me_{G12}$, si elle est 'complète', comporte (par la définition correspondante de *2.4.2*) [une mesure $Mes_1(X)$ d'une grandeur *X*, opérée sur le micro-système *S1* comporté par l'exemplaire considéré du micro-état $me_{G12}$] *et* [une mesure $Mes_2(Y)$ d'une grandeur *Y*, opérée sur le micro-système *S2* comporté par ce même exemplaire considéré du micro-état $me_{G12}$] (en général *X* et *Y* sont différentes). Dénotons par $Mes_{12}(XY)$ une telle mesure complète (par convention, ici l'ordre d'écriture *XY* indique que *X* concerne le micro-système *S1* de l'exemplaire considéré du micro-état $me_{G12}$ et *Y* concerne le micro-système *S2* comporté par ce même exemplaire du micro-état $me_{G12}$).

Soit $\{X1,X2,...Xj,...X\phi\}$ le spectre des valeurs de *X* qui sont prises en considération (cf. la note 15) comme pouvant apparaître par une mesure de *X*, faite sur un micro-état du micro-système *S1* ou sur un micro-état du micro-système *S2*, n'importe[58] ; $\phi$ est un entier qui dénote le nombre total (le cardinal de l'ensemble) des valeurs du spectre de *X*. Soit $\{Y1,Y2,...Yk,...Y\varphi\}$ le spectre des $\varphi$ valeurs de *Y* prises en considération comme pouvant apparaître par une mesure de *Y* (qu'elle ait été faite sur un micro-état du micro-système *S1*, ou du micro-système *S2*, n'importe).

---

[58] Par convention on pose ici que le cardinal du spectre d'une grandeur reste invariant lorsqu'on passe d'une entité qualifiée à l'aide de cette grandeur, à une autre. Cela ne restreint aucun aspect essentiel.

Les événements *élémentaires* de l'espace de probabilité qui coiffe la branche de l'arbre de probabilité $T(G_{12}, V_M)$ du micro-état $me_{G12}$ qui correspond à des mesures complètes $Mes_{12}(XY)$ (formées chacune d'une $Mes_1(X)$ et d'une $Mes_2(Y)$), consistent en toutes les *paires* possibles *(Xj1,Yk2)* d'une valeur du spectre de *X* dénotée *Xj1* enregistrée – dans un exemplaire donné du micro-état $me_{G12}$ – pour le micro-système *S1* qui y est impliqué, *et* d'une valeur de *Y* dénotée *Yk2*, enregistrée – dans le *même* exemplaire du micro-état $me_{G12}$ – pour le micro-système *S2* qui y est impliqué. L'ensemble *{(Xj1,Yk2)}*, *j=1,2,….. φ, k=1,2,….. φ*, des paires *(Xj1,Yk2)* prises en considération comme possibles constitue donc l'univers des événements élémentaires correspondant à une mesure complète $Mes_{12}(XY)$ sur $me_{G12}$.

On peut avoir en particulier *X≡Y*. Mais en général les deux grandeurs considérées sont différentes *et elles peuvent même être mutuellement non-compatibles, dans le cas d'un micro-état d'un micro-système* : Car ces distinctions n'ont pas de conséquences sur le fait suivant.

Pour un micro-état de *deux* micro-systèmes, les grandeurs *X* et *Y* intervenant dans une $Mes_{12}(XY)$ complète, sont *toujours 'compatibles'*, en *ce* sens que, puisqu'elles sont effectuées sur deux micro-*systèmes distincts S1* et *S2*, elles s'incorporent respectivement à un acte de mesure d'indice *1* et une acte de mesure d'indice *2*, et que l'on peut toujours réaliser ces deux actes de mesure *simultanément sur un seul exemplaire du micro-état étudié* $me_{G12}$ [59].

Il en résulte que les deux actes de mesure de la paire *(Mes₁(X), Mes₂(Y))* qui constituent ensemble un acte de mesure complète $Mes_{12}(XY)$ sur $me_{G12}$, appartiennent toujours à *une même branche* de l'arbre de probabilité $T(G_{12}, V_M)$ de $me_{G12}$, celle des $Mes_{12}(XY)$ (pas des $Mes_{12}(ZW)$, etc.) qui est coiffée par l'espace de probabilité fondé sur l'univers des événements – 'élémentaires' au sens des probabilités – constitué de l'ensemble de paires *{X1jY2k}, j=1,2,….. φ, k=1,2,….. φ*, telles qu'elles viennent d'être définies. Donc :

Les deux événements *Xj1* et *Yk2* produits par ces deux actes de mesure *Mes₁(X)* et *Mes₂(Y)* – *observables séparément* – appartiennent néanmoins toujours à *un même* événement 'élémentaire' $Mes_{12}(XY)$ au sens des probabilités, d'*un même* espace de probabilités de l'arbre de probabilité du micro-état $me_{G12}$. Cela *quelle que soit la distance d'espace-temps* qui sépare les deux événements *physiques* [enregistrement de la valeur *Xj1* sur l'exemplaire considéré du micro-état $me_{G12}$] et [enregistrement de la valeur *Yk2* sur ce même exemplaire du micro-état $me_{G12}$] (i.e. quelle que soit la distance qui sépare les deux ici-maintenant au sens de la relativité, des deux événements physiques spécifiés).

Cela en conséquence du fait que les descriptions de microétats sont dépourvues d'une structure définie d'espace-temps. Seules les représentations – classiques – des actions

---

[59] Selon les définitions posées dans le paragraphe *2.4.2* de ce travail, et en généralisant le critère de compatibilité de grandeurs posé dans *2.5.3.2* pour le cas d'un micro-état de *un* microsystème tout en conservant l'essence de ce critère, *c'est bien la possibilité de réaliser simultanément sur un seul EXEMPLAIRE du micro-état étudié, toutes les mesures partielles impliquées dans une mesure complète, qui est l'unique critère de 'compatibilité' des grandeurs impliquées dans les actes de mesure partielle réalisables sur ce micro-état là*. Ce critère unique vaut quel que soit, parmi les différents types de définitions posés dans *2.4.2*, le type auquel appartient le micro-état considéré. *Ainsi se distillent ici et se séparent, d'une part les relativités impliquées dans le concept général de grandeurs compatibles, et d'autre part l'essence invariante du concept de "complémentarité"*.

descriptionnelles du concepteur-observateur macroscopique sont dotées d'une telle structure.

Que peut-on dire concernant la dépendance ou l'indépendance probabiliste mutuelle des événements physiques $Xj1$ et $Yk2$ impliqués dans l'univers de paires $\{X1jY2k\}$, $j=1,2,\ldots\ \phi$, $k=1,2,\ldots\ \varphi$, où l'ici-maintenant d'un événement $Xj1$ peut être arbitrairement éloigné de l'ici-maintenant de l'événement $Y2k$ qui lui est apparié?

Comme il a été déjà remarqué, le calcul des probabilités classique n'offre aucune façon de répondre à cette question à l'intérieur même de l'espace de probabilité, où les paires $X1jY2k$ ont le rôle d'événements *élémentaires*, car *(a)* la définition classique de la dépendance probabiliste ne s'applique *pas* aux événements élémentaires de l'univers qui fonde l'espace, et *(b)* si l'on passe dans l'algèbre d'événements posée dans cet espace, la définition classique de la dépendance est liée au concept de produit ensembliste (intersection ensembliste) de deux événements de l'algèbre, et ce produit, dans le cas considéré, ne contient que des paires *entières*, donc ce concept ne permet pas de séparer une paire $X1jY2k$ dans ces éléments.

Mais on peut utiliser une voie indirecte. Rien n'interdit de considérer *séparément*: *(a)* l'espace de probabilité fondé sur, *exclusivement*, l'ensemble des événements physiques $X1j$ de l'univers (probabiliste) $\{X11, X12,\ldots X1j, \ldots X1\phi\}$ (qui en général est le *seul* accessible à l'observateur qui exécute des mesures sur le micro-système $S1$ (dénotons par $o1$ cet observateur)) et d'assigner cette fois aux événements $X1j$ un caractère d'*'élémentarité'* probabiliste[60] ; et *(b)* l'espace de probabilité fondé sur l'ensemble $\{Y21, Y22, \ldots Y2k, \ldots Y2\varphi\}$ des événements physiques $Y2k$ qui en général sont les seuls perceptibles par l'observateur qui performe les mesures sur le micro-système $S2$ (dénotons cet observateur par $02$) et qui cette fois sont regardés eux aussi comme des événements élémentaires au sens des probabilités. Ainsi l'univers de paires élémentaires $\{X1jY2k\}$, $j=1,2,\ldots\ \phi$, $k=1,2,\ldots\ \varphi$, aura été scindé en deux univers distincts d'événements élémentaires individuels. Si ensuite l'on compare la distribution statistique (dite de 'probabilités') sur l'univers des paires $\{X1jY2k\}$, $j=1,2,\ldots\ \phi$, $k=1,2,\ldots\ \varphi$, aux deux distributions statistiques (de 'probabilités') sur les univers $X11, X12,\ldots X1j, \ldots X1\phi\}$, $j=1,2,\ldots\ \phi$, et $\{Y21, Y22, \ldots Y2k, \ldots\ Y2\varphi\}$, $k=1,2,\ldots$ $\varphi$, considérés séparément, on trouve – *ceci est un fait* – que la probabilité d'un événement $X1jY2k$ selon la distribution trouvée sur $\{X1jY2k\}$, $j=1,2,\ldots\ \phi$, $k=1,2,\ldots\ \varphi$, n'est en général *pas* égale au produit des deux probabilités que les deux événements $X1j$ et $Y2k$ impliqués possèdent, respectivement, à l'intérieur des univers $X11, X12, \ldots X1j, \ldots X1\phi\}$, $j=1,2,\ldots\ \phi$, et $\{Y21, Y22, \ldots Y2k, \ldots\ Y2\varphi\}$, $k=1,2,\ldots\ \varphi$, considérés séparément. Selon les règles du langage de la théorie classique des probabilités, il est permis d'exprimer ce fait en disant qu'*il y a une dépendance probabiliste* entre les événements $X1j$ et les événements $Y2k$ obtenus 'à partir' des mesures complètes $Mes_{12}(XY)$. On peut également exprimer ce fait en disant que les univers $X11, X12, \ldots X1j, \ldots X1\phi\}$, $j=1,2,\ldots\ \phi$, et $\{Y21, Y22, \ldots Y2k, \ldots$ $Y2\varphi\}$, $k=1,2,\ldots\ \varphi$, sont 'corrélés'.

Or, *dans la représentation obtenue, cette dépendance ou corrélation émerge en dehors de toute restriction d'espace-temps*.

---

[60] Le caractère d' "élémentarité" au sens probabiliste, est **foncièrement relatif**, notamment, il est relatif à la manière d'introduire une entité dans le rôle descriptionnel d'une entité-à-qualifier (cf. MMS [1992], [2002A], [2002B], **[2006]**, [2013]).

Et en outre, la strate de conceptualisation radicalement première, en termes de descriptions transférées primordiales, qui est considérée ici, est dépourvue par construction de tout substrat de déjà conceptualisé préalablement, cependant que cette strate elle-même consiste exclusivement en groupes de marques observables transférées sur des enregistreurs d'appareils et codés. Point. Donc *selon l'ordre de constructibilité des conceptualisations qui s'amorce dans le travail présent*, la dépendance constatée se manifeste comme inscrite *primordialement*, sans aucune 'explication' logeable à son intérieur même, ou bien 'en dessous'. Et si l'on veut alors tenter une explication à l'intérieur d'une phase de conceptualisation *subséquentes* selon cet ordre de constructibilité des conceptualisations – une explication qui soit contrôlable en quelque sens et *consensuelle* – il faudrait être en possession de *normes* admises généralement pour élaborer *le passage d'une description transférée primordiale, à une modélisation explicative super-posée* (une première esquisse de telles normes est donnée dans MMS [2002A], [2002B], [2006]). Tant que de telles normes ne sont pas encore élaborées, les manifestations probabilistes liées aux descriptions transférées primordiales de microétats de deux ou plusieurs microsystèmes, ne sont simplement pas 'explicables', elles sont juste des faits d'observation [61].

Ajoutons ceci: Dans un arbre de probabilité absolument quelconque, même les méta-dépendances 'probabilistes' (corrélations statistiques) $p(Yk)=F\{p(G,Xj)\}$ entre des spécifications probabilistes qui coiffent deux branches *distinctes*, ont elles aussi un caractère *primordial* qui juste se manifeste de manière factuelle, mais est dépourvu de toute 'explication' ou 'cause' formulable à l'intérieur de *cette* même phase de conceptualisation *transférée primordiale* (qui est aussi la *seule* prise en considération par le formalisme mathématique de la mécanique quantique (*2.3.2.4.*)). Et notons que dans cette phase transférée primordiale, les 'corrélations statistiques' de méta-dépendance $p(Yk)=F\{p(G,Xj)\}$ émergent elles aussi *libres de toute contrainte d'espace-temps*, juste innées dans la représentation. Tout simplement il en est ainsi. Il s'agit de données premières.

Tout autre trait de conceptualisation, notamment la définissabilité d'une structuration confinante, individuante, et d'un postulat de causalités agissant entre des individualités, ne peut émerger *constructivement* qu'à des niveaux de conceptualisation placés plus ou moins loin au-dessus de la bien connue 'coupure [quantique-classique]' ; en tout cas certainement au-dessus de la modélisation *microphysique* de Broglie-Bohm, dans une zone déjà classique où les 'aspects ondulatoires' peuvent être négligés relativement à des buts descriptionnels qui peuvent être explicités.

### 2.6.2.2. *L'arbre de probabilité d'un micro-état à génération composée et à un ou plusieurs micro-système : L'arbre de probabilité d'un micro-état quelconque*

---

[61] Le 'problème de non localité', on l'aura déjà bien noté, concerne précisément un micro-état de deux micro-systèmes. Et l'on s'y étonne que la mécanique quantique – qui a été construite en tant qu'une représentation *transférée primordiale des microétats*, dépourvue de structure d'espace-temps – implique le type de corrélations mentionné, qui ne s'accorde pas avec certaines exigences des théories de la relativité d'Einstein élaborées dans le cadre de la pensée causale classique, concernant des mobiles macroscopiques. On imagine les réactions qu'a pu produire la situation délinéée plus haut, sur des esprits formés dans la pensée classique, disciplinés par la logique classique et la théorie classique des probabilités, et par la physique *macroscopique* où se trouvent logées les théories de la relativité d'Einstein; et en plus, dans l'absence des concepts de description transférée primordiale et d'arbre de probabilité d'un micro-état de deux micro-systèmes. Ceci dit, la question de non-localité soulève une question qui dépasse les conséquences purement épistémologiques de la seule structure interne des descriptions de microétat. Elle parvient à définir un face-à-face entre cette structure et des faits d'observation. Par cela elle s'étend à, précisément, la question de la *modélisation microscopique adéquate de microphénomènes* ainsi qu'à la relation entre une telle modélisation et d'autre part les modèles classiques d'objets à structure d'espace-temps définie et à interactions causales. Dans le chapitre 5 de cette première partie de ce travail ces questions sont examinées plus en détail, et nous en dirons aussi quelques mots à la fin de la deuxième partie de ce travail.

Considérons d'abord un microétat $me_{G(G1,G2}$ à génération composée, d'*un seul microsystème S*, engendré par l'action sur $S$ d'une opération de génération $G(G1,G2)$ où se composent seulement deux opérations de génération $G1$ et $G2$ (cf. *2.2.4*). Le fait suivant est à noter attentivement. Le microétat à génération composée $me_{G(G1,G2)}$ engendré pour le micro-système $S$ par l'opération de génération composée $G(G1,G2)$ – comme *tout* microétat donné – est lié à *un seul* arbre de probabilité $T(G(G1,G2),V_M)$ (où $V_M$ est une 'vue mécanique définie pour des microétats'). Néanmoins, on s'attend irrépressiblement à ce qu'il y ait une relation définie entre le microétat à génération composée $me_{G1,G2}$ effectivement produit par $G(G1,G2)$ et les deux microétats $me_{G1}$ et $me_{G2}$ qui se *seraient* réalisés, respectivement, *si* soit $G1$ seul, soit $G2$ seul, avait agi sur $S$. Et l'on s'attend également à ce que cette relation actuel-virtuel se reflète de quelque manière dans une relation entre l'arbre $T(G(G1,G2),V_M)$ du microétat effectivement réalisé $me_{G(G1,G2)}$, et d'autre part les arbres $T(G1,V_M)$ et $T(G2,V_M)$ des deux microétats $me_{G1}$ et $me_{G2}$ qui auraient pu se réaliser pour $S$ si $G1$ seul ou $G2$ seul, respectivement, avait agi sur un état de départ quelconque de $S$. Toutefois, dans l'approche présente qui ignore le formalisme mathématique de la mécanique quantique et est exigée strictement qualitative, on ne peut répondre à cette attente qu'en signalant *un fait d'observation directe* exprimé dans les termes qualitatifs *négatifs* suivants.

Imaginons que tous les trois arbres $T(G1,V_M)$, $T(G2,V_M)$, $T(G(G1,G2),V_M)$ ont été effectivement réalisés et que les lois de probabilité qu'ils comportent peuvent être comparées expérimentalement. Soit $X$ une grandeur qui contribue à la vue mécanique $V_M$. Soit $p12(Xj)$ la probabilité assignée dans $T(G(G1,G2),V_M)$ à l'événement qui consiste en l'enregistrement de la valeur $Xj$ de $X$. Et soient $p1(Xj)$ et $p2(Xj)$, respectivement, les probabilités assignées à ce même événement dans les arbres $T(G1,V_M)$ et $T(G,V_M)$ des microétats $me_{G1}$ et $me_{G2}$: La probabilité $p12(G(G1,G2),Xj)$ n'est – en général – *pas* la *somme* des probabilités $p1(G1,Xj)$ et $p2(G2,Xj)$ (cf. MMS [1992B]: En général on trouve que

$$p12(G(G1,G2),Xj) \neq p1(G1,Xj) + p2(G2,Xj)$$

En *ce* sens, le microétat $me_{G(G1,G2)}$ ne peut pas non plus être regardé comme la 'somme' des deux microétats $me_{G1}$ et $me_{G2}$. On peut exprimer cette situation – et on le fait – en imaginant que les microétats $me_{G1}$ et $me_{G2}$ se 'réalisent' tous les deux à l'intérieur du microétat à génération composée $me_{G(G1,G2)}$, mais qu'ils y 'interagissent' ou 'interfèrent' en modifiant mutuellement les effets probabilistes observables que chacun produit lorsqu'il se réalise séparément.

Les caractères d'un arbre de probabilité correspondant à un micro-*état* à génération composée qui, en plus, est aussi un micro-*état* de deux ou plusieurs micro-*systèmes*, découlent facilement des spécifications précédentes.
*Là encore s'introduit une classe de dépendances probabilistes dotée des spécificités non-classiques* [62].

---

[62] On verra dans la deuxième partie de ce travail que lors de la construction du formalisme mathématique de la mécanique quantique, cette situation physique-représentationnelle a suggéré (plus ou moins explicitement) l'utilité de l'introduction d'une représentation *vectorielle* caractérisée par un **axiome** de 'superposition linéaire' **au sens mathématique**. Et que, en outre, cette même situation suggère la possibilité et l'intérêt de la définition explicite **d'un calcul avec des arbres de probabilités considérés globalement** ([MMS [1991] et [ 2009]). La relation entre *l'axiome mathématique* de superposition d'une part et d'autre part *le postulat physique* de composabilité des opérations de génération de microétat de *2.2.4*, est en général très mal comprise. On tend à confondre les natures conceptuelles de ces deux catégories d'éléments descriptionnels. Je pense que cela fait notamment obstacle au développement d'une 'théorie quantique de l'information' bien construite.

*2.6.2.3. La généralité de l'absence de structure d'espace-temps de l'effet observable global*

*des actions cognitives qui engendrent une description de microétat*

L'absence de structure propre d'espace-temps de l'effet observable global $D_M \equiv \{p(G,Xj)\}$, $j=1,2,...J$, $\forall V_X \in V_M$ d'une action descriptionnelle humaine $D_M/G,me_G,V_M/$, qui a été constatée d'abord pour le cas fondamental de l'arbre de probabilité d'un microétat de un microsystème et à opération de génération simple, subsiste pour *toute* sorte d'arbre de probabilité d'un microétat.

Pour tout tel arbre la description finale globale consiste en – exclusivement – des marques observables éparses sur des enregistreurs d'appareils différents et placés dans des endroits d'espace non spécifiés, arbitrairement éloignés les uns des autres. Ces marques émergent à des moments divers et non spécifiés du temps social des observateurs-concepteurs. Le support d'espace-temps de *ce* qui est *observé* au bout de toutes les actions de description primordiale transférée correspondantes à toutes le grandeurs mécaniques considérées, en général *n'est pas connexe*. Et toutes les caractérisations *statistiques* significatives (tendance à convergence, corrélations) que l'on peut trouver par des dénombrements a posteriori de telle ou telle description observable achevée, y sont inscrites d'une manière *indépendante* de toutes les spécifications opérationnelles d'espace ou de temps mises en jeu par l'observateur-concepteur cependant qu'il engendrait la genèse de la description. Les qualifications-cadre d'espace et de temps opérées *par l'observateur* – essentielles afin qu'il puisse *construire* une description transférée de base $D_M/G,me_G,V_M/$ et lui assigner un sens défini – ne laissent aucune trace dans cette description finale $D_M \equiv \{p(G,Xj)\}$, $j=1,2,...J$, $\forall V_X \in V_M$ elle-même : *Car aucune place conceptuelle n'y est aménagée où l'on pourrait loger des qualifications d'espace-temps*.

Seule une éventuelle *modélisation* ultérieure des microétats *et* de leurs interactions de mesure avec des appareils macroscopiques pourrait construire des qualifications d'espace et de temps *liées a posteriori à l'effet final observable* d'une action descriptionnelle $D_M/G,me_G,V_M/$ [63,64].

## 2.7. Conclusion générale
### sur le concept d'arbre de probabilité de l'opération *G* de génération d'un microétat

Les généralisations du concept d'arbre de probabilité d'un seul microétat qui viennent d'être spécifiées dans *2.6.2.1* et *2.6.2.2*, confrontées aux définitions générales de *2.4.2*, entraînent que le concept d'arbre de probabilité s'étend à *toutes* les sortes possibles de microétats, tout en acquérant dans chacun des cas considérés certaines caractéristiques nouvelles, *spécifiques* de ce cas [65]. Donc le symbole $D_M/G, me_G, V_M/$ – où désormais $me_G$ désigne un micro-état absolument quelconque – impliquant un seul micro-système ou plusieurs ou une opération de génération simple ou composée – est doté d'une signifiance non restreinte.

---

[63] La modélisation microscopique de Broglie-Bohm accomplit cela, mais *sans* introduire d'individuations à contours définis et une causalité 'locale' liée à un observateur inertiel, comme le fait la relativité restreinte d'Einstein; ni une causalité locale à consensus parmi des observateurs quelconques, comme le fait la relativité générale d'Einstein.

[64] **Ces remarques sont la première expression d'un leit-motiv qui, dans le chapitre 5 de cette première partie de ce travail, se développera en une conception sur la constructibilité successive des conceptualisations humaines du réel physique en partant du niveau de connaissance zéro concernant *spécifiquement* l'entité-à-décrire particulière considérée.**

Le contenu de chacun des symboles qui interviennent dans la notation $D_M/G, me_G, V_M/$ a déjà été amplement explicité. Pourtant le désigné global de cette notation restait abstrait, non intuitif. Tandis que la représentation d'espace-temps $T(G,V_M)$ de l'entière genèse $D_M/G, me_G, V_M/$ descriptionnelle en rend immédiatement présents à l'intuition tous les contenus, aussi bien que leurs relations. Notamment, elle frappe les yeux avec le fait, essentiel, que la description 'proprement dite' d'un microétat, à savoir l'ensemble $D_M \equiv \{p(G, X_{jM})\}$, $\forall X_M \equiv V_{XM}$ de $V_M$ de toutes les lois de probabilité liées au microétat étudie $me_G$, seulement *coiffe* les branches de l'arbre, et que si la genèse symbolisée par la triade $/G, me_G, V_M/$ est ignorée la description proprement dite ne peut paraître qu'un oiseau sans ailes qui se tient dans l'air mystérieusement. Car ce sont le tronc et les branches de l'arbre qui constituent le méta-phénomène aléatoire qui engendre, explique, soutient, la méta-description proprement-dite de $me_G$, $\{p(G, X_{jM})\} \equiv D_M$, $\forall X_M \equiv V_{XM}$, $V_{XM} \in V_M$ qui coiffe les branches, comme un feuillage.

C'est ce méta-phénomène aléatoire lié à $G - \mathcal{M}\textit{éa}(G)$ – qui, dans sa globalité, est *le méta-générateur de la méta-description globale* $\mathcal{M}dg(G) \equiv \{p(G, X_{jM})\} \equiv D_M$, $\forall (X_M \equiv V_{XM})$, $V_{XM} \in V_M$ du microétat $me_G$, conçu comme engendré *individuellement* par une réalisation de l'opération de génération $G$ définie individuellement et liée à $me_G$ par la relation de un-à-un $G \leftrightarrow me_G$ posée par la décision méthodologique $DM$.

Ceci attire l'attention sur *l'importance cruciale du processus de génération G d'un microétat*. Car c'est l'opération de génération $G$ qui fonde cette unité méta-probabiliste dénotée $T(G, V_M)$, dotée d'autonomie organisationnelle et fonctionnelle, et où se manifestent des dépendances probabilistes *observables* mais que la théorie classique des probabilités ne singularise pas. Or souvent ces dépendances probabilistes sont très inattendues, très contre-intuitives même, surtout dans le cas d'un micro-état de deux micro-systèmes. Et alors c'est encore l'opération de génération $G$ qui guide pour comprendre et accepter ces dépendances, à l'intérieur de l'atemporalité *foncière* du résultat final du résultat final $\mathcal{M}dg \equiv \{p(G, X_{jM})\} \equiv D_M$, $\forall (X_M \equiv V_{XM})$, $V_{XM} \in V_M$ de l'entier processus génétique qui conduit à l'arbre $T(G, V_M)$, géometrisé a posteriori par l'effacement des aspects temporels [66].

---

[65] *Pour un physicien de la mécanique quantique, s'il examine attentivement la situation conceptuelle, cela exprime d'ores et déjà que **le concept d'arbre de probabilité sous-tend l'entier formalisme quantique**.*

[66] **Ce fait est chargé de conséquences qui, une fois perçues dans le cas particulier des microétats, peuvent s'étendre à l'entière conceptualisation d'entités physiques. Notamment, le concept d'arbre de probabilité se transpose à la conceptualisation probabiliste en général, où il** *groupe* **des phénomènes aléatoires distincts mais qui introduisent tous une même entité-objet, en reliant les effets observables de ces phénomènes aléatoires dans un 'pattern' non trivial de méta-dépendances probabilistes qui dans la théorie classique des probabilités ne sont pas singularisées par des définitions qui leur soient spécifiques. Ce 'pattern' revient à caractériser explicitement un type de corrélations probabilistes qui, bien que particulier, est doté d'une grande importance pragmatique parce qu'il singularise ce qu'on peut regarder comme** *les conséquences probabilistes observables d'une genèse commune* **(MMS [2006] pp. 250-256).**

## 2.7. Remarque sur l'évolution d'un microétat

On peut se demander si, en respectant les contraintes imposées ici, il est possible de dire quelque chose sur l'évolution d'un microétat dans des conditions extérieures données [67]. La réponse est positive.

Imaginons une opération $G(t_o)$ de génération d'un microétat $me_G(t_o)$ où $t_o$ est le moment que l'on assigne à la fin de l'opération, donc aussi au début de l'existence de l'exemplaire considéré du microétat $me_G$. Supposons aussi qu'au moment $t_o$ on ne démarre aucune opération de mesure sur cet exemplaire de l'état $me_G$, mais qu'on le laisse subsister pendant une durée $\Delta t=t-t_o$ dans des conditions extérieures, disons $CE$, que l'on aura pu, et choisi, de mettre en place (champs macroscopiques, obstacles). Rien, dans la démarche élaborée ici, ne s'oppose à ce qu'on considère que l'association de l'opération de génération $G(t_o)$, avec les conditions extérieures $CE$ et *le passage de la durée $\Delta t=t-t_o$*, constituent ensemble une nouvelle opération de génération $G'=f(G(t_o),CE,\Delta t)$ qui produit un nouveau microétat $me_G'$ que l'on peut étudier par des mesures, exactement de la même manière que $me_G(t_o)$.

En outre, rien ne s'oppose non plus à ce que la durée $\Delta t$ soit choisie aussi petite (ou grande) qu'on veut. On peut donc étudier un *ensemble* de microétats correspondant à un *ensemble* d'opérations de génération $G'=f(G(t_o),CE,\Delta t)$ où $G(t_o)$ et $CE$ *restent les mêmes*, cependant que $\Delta t$ change de la façon suivante. L'on accomplit un *ensemble* de mesures qui permet de spécifier l'entier arbre de probabilité $T(G',V_M)$ avec $G'=f(G(t_o),CE,\Delta t1)$ et $t1$ dans $\Delta t1=t1-t_o$ très proche de $t_o$ ; puis on accomplit un ensemble de mesures pour $G'=f(G(t_o),CE,\Delta t2)$ et $t2$ dans $\Delta t2=t2-t_o$ plus éloigné de $t_o$; et ainsi de suite, jusqu'à un arbre final correspondant à $G'=f(G(t_o),CE,\Delta tf)$ où $\Delta tf=(tf-t_o)$. De cette manière on peut constituer concernant le microétat *de départ* $me_G(t_o)$, des connaissances qui sont équivalentes en principe à celles qu'offre une 'loi d'évolution'.

Évidemment, jamais une démarche qualitative ne pourra atteindre l'efficacité de représentation d'un système bien construit d'algorithmes mathématiques. Mais ici il ne s'agit que d'une remarque de principe qui fonde la possibilité de parler de l'évolution du microétat $me_G$ dans le contexte de la démarche présente.

Mais notons bien que le concept d'évolution dont il est question ici ne fait que *méta*-qualifier globalement en termes de 'changements', de 'descriptions' déjà accomplies précédemment et dont chacune consiste exclusivement en résultats de dénombrements a posteriori de codages de groupes de marques physique qui émergent épars dans l'espace et le temps. La possibilité de *telles* méta-qualifications temporelles, ne change rien au fait que l'effet observable final $D_M\equiv\{p(G,X_{jM})\}$ de l'action dénotée $D_M/G,me_G,V_M/$ de description *d'un microétat physique*, est dépourvu, lui, d'une structure interne définie d'espace et de *temps* : La scission entre représentations d'espace-temps des actions cognitives déployées par les concepteurs-observateurs afin de qualifier un microétat physique $me_G$, et d'autre part l'organisation d'espace-temps propre aux contenus physiques observables qui sont produits par ces actions, reste inaltérée.

---

[67] Dans le formalisme quantique l'équation d'évolution de Schrödinger *ne s'applique pas directement au microétat étudié*, mais *au descripteur mathématique* |Ψ> à l'aide duquel (avec les autres algorithmes aussi) l'on calcule l'*ensemble* des lois de probabilité qui, ici, a été symbolisé par $\mathcal{M}\equiv\{p(G,X_{jM})\}\equiv D_M$, $\forall(X_M\equiv V_{XM})$, $V_{XM}\in V_M$. Ces lois changent, en général, lorsque du temps passe. Dans les cas les plus 'simples' mais les plus importants d'un point de vue pragmatique le changement est commandé par un hamiltonien d'évolution, i.e. un descripteur de la grandeur mécanique 'énergie totale' qui est impliquée.

## 2.8. L'infra-[mécanique quantique] (*IMQ*)

Le processus de construction de descriptions qualitatives de microétats entrepris dans ce chapitre 2, est accompli. Le résultat sera appelé *l'infra-[mécanique quantique]* et l'on peut le dénoter *IMQ*. Cette dénomination doit être comprise comme: *'dans le substrat du formalisme mathématique de la mécanique quantique et partiellement inséré dans ses algorithmes de manière implicite et cryptique'* (Les parenthèses droites, qui traitent l'expression 'mécanique quantique' comme un tout non dissocié, tentent d'indiquer que l'accent ne tombe nullement sur le mot 'mécanique').

L'*IMQ* s'est constituée comme *une discipline à part entière développée indépendamment du formalisme quantique*. C'est une discipline qualitative d'une nature inusuelle, une discipline *épistémo-physique* qui émane d'une supposition et d'un but corrélés posés à la base librement et qui par la suite ont agi comme des contraintes permanentes :

- Le début de la conceptualisation à développer a été supposé être placé sur le niveau extrême, limite, de connaissance zéro, en ce qui concerne les *spécificités* du microétat *particulier* considéré ; seul le concept général et son nom ont été importés de la conceptualisation classique du réel physique microscopique.

- L'on a voulu mettre en évidence de *A à Z* sous quelles conditions et comment il est possible de construire des connaissances strictement premières concernant spécifiquement un microétat donné, bien que quelconque.

Cette démarche a mis en évidence une forme descriptionnelle qui émerge primordialement statistique et transférée, en ce sens qu'elle consiste en marques physiques observables sur des enregistreurs d'appareils macroscopiques en conséquence d''interactions de mesure', et que ces marques ne sont jamais toutes identiques pour toute interaction de mesure, leurs caractères manifestent des dispersions statistiques.

Les statistiques obtenues sont relatives – d'une manière inamovible – à trois éléments descriptionnels : l'opération de génération $G$ du microétat $me_G$ à étudier, ce microétat $me_G$ lui-même, et la 'vue' (la structure de qualification) utilisée (qui, en particulier, peut être 'mécanique'). Ces trois relativités sont rappelées dans la parenthèse oblique $/G,me_G,V_M/$ du symbole global $D_M/G,me_G,V_M/$ assigné à la forme descriptionnelle trouvée, où le symbole $D_M$ – *isolément* – avec $D_M \equiv \{p(G,X_j)\}$, $j=1,2,...J$, $\forall V_X \in V_M$ représente *seul* la description finale proprement dite produite par les actions cognitives humaines rappelées dans le signe $/G,me_G,V_M/$ : Ces dernières ne constituent que la genèse conceptuelle-opérationnelle de la description finale $D_M$.

La notation globale $D_M/G,me_G,V_M/$ souligne le fait que – d'un point de vue épistémologique, et notamment pour l'intelligibilité – *dans le cas des descriptions de microétats la genèse est aussi importante que le résultat final* : En effet c'est au cours de cette genèse que la situation cognitive où se déroule l'action descriptionnelle associée au but et à la supposition posés à la base, ont imposé avec une nécessité incontournable la *décision méthodologique DM* (*2.2.3*), et ont comporté un problème de codage des effets physiques observables des interactions de mesure, d'une nature particulière (*2.3.2.2,*

*2.3.2.3*) dont la maîtrise a soulevé des problèmes inusuels. Or les deux pas constructifs qui viennent d'être rappelés marquent le processus génétique et ses résultats de caractères foncièrement *non*-classiques.

Au cours de l'élaboration d'une description de microétat s'est installée spontanément une remarquable scission entre : d'une part l'action du concepteur-observateur humain qui – en dehors des deux moments critiques mentionnés plus haut – conçoit et procède selon sa conception *classique* irrépressiblement empreinte de structuration spatio-temporelle et causale ; et d'autre part le produit final $D_M \equiv \{p(G,Xj)\}$, $j=1,2,...J$, $\forall V_X \in V_M$ de la construction qui, lui, en *conséquence* des pas constructifs non-classiques qui se sont imposés au cours de sa genèse, reste dépourvu d'une telle structuration, vide de toute définition d'espace-temps et de causalité.

Il s'ensuit que l'action constructive *du concepteur-observateur humain* peut être considérée séparément. Or cette action s'assemble dans une structure d'espace-temps clairement définie qui, par une élimination finale des traits temporels, se géométrise en une figure arborescente – l'arbre de probabilité d'une opération de génération d'un microétat – doté d'une unité organique qui a été caractérisée d'une manière explicite, détaillée et exhaustive. Cette unité arborescente pointe vers un concept de probabilité beaucoup plus complexe que celui qu'introduit un espace de probabilité au sens classique de Kolmogorov. Dans le travail exposé ici ce concept probabiliste non-classique n'est défini que de manière qualitative et pour le cas particulier qui implique un microétat. Mais d'ores et déjà le concept d'arbre de probabilité d'une opération de génération d'un microétat[68] :

- Élucide le contenu d'un concept de 'phénomène aléatoire-branche' comportant un microétat $me_G$ et une seule grandeur $X$ (*2.6.1.2*).

- Met en évidence un type non-classique de dépendance probabiliste entre des événements produits par deux phénomènes aléatoires-branches distincts de l'arbre $T(G,V_M)$ (*2.6.1.3*).

- *Conduit à un concept de méta-phénomène aléatoire $\mathcal{M}_{pa}(G)$ lié à l'opération de génération G (2.6.1.4) où :*

- Se manifestent des méta-dépendances probabilistes non-classiques qui :

\* Apparaissent d'emblée comme liées au théorème d'incertitude de Heisenberg.

\* Frappent d'emblée comme étant reliées à la 'théorie des *transformations de base*' de Dirac.

\* Suggèrent une définition simple et claire d'une méta-loi de probabilité unique $\{p(G,Xj)\} \equiv D_M \equiv \mathcal{M}_{d\varphi}(G)$ qui *caractérise globalement* le microétat $me_G$ impliqué dans l'arbre considéré (cf. la note 48).

- Indique la voie vers une logique quantique de l'ensemble des *événements* d'un arbre de probabilité organiquement *unifiée* avec les probabilités définies par cet arbre (*2.6.1.5*).

---

[68] L'énumération qui suit est à lier aux notes qui accompagnent les points *2.6.1.2, 2.6.1.3, 2.6.1.4, 2.6.1.5, 2.6.2.1,* (*2.6.2.2*).

- Définit le cadre où s'inscrit le problème de non-localité soulevé par Bell (*2.6.2.1*) et suggère une voie de compréhension liée à l'absence de structure d'espace-temps d'une description proprement dite $D_M$ d'un microétat quelconque.

- Met en évidence un fondement *physique-conceptuel* sur lequel on peut assoir le choix d'une représentation *mathématique* des descriptions de microétats à l'aide d'espaces vectoriels, et corrélativement, suggère l'intérêt de la définition d'un calcul avec des arbres de probabilité de microétats considérés globalement (*2.6.2.2*).

Ainsi l'infra-[mécanique quantique] pointe déjà vers les algorithmes mathématiques de la mécanique quantique. D'ores et déjà il émane de l'infra-[mécanique quantique] comme une force d'interaction explicative avec ces algorithmes.

Il nous semble notable qu'une démarche purement qualitative et si sévèrement contrainte que celle qui vient d'être développée dans ce chapitre, ait pu produire une structure descriptionnelle aussi définie, dotée de caractères et de variantes tellement complexes, et qui paraisse fertile.

Mais avant de considérer cette structure descriptionnelle comme pleinement acquise, nous allons la soumettre dans le chapitre suivant à un examen critique, en analysant les problèmes qu'elle peut soulever dans l'esprit du lecteur et la manière dont elle résiste à ces problèmes.

# 3

# Problèmes d'interprétation qui émergent *à l'intérieur* de *IMQ*, leurs solutions, et les conclusions qui s'en dégagent

### 3.1. Problèmes sur la forme descriptionnelle $D_M/G,me_G,V_M/$ et réponses

Dans ce qui suit[69] on assistera à un processus remarquable. On percevra en pleine lumière la genèse et la nature profonde de quelques problèmes *fondamentaux* d'interprétation que le formalisme mathématique de la mécanique quantique suscite depuis sa constitution: le problème de complétude, le problème ontologique, le sens exact du qualificatif 'essentiel' que l'on associe aux probabilités qui concernent des microétats et la question de la relation entre celles-ci et le postulat de déterminisme, la question de 'la coupure quantique-classique' et celle, corrélative, de la modélisation des microétats. Parce que le formalisme quantique s'est constitué avant la compréhension de son formalisme mathématique, et ensuite pendant très longtemps a constitué l'unique expression disponible de connaissances concernant des microétats, les problèmes énumérés ont été perçus tout d'abord en relation *inextricable* avec ce formalisme mathématique, en tant que problèmes d''interprétation' *du formalisme*. D'ailleurs cette optique perdure à ce jour.

Or il apparaîtra que face à la démarche purement qualitative développée ici, ces problèmes émergent de nouveau, ***indépendamment du formalisme quantique et pourtant dans des termes quasi identiques à ceux qui se sont imposés relativement à ce formalisme***.

Il apparaîtra également que lorsque les problèmes énumérés sont rapportés à la forme descriptionnelle $D_M/G,me_G,V_M/$ établie dans le chapitre **2**, ces problèmes *s'élucident* d'une façon contraignante que l'on pourrait qualifier de 'directe' en ce sens que la solution découle – sans intermédiaire – de la *genèse* de la forme $D_M/G,me_G,V_M/$.

Ainsi, dans le bref texte qui suit, sera dissoute pour la première fois une coalescence erronée, qui a trop duré, entre conceptualisation quantique et formalisme mathématique de la mécanique quantique.

---

[69] Certains passages de ce paragraphe interviennent aussi dans Mugur-Schächter [2006] mais à un niveau conceptuel doté d'une généralité non restreinte, où le cas spécial des microétats se trouve incorporé en tant qu'un cas particulier parmi une foule d'autre cas de descriptions 'de base' transférées.

### 3.1.1. Le désigné du symbole $D_M$ /$G$,$me_G$,$V_M$/ peut-il être regardé comme une "description" du microétat $me_G$?

On vient de voir que la connaissance qu'il a été possible de construire concernant le comportement mécanique d'un microétat, ne consiste en général pas en descriptions individuelles. Elle consiste généralement en distributions de probabilité $p(G,X_j)$ concernant l'émergence des valeurs $X_j$ de telle ou telle grandeur mécanique $X$. En outre, les manifestations observables que ces distributions concernent ne peuvent *pas* être regardées comme étant liées à des propriétés que le microétat étudié $me_G$ engendré par l'opération de génération $G$ aurait 'possédées' d'emblée, avant toute évolution de mesure, d'une façon déjà actuelle, réalisée, et réalisée pour lui isolément, de façon indépendante de tout acte d'observation. Ces lois de probabilité *{p(Xj)}* n'offrent *aucun* renseignement concernant la façon d'être du microétat $me_G$ lui-même, indépendamment de nos actions cognitives sur lui.

Il en est ainsi à tel point qu'il est même possible de reformuler la représentation construite plus haut, en termes strictement opérationnels-observationnels-prévisionnels: Une fois qu'une description $D_M$/$G$,$me_G$,$V_X$/ incluse dans l'ensemble *{$D_M$/$G$,$me_G$,$V_X$/}*, $\forall V_X \in V_M$ a été établie, si d'abord l'on opère de la façon dénotée $G$ et ensuite l'on opère de la façon dénotée *Mes(X)*, on sait à l'avance qu'on a telle probabilité $p(G,Xj)$ d'observer tel groupe *{$\mu_k$}*, *k=1,2,...m* de manifestations (marques) physiques *de l'appareil employé*, codé $Xj$ en termes conceptualisés. En ces conditions peut-on affirmer que $D_M$/$G$,$me_G$,$V_X$/ et $D_M$/$G$,$me_G$,$V_M$/ sont des descriptions *du microétat $me_G$* lui-même?

La question a déjà été formulée et la réponse affirmée a été positive (*2.5.4*). Mais on peut se sentir non satisfait et y revenir en se disant ce qui suit.

«Penser à la manière d'exister du microétat $me_G$ lui-même n'est qu'une intrusion philosophique dans la pensée et le discours scientifique. L'ensemble des lois de probabilité $p(G,X_j)$ liées aux grandeurs mécaniques $X$ et associées à une même opération de génération $G$ donnée, constituent un invariant opérationnel-observationnel-prévisionnel relatif à $G$ et aux processus de *Mes(X)*, et cet invariant est spécifique à $G$ ; *et cela suffit*. On peut même, à la limite, se débarrasser en fin de parcours de toute trace de pensée hypothétique, comme on se débarrasse de tous les éléments de l'échafaudage quand la bâtisse est achevée. Même l'expression 'le microétat correspondant à l'opération de génération $G$' peut être regardée comme un simple appui verbal qui est utile mais qu'il faut se garder de réifier. On se retrouve finalement devant une sorte de pont entre, d'une part des opérations physiques, et d'autre part des observations codées en termes de valeurs d'une grandeur mécanique et des prévisions probabilistes tirées de ces observations. Le microétat étudié ne fait que hanter cette construction comme un fantôme inutile. En *tout* cas, il ne s'agit nullement d'une description de ce microétat lui-même».

Ce problème a plusieurs visages que je vais maintenant spécifier à tour de rôle. D'abord je caractériserai les aspects que, face $D_M$/$G$,$me_G$,$V_M$/, l'on peut désigner par les mêmes dénominations qui ont déjà été employées en relation avec le formalisme mathématique: le problème de *complétude* et le problème *ontologique*.

Une fois que ceci aura été accompli, je montrerai qu'à l'intérieur de la démarche développée ici *la décision méthodologique d'affirmer la relation de un-à-un $G \leftrightarrow me_G$, élimine a priori ces deux problèmes*. Cela permettra de percevoir clairement aussi les réponses à autres deux questions qui, face au formalisme quantique, ont soulevé un grand nombre de débats, à savoir la question de la nature des probabilités quantiques, et la question d'une définition claire de 'la coupure quantique-classique'.

Au bout de ce cheminement, une solution globale à tous les questionnements abordés, apparaîtra avec évidence: Dans tous les cas examinés, uniformément, il s'agit d'incompréhensions engendrées par le fait que la *genèse* qui conduit à une connaissance probabiliste $D_M \equiv \{p(G,Xj)\}$, $j=1,2,...J$, $\forall V_X \in V_M$ et qui est rappelée dans le symbole $D_M/G,me_G,V_M/$, est ignorée. En conséquence de cela – sans s'en rendre compte – l'on reste aveugle face aux exigences spécifiques imposées par les conditions cognitives qui dominent la construction de connaissances concernant des microétats, et l'on remplit ce vide en y déversant des exigences de conceptualisation *classique*, qui tout simplement n'ont pas de *sens* face à des qualifications liées à des microétats.

L'on fabrique ainsi des questions illusoires dans lesquelles on s'enlise.

### 3.1.2. Le problème de 'complétude' de la forme descriptionnelle $D_M/G,me_G,V_M/$

Le débat sur la 'complétude' *du formalisme quantique* – pas sur celle de la forme descriptionnelle qualitative $D_M/G,me_G,V_M/$ – a conduit à des *théorèmes d'impossibilité* dont les plus importants sont le théorème de von Neumann affirmant l'impossibilité de paramètres cachés compatibles avec le formalisme quantique (J. von Neumann [1955]) et le théorème de Wigner affirmant l'impossibilité de définir une probabilité conjointe de position et de quantité de mouvement qui soit compatible avec le formalisme quantique (E.P. Wigner [1971]). Ces théorèmes semblaient trancher ce qu'on avait dénommé le problème d'incomplétude de la mécanique quantique, sans pour autant le faire comprendre [70]. En fait, l'un comme l'autre de ces deux théorèmes faisaient usage, dans la démonstration, du formalisme quantique lui-même, ce qui est circulaire. Finalement ces deux théorèmes ont été invalidés (MMS [1964], Bell [1966], MMS [1977], MMS [1979]). Mais ces invalidations ne font que montrer des vices de traitement du problème, elles n'éliminent nullement le problème considéré, celui d'une représentation plus 'complète' des microsystèmes 'eux-mêmes'.

Or le problème de 'complétude' renaît également face à la forme descriptionnelle qualitative $D_M/G,me_G,V_M/$ qui a émergé à l'*extérieur* du formalisme quantique ainsi que de toute autre formalisation préexistante, et cela suggère que les racines du problème de complétude se trouvent *sous* le formalisme quantique, dans la situation cognitive même qui est impliquée. En effet on se dit:

«Il est clair que si l'on veut se mettre en possession d'une qualification du microétat $me_G$ en termes d'une valeur $Xj$ d'une grandeur mécanique $X$ redéfinie pour des microétats, il faut réaliser une succession *[G.Mes(X)]* où une réalisation de $G$ soit immédiatement suivie d'un acte *Mes(X)* de mesure de $X$ effectué sur l'exemplaire de $me_G$ engendré par cette réalisation de $G$. Mais si, afin de vérifier qu'on a bien procédé, l'on répète la réalisation d'une succession *[G.Mes(X)]*, en général on ne retrouve plus la même valeur $Xj$ de la grandeur $X$. Ceci est un fait d'expérience indéniable qu'on a exprimé en disant que 'la situation est statistique'. Donc afin d'avoir une chance d'accéder à une caractérisation stable de $me_G$ en termes de valeurs $Xj$ de $X$ (alors nécessairement probabiliste), on doit réaliser un très grand nombre de répétitions de la succession *[G.Mes(X)]*. Or après l'enregistrement d'une valeur *Xj* de *X*, l'exemplaire du microétat *me*$_G$ qui avait été mis en jeu par la réalisation de *G*, n'existe plus ; il a transmuté en un *autre* microétat. Donc le microétat étudié doit être recréé pour chaque nouvel acte de mesure. *Et chaque séquence [G.Mes(X)] brise la continuité du processus global de constitution de la connaissance que l'on acquiert sur le microétat me*$_G$. Si je travaillais avec un dé macroscopique je pourrais réutiliser indéfiniment le même dé, sans avoir à le recréer à chaque fois, et le processus de constitution d'une connaissance probabiliste ne serait pas brisé ainsi, ce serait clairement un tout, en ce sens que je pourrais concevoir que la dispersion statistique n'est due qu'à la non identité des *jets*, d'ailleurs permise à l'avance. Tandis que dans le cas d'un microétat il apparaît un problème qui est grave: *Comment peut-on être certain que l'opération de génération G recrée vraiment le même microétat à chaque fois qu'elle est réalisée, comme on l'a admis – peut-être trop vite – par la décision méthodologique de poser une relation de un-à-un, G↔me*$_G$*?* C'est plutôt du contraire de ce que pose cette relation qu'on se sent incliné à être certain. En effet l'opération *G* est définie par des paramètres *macroscopiques* dont il est certainement impossible de dominer tous les aspects *microscopiques*. Donc ce qui selon les paramètres macroscopiques semble être un ensemble de répétitions de la même opération *G*, en fait, au niveau microscopique, est sans aucun doute tout un ensemble d'opérations de génération mutuellement différentes, qui engendrent tout un ensemble de microétats eux aussi mutuellement différents. En plus, un processus de *Mes(X)* est lui aussi défini seulement à l'aide de paramètres macroscopiques en dessous desquels se cache sans doute tout un ensemble de réalisations microscopiques différentes de ce processus. Dans ces conditions, la description des microétats que nous avons élaborée est certainement *incomplète* car elle escamote les différences qui existent entre les exemplaires de microétats engendrés par les répétitions d'une opération de génération *G*, comme elle escamote également les différences qui existent entre les réalisations distinctes de l'opération *G* elle-même, et celles qui existent entre les réalisations distinctes d''une' *Mes(X)*. On affirme des répétitions 'identiques' des séquences *[G.Mes(X)]*, mais en fait celles-ci sont fluctuantes dans *tous* leurs éléments, dans l'élément *G*, dans l'effet *me*$_G$ de *G*, et dans l'acte de *Mes(X)*. Nous avons partout indiqué fallacieusement par un symbole invariant, tout un *ensemble* caché d'entités physiques mutuellement distinctes. Nous aurions dû introduire explicitement tout un *ensemble* d'opérations de génération différentes au niveau microscopique mais correspondant toutes à un même groupement de paramètres macroscopiques, et étiqueter *cela* par le symbole *G*.   Et à l'intérieur de cet ensemble il aurait fallu symboliser les différences, les dénoter. Et de même pour *me*$_G$ et pour *Mes(X)*. Ce n'est que de cette façon qu'on aurait pu espérer de construire une description véritablement complète des microétats»[71].

---

[70] Les conclusions y étaient exprimées en termes absolus et définitifs, comme s'il était concevable de déduire une impossibilité absolue et définitive! (A l'intérieur de quel système formalisé? Un système qui, dès tel moment donné, établirait *tout* ce qui est possible, *à jamais*?).

Toutefois, parvenus en ce point on peut se troubler car on peut se dire également:

«Mais même de cette façon, ce qu'on obtiendrait finalement ne serait toujours pas une description des microétats eux-mêmes ! Car malgré toutes les précautions conceptuelles et notationnelles mentionnées, nous n'apprendrions finalement toujours rien d'*individuellement défini* concernant un microétat donné, ni de défini concernant ce microétat *isolément*, strictement rien. Tout ce que nous apprendrions concernerait toujours seulement les manifestations observables produites par des processus de mesure. Car lors d'une succession *[G.Mes(X)]* donnée nous ne saurons pas comment choisir *factuellement*, dans les ensembles microscopiques que nous venons de *concevoir*, la variante microscopique de l'opération de génération d'un microétat qui s'est réalisé effectivement, ni la variante microscopique de microétat qui s'est réalisée, ni la variante microscopique d'un acte de mesure *Mes(X)* qui s'est réalisée. L'on aura donc alourdi notre façon de dire et de penser, sans avoir enrichi d'un seul iota nos connaissances établies factuellement.

---

[71] Notons qu'un grand nombre de spécialistes de la mécanique quantique parlent effectivement en termes d' 'ensembles'.

Non, il n'y a rien à faire. D'une part, toute la façon de parler et de noter qui a été développée – avec des singuliers partout, *' l'* 'opération de génération *G'*, *'le'* microétat *'correspondant'*, *'le'* processus de *Mes(X)* – induit tout simplement en erreur. Cette façon de parler masque l'incomplétude générale de l'approche. Et d'autre part, si l'on supprime ce masque, si à la place de ces singuliers qui faussent on introduit partout les ensembles microscopiques qui s'imposent à la raison, on reste bloqué dans une impossibilité de choix factuel en conséquence de laquelle toutes les distinctions imaginées ne servent à rien. *Nous sommes condamnés à l'ignorance*. Dans *ce* sens-là, face à nos possibilités de connaissance établies par des faits observables, la description accomplie est effectivement 'complète'.

Nous sommes piégé dans la même difficulté que celle qui nous nargue dans le cas du formalisme quantique. Et la complétude qu'affirme l'orthodoxie concernant le formalisme mathématique de la mécanique quantique est bien vraie, puisqu'elle se reproduit au niveau de la démarche sous-jacente pratiquée ici. C'est cette conclusion qui s'impose. Même si cette complétude n'est pas démontrable déductivement, elle est néanmoins *vraie*.

Mais cette sorte de 'complétude' est une prison insupportable ! Il faut absolument trouver le moyen d'en sortir et d'accomplir une représentation vraiment complète des microétats».

Voilà l'essence du discours qui naît dans les esprits, et comment l'assertion de 'complétude' de la forme descriptionnelle $D_M/G,me_G,V_M/$ s'érigerait comme un mur à la fois inacceptable et indestructible, exactement comme c'est le cas pour le formalisme quantique.

Ces questions d'identité, ou pas, lors des répétitions d'une succession *[G.Mes(X)]*, sont très insidieuses. D'une part elles incluent un noyau dur qu'on ne peut pas ignorer. Elles touchent aux limites de la pensée. L'effet du choc est viscéral. Ce noyau dur est doté d'une grande force de fascination car on supporte mal de véritablement sentir que la pensée se heurte à une limite de ses capacités. Non pas se le dire ou l'entendre dire, mais le sentir. Dans le même temps cette question introduit une foule de glissements que l'on sent être fallacieux. Cela aussi on le sent de façon intime. Et cela aussi inquiète, tout en augmentant la fascination.

C'est ainsi que 'le problème de complétude' qui a tant hanté le formalisme de la mécanique quantique se manifeste aussi, en effet, face à la forme descriptionnelle qualitative $D_M/G,me_G,V_M/$. Et face à celle-ci, il acquiert même une texture plus concentrée et une intensité psycho-intellectuelle plus grande que face au formalisme quantique, parce que, en l'absence d'un formalisme mathématique le regard n'a pas où se disperser en incompréhensions et suspicions adjacentes qui occupent et détendent.

D'autre part, cette fois le problème de complétude apparaît face à une genèse explicite que l'on peut retracer, et cela l'expose à un contrôle critique.

Mais suspendons la progression vers ce contrôle, le temps d'énoncer aussi le problème ontologique.

### 3.1.3. Le problème du contenu 'ontologique' du concept de microétat

On peut suivre aussi un autre cheminement qui est intimement relié au précédent mais où l'accent tombe plutôt sur la question 'ontologique': comment *est* un microétat lui-même, vraiment, indépendamment de toute opération cognitive humaine accompli sur lui? Cependant que cette fois la question de complétude de *notre représentation* reste en retrait. Ce cheminement, lui aussi, a émergé d'abord relativement au formalisme quantique. Mais de nouveau on le retrouve relativement à la forme descriptionnelle $D_M/G,me_G,V_M/$. On se dit .

«Essayons malgré tout d'admettre toutes les 'identités' affirmées par décision méthodologique. Mais alors pourquoi un microétat, toujours le même, identiquement reproduit par des opérations de génération identiques et soumis à chaque fois à une évolution de *Mes(X)* qui est toujours la même, d'une seule et même observable *X*, conduirait-il en général à des valeurs *Xj* différentes, au lieu d'engendrer toujours la même valeur? Ne serait-ce pas parce qu'un microétat est une entité dont la nature est *essentiellement* aléatoire?».

Mais aussitôt on réagit:

«Que peut vouloir dire, exactement, 'une nature essentiellement aléatoire' d'un microétat? Et pourquoi, dans les conditions considérées, un microétat aurait-il une nature plus aléatoire qu'une opération de génération *G* ou qu'un acte de mesure *Mes(X)*? Ne s'agit-il pas en fait exclusivement de l'incapacité opératoire, de notre part à nous, de reproduire, à partir de contraintes macroscopiques, exactement la même opération de génération, le même état microscopique, et aussi, exactement le même acte de mesure? Car l'idée qu'un microétat serait de par lui-même 'essentiellement' aléatoire – ou alors peut-être plutôt *intrinsèquement* aléatoire? Ou bien *aléatoire en soi*? – paraît vraiment très obscure. En outre, s'il ne s'agit en effet que d'une incapacité opératoire humaine, qu'est-ce qui me donne le droit de retourner une telle incapacité de l'homme, en affirmation ontologique de caractères qui seraient – *eux* – essentiellement aléatoires, c'est à dire aléatoires *dans les faits même*? Et d'ailleurs qu'est-ce que cela veut dire 'la même opération *G*' ou 'le même microétat' ou 'le même processus de mesure'? 'Même', *de quel point de vue? Dans l'absolu? Mais n'est-ce pas là un non-sens?*».

Les significations des mots glissent et se tortillent comme des anguilles et elles échappent à l'entendement. La pensée rebondit indéfiniment contre un mur insaisissable qui l'use, la déchire et l'enlise. Alors on renonce à penser. On se tait jusque dans l'âme et on attend, avec une sorte de foi impuissante.

Voilà en quoi consiste 'le problème ontologique' que suscite la forme descriptionnelle $D_M/G,me_G,V_M/$.

Ce même problème – ***tel quel*** – émerge également face au concept de microétat employé dans les exposés du formalisme mathématique de la mécanique quantique.

On voit déjà en quel sens la question de complétude et la question ontologique sont distinctes mais reliées: la question de complétude ne concerne directement que notre représentation des microétats, cependant que la question ontologique ne concerne directement que la façon d'être assignée aux microétats eux-mêmes.

### *3.1.4. Le piège abyssal du réalisme naïf*

Mais ces deux questions, l'une comme l'autre, ***présupposent que cela posséderait un sens définissable*** *de vouloir "savoir comment les microétats sont vraiment en eux-mêmes, indépendamment de toute action cognitive humaine, d'une manière absolue".*

Elles présupposent également toutes deux, bien que de manière plus vague, que cela posséderait un sens spécifiable de vouloir savoir *d'emblée* cela, ***avant*** de se lancer dans le processus de construction exposée ici.

Enfin, elles présupposent aussi que cela posséderait un sens que de vouloir réaliser une sorte de saturation absolue de la description d'une entité-objet; d'arriver à savoir 'tout' ce qui concerne cette entité-objet.

Il est remarquable quels épais brouillards émanent des inerties de la pensée rationnelles et cachent les frontières entre le domaine de la rationalité et celui de la métaphysique. Lorsqu'à la faveur de ces brouillards on passe cette frontière et l'on perçoit les paysages qui s'y déploient, on se sent tout à coup perdu. Saisi d'angoisse l'on voudrait avec urgence retourner dans les pays de la rationalité mais on ne sait plus par quelle voie.

Voilà l'essence multiple, floue et obscure des deux problèmes liés de complétude et ontologique. Dans *IMQ* cette essence agit distillée, rendue indépendante de la formalisation mathématique des descriptions de microétats, et les présuppositions desquelles elle émane apparaissent sans voile. Car ces problèmes, qui ont d'abord été perçus face au formalisme mathématique de la mécanique quantique, émanent en fait d'en dessous de ce formalisme, et le formalisme les obscurcit et les déforme. Cependant que la forme descriptionnelle $D_M/G,me_G,V_M/$ – avec sa genèse – en met en évidence la source profonde qui peut s'énoncer ainsi:

***Être*** et ***description*** *se confondent dans l'esprit en un seul absolu dont on présuppose qu'on peut l'englober entièrement dans la connaissance du réel 'tel qu'il est en lui-même'.*

Le concept de description est alors illusoirement exonéré du tribut inévitable qu'il doit payer à des opérations de *qualification,* qui *seules* peuvent engendrer du *connu,* mais en le *séparant* de l'être-'en-soi', radicalement inconnaissable par les voies de la conceptualisation rationnelle.

Les trompe-l'œil conceptuels coagulés dans cet absolu impossible sont projetés sur l'horizon de la connaissance où ils rejoignent le trompe-l'œil du 'vrai-en-soi'. Ces différentes variantes en trompe-l'œil de notre refus viscéral de réaliser que ***tout CONNU*** *–* qui est description, qui comporte du qualifié *–* ***est confiné à l'intérieur du domaine du relativisé aux qualifications accomplies***, s'agitent vainement dans un tourbillon éternel de néants de sens, comme dans un enfer de Dante des concepts qui ont péché.

### *3.1.5. Retour sur la relation de un-à-un G↔me_G*

Par les questions de complétude et ontologique qu'elle soulève irrépressiblement, la description $D_M/G,me_G,V_M/$ d'un microétat pousse la pensée naturelle à un corps à corps avec l'affirmation kantienne métaphysique de l'impossibilité de connaître le-réel-physique-en-soi. Or les questionnements de complétude et ontologique sont fondés tous les deux sur la mise en doute de la pertinence de la relation de un-à-un dénotée $G↔me_G$. D'autre part, on l'a fortement souligné, seule l'acceptation *méthodologique* de la relation $G↔me_G$ permet de démarrer la construction de la forme qualitative $D_M/G,me_G,V_M/$ de la description d'un microétat en partant – non plus exclusivement de la conceptualisation classique en termes d' 'objets' à contour délimité et à interactions causales – mais en partant d'interactions factuelles avec du réel physique aconceptuel qui n'est pas directement perceptible par l'homme et en rendant compte des résultats de celles-ci en termes aussi purs de mélanges avec la pensées classique et aussi simples et clairs que possible. En outre, la démarche symbolisée $D_M/G,me_G,V_M/$, avec son résultat observable $D_M \equiv \{p(G,Xj)\}$, $j=1,2,...J$, $\forall V_X \in V_M$, permettent à un œil averti d'y discerner un équivalent intégré et à genèse épistémologique explicite, des descriptions de microétats offertes par les algorithmes quantiques, qui ont donné des preuves de leur efficacité : $IMQ$ et le formalisme quantique – de commun accord et à l'unisson avec l'histoire de la physique – clament que les toutes premières descriptions efficaces de microétats sont des descriptions transférées primordiale qui émergent statistiques. Cependant que $G \leftrightarrow me_G$ absorbe la statisticité primordiale de ces descriptions dans une manière de dire et de penser résolument constructive et opérationnelle, maximazlement simple, et au cours de laquelle la factualité et les problèmes qu'elle soulève sont épurés de toute hypothèse qui n'est imposée ni par les contraintes locales qui agissent, ni par le but final.

En ces conditions que faut-il finalement penser de la relation $G \leftrightarrow me_G$?

Il me semble que la question peut être tranchée définitivement par le dialogue imaginé qui suit [72].

**L (le lecteur)**. Malgré tout, je me demande si vraiment on est *obligé* de poser la relation de un-à-un $G \leftrightarrow me_G$. Il existe peut-être une autre solution.

**M (moi)**. Tout d'abord, rien n'est obligatoire dans une construction. Ceci est convenable, cela ne l'est pas. Point. En l'occurrence, si l'on imaginait au départ que l'opération $G$ peut produire tantôt une chose et tantôt une autre, on aurait des difficultés pour parler de ce que $G$ produit. Et aussi pour y réfléchir, ce qui est beaucoup plus grave. Alors pour quelle raison devrait-on éviter d'introduire une organisation de langage-et-concepts qui évite ces difficultés?

**L**. Pour ne prendre aucun risque de découvrir plus tard que l'on a affirmé quelque chose de faux.

**M**. Faux? Mais face à quoi? La question est là: ***face à quoi?*** Forcément face à quelque *examen* futur pour qualifier le microétat $me_G$, n'est-ce pas? Or ici, *il ne s'agit pas d'une relation entre le microétat $me_G$ et les résultats d'examens futurs pour le qualifier. Il s'agit* **exclusivement** *de* **la relation entre l'opération de génération $G$ et son effet $me_G$.** Lorsqu'on glisse subrepticement d'un problème à un autre, on étrangle l'entendement dans un nœud.

**L**. D'accord, mais ce qu'on admet maintenant peut entraîner des effets concernant ce qui se manifestera plus tard.

**M**. Magnifique! Finalement je trouve que cet échange est le plus utile que l'on ait pu concevoir afin de rendre intuitive la relation $G \leftrightarrow me_G$. Vous êtes en train de m'offrir l'occasion d'étaler sans un pli devant les yeux publics, l'un de ces glissements incontrôlés qui sécrètent des faux absolus et faux problèmes où l'entendement reste piégé comme une mouche dans une toile d'araignée.

Donc exprimons-nous jusqu'au bout: vous craignez que le fait de poser d'emblée une relation de un-à-un entre l'opération de génération $G$ et le microétat $me_G$ qu'elle produit, puisse avoir des implications qui se révéleront fausses face aux résultats de quelque examen futur de $me_G$. Et cette crainte vous fait préférer de laisser ouverte la possibilité que cette relation ne soit pas de un-à-un, plutôt que de l'exclure prématurément par une assertion dictatoriale qui pourrait se trouver démentie par la suite. C'est bien cela?

---

[72] Cf. aussi la définition $D4$ dans l'exposé du noyau de la méthode de conceptualisation relativisée, dans Mugur-Schächter [2006], p.64: là ce même dialogue intervient à un niveau *général*, non lié au cas particulier des microétats.

**L**. Tout à fait cela.

**M**. Alors faisons une expérience de pensée. Imaginons un examen dénoté *Ex.1* du microétat *me$_G$* qui serait tel que, chaque fois que l'on réalise *G* et l'on soumet l'effet *me$_G$* de *G* à l'examen *Ex.1*, l'on obtienne invariablement le même résultat. Que diriez-vous dans ce cas *concernant la relation entre G et me$_G$*? Qu'il est désormais *démontré* qu'il s'agit en effet d'une relation de un-à-un? Vous pouvez répondre « oui », vous pouvez répondre « non », ou bien vous pouvez répondre « pas encore démontré ». Cela épuise les possibilités.

Supposons d'abord que vous répondiez « oui ». En ce cas, imaginons maintenant un autre examen dénoté *Ex.2* qui est différent de *Ex.1* et qui est tel que lorsqu'on répète *G* plusieurs fois et à qu'à chaque fois on soumet l'effet *me$_G$* obtenu, à l'examen *Ex.2*, l'on constate tantôt un résultat, tantôt un autre, donc en fin de compte tout un ensemble de résultats différents. Cela vous paraît-il impossible, étant donné que l'effet du premier examen *Ex.1* s'est avéré être stable?

**L**. Non, pas nécessairement, en effet.... On peut imaginer par exemple que l'opération *G* est définie de façon à produire à chaque fois une bille sphérique de dimensions données, mais dont on laisse la matière varier d'une réalisation de *G* à une autre. Alors en répétant *G* et en soumettant à chaque fois le produit de *G* à un examen de forme, on obtiendrait un ensemble de résultats identiques, cependant qu'avec un examen de poids on obtiendrait un ensemble de résultats dispersés…….. Si l'on n'essaie pas de restreindre *G* à l'avance convenablement, on ne peut pas éliminer la possibilité que vous venez d'envisager.

**M**. Restreindre *G à l'avance* pour que ***tout*** examen futur, disons *Ex.j, j=1,2,......,…* conduise à un ensemble de résultats identiques si l'on répète des séquences *[G.Ex.j]*? Cela avec un *j* quelconque, de l'ensemble d'examens quelconques considéré? Cela vous paraît-il concevable? Il me semble que vous ne distinguez pas clairement entre *une restriction qui pèserait **sur l'opération de génération G**, et une restriction concernant les examens futurs que l'on pourrait accomplir **sur les résultats de G***. Mais progressons systématiquement. Donc vous admettez que lorsqu'on répète l'opération de génération *G*, telle qu'elle a été spécifiée de par sa définition, le microétat qui en résulte pourrait manifester à chaque fois des résultats identiques lorsqu'il est soumis à l'examen *Ex.1*, cependant que l'examen *Ex.2*, lui, produirait des résultats non-identiques. Que diriez-vous en ce cas *concernant la relation entre G et me$_G$*? Qu'il est désormais *démontré* qu'elle n'est pas une relation un-à-un?……

J'ai l'impression que vous hésitez? Pourquoi?

**L**. Parce que je commence à concevoir qu'il se pourrait que le comportement du microétat produit par *G*, face à des examens futurs *sur ce **résultat**,* ne puisse jamais imposer une conclusion quant à la relation entre *G* et le microétat produit par *G*.

**M**. Donc finalement nous sommes en train de converger. Néanmoins allons jusqu'au bout systématiquement. Examinons maintenant la troisième réponse possible de votre part. Supposons donc qu'à ma première question concernant l'examen *Ex.1* vous ayez répondu « non, cela ne démontre pas encore que la relation entre *G* et son effet dénoté *me$_G$* soit une relation de un-à-un ». Dans ce cas, je vous demanderais: *Quand* admettrez-vous qu'il *est* démontré que la relation entre l'opération *G* de génération du microétat et le microétat que *G* produit, est une relation de un-à-un? Quand vous aurez vérifié l'identité des résultats pour tous les examens futurs? Mais que veut dire 'tous' ici? Tous les examens que l'on connaît, ou bien tous ceux que l'on connaît plus ceux que l'on imaginera jusqu'à la fin des temps? Sur quelle base pourrait-on affirmer quoi que ce soit concernant cette 'totalité' ouverte, indéfinie d'effets d'examens futurs?……

Je prends la liberté de considérer que le dialogue imaginaire qui précède a valeur d'une preuve ; qu'il a pu imposer désormais la nécessité, en général, de décisions méthodologiques lorsqu'on entreprend une construction ; et aussi, en l'occurrence, la nécessité de la décision méthodologique de poser la relation de un-à-un $G \leftrightarrow me_G$. La justification – pas la *preuve*, mais la *justification* – ne pourra venir qu' *a posteriori*. Car cette relation de un-à-un non seulement est nécessaire *a priori* pour pouvoir construire une représentation qualitative de la description d'un microétat, mais en outre la nier, on vient de le voir, serait dépourvu de toute conséquence acceptable d'un point de vue conceptuel, comme aussi de toute conséquence utile.

Or la relation de un-à-un $G \leftrightarrow me_G$ **élimine a priori** les deux problèmes reliés de la complétude de la forme descriptionnelle $D_M/G, me_G, V_M/$ et du contenu ontologique du concept de microétat.

Ce qui a fait obstacle à la compréhension de la situation conceptuelle lorsque les problèmes de complétude et ontologique ont été soulevés en relation avec la formalisme mathématique de la mécanique quantique, a été l'idée fausse induite par la présence du formalisme mathématique, qu'il s'agirait de questions de nature formelle, à résoudre par des démonstrations de théorèmes. Cependant que lorsqu'ils sont rapportés à la genèse comportée par la forme descriptionnelle $D_M/G, me_G, V_M/$ – avec la relation de un-à-un $G \leftrightarrow me_G$ qui y joue un rôle *méthodologique* fondamental – ces célèbres et persistantes questions de la complétude descriptionnelle et du contenu ontologique d'un 'microétat', qui ont fait couler tant d'encre, simplement s'évanouissent. Lorsqu'on tient compte de la genèse de la forme descriptionnelle qualitative $D_M/G, me_G, V_M/$, lorsqu'on s'imprègne du **néant** conceptuel duquel la forme $D_M/G, me_G, V_M/$ a pu émerger *via* des contraintes cognitives et méthodologiques inévitables qu'elle a incorporées pas à pas, on comprend enfin *intuitivement* qu'**avant** les descriptions *transférées* des microétats – par construction – il n'y a **RIEN** en tant que connaissances sur des microétats spécifiés. On heurte ainsi le **sol-limite** en dessous duquel aucun connu spécifique d'un microétat donné ne peut trouver place. Seule une *construction* **super-posée** reste concevable.

L'on réalise ainsi *qu'un processus d'élaboration de connaissances est inéluctablement marqué par le choix d'une* **origine** *sur la verticale des conceptualisations et par* **l'ordre** *qui en découle dans l'action épistémologique*. En certains cas les conséquences d'un tel choix peuvent dominer tout autre caractère. Notamment ils peuvent évaporer les prudences logiques, comme dépourvues à la fois de sens, d'utilité et – radicalement – de possibilité : c'est le cas pour la relation de un-à-un $G \leftrightarrow me_G$.

### 3.1.6. Probabilités primordiales versus *probabilités classiques* [73]

Éclaircissons ces question d'origine et d'ordre sur la verticale des phases de conceptualisation, de 'haut' et 'bas', d' 'au-dessus' et 'au-dessous' d'une phase donnée de conceptualisation, qui peuvent paraître obscures et étranges.

Dans le cas de *IMQ* le but posé était d'identifier et traiter toutes les contraintes factuelles et épistémologiques qu'imposent – spécifiquement – les conditions cognitives comportées par la conceptualisation des microétats. Face à ce but, le choix de placer l'origine du processus de conceptualisation sur le niveau de connaissance zéro concernant tel ou tel microétat particulier, était incontournable, car les contraintes qu'imposent les conditions cognitives comportées par la conceptualisation d'une entité physique, ne peuvent pas être inventées, elles ne peuvent qu'être constatées, or ceci s'est avéré impossible en partant de la conceptualisation classique qui est fondée sur la notion d''objets' au sens courant (MMS [2006 pp.118-127]) [74]. En effet, ces 'objets' classiques sont quasi généralement conçus comme dotés 'vraiment' de 'propriétés' propres – si l'on peut dire – notamment, comme étant délimités spatialement par des contours intrinsèques et absolus et comme étant soumis à des 'causes' de changement, intrinsèquement vraies elles aussi et qui agissent avec un détail et une rigueur sans bornes.

La physique classique est elle aussi dominée par les caractéristiques de la conceptualisation classique courante. Elle aussi s'appuie directement sur des modèles-'objets'. Et selon la physique classique le caractère probabiliste d'une description (par exemple celle des effets des jets d'un dé, où celles de la théorie cinétique des gaz) peut *toujours* être *éliminé* par l'application en toute rigueur de la théorie classique fondamentale utilisée (la mécanique newtonienne, ou celle d'Einstein, ou l'électromagnétisme de Maxwell, ou quelque composition de ces théories). Toutes les théories de la physique classique sont absolument déterministes dans leurs principes. Les caractères statistiques ou probabilistes ne s'imposent qu' "en pratique" et l'on postule qu'ils sont explicables par quelque ignorance ou quelque abstraction faite de données qui sont disponibles dans la représentation rigoureuse placée en-dessous de ce caractère, dans les théories de base qui interviennent.

Le fait que ce postulat causal explicatif est entaché de non effectivité au sens de l'impossibilité à la fois factuelle et théorique d'empêcher des effets de 'chaos déterministe', n'a fait qu'un scandale de salon. Il n'a pas épuré les mentalités (Longo [2002], MMS [2002C])). L'univers conceptuel de la science classique fonctionne sur un substrat lissé par un déterminisme foncier *décrété* dont le caractère faussement absolu et universel se cache dans le flou de la pensée courante.

*La physique classique opère donc à l'intérieur du piège du réalisme naïf* (**3.1.4**).

---

[73] J'ai montré ailleurs que **le concept de 'probabilité', comme celui d''objet', est un concept foncièrement classique** (MMS [2009], [2013]). L'utiliser dans le cadre d'une description purement transférée, et avant d'avoir explicitement élaboré un mode *normé* de *passage* d'une description transférée donnée, à une modélisation correspondante, revient à un mélange non contrôlé de niveaux descriptionnels essentiellement distincts. Plus tôt ou plus tard, mais toujours, de tels mélanges engendrent des paradoxes et des faux problèmes qui polluent et finalement bloquent le développement des processus de construction de connaissances. Ce problème est rappelé dans la deuxième partie de ce travail. Mais il n'est traité que dans MMS [2013].

[74] Husserl a montré dans sa Phénoménologie, "mais un peu tard", que – très paradoxalement – les 'objets' classiques qui sont communément considérés comme le paradigme de matérialité, sont en fait des construits abstraits, des *modèles* tirés des résultats d'interactions des appareils sensoriels humains, avec du réel physique, via ce que l'on peut appeler des descriptions transférées macroscopiques produisant des enregistrements qui s'accompagnent de qualia. Mais ce n'est *pas* de ces descriptions transférées que part la conceptualisation classique *explicite*. Elle part des 'objets' préconstruits de façon implicite, souvent réflexe ou même câblée génétiquement dans le système nerveux humain ; la genèse des modèles-'objet' se perd souvent dans les brumes de l'évolution biologique qui l'a enfermée dans les boites noires des réflexes neurosensoriels.

Il n'est donc pas étonnant que les modélisations initiales *classiques* successives – moléculaire, atomique, nucléaire, et même celle en termes de particules élémentaires – n'aient pas véritablement abouti. Les maillons de cette chaîne de modèles en termes d''objets' classiques à des échelles d'espace-temps de plus en plus petites ont laissé la pensée physique en panne. Les éléments de cette chaîne suspendue au plafond de la conceptualisation classique, *ne s'implantent pas dans la factualité physique qu'ils concernent*. Ils ne l'atteignent même pas. Ils restent comme ballants au dessus d'un non-fait conceptuel où tourbillonnent les malentendus métaphysiques du réalisme naïf, cependant que de leur intérieur leur nature de répliques arbitraires de la croûte classique de nos modes de conceptualisation ancestraux les rend friables.

Seule la mécanique quantique *fondamentale* – celle qui s'est forgée entre 1900 et 1930 – a trouvé moyen de véritablement enraciner les descriptions de microétats, dans la factualité physique qu'elles concernent, en introduisant pour la première fois dans l'histoire de la pensée ce qui – dans l'*IMQ* – a été dénommé des *descriptions primordialement transférées*. Par cela elle a vitalisé a posteriori la physique classique des noyaux, des atomes, des molécules, et aussi, la physique encore inachevée des champs et des particules élémentaires. Mais elle a accompli ces performances via un formalisme cryptique qui soulève des problèmes bien connus.

C'est cela qui a conduit à l'approche de *IMQ* dont le premier but était d'identifier explicitement et jusqu'au bout les exigences d'une théorie des microétats débarrassée de tout dogme ou ambiguïté épistémologique. A son tour ce but a conduit à enraciner explicitement la démarche *IMQ* dans la factualité microphysique aconceptuelle, afin qu'aucun obstacle sur le trajet du développement ne puisse échapper *non rencontré* et a fortiori non traité. Or *de là*, de la factualité microphysique aconceptuelle, on ne peut développer la conceptualisation *que* vers le 'haut' classique de la verticale des phases de conceptualisations humaines du réel physique. Car, *par construction*, en 'dessous' il ne reste aucun espace abstrait où l'on puisse loger de façon non-contradictoire la *notion* de quelque conceptualisé ou conceptualisable qui puisse 'expliquer' les résultats obtenus d'une manière plus définie, plus 'complète', plus 'singulière' (au sens grammatical). Voilà les significations qu'exprime ce langage de 'haut' et 'bas', en-dessous…, etc.

En ces conditions, si par une application inertielle du principe déterministe l'on éliminait le caractère factuel primordialement statistique-'probabiliste' de la description de microétat qui émerge, qu'en resterait-il? ***Rien***. Cette description toute entière n'est rien de plus que les lois de 'probabilité' primordiales construites en relation directe avec des fragments de substance physique a-conceptuelle non-perçus. *Point*.

Donc l'explication des probabilités comportées par les descriptions transférées primordiales de microétats construites dans *IMQ* ne peut être construite que *sur la base* de ces descriptions et en cohérence constamment contrôlée avec elles.

*Mais ce renversement d'ordre de constructibilité – qui découle du choix délibéré de l'origine de la démarche constructive de IMQ – désigne un **statut méthodologique-épistémologique**. Il est **pur de toute connotation ontologique***.

(Je refuse l'expression courante de 'probabilités essentielles' précisément parce qu'elle comporte une forte connotation ontologisante qui suggère des 'propriétés intrinsèques' des entités représentées elles-mêmes)[75].

---

[75] Tout ce qui vient d'être rappelé n'est qu'une manifestation particulière du fait que, *toujours*, nos conceptualisations du réel physique sont nécessairement marquées d'un caractère méthodologique-épistémologique, même dans les sciences

Toutefois, de ce qui vient d'être souligné il découle le fait de connaissance tout à fait réel que les contenus des représentations de microétats construites sur la base des descriptions primordialement transférées statistiques-probabilistes et en montant vers la conceptualisation classique, sont certainement différents des contenus des représentations d'entités physiques microscopiques accomplies dans l'ordre *historique* 'descendant', car ces deux démarches en sens opposées incorporent des contraintes mutuellement distinctes et dans des successions structurantes opposées.

### 3.1.7. Descriptions 'primordialement transférées' versus *réalisme scientifique*

Notamment, des chaînes de conceptualisation construites sur la base des descriptions de microétats primordialement transférées, désintègrent le piège du réalisme naïf où se débattent les esprits torturés par les problèmes de complétude et ontologique.

En effet, d'une manière évidente et irrépressible, le processus de construction de connaissances absolument premières concernant des microétats a conduit dans *IMQ* à des descriptions primordialement transférées qui *éjectent d'elles explicitement tout réalisme naïf* : Ces descriptions sont foncièrement, inextricablement marquées de *relativités descriptionnelles*. et cela – tout aussi foncièrement et inextricablement – élimine toute possibilité de retour à un réalisme naïf en ce qui concerne le contenu de, également, toutes les *modélisations* qui peuvent se constituer sur la base de ces descriptions marquées de relativités descriptionnelles. Ces modélisations, désormais, ne peuvent plus être conçues elles non plus autrement que foncièrement relatives à nos actions cognitives, même si les relativités qui y interviennent génétiquement n'atteignent la conceptualisation classique que *cachées* dans des boules de modèles-'objets' qui roulent causalement sur un terrain conceptuel lissé par un postulat déterministe.

### 3.1.8. Descriptions transférées primordiales versus *le postulat déterministe*

Notamment, sous les effets descriptionnels infusés à partir de la strate initiale des descriptions de microétats primordialement transférées et statistiques-'probabilistes', ce postulat déterministe ne peut plus être admis comme rigoureusement 'vrai' et comme universel. Au niveau de conceptualisation où il devient possible de construire de la causalité et du déterminisme – via des abstractions fondées sur des considérations concernant une certaine hiérarchie, sur ce niveau là, des effets des ordres de grandeur des distances, durées, fonctions d'action, énergies, qui y interviennent – *le postulat déterministe lui-même émerge marqué à la fois de relativités ET D'APPROXIMATIONS*.

Dans ces conditions, tirons les conséquences jusqu'au bout. Lorsque la microphysique fondamentale actuelle n'offre que des descriptions primordialement transférés et statistiques, cependant que la physique macroscopique postule directement un déterminisme *ontologique*, assigné aux faits mêmes, qui se trouve en difficulté à la fois épistémologique et métaphysique, il paraît approprié d'introduire le postulat suivant.

---

classiques. La croyance qu'il serait possible d'échafauder un système logiquement cohérent de représentations d'un domaine de réel physique, exclusivement à coup de purs constats, de pures découvertes, se révèle avec évidence comme illusoire dès qu'on y pense véritablement. Il *faut* commencer par vouloir *ceci* ou *cela*, opérer 'afin que', *poser*, définir et dénommer, tout cela délibérément, et bien sûr en relation constante avec du factuel et avec les exigences cognitives et logiques. En ces conditions il convient de déclarer en chaque phase *explicitement* en fonction de quelles contraintes on agit, afin d'offrir une perceptibilité claire aux regards critiques.

*Toute* phase absolument initiale d'un processus de construction de connaissances nouvelles – que ce processus soit conscient ou réflexe, physique ou abstrait, à entités-objet-de-qualification microscopiques ou macroscopiques ou cosmiques – comporte toujours un caractère primordialement statistique.

Selon un tel postulat chaque discipline déterministe de la physique classique est à regarder d'emblée comme *un construit modélisant et relatif qui est tiré de façon constructive, directement ou de manière médiate, de la phase absolument initiale du processus de construction de connaissances qui est impliqué, qui existe nécessairement et – toujours – est primordialement transférée et statistique.*

Ces déterminismes conçus comme des construits à partir de descriptions initiales primordialement transférées, statistiques et foncièrement relativisées, préserveraient tous les avantages pragmatiques qui découlent du postulat déterministe classique universel et absolu, sans pour autant assigner un caractère déterministe au 'mode d'être des faits physiques eux-mêmes'. Et sans que, sur cette base insoutenable, l'on soit conduit à incriminer comme 'incomplètes' les représentations d'un domaine du réel physique qui ne possèdent pas un caractère déterministe, notamment parce que tout simplement elles ne disposent pas encore d'une matière descriptionnelle de laquelle forger un tel caractère, ni d'une place conceptuelle aménagée où le loger.

Cela – associé à une unification des conceptualisations classiques, avec leurs sources premières factuelles-épistémologiques-méthodologiques, via des chaînes de connaissance qui débuteraient systématiquement avec des descriptions primordialement transférées et statistiques – nous mettrait dans un schéma global de pensée entièrement cohérent.

### 3.1.6. Une 'coupure' classique-IMQ

Nous avons mis en évidence que les descriptions transférées $D_M/G,me_G,V_M/$ de microétats appartiennent à une strate primordiale de conceptualisation dont la structure descriptionnelle est radicalement distincte de celle des descriptions classiques.

D'autre part nous avons rappelé que les descriptions au sens classique comportent des modèles-'objets' (élaborés sur la base des descriptions transférées primordialement probabilistes qui se sont construites de manière réflexe ou bien seulement implicite). Ces modèles-'objets' préexistants sont dotés par la modélisation qui les a générés, de 'propriétés' indépendantes de toute action cognitive. Et – selon la pensée courante, selon les langages naturels avec leurs **grammaires**, selon la *logique* et les probabilités classiques, et toutes les disciplines 'dures' de la science classique – ces 'propriétés' indépendantes permettraient de sélectionner des objets-d'étude parmi les modèles-'objets' préexistants et *dans une même foulée* de qualifier les objets-d'étude sélectionnés.

On a appréhendé les distances qui, selon diverses directions, séparent les descriptions au sens classique, des descriptions $D_M/G,me_G,V_M/$ primordialement transférées et statistiques: Il en ressort clairement qu'il s'agit de deux phases radicalement distinctes des processus de conceptualisation.

Ces remarques conduisent trivialement à la conclusion suivante.

Entre: *(a)* les descriptions de microétats $D_M/G,me_G,V_M/$ primordialement transférées et statistiques et *(b)* les descriptions au sens classique, fondées sur des modèles-'objets' conçus comme dotés de propriétés intrinsèques, il existe une *'coupure [IMQ-classique]'* qui ne peut être comblée que par des modélisations fondées sur les caractéristiques des descriptions $D_M/G,me_G,V_M/$ [76].

La mystérieuse *'coupure [quantique-classique]'* s'inscrit dans la coupure *[IMQ-clasique]*, puisque *IMQ* inclut par construction les descriptions transférées élaborées dans la mécanique quantique (*2.3.2.3, 2.3.2.4*).

La raison pour laquelle la définition du contenu de cette coupure oppose à ce jour même des résistances, est que la forme qualitative intégrée $D_M/G,me_G,V_M/$ des descriptions de microétats primordialement transférées et statistiques, est restée non connue d'une manière explicite.

## 3.2. La source commune des problèmes suscités par la forme descriptionnelle $D_M/G,me_G,V_M/$

Dans les problèmes qui émergent concernant la forme descriptionnelle $D_M/G,me_G,V_M/$, ce qui agit à la base est – uniformément – une compréhension insuffisante du fait que la spécificité majeure d'une description de microétat est d'être une description toute première, primordialement transférée et statistique. Malgré plus de 70 années d'existence de la mécanique quantique et d'utilisation du formalisme mathématique de cette théorie, la forme *intégrée* des descriptions de microétats symbolisée ici $D_M/G,me_G,V_M/$, avec sa genèse, ses caractéristiques et ses conséquences, sont restées foncièrement étrangères à nos esprits éduqués par des interactions cognitives directes avec le réel factuel macroscopique, via nos sens biologiques. Mais dès que la forme descriptionnelle $D_M/G,me_G,V_M/$ est connue et véritablement comprise *via* sa genèse, la situation change. Les traits méthodologiques et les relativités que cette forme incorpore dissolvent d'ores et déjà le problème de complétude, le problème ontologique, le problème de la spécificité de l'indéterminisme quantique face au 'hasard' classique, et le problème de la définition de 'la coupure quantique-classique'.

### 3.3. *IMQ* versus consensus

Enfin, examinons la question des conditions de constructibilité de consensus concernant les assertions observationnelles faites à l'intérieur de l'*IMQ*. Ces conditions s'expriment rapidement, mais leurs conséquences sont importantes en ce qui concerne le 'problème' de l'unification entre mécanique quantique et relativité einsteinienne, examiné dans la deuxième partie de ce travail,.

L'*unique* type de consensus que l'on puisse exiger dans le cas d'une théorie à descriptions primordialement transférées et statistiques, est celui obtenu par la comparaison directe des distributions statistiques élaborées *séparément* dans les différents référentiels *propres* des laboratoires. En effet chacun des observateurs-concepteurs, dans son laboratoire, *ne peut faire rien d'autre* afin d'entrer en possession de données factuelles, que (*2.3.2.2, 2.3.2.3*):

---

[76] Dans MMS [2006] – à l'intérieur de la méthode de conceptualisation relativisée dénotée *MCR* – cette coupure apparaît explicitement comme un cas *particulier* incorporé dans une coupure *générale* [(descriptions primordialement transférés et statistiques)-(descriptions classiques)], *libérée de coalescence avec la coupure micro-macro* : Au niveau macroscopique, comme au niveau microscopique, comme aussi au niveau cosmique, le seul ordre de construction de connaissances qui assure la capacité d'une intégration exhaustive de toutes les contraintes qui agissent (cachées ou explicitées) dans la conceptualisation d'une modèle-'objet' donné, place à la base des descriptions primordialement transférées et statistiques qui peuvent y conduire. Cependant que l'ordre explicatif-logique-déductif n'intervient qu'*a posteriori* pour des buts d'unification syntaxique d'îlots de descriptions préconstituées).

- Effectuer un grand nombre de successions *[G.X]* ou *[G.Mes(X)]*.

- Prendre connaissance des groupes de marques enregistrées sur les enregistreurs de son appareil, à la suite des opérations d'interaction *X* ou *Mes(X)* entre l'appareil et un exemplaire de l'entité etudiée '*me_G*', les marques enregistrées étant dépourvues de toutes qualia *associable à l'entité physique étudiée 'me_G'*, cependant que cette entité elle-même est inobservable isolément en conséquence des conditions cognitives dans lesquelles se déroule l'investigation.

- Classer ces groupes de marques selon quelque critère codant.

- *Dénombrer* les groupes de marques de chaque classe et en établir la répartition statistique.

Par construction, les seuls invariants requis lorsqu'on passe d'un observateur à un autre, sont le procédé et ses résultats, qui, en leur essence, doivent être identiques à ceux de tout autre observateur qui opère séparément dans le référentiel de son propre laboratoire.

Notamment, le cas de deux ou plusieurs observateurs liés à des référentiels en mouvement (inertiel, ou non) les uns par rapport aux autres, qui tous observent directement via des signaux lumineux un mobile extérieur à tous les laboratoires et, pour s'assurer qu'il s'agit bien d'un 'même' mobile, doivent spécifier et respecter telle ou telle loi de transformation des coordonnées d'espace et de temps, *tout cela simplement ne se présente pas, et n'a même pas de sens*, puisque les concepts même de mobile et de trajectoire n'ont aucun sens dans la situation cognitive liée à des descriptions transférées primordialement statistique. Des concepts de cette sorte ne peuvent (éventuellement) être associés à des microétats que via des modélisations accomplies dans des phases de conceptualisation subséquentes.

Or, selon les principes même de la relativité restreinte :

« Le déroulement de phénomènes analogues liés de la même façon à divers systèmes de référence, ne dépend pas du mouvement rectiligne et uniforme du système de référence »[77]

Bref, en ce qui concerne le type de consensus exigible, la situation est du même genre que celle qui se réalise dans l'entière physique classique: L'assertion d'un résultat expérimental donné doit être *vérifiable* dans chaque référentiel propre. *Point*.

### 3.4. Conclusion sur le chapitre 3

On peut espérer que les résultats obtenus dans ce chapitre concernant les problèmes soulevés par *IMQ* se montreront utiles pour libérer également le formalisme mathématique de la mécanique quantique des problèmes d'interprétation qui l'entachent. Car d'ores et déjà on peut percevoir que les problèmes examinés plus haut, tels qu'ils se manifestent face à la forme descriptionnelle *D_M/G,me_G,V_M/* établie dans *IMQ*, concentrent en eux, distillé, l'essence des problèmes qui portent les mêmes dénominations et qui ont *d'abord* été formulés relativement au formalisme mathématique de la mécanique quantique, parce que ce formalisme s'est constitué et a été utilisé avant que sa manière de signifier ait été élucidée.

---

[77] Marie-Antoinette Tonnelat [1971], p. 152.

*Troisième partie*

**L'infra-mécanique quantique
regardée du dehors de manière globale,
versus
le problème de localité**

# *Mot d'introduction à la troisième partie*

L'infra-mécanique quantique étant maintenant délinéée, il est possible de l'examiner dans sa globalité, de l'extérieur, afin de repérer la place qu'elle occupe face à trois aspects d'importance majeure: la façon de laquelle y interviennent l'espace, le temps, et les géométries d'espace-temps; le type de consensus qu'elle permet; et enfin, ce qu'elle implique sur les relations entre la microphysique actuelle et les théories de la relativité d'Einstein.

Les 'problèmes' soulevés par le type de description qualitative d'un microétat qui s'est constitué dans l'infra-mécanique quantique, sont internes à cette démarche. Ceci a permis de les aborder dans le courant même de la construction de la démarche. Mais le problème de localité est extérieur à l'infra-mécanique quantique. Il concerne la relation entre les descriptions de microétats et les théories relativistes einsteiniennes. Pour cette raison, afin de pouvoir l'aborder non pas sur la base de non dits, mais sur la base de formulations explicites, nous avons relégué son examen à la fin de cette troisième partie. La conclusion qui apparaîtra concernant le théorème de Bell est la suivante.

La preuve mathématique de 'l'inégalité de Bell' n'est pas réfutable. Mais sa signification se précise et il est important pour l'évolution immédiate des recherches en physique théorique d'en prendre connaissance.

Quant au théorème considéré globalement, la conclusion énoncée par Bell *ne découle nullement de la preuve mathématique de l'inégalité*. C'est la mise en évidence de cette absence de conséquence logique qui équivaut à ce que j'appelle une 'invalidation conceptuelle' du théorème. Cette sorte d'invalidation est liée foncièrement à la conceptualisation générale qui sous-tend la démarche démonstrative et qui conduit à la signification précisée de la preuve mathématique considérée isolément.

Finalement est esquissée une variante des expériences concernant la question de localité, modifiée de manière à se focaliser spécifiquement sur la signification nouvelle de la preuve mathématique de l'inégalité.

Les considérations qui suivent n'expriment ni des vues actuellement consensuelles, ni même des vues personnelles qui seraient toutes bien stabilisées. Il ne s'agit que d'une rapide exploration du terrain conceptuel introduit par l'infra-mécanique, encore mal affermi par endroits. Cette exploration est soumise aux lecteurs d'une manière informelle, et souvent elle se limite à des remarques et de *questions*.

# Chapitre 4
# *Théories physiques,* versus *espace, temps, géométrie, consensus*

## 4.1. Remarques générales sur espace, temps et géométrie

Un individu humain normal ne peut percevoir, ni même seulement concevoir, une entité *physique* – objet, événement, substance – sans la placer dans l'espace et le temps. Ceci est un fait psychique qu'il paraît difficile de contester. Kant a exprimé ce fait en posant que l'espace et le temps sont deux formes *a priori* de l'intuition.

Dans ce qui suit j'adopte ce postulat.

J'ajouterai un aveu épistémologique: réciproquement, je n'arrive pas à concevoir de l'espace ou du temps en l'absence – strictement – de *toute* existence physique, ou au moins d'une émanation d'une existence physique, comme mon attention cachée quelque part pour surveiller, et mon souffle qui en quelque sorte dénombre qualitativement du passage de temps.

Le postulat kantien rappelé plus haut n'implique aucune *structure* d'espace, ou de temps, ou d'espace-temps. Il n'affirme qu'un fait concernant le psychisme des individus humains. Comment, alors, s'engendrent des 'géométries' d'espace, ou de temps, ou d'espace-temps?

Henri Poincaré [1898] a notablement élaboré l'idée que l'assignation à l'espace – considéré comme une donnée première non structurée[78] – d'une structure *géométrique* euclidienne, émerge par l'intégration dans un système unique, stable et cohérent, de toute la diversité des aspects spatiaux qui sont impliqués dans les *interactions* naturelles kinésiques et sensorielles entre les individus humains et du réel physique. Cette intégration, la géométrie euclidienne, s'exprime par un système de relations entre, exclusivement, des concepts spatiaux abstraits, points, lignes, figures (cf *1.5.3*), où les relativités à tel ou tel 'point de vue' sont *évacuées*[79], comme aussi les éléments physiques et biologiques sensoriels qui ont participé aux interactions.

Einstein (peu après Poincaré, mais sans faire référence à lui) a fait une assertion similaire dans son exposé de la relativité restreinte. Il y affirme que les interactions *de mesure* que des observateurs inertiels réalisent avec des mobiles macroscopiques, *via* des signaux lumineux, conduisent à la géométrie d'espace-*temps* 'Minkowski-Einstein'. Mais notons que, à la différence de la géométrie d'espace euclidienne, la géométrie d'espace-temps de Minkowski-Einstein n'est *pas* intégrative jusqu'au bout. Elle n'intègre (dans un schéma à deux cons de lumière et deux ailleurs) que l'ensemble des interactions de mesure d'*une seule classe* d'observateurs inertiels ayant tous des états identiques de mouvement inertiel. Une méta-intégration de toutes ces intégrations dans une synthèse unique de toutes les interactions de mesure de tous les observateurs inertiels, n'est pas opérée dans la relativité restreinte. C'est pour cette raison que la géométrie Minkowski-Einstein ne permet pas de définir une causalité générale cohérente.

---

[78] Poincaré ne fait pas référence à Kant, pour autant que je sache.
[79] Elles constituent un autre système, annexe, la géométrie euclidienne projective.

Quelques trente années plus tard Husserl a développé sa célèbre phénoménologie où il décrit les processus de *constitution transcendantale des "objets" physiques*[80]. Une entité physique donnée est toujours perçue exclusivement de tel ou tel point de vue particulier; elle n'est jamais perçue de tous les points de vue possibles. Une perception relative à un point de vue donné, conduit à une description correspondante *particulière* de cette entité. Mais l'ensemble de toutes les différentes descriptions particulières possibles d'une entité donnée, est intégré par l'esprit dans un concept abstrait d'un "objet" conçu comme l'invariant de cet ensemble, qui transcende la perceptibilité sensorielle (c'est ce qui, dans ce travail, a été à plusieurs reprise désigné par l'expression 'objet'-modèle).

Il paraît naturel (sinon même inévitable) de regarder le concept de géométrie euclidienne de Poincaré comme *un 'objet-**cadre'** général* construit comme l'invariant de *tous* les ensembles d'interactions humaines courantes avec des entités physiques, chacun de ces ensembles engendrant un "objet" au sens de Husserl.

Dans les sciences cognitives actuelles une vue de cette sorte commence à poindre chez des neurobiologistes et elle est étayée par des philosophes (Berthoz et Petit [2007])[81].

Quant à la relativité générale, ce qu'elle introduit n'est *pas* ce que j'accepte d'appeler une géométrie d'espace-temps. Selon ma vue c'est une *représentation géométrique* où une certaine géométrie d'espace-temps qui n'est pas explicitée de façon isolée, est d'emblée mise en coalescence avec un codage géométrique de distributions variables de masses et de champs, cela sous la contrainte d'un but descriptionnel: le but que la loi de mouvement de tout mobile macroscopique donné observé par tout ensemble d'observateurs – *via* des signaux lumineux –, soit construite par tous ces observateurs comme une géodésique de la représentation géométrique mentionnée.

De ces considérations, retenons ce qui suit.

Une géométrie d'espace ou d'espace-temps émerge comme une 'constitution transcendantale', i.e. comme la constitution d'une structure intégrative abstraite fondée sur un ensemble d'interactions d'un type donné mais qui transcende ces interactions (des interactions naturelles, ou de mesures macroscopiques opérées par des observateurs physiciens *via* des signaux lumineux, ou encore, sans doute, des interactions d'autres catégories possibles).

*Lorsque le type considéré d'interactions change, la géométrie qui en émerge change elle aussi.*

---

[80] Rappelons que chez Husserl 'transcendantal' veut simplement dire 'par interaction'.

[81] On pourrait regarder la description d'une entité physique macroscopique donnée, fondée exclusivement sur telle ou telle structure perceptive particulière éprouvée par un individu humain à la suite d'interactions sensorielles avec cette entité physique, comme une description de l'entité physique qui est *transférée* sur les enregistreurs d'appareils sensoriels biologiques de l'individu humain. Mais dans la mesure où l'entité-objet-de-description qui intervient est connue à l'avance, il s'agirait d'une hybridation du concept de description transférée primordiale, avec le concept d'"objet"-modèle-macroscopique.

## 4.2. Infra-mécanique quantique et mécanique quantique
### *versus*
### espace, temps et géométrie

Comment interviennent l'espace et le temps dans l'infra-mécanique quantique?

Ils y interviennent d'abord fondamentalement, en tant que formes *a priori* de l'intuition des concepteurs humains qui ne peuvent concevoir des entités physiques sans les loger dans de l'espace et du temps.

Ils y interviennent également dans les assignations de coordonnées d'espace et de temps impliquées dans les définitions des opérations *G* de génération d'un microétat et des opérations de qualification par des *Mes(X)*, où elle conduisent à la 'condition-cadre' qui est cruciale pour la qualifiabilité d'un microétat (cf. *3.3.2.2*).

Ces assignations de coordonnées dépassent l'appartenance aux formes *a priori* de l'intuition humaine. Elles appartiennent à une activité scientifique qui exige l'incorporation dans une structure géométrique.

Mais *laquelle*? La géométrie d'espace euclidienne associée au temps absolu de Galilée et Newton, ou bien des géométries d'espace-temps de Minkowski-Einstein?

Examinons la situation de plus près. Selon l'infra-mécanique quantique, l'algorithme qualitatif de construction de connaissances concernant un microétat $meG$ conduit à des descriptions primordiales transférées $D_M/G,meG,V_M/$. La genèse et le résultat de cette sorte de descriptions sont explicités dans la structure $T(G,V_M/)$ d'arbre de probabilité d'un microétat. Le protocole de construction de cet arbre comporte la répétition un très grand nombre de fois, de la succession *[G.Mes(X)]* où $X$ varie sur l'ensemble des 'grandeurs mécaniques' (tests liés à un signification mécanique) redéfinies pour des microétats. Chaque succession *[G.Mes(X)]* implique les données d'espace-temps $d_G.(t_G-t_o)$ et $d_X.(t_X-t_G)$ (voir la *Fig.1* du chapitre **2**) ainsi que – en général – les coordonnées d'espace et de temps des marques physiques observables $\mu_j$, $j=1,2,...m$ observées sur les enregistreurs de l'appareil $A(X)$ mis en jeu. Que peut-on dire concernant ces différentes qualifications d'espace et de temps qui interviennent?

La première remarque qui vient à l'esprit est que toutes ces qualifications s'appliquent *directement*, non pas à des interactions entre l'expérimentateur et les microétats $meG$ étudiés, mais aux interactions de l'expérimentateur avec les appareils macroscopiques utilisés pour réaliser les opérations *G* et *Mes(X)*. Chaque expérimentateur accomplit des opérations *G* et *Mes(X)* dans son référentiel propre lié à son laboratoire, sans observer rien d'autre que ses appareils, leur état et les marques physiques qui s'y affichent. Nulle part n'interviennent ni des 'mobiles' observés par plusieurs observateurs à la fois, ni des 'signaux lumineux' pour accomplir les observations et notamment pour assigner des coordonnées d'espace et de temps. En outre, le but n'est pas d'établir, pour un 'mobile' donné qui serait observé à partir d'états d'observation différents, une loi de mouvement dont la forme soit invariante aux changements d'états d'observation inertiels. Le but, dans ce cas, est d'établir des distributions de probabilité fondées sur des dénombrements de marques physiques amassées sur des enregistreurs d'appareils, où elles peuvent attendre leur lecture et leur dénombrement aussi longtemps qu'on veut.

Tout cela paraît entièrement étranger aux géométries de Minkowski-Einstein et d'autant plus à le géométrie riemanienne de la relativité générale. Tout ce qui, dans l'élaboration des descriptions primordiales transférées des microétats, se rapporte aux interactions entre l'expérimentateur et des entités physiques, semble s'intégrer à la géométrie euclidienne et au temps social conventionnel utilisés dans la physique classique.

Toutefois on peut encore hésiter. La forme descriptionnelle qualitative $D_M/G, me_G, V_M/$ mise en évidence par l'infra-mécanique quantique pourrait induire en erreur, peut-on se dire, par sa généralité, par son caractère qualitatif, par l'abstraction qui y est faite des définitions des opérations de mesure. En effet *dans le formalisme mathématique de la mécanique quantique* on spécifie entièrement la définition de chaque opération de mesure et cette définition est construite sur la base de modélisations de prolongement de la mécanique macroscopique classique. Celles-ci – sans supprimer le caractère primordial transféré des marques physiques observables – peuvent notamment introduire des mesures de durées et de distances que l'on assigne *au microétat étudié lui-même*, conçu subrepticement comme un mobile microscopique qui, du point de vue des déplacements et aux dimensions près, serait assimilable à un mobile macroscopique; ou au moins, qui impliquerait un élément assimilable à un tel mobile. C'est effectivement le cas pour la méthode du temps de vol discutée dans *3.3.2.1*. En ces circonstances, est-il toujours tout à fait pertinent de dire que les données d'espace et de temps qui interviennent dans une description de microétat primordiale transférée, concernent toujours exclusivement les interactions entre l'expérimentateur et ses appareils?

Notons tout d'abord que cette objection dépasse l'infra-mécanique quantique; elle concerne les descriptions primordiales transférées de la *mécanique quantique*. Mais examinons-la tout de même en prenant appui sur la vue exprimée ci-dessus concernant les géométries. Admettons que dans le cas des enregistrements d'une durée et d'une distance lors d'un acte de mesure par la méthode du temps de vol, il s'agirait d'une interaction entre l'expérimentateur et le microétat étudié. Quel sens y aurait-il de faire usage, pour cette raison, du formalisme de la relativité restreinte qui a été construit sur la base d'interactions où les signaux lumineux tiennent une place centrale? Et de l'idée d'interactions accomplies par plusieurs observateurs qui tous observent un même mobile à partir de référentiels inertiels distincts? Des interactions dont les résultats seraient ensuite organisés (*via* de nouvelles variances assignées aux valeurs des grandeurs mécaniques) de telle manière que tous les observateurs en tirent une loi de mouvement de la même forme? *Rien* de tout cela ne correspond au cas d'une mesure de la quantité de mouvement d'un microétat par la méthode du temps de vol. Il n'y a dans ce cas ni observation par plusieurs observateurs inertiels différents, ni signaux lumineux, ni le but de construire une loi de mouvement invariante aux changements de l'état d'observation. Il n'y a qu'enregistrements sur les enregistreurs du laboratoire, de marques codées qui y persistent indéfiniment et que l'on dénombre quand on veut afin d'établir leurs fréquences relatives.

## 4.3. La microphysique actuelle et les relativités d'Einstein

Mais alors, peut-on se dire, pourquoi dans le calcul célèbre de la thèse de Louis de Broglie, l'application des lois de transformation de Lorentz-Einstein a conduit à la relation fondatrice $p=h/\lambda$? Pourquoi dans le calcul des durées de vie de microsystèmes radioactifs les transformations de Lorentz-Einstein sont pertinentes? Pourquoi l'équation de Dirac a-t-elle conduit à une découverte nouvelle?

Ces questions, sans doute, méritent beaucoup de réflexion. Je n'essaierai pas de les examiner ici car cela dépasserait trop le but de ce livre. Néanmoins j'avance les remarques suivantes.

La thèse de Louis de Broglie a introduit au départ un **modèle** de microétat; un modèle d'un micro-état particulier, à savoir d'un *électron* libre. Ce n'est pas un modèle fantôme consistant dans une écriture mathématique de prolongement d'une qualification mécanique classique, comme dans le cas des implications conceptuelles des opérateurs dynamiques du formalisme quantique. C'est un modèle en chair et en os, d'un type particulier de microétat, mais un modèle radicalement novateur, celui d' 'onde-particule' (en fait un modèle onde *tout court* mais comportant une singularité de l'amplitude localisée dans l'espace-temps quasi ponctuellement qui concentre en elle presque l'entière énergie du microétat et qui serait ce qui produit les marques observables). De Broglie a ensuite généralisé ce modèle pour des microsystèmes quelconques se trouvant dans un microétat quelconque évoluant dans des milieux quelconques, meublés par des obstacles ou investis par des champs.

Par cela le traitement de Louis de Broglie s'est placée d'emblée sur le niveau d'une conceptualisation modélisante, pas sur le niveau d'une description primordiale transférée. Ce niveau là, celui de la théorie quantique, s'est constitué après.

Cela explique sans doute l'impossibilité qui s'est manifestée par la suite, d'inclure le concept de Broglie, tel quel, dans la construction subséquente du formalisme mathématique de la mécanique quantique, qui, elle, s'est placée sur le plan de conceptualisation radicalement primordiale. Le modèle de Broglie a fortement influencé le processus de construction du formalisme quantique, d'une manière et à un degré qui mériteraient un examen approfondi; et en outre il s'est installé également dans le langage qui introduit le formalisme et dans le langage qui accompagne l'utilisation du formalisme. Toutefois ce modèle *n'est pas représenté **dans** le formalisme et il s'est avéré impossible de l'y incorporer*.

Or on *peut* concevoir que, pour des modélisations de microétats libres d'*électrons* dotés de charge électrique, qui donc interagissent *via* des champs électriques beaucoup plus fortement que *via* des champs gravitationnels, l'emploi des transformations Lorentz-Einstein pour les coordonnées d'espace et de temps se soit montré pertinent, c'est-à-dire, qu'il conduise à des résultats que l'expérience vérifie. En effet, même si les observateurs humains et leurs signaux lumineux sont absents au niveau microscopique, les modèles d'états d'électrons prennent en quelque sorte le relais, car ils interagissent *via* des champs qui se propagent avec la vitesse de la lumière. Mais trouve-t-on *strictement* le même degré de confirmation expérimentale de la relation $p=h/\lambda$ lorsqu'on mesure, pour ce but, l'interfrange d'une interférence neutronique? A-t-on même essayé de s'en assurer? A-t-on examiné à fond l'hypothèse – ou plutôt le postulat plus ou moins tacite – que tout champ de toute nature se propage avec la vitesse *de la lumière*?

Chaque exemple de clair succès d'un traitement relativiste appliqué à des microétats, devrait être examiné sévèrement en explicitant le type d'interactions impliqué.

En tout cas, une théorie comme la mécanique quantique qui, se place toute entière sur le niveau de conceptualisation *primordiale* transférée, ne peut *pas* incorporer une géométrie fondée sur des interactions entre des **modèles** de micro-systèmes et de micro-états qui sont élaborés sur *la base* de descriptions primordiales transférées, dans une phase de conceptualisation qui est subséquente dans l'ordre génétique des modes de conceptualisation.

On ne peut pas réaliser une cohérence logique-formelle en mélangeant aveuglément des niveaux de conceptualisation distincts, comme on ne peut pas avoir une équation correcte si les deux membres ne sont pas identiques du point de vue des dimensions sémantiques qui interviennent.

La résistance du formalisme quantique à l'intégration du modèle de Louis de Broglie, bien qu'il ne s'agisse là que d'une modélisation *immédiate, microphysique*, d'ores et déjà manifeste cet impératif d'homogénéité conceptuelle.

Mais lorsqu'il s'agit d'une conceptualisation grossièrement distincte de celle de la mécanique quantique, si éloignée de la conceptualisation primordiale transférée de la mécanique quantique que l'est la conceptualisation macroscopique du mouvement de 'mobiles' selon la théorie de la relativité restreinte d'Einstein, on comprend que la résistance à une 'unification' globale soit une manifestation insurmontable de rejet conceptuel. Une telle unification ne pourrait être envisagée qu'entre une *modélisation* de la mécanique quantique fondamentale achevée en termes de micro-'mobiles', et la relativité restreinte. Et encore, même dans ces conditions modifiées, il n'y aurait *lieu* d'envisager une telle unification que si *(a)* les interactions entre les microétats modélisés *justifiaient* toutes, uniformément, l'emploi des transformations de Lorentz-Einstein dans la représentation des façons dont ces microéts modélisés 'ressentent' les 'influences' qui leur parviennent *via* des champs, en fonction de leur état de mouvement mécanique ; *(b)* si l'on peut justifier *jusqu'au bout* une métaphore en termes microphysiques, des conditions d'observation qui sous-tendent la relativité restreinte (observateurs *inertiels* et le *but* d'établir une loi de mouvement invariante aux changements de l'état de mouvement mécanique). Ce qui est loin de paraître évident.

Quant à une unification de la mécanique quantique *fondamentale*, à descriptions primordiales transférées, avec la relativité générale, on perçoit immédiatement toutes les réticences qui émanent du point de vue qui vient d'être esquissé, tout autant quant à la possibilité d'un tel concept d'unification, qu'en ce qui concerne *sa valeur en tant qu'un but*.

La mécanique quantique fondamentale est l'*unique* théorie mathématique transférée, primordialement probabiliste, construite à ce jour. Cette théorie a réussi à incorporer dans ses algorithmes des **racines de la connaissance** humaine, et à les dominer formellement. C'est pour cette raison – même si celle-ci reste implicite – que la mécanique quantique apparaît à tous comme une révolution majeure dans la pensée scientifique. Et c'est parce que la percée cognitive impliquée dans la mécanique quantique s'est réalisée directement en termes mathématiques, sans que l'on ait pu discerner d'une manière *intégrée* les contenus épistémologiques et méthodologiques si radicalement nouveaux captés dans les algorithmes mathématiques, que cette théorie apparaît comme cryptique: *la méthodologie de prise en compte explicite des contraintes qui émanent de la situation cognitive à supposer, et du but de chaque étape descriptionnelle, manque*. C'est un manque à la base qui empêche de doter les modélisations microphysiques, de fondations élaborées avec précision.

L'on a d'abord, presque miraculeusement, réussi à capter – sans les identifier explicitement – certaines racines de la conceptualisation humaine desquelles a poussé la mécanique quantique fondamentale. Mais ensuite la liaison avec ces racines s'est brisée et nous sommes remontés en vol non contrôlé vers le ciel conceptuel des modèles classiques, comme accrochés à des cerfs volants dont la ficelle a été coupée.

Le modèle standard a sans doute réussi dans une large mesure à compenser cette brisure d'une véritable continuité avec la mécanique quantique fondamentale. Il serait intéressant d'analyser les causes, les modalités et le degré de cette réussite.

Mais la théorie des cordes s'est énormément éloignée de la mécanique quantique fondamentale. On n'y perçoit plus aucune connexion conceptuelle avec cette théorie de base, seulement quelques signes mathématiques qui la rappellent. On y perçoit par contre une véritable fascination pour les démarches relativistes qui se sont forgées dans le domaine de la physique macroscopique, sous l'empire de situations observationnelles et de buts descriptionnels qui n'ont un sens *défini* que dans la physique macroscopique. La transposition acritique de ces démarches afin de représenter des entités et des interactions microscopiques, semble vouée à rester non pertinente. Car quelle chance a-t-on de tomber en accord avec les manières de se manifester à nous, de ces entités et interactions, lorsqu'on recherche des invariants formels sans se soucier à chaque pas de savoir pour quel *but* on recherche telle ou telle invariance et en cohérence avec *quelle* méthode de transposition à une représentation microscopique, de *quels* traits de *quelle* situation cognitive? Or en physique mathématique la non pertinence épistémologique et méthodologique se manifeste comme un rejet conceptuel d'une greffe d'un organe abstrait qui n'est pas compatible avec l'organisme du savoir constitué précédemment.

# Chapitre 5

# Le théorème de non localité de Bell dans la perspective créée par l'infra-mécanique quantique

La situation conceptuelle mise en évidence dans le chapitre **4** modifie la perception du problème de localité. Elle permet de préciser la signification de l'inégalité prouvée par Bell en faisant quelques remarques critiques concernant l'exposé. (Afin de miex suivre le texte qui suit il pourrait être utile de se remettre dans l'esprit le contenu du paragraphe *2.3.3*, en tant qu'une introduction approfondie).

## 5.1. Le théorème de Bell

### *5.1.1. La structure du théorème*

Comme il est bien connu, parmi les divers types de microétats décrits par la mécanique quantique, Einstein, Podolski et Rosen (EPR, [1935]) ont sélectionné un micro-état de deux micro-systèmes du type qui, à l'intérieur de l'infra-mécanique quantique, a été caractérisé dans *3.9.1*. Selon l'argument EPR, ce cas conduirait à conclure que (Bell [1964], l'abstract):

> «…quantum mechanics could not be a complete theory but should be supplemented by additional variables. These additional variables were to restore to the theory causality and locality. »

Or Bell ([1964], en continuation dans l'abstract) a affirmé avoir démontré que ceci n'est pas possible:

> « In this note that idea will be formulated mathematically and shown to be incompatible with the statistical predictions of quantum mechanics. It is the requirement of locality, or more precisely that the result of a measurement on one system be unaffected by operations on a distant system with which it has *interacted in the past*[82], that creates the essential difficulty. »

Le travail de Bell se place donc *à l'extérieur* de la mécanique quantique (et aussi de l'infra-mécanique quantique): l'on y examine la possibilité de 'compléter' le formalisme quantique par *une modélisation au niveau microphysique* à l'aide de paramètres additionnels cachés face au formalisme quantique, qui rende ce formalisme conforme à la condition de 'localité' exigée par la théorie de la Relativité d'Einstein: il s'agit donc d'un méta-problème de relation entre des théories différentes.

---

[82] Nos italiques. Elles soulignent ce qui suit. Nulle part dans le travail de Bell, ni dans les innombrables travaux sur le problème de localité qui ont été publiés depuis, on ne parle d'une *opération de génération* $G_{12}$ qui engendre le micro-état de deux micro-systèmes que l'on étudie. On ne parle que d' "interaction" et d' "entanglement" ("intrication"). On s'exprime comme si les micro-*systèmes* considérés n'étaient pas souvent *eux-mêmes* (pas seulement leurs micro-états) **engendrés** de toutes pièces par cette "interaction", ce qui module le sens du mot 'interaction' qui incite à présupposer la *préexistence pérennisée* des éléments interagissants, cependant que dans ce cas il s'agit d'intéoction d'*événements* (d'enregistrements observables), au sens des probabilités. L'adéquation *générale* du terme d' 'opération de génération de micro-état' – et *uniquement* de ce terme (ou d'un équivalent) – reste systématiquement occultée tout au cours de l'exposé de Bell, comme dans tout exposé concernant des microétats, *à ce jour même* (dans le paragraphe 'V. Generalization' on trouve une suite de périphrases qui remplacent le vide d'une définition du concept d' 'opération de génération de micro-état'). C'est *à ce point là* que l'importance conceptuelle et opérationnelle foncière et générale, pour la **possibilité** même des conceptualisations primordiales transférées, a été et continue d'être occultée à ce jour même, autant à l'intérieur de la mécanique quantique, que lorsqu'il s'agit de sa modélisation au niveau microphysique par des paramètres supplémentaires.

Bell a procédé de la façon suivante (très bien connue, mais que je rappelle pour autosuffisance de ce chapitre).

Il a d'abord exprimé mathématiquement, pour le micro-état particulier de deux micro-systèmes considéré, l'exigence de localité au sens d'Einstein transposée dans une modélisation microphysique à l'aide de paramètres additionnels face au formalisme quantique. La transposition est accomplie de la façon suivante. Bell a isolé les effets possible de l' "interaction passée" entre les deux micro-systèmes *S1* et *S2*, en les incorporant *tous* par définition à un paramètre caché unique dénoté $\lambda$ – et en *permettant* ces effets – mais tout en imposant par ailleurs aussi sa 'condition de localité' exprimée par l'exigence d'indépendance du résultat *B* d'une mesure sur *S2*, de la position de l'enregistreur *a* où s'inscrit le résultat *A* de la mesure opérée sur la paire *S1* de *S2*, et *vice versa* pour le résultat *A* et la position de l'enregistreur *b* du résultat *B* de la mesure sur *S2* (cf. les bien connues conditions (1) et (2) du travail de Bell[83]).

Ensuite Bell a démontré son inégalité qui exprime que, dans le cas considéré, sa transposition de l'exigence einsteinienne de localité, en termes d'une modélisation à paramètres cachés, n'est pas compatible avec les prédictions quantiques numériques correspondantes.

Or, dans le cas considéré, il est possible de trancher expérimentalement entre les deux représentations.

Bref, Bell a organisé, pour le cas considéré, une opposition décidable par l'expérience, entre *(a)* [les prédictions comportées par une modélisation du formalisme quantique au niveau microphysique et qui est supposée 'locale' au sens d'Einstein], et *(b)* [les prédictions quantiques].

Ici finit ce qui, dans le théorème de localité, est représenté mathématiquement.

A cette représentation mathématique Bell *juxtapose* l'affirmation du paragraphe suivant intitulé "Conclusion":

«In a theory in which parameters are added to quantum mechanics to determine the results of individual measurements, there must be a mechanism whereby the setting of one measuring device can influence the reading of another measurement, however remote. Moreover, the signal involved must propagate instantaneously so that such a theory could not be Lorentz invariant.»

Est-ce à dire que la corrélation statistique prédite pour le cas d'un microétat de deux microsystèmes, par une modélisation des descriptions transférées quantiques qui serait 'locale' au sens d'Einstein – si elle se vérifiait – entraînerait *nécessairement* l'existence d' 'influences' par des 'signaux' à 'propagation instantanée' ?

J'argumenterai que ceci n'est *pas* le cas.

Il ne s'agit pas ici de la conclusion de la preuve mathématique, mais de la conclusion globale tirée par Bell. Or à la lumière de l'infra-mécanique quantique *cette conclusion globale n'est pas recevable*.

### 5.1.2. La situation expérimentale actuelle

Les expériences accomplies par Alain Aspect ([1980], [1982]) ont établi déjà depuis plus de 25 ans et avec un grand degré de certitude, que les prédictions du formalisme quantique dans le cas considéré, se *vérifient*. Autrement dit, ces expériences ont établi que – dans le cas considéré – l'expression mathématique à l'aide de paramètres additionnels donnée par Bell à l'exigence de localité au sens d'Einstein, n'est *pas* réalisée dans le cas considéré.

---

[83] $A=A(a,\lambda)$, $B=B((b,\lambda)$, $P(a,b)=\int d\lambda \, \rho(\lambda) \, A(a,\lambda)B(b,\lambda)$.

Les expériences ultérieures de Weihs, Jennewein, Simon, Weinfurter, Zeilinger ([1998]), et actuellement les expériences de Salart, Baas, Branciard, Gisin & Zbinden [2008], montrent en outre la quasi 'instantanéité' des corrélations constatées.

On se trouve devant ce *fait*: les prévisions quantiques se vérifient, celle d'une modélisation microscopique 'locale' ne se vérifient pas.

### 5.1.3. Les réactions actuelles à la situation expérimentale

Et 'donc', dit-on, une modélisation des descriptions quantiques des microétats – au niveau microphysique et à l'aide de paramètres cachés face au formalisme quantique – qui soit 'locale' au sens d'Einstein, semblerait exclue[84].

Actuellement la plupart des physiciens se soumettent plus ou moins sereinement et passivement à cette opinion, même si dans leurs manières de dire l'on perçoive uniformément un certain étonnement.

Mais d'autre part cette soumission conceptuelle ne suspend, ni ne ralentit, la recherche d'une 'unification' de la mécanique quantique avec les relativités d'Einstein, où donc les comportements des micro-états soient contraints par les conditions relativistes, notamment celles de séparabilité et de causalité locale.

### 5.1.4. La situation conceptuelle révélée par l'infra-mécanique quantique

Les examens accomplis dans les paragraphes **3.7**, **3.8** et dans le chapitre **4** ont mis en évidence que sur la verticale des niveaux de nos conceptualisations d'entités physiques, il existe un *ordre* de reconstructibilité des *structures internes de nos représentations* en partant du niveau le plus bas de conceptualisation zéro et en s'avançant vers le haut de cette verticale. Cet ordre commence avec une structure de descriptions *transférées* absolument premières. Il conduit ensuite à la question de la structure d'une *modélisation microscopique* (désontologisée et relativisée) des descriptions transférées primordiales. Puis cet ordre *s'arrête* (à ce jour) dans une 'coupure' [quantique-classique] qui n'a été comblée d'une façon accomplie que très récemment (Mugur-Schächter [2002B] et [2006]) (cf. aussi **3.7**). Et il reprend au-dessus du hiatus qui entoure cette coupure, avec des conceptualisations modélisantes en termes d'*'objets' classiques*. Cet ordre doit en principe permettre d'acquérir progressivement une certaine 'compréhension' des relations entre les structures internes de toutes deux strates de conceptualisations successives, en s'avançant du 'bas' (le niveau zéro des descriptions de base transférées) vers le 'haut' (les niveaux modélisants en général plus complexe) sur la verticale des phases de nos conceptualisations. Mais il s'agit là de 'compréhension' en un sens génératif et tâtonnant, pas déductif et certain. Cette sorte de compréhension devrait néanmoins, en principe, permettre de parfaire de proche en proche la reconstruction d'une représentation pleinement accomplie de la relation entre les descriptions primordiales transférées des microétats, et d'autre part les concepts de 'localité', 'séparabilité', 'causalité', 'déterminisme', qui ont été élaborés d'abord sur le niveau des conceptualisations macroscopiques du réel physique[85].

---

[84] J'utilise le conditionnel parce que pendant un certain temps je m'étais attachée à construire un contre exemple à cette dernière affirmation. Et j'ai effectivement réussi à construire un modèle qui est *local* au sens d'Einstein et dans le même temps est aussi *compatible* avec les prévisions quantiques (Mugur-Schächter [1987]). Cela, par sa *seule possibilité*, indépendamment de sa vérité factuelle, nie la généralité de l'affirmation de la conclusion de Bell (Mugur-Schächter [1988]). En outre le fait que ce contre-exemple est en accord avec les prévisions quantiques a été *vérifié* par un calcul explicite de Bordley ([1989]). Mais dans ma perspective actuelle concernant le problème de localité, la question du degré de généralité *logique* de la conclusion de la preuve de Bell, a perdu tout intérêt. Car à la lumière de l'infra-mécanique quantique l'entier problème de localité tout simplement se dissout pour des raisons radicales d'ordre conceptuel, comme il ressort dans ce qui suit.

***Mais la démarche inverse n'est pas possible.***

Si on la tente, on s'enlise inévitablement dans des faux problèmes et des paradoxes de la même nature que ceux qui ont été discutés à fond dans **3.7**. Car l'ordre de constructibilité des structures internes des représentations à partir du niveau zéro de conceptualisation, comme tout ordre, comporte une asymétrie foncière entre les deux sens de circulation sur la verticale des niveaux de conceptualisation. Et cette asymétrie est telle qu'en général elle s'oppose aux transports de concepts incorporés à des conceptualisations macroscopiques, vers des conceptualisations microscopiques. (Bohr l'a assez dit. C'est d'ailleurs ce qui a fait barrage à une physique statistique fondée sur un modèle de 'mobile' au sens de Newton). En effet, dès qu'on tente un transport direct de concepts macroscopiques vers les représentations primordiales transférées, on se heurte sur le niveau d'arrivée ***à une absence de 'volume conceptuel' déjà aménagé où l'on puisse déposer les concepts transportés***. Car en suivant l'ordre de *constructibilité* de la structure interne des représentations à partir du niveau zéro de conceptualisation, les volumes conceptuels (désontologisés et relativisés) ne se laissent aménager qu'au fur et à mesure qu'émergent les concepts spécifiques du niveau nouvellement atteint par ce processus de construction là. Si néanmoins l'on tente le dépôt brutal sur le niveau de conceptualisation primordial transféré, d'un concept *spécifique* du niveau de conceptualisation classique, il s'engendre une sorte de structure chimérique où les modélisations classiques se mêlent aux descriptions primordiales transférées d'une façon non construite, comme dans les ombres projetées sur un mur par une lanterne magique, les images d'objets divers placés à des distances et dans des angles de vue différents, se superposent dans un même plan en perdant la définition des structures et des contours originels.

Cela crève maintenant les yeux, lorsqu'on reconsidère les descriptions primordiales transférées des microétats à la lumière des considérations du chapitre **4**, que les concepts de séparabilité et de causalité locale ne peuvent pas y être insérés. Ces concepts ont été conçus initialement concernant des 'mobiles' ('objets') macroscopiques, supposés préexistants et munis de propriétés qu'ils contiendraient *dans* eux-mêmes, des 'mobiles' observés d'une manière *directe*, à l'aide de *signaux lumineux*, par des observateurs humains munis de *référentiels* (mutuellement inertiels, ou pas). Ces concepts, en outre, sont nés sous la contrainte de consensus parmi de *tels* observateurs, un consensus imposé spécifiquement et exclusivement concernant la *loi de mouvement des 'mobiles'* considérés. Il serait pour le moins gratuit, sinon carrément déplacé et absurde, d'imposer des concepts ayant une telle genèse, à des descriptions primordiales transférées $D_M/G,me_G,V_M/=\{p(G,Xj)\}$ qui, à partir d'une factualité physique jamais encore conceptualisée, tout juste émergent sous la forme d'un ensemble de marques observables *éparses dans l'espace et le temps de l'observateur* ; des marques *transférées* sur des récepteurs d'appareils à la suite d'interactions produites par des concepteurs observateurs humains, oui, mais qui ne sont *pas perceptibles* par ces humains; des marques distribuées en *statistiques*; des marques dont l'appréhension est *vide de toute qualia incorporée* assignable isolément aux microétats 'représentés', qui sont *séparées de tout concept de propriété 'possédée' par ces microétats* et étrangères au concept classique de 'trajectoire'. Les 'descriptions' transférées sont encore foncièrement mélangées à la manière de les faire émerger, encore dépourvues même d'un support *connexe* d'espace-temps à contour définissable, elles sont encore si éparpillées qu'on a du mal à même arriver à leur associer un être et un nom et une relation spécifiable avec les microétats qu'elles sont censées impliquer.

---

[85] A l'intérieur de la méthode générale de conceptualisation relativisée accomplie par généralisation de la démarche développée dans l'infra-mécanique quantique, "la coupure quantique-classique" se trouve englobée dans un "passage" général et réglementé du point de vue constructif, des descriptions transférées aux modélisations classiques, et cela permet effectivement d'expliciter la 'compréhension' spécifiée.

*Aucune* des caractéristiques de la conceptualisation einsteinienne ne s'applique à des descriptions d'un type tellement débutant, naissant, inaccompli.

## 5.1.5. Remarque sur la "conclusion" de Bell

Toutefois on peut rétorquer que l'inégalité de Bell se réfère à une *modélisation* microscopique de descriptions transférées, pas directement à des descriptions transférées. Focalisons alors l'attention plus exactement sur une telle modélisation.

La preuve mathématique de l'inégalité de Bell n'est pas critiquable en tant que telle. Mais globalement l'exposé de Bell induit confusion, pour la raison suivante:

Ce que Bell appelle "conclusion" ne se rapporte pas à la preuve mathématique de l'inégalité, cela se rapporte exclusivement à la pertinence de l'application de la relativité d'Einstein, aux descriptions de microétats, transférées *ou modélisées*.

Il n'est pas inutile de souligner cette distinction car, à ce jour même, la communauté scientifique, plus ou moins explicitement, prend le travail de Bell sur la non-localité comme base de la croyance en la nécessité d'***unifier*** les descriptions des microétats, avec la relativité d'Einstein, quant ce serait exactement le contraire qu'il conviendrait d'en conclure.

Si l'on admet que les prévisions quantiques sont vraies expérimentalement et si l'on veut expliquer ce fait, alors rien*, mais strictement rien, ni en 1964 ni actuellement, n'impose pour autant avec nécessité logique une explication en termes d' 'influences' par des 'signaux' à 'propagation instantanée'[86]. Au *contraire*, la hiérarchie de la constructibilité des structures des représentations à partir du niveau de conceptualisation zéro, impose d'*exclure* une explication dans ces termes là. Car entre une modélisation microphysique à l'aide de paramètres cachés, des descriptions primordiales transférées impliquées dans la mécanique quantique, et un niveau de conceptualisation où l'affirmation ou la négation des qualifications einsteiniennes de séparabilité et de causalité locale soient douées d'une signification *définie*, il s'interpose toute une large zone de construction progressive des structures de conceptualisation.

La situation conceptuelle peut être précisée plus, de la façon suivante.
La preuve mathématique de l'inégalité de Bell se rapporte à une dichotomisation: *(a)* on peut modéliser les descriptions quantiques primordiales transférées, au niveau microphysique, à l'aide de paramètres cachés, en termes de causalité séparable et locale; *(b)* on ne peut pas faire cela.
Bell a réussi à créer cette dichotomisation et à la placer sur un seul plan conceptuel où l'on peut effectuer une comparaison.
Mais *l'explication* d'une éventuelle violation expérimentale de cette preuve mathématique dichotomique, elle, n'est *pas **définissable***. Elle mobilise virtuellement un possible factuel-conceptuel inconnu et infini. Ce possible inconnu ne peut pas être réduit à du dicible en termes logiques, il ne peut pas être inclus dans la conclusion déductivement établie d'une démarche déductive [87], [88]. Pourtant Bell, dans sa "conclusion", s'exprime comme si sa preuve mathématique *imposait par voie logique* un nouveau choix dichotomique: ou bien *(a)* les faits microphysiques obéissent à la modélisation 'locale' spécifiée dans la preuve mathématique; ou bien *(b) dans le cas contraire il doit y avoir 'un mécanisme d'influence' ou des 'signaux' à 'propagation instantanée'* (« …the signal involved must propagate instantaneously »). Quand en fait un tel choix, avec le langage et les *concepts* qui l'expriment, ne découle nullement de la preuve mathématique. Rien n'impose ni n'assure qu'une modélisation microscopique des descriptions transférées des microétats, doive assurer quelque pertinence des notions d''influence' ou de 'signal' tels que celles-ci se sont formées dans la physique macroscopique. Et même si c'était le cas, pour quelle raison la vitesse d'un 'signal' comporté par un phénomène microphysique devrait-elle être précisément celle de la "lumière", quand d'autre part bon nombre de microsystèmes lourds ne comportent même pas de charge électrique, donc n'impliquent rien d'électromagnétique ?

---

[86] Notons la contradiction dans les termes.

La "conclusion" de Bell ne *peut* pas être comprise, ni surtout admise, comme une conséquence logique de la preuve mathématique de l'inégalité. *Elle n'est admissible que comme* **une opposition à l'attitude d'Einstein**.

Pourtant, nous l'avons déjà souligné, la communauté scientifique, à ce jour même, reste encore enfoncée dans l'idée que la microphysique devrait être unifiée avec la relativité d'Einstein.

Cela montre que le contenu du travail de Bell a donné lieu à une mésinterprétation.

### 5.1.6. Une conjecture

Einstein a exigé que l'on 'complétât' la mécanique quantique à l'aide de paramètres cachés introduits dans une modélisation immédiate microphysique qui assure une causalité 'séparable' et 'locale'. Cette exigence montre qu'Einstein n'avait certainement pas imaginé qu'une telle modélisation pourrait impliquer une contradiction avec certaines prévisions observables de la mécanique quantique.

Quant à Bell, avait-il réalisé que l'ordre de constructibilité explicite des conceptualisations permet le but d'expliquer' ou de 'justifier' la condition macroscopique de localité *sur la base* des description quantiques, de leur modélisation immédiate au niveau microscopique, et de la définition de procédés descriptionnels de *passage* des descriptions primordiales transférées impliquées dans la mécanique quantique, à des descriptions classiques en termes d' "objets"? Cependant que la démarche inverse n'est pas possible? Etait-il averti que les concepts de séparabilité, de causalité, de localité, de déterminisme, qui fondent l'entière physique macroscopique, ne pouvaient être

---

construits qu'*à une grande distance conceptuelle des descriptions quantiques*, bien au-dessus de la coupure qui sépare la conceptualisation primordiale des micro-états, des conceptualisations classiques?

Je prends la liberté de formuler la conjecture que – comme nous tous à l'époque – Bell n'en était pas averti; qu'il ne percevait pas la distance conceptuelle qui vient d'être mentionnée; qu'il projetait les descriptions 'locales' macroscopiques et les descriptions 'locales' microscopiques sur un même plan, à l'horizon des connaissances d'alors.

Car je n'arrive pas à trouver une autre explication au fait que Bell ait juste accolé sa "conclusion", à sa preuve mathématique, exactement comme si *avec évidence* l'explication contenue dans sa "conclusion" imposait déductivement le nouveau choix dichotomique signalé plus haut (ou bien *(a)* les faits microphysiques obéissent à la modélisation 'locale' spécifiée dans la preuve mathématique; ou bien *(b)* dans le cas contraire il doit y avoir des 'signaux' à 'propagation instantanée').

Et il semble assez clair qu'*en plus* Bell pensait que l'expérience qu'il avait conçue allait *infirmer* la prévision quantique. Car c'est bien cela qui est suggéré par le fait que Bell ait exprimé l'explication contenue dans sa "conclusion" dans les termes même de la conceptualisation d'Einstein, d' 'influences' portées par des 'signaux', mais en utilisant ces termes *par absurde* en quelque sorte, car liés à une 'propagation instantanée': il voulait, tout en poussant jusqu'à leur limite les conséquences de l'attitude d'Einstein, mettre d'autre part en évidence un caractère *de la conceptualisation quantique* qui lui semblait *faux*, bien que par ailleurs cette conceptualisation quantique soit efficace "for all practical purpose" (expression qu'il a souvent employée en relation avec le formalisme quantique).

Ces traits des formulations de Bell, mis ensemble, me paraissent presque démontrer qu'il s'apprêtait à assister à une mise à mort des principes incorporés aux descriptions quantiques, et 'donc', dans sa vue dichotomique, il s'attendait à un triomphe de la relativité d'Einstein.

Mais dans le même temps il voulait laisser ouverte la possibilité contraire aussi, et s'assurer que dans le cas, qui lui semblait improbable, où celle-ci néanmoins s'avérerait, il allait encore avoir eu raison.

Il y avait dans l'esprit de Bell, une coalescence de buts distincts et cela a engendré une ambiguïté à la fois psychologique et stratégique de sa manière d'exposer.

Évidemment, ceci restera à jamais une simple conjecture.
Mais les remarques critiques formulées plus haut n'en dépendent nullement.

Enfin, quoi qu'il en soit, d'un point de vue heuristique le rôle joué par le travail de Bell a été proprement énorme.

### 5.1.7. *Une autre approche possible*

Revenons maintenant au point **3.4**. Nous y avons explicité des définitions qui sont cohérentes avec les manières de parler qui sont courantes dans la microphysique. Mais rien n'impose précisément ces définitions là. On peut en tenter d'autres, avec une autre manière de les mettre en relation avec la définition d'une opération de génération composée *Gn*. Par exemple, on peut explorer les répercussions qu'auraient un ensemble de définitions selon lequel "un microétat de un microsystème" serait toujours, par définion, *ce* qui est produit par *une* opération de génération, cependant qu'un tel microétat ne peut produire, via *une* succession *[G.Mes(X)]*, qu'*une* seule marque (effet) observable si son opération de génération est "simple", mais produit au maximum *n* marques (effets) observables si son opération de génération est "composée", i.e. s'il s'agit d'une succession *[Gn.Mes(X)]*. Cela modifierait les façons de penser et de parler concernant "un arbre probabilité à opération de génération composée" (cf. **3.9**).

  *Quelles seraient les conséquences sur ce qu'on appelle 'le problème de non-localité' ?*

  Cette simple question attire l'attention sur la *relativité* du tout qui se constitue dans la pensée en ce qui concerne un ensemble donné d'opérations et observations, face aux concepts-et-langage que l'on associe à cet ensemble[89].

## 5.2. Conclusion

  Ainsi la représentation qualitative des microétats construite ici confirme les réticences exprimées dès 1979 face au 'problème de localité' soulevé par Bell (cf. *2.3.3*).

  Depuis presque 30 années accomplies l'entière communauté des physiciens reste hypnotisée par le dogme *a priori* selon lequel la mécanique quantique devrait être relativiste, et cela conduit à une véritable cécité face à la hiérarchie génétiques des niveaux de constructibilité des conceptualisations, et face aux conséquences de cette hiérarchie.

  On raisonne sur une *projection* illusoire sur un plan de conceptualisation unique qui est factice et où se superposent de façon arbitraire, non construite, *dépourvue de cohérence interne*, trois niveaux distincts de représentation d'entités physiques:
  - le niveau *primordial* de représentation transférée des microétats impliqué dans la mécanique quantique fondamentale;
  - le niveau *second* d'une modélisation microphysique immédiate des descriptions transférées de microétats, à l'aide de paramètres supplémentaires, cachés face à la représentation transférée primordiale des microétats;
  - le niveau *classique* de modélisation du réel physique en termes d' "objets" macroscopiques directement perceptibles par l'homme, qui préexisteraient à toute perception et seraient investis de "propriétés" qui préexisteraient elles aussi indépendamment de tout action cognitive réflexe ou délibérée[90].

  Les spécificités du niveau classique de modélisation dominent irrépressiblement les élaborations 'descendantes' des représentations placées sur les deux autres niveaux sous-jacents mentionnés. Car, dans la chronologie *historique* de nos constructions de connaissances, le niveau classique de conceptualisation macroscopique est antérieur, et il infuse aussi bien dans les actions cognitives de première construction des niveaux sous-jacents, que dans les représentations des données observables que ces actions cognitives produisent. Mais à terme , cette domination est vouée à être effacée – à être *niée sur le plan conceptuel* – parce que la *re*construction de nos modélisations d'entités physiques en partant des descriptions de base transférées des microétats accomplies sur le niveau zéro de nos conceptualisations d'entités physiques et en s'avançant vers le haut de la verticale de nos conceptualisations d'entités physiques, incorporent des contraintes factuelles et épistémiques relativisantes et désontologisantes qui étaient restées absentes des modélisations accomplies précédemment, nourries exclusivement par notre pensée classique. D'autant plus que les concepts-et-langage associés aux descriptions transférées des microétats cachent encore des flous (cf. *5.1.6*).

---

[89] En mécanique quantique fondamentale l'opération de génération d'un microétat est dépourvue de représentation mathématique, cependant que les deux concepts envisagés ici – de "un microétat de *n* microsystèmes" (de 3.4) ou celui de "un microétat à opération de génération composée *Gn*" suggéré ici, correspondent tous les deux à une circonstance qui, selon le formalisme quantique, n'introduit qu'une seule fonction d'état. Cela renforce la pertinence de la question formulée plus haut.

[90] En physique ce niveau "classique" possède une structure fine qui distingue entre la structuration newtonienne des formes *a priori* d'espace et le temps, celle de la relativité restreinte d'Einstein, et celle de la relativité générale d'Einstein.

# Conclusion sur la  troisième partie

Lorsqu'on a *vu*, comme au cours de l'humble cheminement développé ici afin de construire l'infra-mécanique quantique, à quel point chaque pas sur un bref trajet de conceptualisation épistémologique-méthodologique d'un domaine particulier du réel physique, dépend des contraintes qui émanent de la situation cognitive impliquée, du but de la représentation recherchée, des *décisions* méthodologiques qui s'imposent face à *cette* situation cognitive et *ce* but, spécifiquement – et des décisions qu'on ne perçoit que successivement lorsqu'on se trouve nez à nez avec la nécessité de l'une d'elles lors d'une impasse dans la progression – ce n'est qu'alors qu'on réalise vraiment l'importance des aspects épistémologiques et des aspects de *méthode*.

*La microphysique actuelle souffre d'une crise d'absence d'une méthode* ***épistémologique***. *Elle souffre plus encore de l'ignorance de l'existence de cette crise.*

L'entière physique moderne est viscéralement travaillée par la nécessité d'engendrer à partir d'elle-même une méthode épistémologique élaborée en concordance avec son propre niveau de détail, de rigueur et de synthèse. Une méthode qui, dans chaque phase descriptionnelle du développement d'une représentation, puisse dicter explicitement les règles à respecter afin d'achever cette phase là d'une manière consistante avec tous les présupposés qui y interviennent.

Il ne me semble pas exclu qu'une 'unification' de la physique actuelle ne puisse s'accomplir *que* dans un sens très différent de celui qu'on poursuit depuis des dizaines d'années.

À savoir, par une *méthode* unitaire de construction des représentations quelconques de faits physiques.

Une méthode qui, lors de chaque action descriptionnelle, distingue clairement tous les éléments auxquels cette action est relative, tous les éléments qui la conditionnent localement; et qui sépare nettement les uns des autres les actes successifs de conceptualisation, ainsi que, globalement, les niveaux de conceptualisation, selon une hiérarchie dictée par la générativité des processus successifs de conceptualisation.

C'est dans la méthode de décrire que pourrait s'accomplir une unification. Dans le cadre d'une méthode descriptionnelle générale, chaque description particulière pourrait conserver pleinement les spécificités de son propre contenu, tout en se trouvant en relation explicitement définie avec les autres descriptions, *via* ses caractères de forme descriptionnelle.

*Quatrième partie*

*De
l'infra-mécanique quantique
vers
une méthode générale
de conceptualisation relativisée*

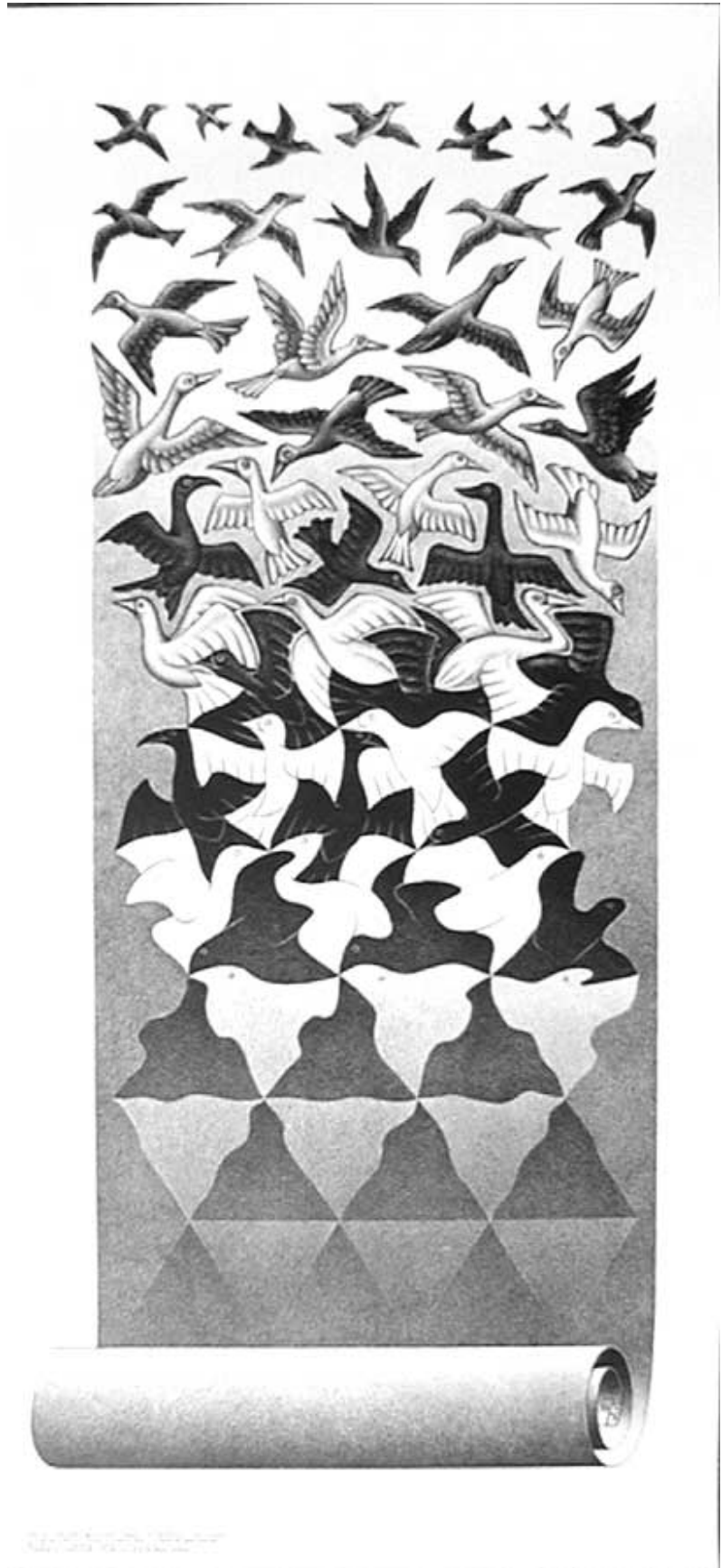

**ESCHER  LIBERATION**

# *Mot d'introduction à la quatrième partie*

La construction de l'infra-mécanique quantique a été entreprise dans un but double. Le premier but, tourné vers le futur, était de se doter d'une structure de référence, construite explicitement et indépendamment, qui permette d'étudier l'ensemble de tous les problèmes d'interprétation du formalisme quantique par comparaison avec cette structure unique, sous le *contrôle* de cohérence assuré par cette unicité. Le deuxième but était tourné vers le passé: doter *a posteriori* la méthode générale de conceptualisation relativisée – déjà construite – d'une fondation plus explicite, plus détaillée et approfondie que toutes celles construites dans les exposés précédents de cette méthode. Car la méthode s'impose véritablement lorsqu'on comprend comment elle émane de la mécanique quantique fondamentale. Ce deuxième but – purement épistémologique et méthodologique – pourrait s'avérer beaucoup plus fertile et d'importance beaucoup plus générale que celui d'utiliser l'infra-mécanique quantique pour des élucidations dans le cadre strict de la physique.

En ce point de notre exposé ces deux buts se séparent. Le premier sera poursuivi dans un autre travail. Le deuxième sera mené à son terme ici.

En effet le bref chapitre qui suit mettra en évidence la direction de pensée qui conduit de l'infra-mécanique quantique à la méthode générale de conceptualisation relativisée[91]. Il apparaîtra ainsi clairement sur quelle base il a été affirmé dans l'introduction générale que les descriptions de microétats, lorsqu'elles sont purifiées de leurs aspects liés *spécifiquement* aux microétats, contiennent les principes d'une radicale révolution de l'épistémologie.

---

[91] Le contenu de la quatrième partie de ce livre se retrouve dans Mugur-Schächter [2006], mais distribué dans d'autres contextes. Comme il a été déjà souligné dans l'introduction générale, cette redite est due au fait que, bien que le livre présent et le livre cité qui contient l'exposé de la méthode générale de conceptualisation relativisée, soient intimement reliés, il a fallu néanmoins chercher à rendre autosuffisant chacun de ces deux écrits, qui s'adressent à des publics différents.

# Chapitre 6

# Les descriptions de microétats
# comme prémisses d'une révolution de l'épistémologie

« When a thing is funny
search it carefully for a hidden truth. »

George Bernard Shaw

## 6.1. Des 'phénomènes'

### 6.1.1. Au sens de l'épistémologie philosophique classique

L'épistémologie philosophique dominante prolonge celle de Kant, principalement par les développements de la phénoménologie de Husserl. Selon l'approche kantienne-husserlienne un 'phénomène' est un événement qui ne peut exister qu'à l'intérieur d'une conscience individuelle où il émerge involontairement *via* des actions réflexes du corps qui loge la conscience ; cet événement, qui originellement est donc essentiellement subjectif, peut être *'légalisé'* après coup de façon à réaliser à son égard de la communicabilité et, éventuellement, du consensus inter-subjectif.

Husserl et ses élèves se sont lancés dans l'analyse de la structure psycho-bio-physique des processus d'émergence d'un phénomène. Mais ils l'ont fait d'un point de vue psychologique qui n'atteint pas la structure biophyslogique profonde du processus de cette émergence, qui est réflexe. Celle-ci est laissée en blanc.

### 6.1.2. Au sens des sciences cognitives

Dans le cadre des sciences cognitives les biologistes – souvent en collaboration avec des philosophes – prolongent la définition d'un phénomène au sens de l'épistémologie philosophique en entreprenant l'étude, dans toute sa profondeur, de la structure physiologique des processus d'émergence qui entrent en jeu (Changeux [1983] ,Edelman [1992], Damasio [1999], Berthoz [1997], Berthoz et Petit [2006], et d'autres). Les études des 'cognitivistes' mettent en relation avec précision et détail, d'une part tel ou tel impact produit par du réel extérieur au corps des hommes, sur un organe sensoriel d'un corps humain (ou d'un corps d'un animal différent de l'homme), avec d'autre part la séquence de processus biologiques *réflexes* que cet impact déclenche à l'intérieur du corps considéré et qui finalement produit un état psychique conscient.

Ces recherches dotent un phénomène au sens de l'épistémologie philosophique classique, de toute une épaisseur de connu concernant les processus biologiques d'émergence de ce phénomène. Cette épaisseur nouvelle est placée dans le temps entièrement après la production par un stimulus externe, d'un impact sur un organe sensoriel, cet impact lui-même étant traité comme une donnée première. Dans l'espace cette épaisseur est logée entièrement à l'intérieur du corps (ceci s'associe au fait que l'on admet que le déclenchement de l'état conscient comporté par un phénomène est entièrement biologique et réflexe). Enfin, l'on admet que l'état conscient comporté par un phénomène constitue son point final.

Bref, la découpe de l'entité-objet-d'étude qui correspond au concept de phénomène au sens des sciences cognitives est essentiellement biologique, les éléments psychiques conscients qui y sont liés y tiennent une place périphérique de témoignage d'effets ressentis *passivement*.

### 6.1.3. Au sens de l'infra-mécanique quantique

Le processus d'élaboration de l'infra-mécanique quantique a mis en évidence que certains processus de construction de connaissances peuvent obliger à concevoir consciemment et à opérer *délibérément*, des changements physiques sur du réel physique *extérieur* au corps de l'observateur-concepteur, *avant* qu'un impact physique ne se produise sur les appareils sensoriels biologiques du corps de l'observateur-concepteur et *afin* de le produire sous une forme préconstruite selon un *but*. Il révèle que ces changements peuvent être conçus de manière à forger pas à pas le processus de déclenchement d'un phénomène conformément à un projet dont chaque étape comporte un *but local* qui porte à tour de rôle sur: *(a)* ce qu'on examine ; *(b)* la façon d'examiner ; *(c)* l'effet observable produit par l'interaction entre ce qu'on examine et ce à l'aide de quoi on examine, qui est distingué foncièrement de ce qu'on examine ; *(d)* l'impacte sensoriel biologique qui déclenche les processus réflexes de l'intérieur du corps de l'observateur-concepteur et produit finalement l'état conscient que comporte le phénomène au sens subjectif du mot ; *(e)* et *après* l'émergence de cet état conscient, dans la même foulée, la signification légalisée assignée à ce qu'on observe de façon consciente, par des codages convenables établis à l'avance et qui transforment ce qu'on observe en phénomène communicable *et* apte à engendrer du consensus.

Les phénomènes au sens de l'infra-mécanique quantique sont délibérés et, comme Athèna qui est née de la tête de Zeus habillée de pied en cap, ils naissent décrits, communicables, et légalisés de manière à pouvoir assurer du consensus intersubjectif.

Les recherches accomplies dans le cadre des sciences cognitives associent donc une épaisseur nouvelle aux phénomènes au sens psycho-philosophique, mais une épaisseur réflexe et interne au corps, qui, entièrement, suit dans le temps un impact sur un organe sensoriel, et qui finit avec l'état conscient déclenché. L'infra-mécanique quantique ajoute d'autres épaisseurs encore, créées délibérément et en dehors du corps, partiellement placées dans le temps *avant* l'impact sensoriel qui déclenche le phénomène conscient et afin de régler ce phénomène conscient sur des buts explicites, et partiellement *après* l'émergence du phénomène conscient, afin de le rendre communicable et capable d'engendrer du consensus.

Ceci est une grande nouveauté épistémologique. Non pas biologique, ni psychologique, mais proprement épistémologique.

*Le concept de phénomène au sens de l'infra-mécanique quantique relie les sciences cognitives à une épistémologie fondée dans la physique moderne.*

Cela ouvre la voie vers une certaine unification entre philosophie, biologie, et physique.

Cependant que l'épistémologie philosophique – tout autant que l'épistémologie au sens des sciences cognitives – au contraire, occultent l'entière structure fine d'un processus de construction délibérée d'un phénomène au sens de l'infra-mécanique quantique.

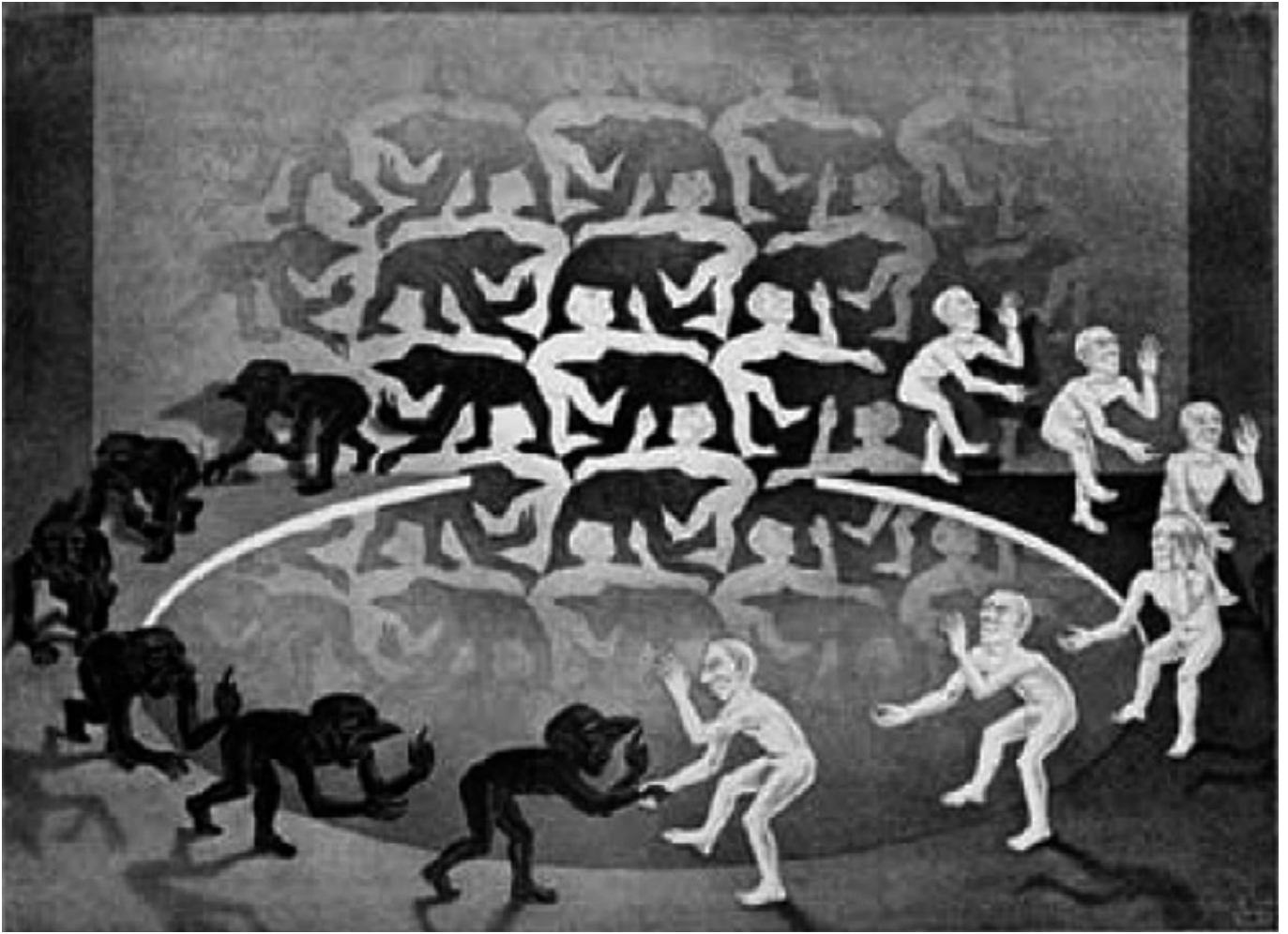

**ESCHER**
**ENCOUNTER**

## 6.2. Phénomènes et descriptions: l'idée d'un canon descriptionnel général

Toute transposition d'un phénomène en termes communicables équivaut en essence à la construction d'une *description*: rien d'autre que des descriptions ne peut être connu d'une manière intersubjective non restreinte. Ni des entités factuelles extérieures à toute conscience, ni des phénomènes non décrits, ne peuvent être connus d'une manière intersubjective non restreinte. Ce qui n'est pas 'décrit' peut être ressenti en un certain sens, mais ne peut pas être communiqué à toute distance et sans limitations, dans tous les détails et sans ambiguïtés (pas juste montré du doigt ou dessiné, dansé, crié ou gémi, etc.).

Cette spécification est loin d'être triviale. Elle focalise l'attention sur l'importance primordiale du contenu et de l'agencement d'une description quelconque et notamment sur le degré et la qualité de la communicabilité que ce contenu et agencement assurent. Et tout à coup, l'on est frappé par l'intérêt que présenterait un moule descriptionnel *normé*, une forme *canonique* d'une description quelconque: une fois un tel canon établi, toutes les procédures de description pourraient être qualifiées face à lui, comparées, comprises, à l'intérieur d'un cadre commun où une certaine unité entre les spécificités de tel ou tel mode particulier de décrire serait aménagée à l'avance.

Mais selon quels critères pourrait-on définir un canon descriptionnel général?

Il est remarquable que, par une certaine voie esquissée ci-dessous, la question formulée ramène l'attention sur la forme descriptionnelle $D_M/G,me_G,V_M/$ de la description d'un microétat.

### 6.2.1. Les sources de l'apparente singularité de la forme $D_M/G,me_G,V_M/$

La forme $D_M/G,me_G,V_M/$ exprime une façon d'accomplir une description qui est créative au maximum et déployée au maximum: chaque étape, chaque élément d'étape, est entièrement à créer, et à créer à son tour, séparément. A la faveur de ce degré de créativité et de déploiement extrêmes, toutes les relativités impliquées agissent en pleine lumière, sous nos yeux. Au premier abord, ce degré de créativité et ce déploiement extrêmes, avec la parfaite visibilité de toutes les relativités descriptionnelles impliquées, paraissent vraiment très singuliers. Les langages naturels avec leurs grammaires nous ont profondément conditionnés à supposer plus ou moins explicitement que ce qui, dans les descriptions, joue le rôle d'entité-objet-de-description, préexiste aux descriptions en tant qu''objets' tout court, qualifiés à l'avance par des 'propriétés' qu'ils 'possèderaient' à l'état déjà actuel et indépendamment de tout examen. Ces suppositions n'ont jamais soulevé des difficultés avant d'avoir été confrontées au but particulier de décrire des microétats. Pas même la description préquantique de micro-*systèmes* (électrons, protons, neutrons, etc.) donnée dans la physique atomique n'a soulevé des difficultés très spécifiques.

> *A ce jour même, la question de savoir comment l'entité-objet d'une description est introduite, est entièrement occultée en tant que question explicite et générale*, non seulement dans la pensée courante, mais aussi dans les sciences, et même dans les sciences les plus modernes.

La théorie de la relativité d'Einstein, par exemple, ignore la question. Nonobstant ses analyse révolutionnaires des mesures de distances spatiales et de durées, cette théorie, parce qu'elle a été conçue initialement au niveau macroscopique, travaille partout – *même dans ses extensions au microscopique* – directement avec de 'objets' préexistants définis par leurs 'propriétés', comme dans la logique classique. La relativité d'Einstein est une théorie presque classique.

Quant à la mécanique quantique elle même (qui selon la reconstruction qualitative exposée ici est la source de la révélation du rôle central que peut jouer l'opération de génération de l'entité-objet-de-description en tant qu'opération physique délibérée et indépendante de toute qualification), la question de la génération des entités-objet-d'étude n'y est mentionnée que dans un contexte qui introduit une confusion. A savoir, en relation avec *le concept de 'préparation d'état' qui intervient dans la définition d'un acte de MESURE à accomplir sur le microétat à étudier*. Ce concept présuppose que le microétat *me_G* à étudier est *déjà disponible en tant qu'entité-objet de la qualification à accomplir* ; mais afin d'être *qualifié* ce microétat doit être *changé* d'une manière *telle* qu'elle permette de parler d'une 'mesure' d'une grandeur mécanique bien définie, produisant des effets observables que l'on puisse coder en termes de valeurs de cette grandeur mécanique bien définie. C'est ce processus bien particulier appropié du microétat-objet-d'étude, supposé déjà existant, lors d'un acte de mesure accompli sur lui, qui est le désigné de l'expression 'préparation d'état' (préparation du microétat en vue de mesure). Dans cette circonstance-là il ne s'agit donc nullement (en général) de la génération de l'entité-objet à décrire – le microétat *me_G* – mais de l'opération de qualification de cette entité. Pourtant, lorsqu'on prononce l'expression nouvelle 'opération de *génération* de microétat', la force d'attraction qui, pour un physicien, émane du sens de *l'autre* expression de '*préparation* d'état', phagocyte le sens de l'expression nouvelle[92]. Le fait qu'il soit clair qu'afin de qualifier un microétat il faut d'abord qu'il soit là 'fixé' en tant qu'objet-détude, disponible pour l'étude, nommable et reproductible, n'y change rien. Telle est la force que les habitudes de langage ont sur la pensée.

Quant aux opérations de qualification, selon la pensée classique telle qu'elle est reflétée par les grammaires et par la logique, tout le processus qui d'abord crée un qualificateur et ensuite crée les qualifications correspondantes est rétréci en un seul acte statique, presque passif, de simple détection sur une entité-objet préexistante, d'une propriété préexistante. C'est précisément ce rétrécissement qui a fait contraste avec l'analyse d'Einstein des mesures de longueur et de durée. Le scandale soulevé par la relativité restreinte a consisté dans la prise de conscience (1905-1925) du fait que les qualifications d'espace et de temps se *construisent* par les processus physiques de mesure, et que les résultats des processus de mesure d'espace et de temps comportent des *relativités* à ces processus de construction. Cette prise de conscience a clairement influencé plus tard la démarche active et relativisante adoptée pour construire des qualifications de microétats. Dans l'élaboration des processus de mesure sur des microétats, cette prise de conscience déclenchée par la relativité restreinte s'est prolongée, généralisée et précisée. A l'opposé de ce qui s'est passé concernant l'opération de génération d'une entité-objet – qui a été occultée – le schéma d'un processus actif et relativisant de qualification d'un microétat par des mesures qui créent la qualification obtenue, s'est royalement installé vers 1935 dans la nouvelle pensée scientifique et philosophique officielle. Néanmoins aujourd'hui encore ce schéma continue de surprendre. On a beaucoup de mal à se débarrasser de la contraction classique d'un processus de qualification en une simple détection d'une propriété posée comme préexistante, absolue et 'possédée' par l'entité-objet en état *actuel*, indépendamment de tout acte cognitif opéré sur elle. Ce sont toujours la grammaire et la logique classique, avec leurs objets et prédicats hypostasiés, qui mènent la danse dans la pensée courante.

Bref, le schéma descriptionnel $D_M/G,me_G,V_M/$ qui vient d'être explicité pour le cas des microétats paraît tellement singulier parce que l'étape de génération de l'entité-objet reste quasi entièrement méconnue à ce jour, tandis que la structure de l'étape de qualification et les conséquences de cette structure, bien que largement discutées, ne sont pas encore vraiment assimilées.

### *6.2.2. L'universalité de la forme descriptionnelle $D_M/G,me_G,V_M/$*

Dès qu'on a clairement perçu la situation cognitive à laquelle est liée la forme descriptionnelle $D_M/G,me_G,V_M/$ et qu'on l'a assimilée à fond, il émerge une tendance vers une certaine inversion. On subit une variation comparable à celle qui fait apparaître certains dessins d'un cube tantôt comme concaves et tantôt comme convexes: ce qui au premier abord frappait comme foncièrement nouveau et très singulier, tout à coup apparaît au contraire comme doué d'une certaine sorte de normalité, donc d'universalité. Il saute aux yeux que la phase strictement première, primordiale, de toute chaîne de descriptions – qui nécessairement existe toujours – comporte souvent tout à fait explicitement des caractères du même type que ceux que l'on peut identifier dans le mode de conceptualisation des microétats. Pas toujours tous ces caractères à la fois, ni toujours séparés l'un de l'autre. Mais on reconnaît clairement la présence d'un équivalent de chacune des phases de la forme descriptionnelle $D_M/G,me_G,V_M/$, même s'il est plus ou moins implicite, ou en état de coalescence avec l'équivalent d'une autre phase. J'indiquerai maintenant cette présence sans aucune prétention de rigueur, comme si j'indiquais du doigt un paysage.

---

[92] J'ai pu constater une tendance très forte à simplement ne *pas* enregistrer la distinction entre 'préparation d'état pour mesure' et 'génération du microétat à étudier''.

Pensons par exemple à un détective qui doit susciter un début de chaîne de connaissances concernant un cas qu'il veut résoudre. Que fait-il alors? D'habitude il focalise son attention sur tel ou tel endroit convenable du réel physique et il en extrait un fragment (il découpe un morceau de tissus, décolle du sang coagulé, suit des gens, prend des photos, etc.). Il opère ainsi certaines 'générations d'entités-objet'. Il s'agit d'actes de génération d'entités-objet qui sont bien moins radicalement créatifs qu'un acte de génération d'un microétat. Néanmoins ce sont bien des actes de génération d'entités-objet en vue d'une description à accomplir qui n'est que future. Des actes qui ne sont pas identifiables avec les actions qui, à proprement parler, établiront des qualifications: celles-là vont suivre, elles seront à accomplir séparément. Quelquefois il peut même arriver que le détective crée une entité-objet en forgeant de toutes pièces une situation-test qui implique les suspects. Les réactions de ceux-ci peuvent se produire bien plus tard, séparément et tout à fait indépendamment de l'acte de génération de la situation-test: elles agiront comme des processus de 'mesure' subséquents, aménagés à l'avance et qui produiront des qualifications planifiées.

On peut également penser à un prélèvement de tissu pour analyses médicales ultérieures, ou aux échantillons arrachés par un robot lunaire qui devront revenir sur terre afin d'être examinés, ou à l'effort par lequel tout élève ou étudiant s'engendre lui-même comme objet futur d'examen par des 'examinateurs', ou à des élections (politiques, législatives, municipales) qui engendrent certaines personnes en tant qu' 'élus' que l'on qualifiera par la suite selon leurs actions, etc., etc..

Le langage courant ne favorise pas la perception des paires *[(opération de génération d'une entité-objet), (opération de qualification de l'entité-objet créée)]*. Il les occulte. Ce sont des cas trop complexes et trop rares pour que des moules de dire-et-penser publics se soient moulés sur leur structure. Le langage courant ne favorise que la vue fictive, statique, d'entités-objet qui préexistent, disponibles pour être qualifiées, et de prédicats qui eux aussi préexistent, disponibles pour être mis à l'œuvre: c'est simple, économique, et cela se prête à toutes les combinaisons, comme les pièces d'un lego.

Pourtant, dans tous les cas mentionnés plus haut l'observateur-concepteur crée une entité-objet qui ne préexistait pas, même si moins radicalement que dans le cas des microétats. Souvent il la crée tout à fait indépendamment des opérations de qualification, qui ne sont réalisées qu'ensuite. Et dans certains cas, l'opération conçue afin de qualifier l'entité-objet selon les buts descriptionnels voulus, peut changer cette entité si radicalement que, si plusieurs examens sont nécessaires, il faut produire plusieurs exemplaires de l'entité-objet, comme dans le cas des microétats (pensons aux analyses médicales). En outre, toujours comme dans le cas des microétats, les qualifications qui émergent à la fin sont marquées d'une manière indélébile d'une *triple* relativité: une relativité au mode de génération de l'entité-objet (ce mode peut tout simplement exclure certains examens, ou en favoriser d'autres, et en tout cas c'est lui qui produit le fragment de matière première pour extraction de connaissances, donc les connaissances obtenues en dépendront) ; une relativité à l'entité-objet elle-même (c'est à elle que sont directement reliées les  manifestations qualifiantes observables) ; et une relativité au modes d'examen de cette entité. Ainsi la stratégie de conceptualisation explicitée dans le cas des microétats apparaît maintenant comme ayant incorporé une certaine *universalité*.

La description des microétats introduit une instance particulière des traits universels de la toute première phase de tout processus par lequel l'homme extrait du réel où il est plongé et auquel il appartient, un début de chaîne de connaissances.

C'est en cela que consiste l'universalité de la science des microétats. Elle n'émane nullement du fait que tout système matériel est constitué de microsystèmes, comme on l'affirme souvent. C'est le contenu universel qui tient à la situation cognitive caractéristique de la phase absolument première d'une chaîne de conceptualisation qui est la source de l'impression d'*essentiel* que produit la mécanique quantique. Car cette toute première phase de conceptualisation existe nécessairement à la base de toute chaîne de conceptualisation.

Aucune autre théorie d'un domaine du réel, pas même la relativité d'Einstein, n'a capté des connexions cognitives premières et universelles entre la pensée de l'homme et le réel physique, aussi complexes que celles enfermées dans les algorithmes quantiques. Mais là ces connexions restent cachées.

Tandis que dans l'infra-mécanique quantique elles s'étalent sous les yeux de tous. C'est pour cette raison que l'infra-mécanique quantique offre les bases d'une révolution majeure de l'épistémologie, autant de l'épistémologie philosophique classique que de celle qui s'appuie sur les sciences cognitives.

### 6.2.3. Vers une généralisation de la structure épistémologique révélée par l'infra-mécanique quantique

L'épistémologie philosophique doit se reconstruire en se fondant sur une généralisation attentive de l'infra-mécanique quantique, au lieu de s'appuyer sur la mécanique newtonienne. Et l'épistémologie au sens des sciences cognitives doit se reconstruire elle aussi en se connectant aux phénomènes liés aux descriptions primordiales, transférées. Les deux sortes d'épistémologies doivent développer en accord mutuel une représentation du *passage* de la strate primordiale de conceptualisation transférée à la strate de conceptualisation classique[93]. Il faut élaborer aussi les connexions qui, dans la 'coupure' entre descriptions transférées et descriptions classiques, construisent les tissus qui *lient* ces deux strates de la conceptualisation.

### 7.2.4. Le concept d'un canon descriptionnel général

Ce qui convient comme forme descriptionnelle canonique utilisable en tant qu'un étalon général, est précisément le moule vide le plus complet imaginable, capable d'offrir à *toute* étape possible d'un processus de description donné mais *absolument quelconque*, une location disponible et spécifique, une case réservée. Sous l'influence du symbole $D_M/G, me_G, V_M/$ d'une description de microétat, je vois ce moule avec des cases déployées à l'horizontale, une case pour chaque étape distincte possible au cours d'un processus de description – une case de génération d'entité-objet, une case d'entité-objet, une case de grille de qualification –, chaque case offrant par sa *profondeur* verticale une place maximale au degré de créativité. Dans tel ou tel cas donné, certaines cases du moule général canonique de référence pourront rester inutilisées, ou partiellement vides de créativité. Mais on le saura, car la comparaison avec la forme canonique y montrera un vide étiqueté, ou un degré d'absence de créativité (de passivité) qui pourra être évalué. Ainsi les conséquences du vide en question pourront être clairement explicitées. Par exemple imaginons que je dise « je considère ce que je perçois devant mes yeux et c'est une surface rouge ». Par référence au moule 'extrémal' tiré du processus de description de microétats il apparaît qu'en ce cas les deux actions descriptionnelles, canoniquement distinctes, celle de génération de l'entité-objet et celle de génération des qualifications de celle-ci, ont subi une coalescence dans l'acte unique « je considère ce que je perçois devant mes yeux » qui, à la fois, délimite l'entité-objet et accomplit l'action de qualification. Dans ce cas, la location qui, dans le moule canonique, est réservée à l'étape initiale de génération *indépendante* de l'entité-objet, reste donc entièrement vide. Ou bien imaginons l'assertion «j'ai cueilli cette fleur, je la regarde et je la trouve très belle». Une comparaison avec le moule canonique met en évidence que dans ce cas l'entité-objet (la fleur) est introduite par une action de génération (cueillir) qui n'est que très partiellement créative, cependant que (dans l'immédiat) l'examen ne change que très peu l'entité-objet introduite de cette façon. Dans ce cas, toutes les cases du moule canonique sont donc peuplées. Mais la case d'une action de génération de l'entité-objet, et celle d'un examen de qualification sont presque vides de créativité. Il s'ensuit qu'un traitement classique (qui suppose la préexistence de l'entité-objet et son invariance face à l'opération de qualification, donc aussi la préexistence de propriétés que l'entité-objet possèderait de par elle-même) peut être posé comme produisant une description qui constituera une très bonne approximation du résultat que fournirait un traitement canonique complet et rigoureux.

---

[93] La phénoménologie de Husserl – le processus de 'constitution transcendentale d'objets' – doit être re-conçu sur les bases élargies tirées de la mise en évidence des contenus épistémologiques de la microphysique actuelle. Les philosophes n'en sont probablement pas conscients, cependant que les physiciens ignorent que la question de la coupure quantique-classique s'incorpore à la question centrale de la phénoménologie husserlienne. Dans la méthode de conceptualisation relativisée (Mugur-Schächter [2006]) cette question centrale a été traitée de *A* à *Z*, dans des termes enracinés dans la microphysique et pourtant tout à fait généraux. Mais cela a été fait sans qu'à l'époque l'auteur ait été conscient que cette question avait déjà été énoncée et, en un certain sens restreint, examinée à fond! Tel est aujourd'hui le cloisonnement de la pensée dans des domaines spécialisés.

## 6.3. Vers une méthode générale de conceptualisation relativisée

C'est ainsi que, en partant de la mécanique quantique, s'est imposée progressivement la visée de construire une méthode générale de conceptualisation relativisée. Cette méthode (Mugur-Schächter [2002], [2006]) incorpore explicitement à sa base la strate universelle des descriptions primordiales transférées, qui marque explicitement [la coupure (descriptions primordiales transférées)-(descriptions classiques modélisantes)]. Elle met en évidence comment on peut conceptualiser d'une manière formalisable, en se libérant de l'emprise de la pensée classique et des langages usuels qui l'expriment. Et – tout en marquant les *limites* de la connaissance inter-subjective consensuelle – la méthode de conceptualisation relativisée permet à chacun de rattacher d'une manière cohérente à cette sorte de connaissance limitée, la clôture métaphysique qui convient à son tempérament et ses intuitions.

# *Conclusion générale*

Dans la première partie de cet ouvrage nous avons repéré la place qu'occupe la mécanique quantique dans l'évolution des relations entre, d'une part, la philosophie et notamment l'épistémologie philosophique, et d'autre part les modes de pensée qui agissent dans les disciplines de la physique.

Dans la deuxième partie du livre nous avons entrepris de spécifier en détail les contenus épistémologiques impliqués dans la mécanique quantique. Le résultat peut surprendre. Tout à fait indépendamment du formalisme de la mécanique quantique, il s'est constitué une discipline qualitative épistémo-physique, l'infra-mécanique quantique, dont le contenu consiste en l'organisation du substrat d'opérations cognitives et de significations encryptées dans le formalisme quantique. Dans l'infra-mécanique quantique cette organisation sous-jacente se montre débarrassée de tout élément mathématique, concentrée en elle-même et dotée d'un contour propre. Sa genèse, explicitée pas à pas, la marque d'un caractère de nécessité. Le cœur de l'infra-méacanique quantique est un type descriptionnel 'transféré' sur des enregistreurs d'appareil, primordialement probabiliste, foncièrement différent du type descriptionnel classique et qui, auparavant, n'avait jamais encore atteint le niveau d'une connaissance exprimée et intégrée. Désormais ce type descriptionnel primordial est défini avec rigueur et détail – pour le cas spécial des microétats – et il élucide la nature de la fameuse 'coupure quantique-classique'. Le brouillard lourd, comme solide, qui cachait la manière de signifier de la mécanique quantique, est entièrement dissipé.

La troisième partie de l'ouvrage a brièvement indiqué que l'infra-mécanique quantique, malgré sa forte singularité face à nos habitudes de pensée telles qu'elles se manifestent dans les langages courants et dans l'entière pensée classique, incorpore une certaine sorte d'universalité qui, plus ou moins clairement, se manifeste souvent dans nos pratiques de tous les jours. C'est cette universalité qui, explicitée et épurée, a ouvert la voie vers une épistémologie générale enracinée en dessous des langages usuels, dans la pure factualité physique.

Cette épistémologie générale – la méthode de conceptualisation relativisée – a été largement exposée dans d'autres ouvrages. Mais l'infra-mécanique quantique est exposée ici pour la première fois. Elle rend visible le développement fœtal de la méthode générale de conceptualisation relativisée, au sein du cas particulier des descriptions de microétats. Le caractère de nécessité qui marque ce développement est beaucoup plus concret et intuitif dans l'infra-mécanique quantique que dans sa réplique abstraite qui se constitue dans le processus de construction de la méthode générale de conceptualisation relativisée. Il pourra désormais irriguer cette réplique aussi d'une nuance d'évidence intuitive.

Je finirai cet ouvrage par un aveu d'admiration. La mécanique quantique cache en elle quelque chose de merveilleux. Chacun est assailli de temps à autre par l'impression que ceci ou cela est merveilleux. Ou que tout est merveilleux. Mais, comme diraient les cochons d'Orwell [1945], certaines merveilles sont plus merveilleuses que les autres. Et la merveille de la mécanique quantique est vraiment très merveilleuse.

———

__________